\documentclass{aa}  

\usepackage{graphicx}
\usepackage{txfonts}

\usepackage[colorlinks=true,linkcolor=blue, citecolor=blue, filecolor=blue, urlcolor=blue]{hyperref}
\newcounter{myenum}

\begin{document}

   \title{Reexploring Molecular Complexity with ALMA:\\
    Insights into chemical differentiation from the molecular composition of 
    hot cores in Sgr~B2(N2)
    }
   \subtitle{}
   \titlerunning{ReMoCA:
    Insights into chemical differentiation from the molecular composition of 
    hot cores in Sgr~B2(N2)
    }

   \author{A. Belloche \inst{1}
          \and R.~T. Garrod \inst{2}
          \and H.~S.~P.~M{\"u}ller \inst{3}
          \and N.~J. Morin\inst{4}
          \and S.~A. Willis\inst{4}
          \and K.~M. Menten \inst{1}\thanks{This article is dedicated to the 
memory of Karl Menten who suddenly passed away shortly before the completion 
of this work. He originally motivated the Sgr~B2 spectral line surveys 
performed at the MPIfR which started with the IRAM 30\,m telescope in 
2004. We are deeply grateful for the numerous inspiring discussions that we 
had the privilege to have with him over the past two decades, in particular 
about star formation and astrochemistry. We miss his enthusiasm, his 
scientific curiosity and creativity, as well as his open and generous 
personality.}
          }

   \institute{Max-Planck-Institut f\"{u}r Radioastronomie, 
              Auf dem H\"{u}gel 69, 53121 Bonn, Germany\\ 
              \email{belloche@mpifr-bonn.mpg.de}
         \and Departments of Chemistry and Astronomy, University of Virginia, 
              Charlottesville, VA 22904, USA        
         \and Astrophysik/I. Physikalisches Institut, Universit{\"a}t zu 
              K{\"o}ln, Z{\"u}lpicher Str. 77, 50937 K{\"o}ln, Germany
         \and Department of Chemistry, University of Virginia, 
              Charlottesville, VA 22904, USA
             }

   \date{Received 7 March 2025 / Accepted 29 April 2025}

% \abstract{}{}{}{}{}
% 5 {} token are mandatory
 
  \abstract
   {Hot molecular cores correspond to the phase of star formation during which 
many molecules, in particular complex organic molecules (COMs), thermally 
desorb from the surface of dust grains. Sophisticated kinetic models of 
interstellar chemistry describe the processes that lead to the formation and 
subsequent evolution of COMs in star forming regions.}
   {Our goal is to derive the chemical composition of hot cores in order to
improve our understanding of interstellar chemistry. In particular, we want to 
test the models by comparing their predictions to the observed composition of 
the gas phase of hot cores.}
   {We used the Atacama Large Millimeter/submillimeter Array (ALMA) to perform
an imaging spectral line survey of the high-mass star forming region
Sagittarius~B2(N) at 3~mm, called Reexploring Molecular Complexity with ALMA 
(ReMoCA). We modeled under the assumption of local thermodynamic equilibrium 
the spectra obtained with this survey toward the sources embedded in the 
secondary hot core Sgr~B2(N2). We compared the chemical composition 
of these sources to that of sources from the literature and to predictions of
the chemical kinetics model MAGICKAL.}
   {We detected up to 58 molecules toward Sgr~B2(N2)'s hot cores, including up 
to 24 COMs, as well as many less abundant isotopologs. The compositions of 
some pairs of sources are well correlated, but differences also exist in 
particular for HNCO and NH$_2$CHO. The abundances of series of homologous 
molecules drop by about one order of magnitude at each further step in 
complexity. The nondetection of radicals yields stringent constraints on the 
models. The comparison to the chemical models confirms previous evidence of 
a high cosmic-ray ionization rate in Sgr~B2(N). The comparison to sources from 
the literature gives new insight into chemical differentiation. The 
composition of most hot cores of Sgr~B2(N2) is tightly correlated to that of 
the hot core G31.41+0.31 and the hot corino IRAS~16293--2422~B after 
normalizing the abundances by classes of molecules (O-bearing, N-bearing, 
O+N-bearing, and S-bearing). There is no overall correlation between 
Sgr~B2(N2) and the shocked region G+0.693--0.027 also located in Sgr~B2, and 
even less with the cold starless core TMC-1. The class of N-bearing species 
reveals the largest variance among the four classes of molecules. The 
S-bearing class shows in contrast the smallest variance.}
   {These results imply that the class of N-bearing 
molecules reacts more sensitively to shocks, low-temperature gas phase 
chemistry after non-thermal desorption, or density. The overall abundance 
shifts observed between the N-bearing and O-bearing molecules may indicate how 
violently and completely the ice mantles are desorbed.}

   \keywords{astrochemistry -- line: identification -- 
             radio lines: ISM --
             ISM: molecules -- 
             ISM: individual objects: \object{Sagittarius B2(N)}}

   \maketitle
%
%________________________________________________________________

\section{Introduction}
\label{s:introduction}

Molecular clouds are the birth place of stars and their planetary systems. 
In these interstellar environments, star and planet formation results from an 
intricate interplay between physical and chemical processes. This work focuses 
on the chemistry that occurs in star forming regions, in particular on the 
formation and evolution of complex organic molecules (COMs), which are defined
as carbon bearing molecules that contain at least six atoms 
\citep[][]{Herbst09}. Interstellar chemistry proceeds in both the gas phase 
and the solid phase, the latter corresponding to the surface and icy mantles 
of dust grains. While both phases are involved in the formation and evolution 
of COMs \citep[e.g.,][]{Herbst86,Charnley92,vanDishoeck98,Garrod08}, the major 
role of the solid phase in the production of COMs has become more obvious in
the past decade, in particular since the detection of gas-phase saturated COMs 
at low temperatures in prestellar cores and protostellar envelopes
\citep[e.g.,][]{Oberg10,Bacmann12,Vastel14} and, more recently, with
the identification of several COMs in interstellar ices with the James Webb
Space Telescope \citep[JWST, e.g.,][]{Yang22,Rocha24,Nazari24}. While earlier 
kinetic models of interstellar chemistry relied on the increased mobility of 
radicals at the surface of dust grains during the warm-up phase of protostellar 
evolution to explain the production of many COMs 
\citep[e.g.,][]{Garrod06,Garrod08}, additional nondiffusive processes 
efficient at low temperatures on dust grains have been added to these models 
in order to account for the early formation of COMs in star forming regions 
\citep[e.g.,][]{Jin20,Garrod22}. 

One of our goals is to obtain observational constraints on the chemical and 
physical processes that play a role in the emergence of molecular complexity 
in star forming regions. One way to achieve this is to test chemical models 
such as those mentioned above by comparing their predictions to the 
post-desorption chemical composition of the gas phase of hot molecular cores 
in star forming regions. 

Hot cores correspond to the dense and compact regions around nascent high-mass 
protostars where spectral emission from COMs is detected, which is associated 
with molecular desorption from dust grains upon heating by the central object
\citep[e.g.,][]{Walmsley95}. They have typical kinetic 
temperatures of 150--200~K. Their lifetime is on the order of 
$6 \times 10^4$~yr \citep[e.g.,][]{Bonfand17,Nony24}. In low-mass star forming 
regions, these objects are called hot corinos \citep[][]{Ceccarelli04}. The 
chemical composition of hot cores can be determined by analyzing their 
spectrum in the millimeter/submillimeter wavelength range where molecules have 
most of their rotational transitions. With the advent of broadband 
instrumentation at single-dish telescopes and interferometers over the past 
two decades, spectral line surveys of multiple sources covering a broad 
frequency range have become more affordable 
\citep[see, e.g., the review by][]{Jorgensen20}. This means that the 
predictions of chemical models can be tested by determining the abundances of 
a large number of COMs for sample of sources that are located in different 
environments, have various masses and luminosities, or are in different 
evolutionary stages.

This work focuses on hot cores embedded in Sagittarius (Sgr) B2. This
molecular cloud complex is located in the central molecular zone of our 
Galaxy, at a projected distance of about 100~pc from the galactic center, 
Sgr~A*. Sgr~B2 consists of several protoclusters, in particular the high-mass 
star forming region Sgr~B2(N) that contains a population of hot cores and more 
evolved \ion{H}{ii} regions 
\citep[e.g.,][]{Gaume95,Schmiedeke16,Bonfand17,SanchezMonge17}.
Since the 1970s, thanks to its high H$_2$ column density, Sgr~B2 has been a 
place of choice to search for interstellar molecules, in particular COMs 
\citep[e.g.,][]{Ball70,Rubin71,Hollis00,Belloche08,Belloche22}. It has been 
the target of multiple spectral line surveys, starting with single-dish 
telescopes in the 1980s and 1990s 
\citep[e.g.,][]{Cummins86,Turner89,Nummelin98}. Our own spectral line survey 
of Sgr~B2(N) with the IRAM 30~m telescope at 3~mm, 2~mm, and 
1.3~mm led to the detection of several new COMs 
\citep[][]{Belloche08,Belloche09}. Despite the limited angular resolution 
(half-power beam width, HPBW, $\sim$ 25$\arcsec$), we were able to identify 
the presence of two hot cores in Sgr~B2(N) within the single-dish beam thanks 
to their different systemic velocities  \citep[][]{Belloche13}. Our analysis 
of the IRAM 30~m spectrum showed that the spectral confusion, exacerbated by 
the presence of these two velocity components, was much higher at 1.3~mm than 
at 3~mm and that the opacity of the dust made it more difficult to derive 
reliable COM column densities at 1.3~mm in this source. 

In order to make further progress with the determination of the
chemical composition of Sgr~B2(N)'s hot cores, we took advantage of the
advent of the Atacama Large Millimeter/submillimeter Array (ALMA) in the early
2010s to perform an imaging spectral line survey of Sgr~B2(N) at 3~mm. Thanks
to its angular resolution of $\sim$1.6$\arcsec$, this survey, called Exploring 
Molecular Complexity with ALMA (EMoCA), resolved the two main hot cores into 
separate sources which we called Sgr~B2(N1) and Sgr~B2(N2) 
\citep[][]{Belloche16}. This gain in angular resolution reduced the spectral 
confusion considerably. The resolution was however still insufficient to 
resolve the internal structure of both hot cores, and the high dust optical
depth of the main hot core, Sgr~B2(N1), still made it difficult to explore its
chemical composition even at 3~mm. This motivated a second high-sensitivity, 
high-angular-resolution ($\sim$0.5$\arcsec$) survey with ALMA, called 
Reexploring Molecular Complexity with ALMA (ReMoCA) \citep[][]{Belloche19}.
This survey allowed us to resolve the thermal structure of Sgr~B2(N1), probe 
the transition between non-thermal and thermal desorption, and establish
that a number of COMs co-desorb thermally with water in this source 
\citep[][]{Busch22}. Our series of spectral line surveys of Sgr~B2(N) has
demonstrated that a high angular resolution is key to reduce the spectral
confusion of high-mass star forming regions by probing individual sources and 
pockets of gas with narrower line widths \citep[][]{Belloche22}.

In this article, we focus our attention on the secondary hot core 
Sgr~B2(N2), which ReMoCA resolves into several sources, and we derive their 
individual chemical compositions. \citet{Moeller25} recently used ALMA at 
1.2~mm to study the whole population of hot cores in Sgr~B2(N) and Sgr~B2(M). 
They derived the chemical composition of several dozen sources, albeit on 
the basis of a limited number of molecules, and focused their analysis on 
evaluating the evolutionary stages of the sources. Thanks to the lower 
degree of spectral confusion at 3~mm, we report here the detection of a much 
larger set of molecules and investigate the constraints that these 
observational results impose on chemical models. This article is structured as
follows. We describe the observational setup and the method used to identify
the molecules and measure their column densities in Sect.~\ref{s:observations}.
Section~\ref{s:results} presents the results of this analysis and a comparison
to other sources from the literature. In Sect.~\ref{s:chemical_models} we 
compare the observational results with the predictions of chemical models.
We discuss the results obtained in this work in Sect.~\ref{s:discussion}. Our 
conclusions are reported in Sect.~\ref{s:conclusions}.

\section{Observations and radiative transfer modeling}
\label{s:observations}

\subsection{ALMA observations}
\label{ss:alma}

\begin{table}
\caption{Beam sizes and noise levels.}
\label{t:beam_noise}
\centering
\begin{tabular}{cccccccc}
\hline
\hline
\noalign{\smallskip}
S\tablefootmark{(a)} & \hspace*{-3ex} W\tablefootmark{(b)} & \hspace*{-3ex} Freq. range\tablefootmark{(c)} & \multicolumn{2}{c}{Synthesized beam} & \hspace*{-2ex} & \multicolumn{2}{c}{rms\tablefootmark{(e)}} \\
%\noalign{\smallskip}
\cline{4-5} \cline{7-8}
\noalign{\smallskip}
      & & \hspace*{-3ex} {\small (MHz)} & HPBW & \hspace*{-2ex} PA\tablefootmark{(d)} & \hspace*{-2ex} & \hspace*{-2ex} {\small (mJy} & \hspace*{-2ex} {\small (K)} \\
      & &     & {\small ($''\times''$)} & \hspace*{-2ex} {\small ($^\circ$)}    & \hspace*{-2ex} & \hspace*{-2ex} {\small beam$^{-1}$)} & \\
\hline
\noalign{\smallskip}
S1 & \hspace*{-3ex} 0 &  \hspace*{-3ex} 84\,112--85\,990  & $0.86 \times 0.67$ & \hspace*{-2ex} $-84$ & \hspace*{-2ex} & \hspace*{-2ex} 0.96 & \hspace*{-2ex} 0.28 \\
   & \hspace*{-3ex} 1 &  \hspace*{-3ex} 85\,938--87\,815  & $0.90 \times 0.66$ & \hspace*{-2ex} $-87$ & \hspace*{-2ex} & \hspace*{-2ex} 0.94 & \hspace*{-2ex} 0.26 \\
   & \hspace*{-3ex} 2 &  \hspace*{-3ex} 96\,116--97\,993  & $0.74 \times 0.59$ & \hspace*{-2ex} $-84$ & \hspace*{-2ex} & \hspace*{-2ex} 0.91 & \hspace*{-2ex} 0.27 \\
   & \hspace*{-3ex} 3 &  \hspace*{-3ex} 97\,941--99\,818  & $0.72 \times 0.57$ & \hspace*{-2ex} $-86$ & \hspace*{-2ex} & \hspace*{-2ex} 0.94 & \hspace*{-2ex} 0.29 \\
S2 & \hspace*{-3ex} 0 &  \hspace*{-3ex} 87\,763--89\,640  & $0.85 \times 0.62$ & \hspace*{-2ex}  $87$ & \hspace*{-2ex} & \hspace*{-2ex} 1.01 & \hspace*{-2ex} 0.30 \\
   & \hspace*{-3ex} 1 &  \hspace*{-3ex} 89\,588--91\,465  & $0.93 \times 0.61$ & \hspace*{-2ex}  $83$ & \hspace*{-2ex} & \hspace*{-2ex} 1.06 & \hspace*{-2ex} 0.28 \\
   & \hspace*{-3ex} 2 &  \hspace*{-3ex} 99\,766--101\,643 & $0.72 \times 0.60$ & \hspace*{-2ex}  $83$ & \hspace*{-2ex} & \hspace*{-2ex} 0.82 & \hspace*{-2ex} 0.23 \\
   & \hspace*{-3ex} 3 & \hspace*{-3ex} 101\,591--103\,468 & $0.72 \times 0.62$ & \hspace*{-2ex}  $66$ & \hspace*{-2ex} & \hspace*{-2ex} 0.73 & \hspace*{-2ex} 0.19 \\
S3 & \hspace*{-3ex} 0 &  \hspace*{-3ex} 91\,403--93\,280  & $0.70 \times 0.62$ & \hspace*{-2ex}  $89$ & \hspace*{-2ex} & \hspace*{-2ex} 0.83 & \hspace*{-2ex} 0.27 \\
   & \hspace*{-3ex} 1 &  \hspace*{-3ex} 93\,228--95\,105  & $0.69 \times 0.59$ & \hspace*{-2ex} $-86$ & \hspace*{-2ex} & \hspace*{-2ex} 0.80 & \hspace*{-2ex} 0.27 \\
   & \hspace*{-3ex} 2 & \hspace*{-3ex} 103\,405--105\,282 & $0.63 \times 0.53$ & \hspace*{-2ex} $-85$ & \hspace*{-2ex} & \hspace*{-2ex} 0.87 & \hspace*{-2ex} 0.29 \\
   & \hspace*{-3ex} 3 & \hspace*{-3ex} 105\,230--107\,107 & $0.61 \times 0.52$ & \hspace*{-2ex} $-86$ & \hspace*{-2ex} & \hspace*{-2ex} 0.91 & \hspace*{-2ex} 0.31 \\
S4 & \hspace*{-3ex} 0 &  \hspace*{-3ex} 95\,062--96\,939  & $0.57 \times 0.46$ & \hspace*{-2ex} $-53$ & \hspace*{-2ex} & \hspace*{-2ex} 0.35 & \hspace*{-2ex} 0.18 \\
   & \hspace*{-3ex} 1 &  \hspace*{-3ex} 96\,887--98\,764  & $0.56 \times 0.45$ & \hspace*{-2ex} $-53$ & \hspace*{-2ex} & \hspace*{-2ex} 0.35 & \hspace*{-2ex} 0.18 \\
   & \hspace*{-3ex} 2 & \hspace*{-3ex} 107\,064--108\,942 & $0.51 \times 0.41$ & \hspace*{-2ex} $-54$ & \hspace*{-2ex} & \hspace*{-2ex} 0.41 & \hspace*{-2ex} 0.21 \\
   & \hspace*{-3ex} 3 & \hspace*{-3ex} 108\,890--110\,767 & $0.50 \times 0.40$ & \hspace*{-2ex} $-54$ & \hspace*{-2ex} & \hspace*{-2ex} 0.43 & \hspace*{-2ex} 0.22 \\
S5 & \hspace*{-3ex} 0 &  \hspace*{-3ex} 98\,714--100\,591 & $0.43 \times 0.30$ & \hspace*{-2ex} $-78$ & \hspace*{-2ex} & \hspace*{-2ex} 0.68 & \hspace*{-2ex} 0.65 \\
   & \hspace*{-3ex} 1 & \hspace*{-3ex} 100\,539--102\,417 & $0.42 \times 0.29$ & \hspace*{-2ex} $-78$ & \hspace*{-2ex} & \hspace*{-2ex} 0.67 & \hspace*{-2ex} 0.65 \\
   & \hspace*{-3ex} 2 & \hspace*{-3ex} 110\,717--112\,594 & $0.39 \times 0.26$ & \hspace*{-2ex} $-77$ & \hspace*{-2ex} & \hspace*{-2ex} 0.83 & \hspace*{-2ex} 0.80 \\
   & \hspace*{-3ex} 3 & \hspace*{-3ex} 112\,542--114\,419 & $0.38 \times 0.25$ & \hspace*{-2ex} $-77$ & \hspace*{-2ex} & \hspace*{-2ex} 0.99 & \hspace*{-2ex} 0.99 \\
\hline
\end{tabular}
\tablefoot{
\tablefoottext{a}{Setup.}
\tablefoottext{b}{Spectral window.}
\tablefoottext{c}{The frequencies correspond to rest frequencies at a systemic 
velocity of 62~km~s$^{-1}$.}
\tablefoottext{d}{Position angle East from North.}
\tablefoottext{e}{Median rms noise level measured in the channel maps of the 
continuum-removed data cubes.}
}
\end{table}

The imaging spectral line survey ReMoCA was performed with ALMA toward 
Sgr~B2(N) at high angular resolution in the 3~mm atmospheric window between 
84.1 and 114.4~GHz. The observations and data reduction were described in 
detail in \citet{Belloche19}. We summarize here the main features. The survey 
was performed in five parts that were tuned to different frequencies with a 
spectral resolution of 488.3~kHz (1.7--1.3~km~s$^{-1}$). We called these five
spectral setups S1--S5. Each setup consists of four 1.88-GHz wide spectral 
windows labeled W0--W3 that together cover both sidebands. The five setups were 
observed on different days with various antenna configurations 
\citep[see Table~1 of][]{Belloche19}, which 
resulted in different angular resolutions. The spectral coverage, angular 
resolution, and noise level of each spectral window are listed in 
Table~\ref{t:beam_noise}, which is slightly updated compared to the initial 
table published in \citet{Belloche19}. The survey has a median angular 
resolution of 0.6$\arcsec$, which corresponds to 4900 au at the distance of 
Sgr~B2 \citep[8.2~kpc,][]{Reid19}. The median rms noise level is 
0.8~mJy~beam$^{-1}$, which corresponds to a brightness temperature noise level 
of 0.27~K for a half-power beam width (HPBW) of 0.6$\arcsec$ at a frequency of 
100~GHz.

Because of the high spectral line density, which translates into spectra close 
to the confusion limit, and the presence of several sources with different 
systemic velocities in the field of view, splitting the line and continuum 
emissions in the Fourier plane was not possible. This was done instead in the 
image plane as described in \citet{Belloche19}. This splitting was 
subsequently slightly improved, as we reported in \citet{Melosso20}.

The primary beam of the ALMA 12\,m antennas has a HPBW that varies between 
69$\arcsec$ at 84~GHz and 51$\arcsec$ at 114~GHz \citep[][]{Remijan15}.
The data cubes of the ReMoCA survey were not corrected for the primary beam
response. However, this correction was applied to all spectra used to derive
column densities and those displayed in the figures of this article. Figures 
showing maps are not corrected for the primary beam response. This has the
advantage of keeping the noise level uniform. In these maps, the longest 
(shortest) angular distance to the phase center is 6.8$\arcsec$ (0.5$\arcsec$), 
which corresponds to a correction factor of 1.05 (1.0003) at 84~GHz and 1.03 
(1.0001) at 114~GHz. The distortion of the maps is thus marginal (smaller than
5\%).

\subsection{Radiative transfer modeling}
\label{ss:weeds}

We modeled the ReMoCA spectra under the assumption of local thermodynamic
equilibrium (LTE), which is well justified given the high densities that 
characterize the hot cores in Sgr~B2(N) 
\citep[$>1 \times 10^7$~cm$^{-3}$,][]{Bonfand19}. We used the astronomical 
software Weeds \citep[][]{Maret11}, which is part of the CLASS program of the 
GILDAS package\footnote{See http://www.iram.fr/IRAMFR/GILDAS}, to compute 
synthetic spectra. Weeds accounts for the finite angular resolution of the 
observations, the line optical depth, and the contribution of the background 
continuum emission to the equation of radiative transfer (see the Weeds 
documentation\footnote{https://www.iram.fr/IRAMFR/GILDAS/doc/html/weeds-html/weeds.html} 
for more details). We emphasize that neglecting the strong continuum emission 
of hot cores in the equation of radiative transfer would lead to 
underestimating the molecular column densities.

We modeled the contribution of each molecule separately before adding them 
together. The synthetic spectrum of each species is defined by five 
parameters: size of the emitting region ($\theta_{\rm s}$), column density 
($N$), temperature ($T_{\rm rot}$), line width ($\Delta V$), and velocity 
offset ($V_{\rm off}$) with respect to the assumed systemic velocity of the 
source. Many molecules contribute to the ReMoCA spectra with rotational 
transitions that originate not only from their vibrational ground state but 
also from various vibrationally excited states. In those cases, we modeled 
each vibrational state separately in order to be able to account for potential 
differences between vibrational and rotational temperatures. 

When the vibrational and rotational temperatures of a given molecule are 
equal, the column density parameters $N$ of all vibrational states of that 
molecule are identical in our Weeds models. This value corresponds to the 
total column density of the molecule, that is it includes the populations of 
the ground state and all vibrationally excited states. On the contrary, if the 
vibrational temperature of a given vibrational state $V$ is higher than the 
rotational temperature because of, e.g., radiative pumping, then, in order to 
fit the observed spectrum, the column density parameter $N_V$ of that state 
would have to be artificially increased to account for the overpopulation in 
that state. In such a case, the calculation of the total column density of the 
molecule would be tricky. In practice, we did not find any difference between 
the vibrational and rotational temperatures of the detected molecules.

We decided to keep the original angular resolutions of the 20 ReMoCA spectral 
windows despite their differences rather than degrade the data to the coarsest 
angular resolution (0.8$\arcsec$). The reason for this is that the shortest 
separation between the sources studied in this article is 0.55$\arcsec$. Such
a small angle is resolved by setups S4 and S5 but not completely by the other 
setups. The LTE parameters derived for each source were optimized as much as 
possible on S4 and S5. Part of the discrepancies between the synthetic and 
observed spectra of setups S1--S3 result from their coarser angular 
resolution that leads to partial contamination between neighboring sources, an 
effect that cannot be accounted for by our simple modeling framework.

\subsection{Spectroscopy}
\label{ss:spectro}

Table~\ref{t:spectrobib} provides the list of the 448 spectroscopic entries 
that were used in this work to model the observed spectra or derive upper 
limits to the column densities of nondetected molecules. These spectroscopic
entries were downloaded from the Cologne Database for Molecular Spectroscopy
\citep[CDMS\footnote{https://cdms.astro.uni-koeln.de/},][]{Mueller01,Mueller05,Endres16}, 
the Lille Spectroscopic Database (LSD\footnote{https://lsd.univ-lille.fr/}),
and the database for molecular spectroscopy of the Jet Propulsion Laboratory  
\citep[JPL\footnote{https://spec.jpl.nasa.gov/},][]{Pickett98}. They were all
inserted into a local database connected to the radiative transfer software 
Weeds. In addition to these publicly available entries, our database also 
includes predictions that were provided to us by various collaborators over 
the past two decades and a few entries that we prepared ourselves on the basis 
of published studies.

Spectroscopic entries in all three public databases mentioned above have a 
specific 5- or 6-digit index called the tag. The third-to-last digit of the tag
codes for the database (0 for JPL, 5 for CDMS, and 8 for LSD). The two or three 
digits that precede the database code represent the molecular weight of the 
molecule (e.g., 32 for CH$_3$OH or 103 for c-C$_6$H$_5$CN). The last two 
digits distinguish molecules in the same database that have the same 
molecular weight. For a given molecule, we modeled the emission of vibrational 
excited states separately in order to account for potential deviations from 
LTE in the event that the vibrational temperature would be different from the 
rotational one (see Sect.~\ref{ss:weeds}). To do that, we had to split the 
original entries that contain several vibrational states in the public 
databases into multiple entries containing one state each, and we assigned new 
tags to the individual entries, with a database code different from 0 or 5. 
Unfortunately, we had used the database code 8 for some of our local entries 
before the LSD database was created. We have started to change these tags when 
new LSD entries are in conflict with our existing local entries. Some CDMS 
entries are provided with hyperfine structure in the documentation area of the 
CDMS website with no specific tag. Here again, we had to assign new tags to 
these CDMS entries with hyperfine structure.

We made a considerable effort to estimate a posteriori corrections to the 
column densities of the analyzed molecules when their partition functions 
did not include, or only partially, the contribution of vibrationally excited 
states or higher energy conformers. In most cases, the vibrational correction 
($C_{\rm vib}$) was computed in the harmonic approximation using the following 
equation:
\begin{equation}
C_{\rm vib} = \prod_{i=1}^{N} \frac{1}{1-e^{-E_i/kT}}
\end{equation}
with $k$ the Boltzmann constant, $T$ the temperature, and $E_i$ the energies of
the $N$ fundamental modes of vibration. In most cases, the conformational 
correction 
($C_{\rm conf}$) was computed assuming that a conformer $j$ of energy $E_j$ 
with respect to the energy of the lowest-energy conformer contributes to the 
total partition function of the molecule with a population equal to 
$e^{-E_j/kT}$ times the population of the lowest-energy conformer. We also
accounted for the degeneracies of the conformers when they differed.

For each spectroscopic entry, Table~\ref{t:spectrobib} provides the chemical 
formula of the molecule, along with the vibrational state or name of the 
conformer when relevant, the tag, the database where it was downloaded from, 
and a list of relevant references (``Spectro'' for the frequencies and dipole 
moment, ``$C_{\rm vib}$'' for the vibrational energies, and ``$C_{\rm conf}$'' 
for the energies of the conformers).

\section{Results}
\label{s:results}

\subsection{Source selection and description}
\label{ss:sources}

\begin{table}[!t]
 \begin{center}
 \caption{
 Properties of the positions analyzed within Sgr~B2(N2).
}
 \label{t:sources}
 \vspace*{-1.2ex}
 \begin{tabular}{llcrrcc}
 \hline\hline
 \multicolumn{1}{c}{Source} & \multicolumn{2}{c}{Coordinates\tablefootmark{a}} & \multicolumn{2}{c}{Offsets\tablefootmark{b}} & \multicolumn{1}{c}{\hspace*{-1ex}$\alpha_{\mathrm{1.2mm}}$\tablefootmark{c}} & \multicolumn{1}{c}{\hspace*{-1ex}$V_{\rm lsr}$\tablefootmark{d}} \\ 
  & \multicolumn{1}{c}{$\alpha$} & \multicolumn{1}{c}{$\delta$} & \multicolumn{1}{c}{$\Delta\alpha$} & \multicolumn{1}{c}{\hspace*{-1ex}$\Delta\delta$} & & \hspace*{-1ex}(km \\ 
  & \multicolumn{1}{c}{17h47m} & \multicolumn{1}{c}{$-$28$^\circ$22$\arcmin$} & \multicolumn{1}{c}{($\arcsec$)} & \multicolumn{1}{c}{\hspace*{-1ex}($\arcsec$)} & & \hspace*{-1ex}s$^{-1}$) \\ 
 \hline
 AN06 & 19.96s & 13.94$\arcsec$ & $ 1.19$ & \hspace*{-1ex}$ 2.06$ & \hspace*{-1ex}2.4  & \hspace*{-1ex}68.3  \\ 
 AN03 & 19.90s & 13.52$\arcsec$ & $ 0.40$ & \hspace*{-1ex}$ 2.48$ & \hspace*{-1ex}3.2  & \hspace*{-1ex}73.2  \\ 
 AN02 & 19.86s & 13.17$\arcsec$ & $-0.13$ & \hspace*{-1ex}$ 2.83$ & \hspace*{-1ex}3.1  & \hspace*{-1ex}72.9  \\ 
 N2b & 19.834s & 13.60$\arcsec$ & $-0.48$ & \hspace*{-1ex}$ 2.40$ & \hspace*{-1ex}$\sim$3  & \hspace*{-1ex}74.2  \\ 
 \hline
 \end{tabular}
 \end{center}
 \vspace*{-2.5ex}
 \tablefoot{
 \tablefoottext{a}{J2000 equatorial coordinates of the ALMA continuum peak measured at 1.2~mm by \citet{SanchezMonge17} except for N2b. The coordinates of the latter were taken from \citet{Belloche22}.}
 \tablefoottext{b}{Equatorial offsets from the ReMoCA phase center located at $17^{\rm h}47^{\rm m}19{\fs}87, -28^\circ22'16{\farcs}0$.}
 \tablefoottext{c}{Spectral index of the ALMA 1.2~mm continuum emission reported by \citet{SanchezMonge17}. The spectral index of N2b was visually extracted from Fig.~7 of \citet{SanchezMonge17}.}
 \tablefoottext{d}{Systemic velocity assumed for the LTE modeling.}
 }
 \end{table}

\begin{figure*}[!ht]
\centerline{\resizebox{\hsize}{!}{\includegraphics[angle=0]{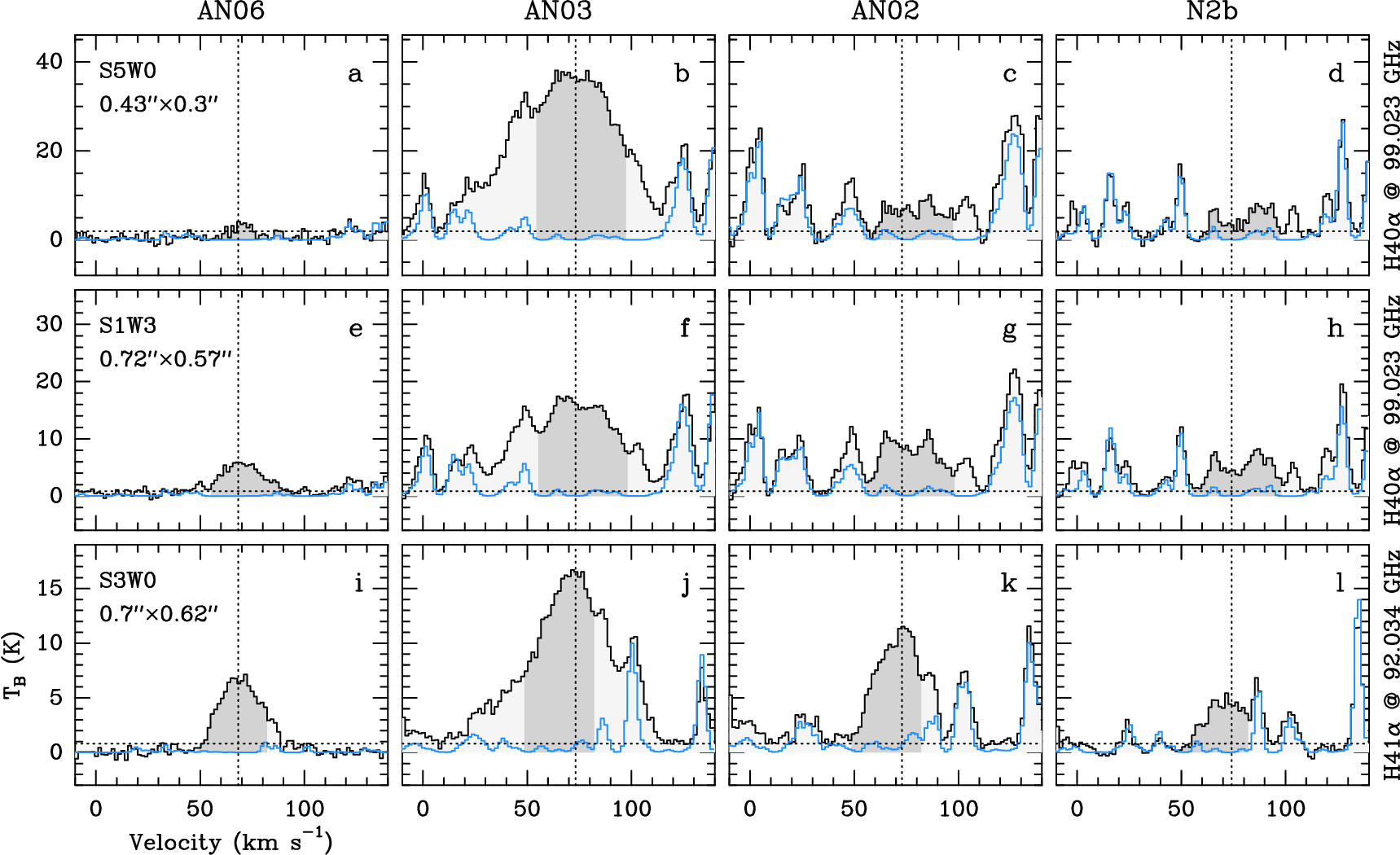}}}
\caption{ALMA continuum-subtracted spectra toward the hot core positions AN06, 
AN03, AN02, and N2b (from left to right) at the frequencies of the hydrogen 
recombination lines H40$\alpha$ (top and middle rows, for two setups with 
different angular resolutions) and H41$\alpha$ (bottom row). In each panel, the 
vertical dashed line marks the systemic velocity of the source adopted for the 
LTE modeling of the molecular emission. The velocity axis refers to the rest 
frequency of the recombination line. The horizontal dashed line indicates the 
3$\sigma$ noise level. The blue spectrum represents the LTE model that includes 
the contribution of all molecules identified so far. The velocity range 
highlighted in dark gray is specific to each recombination line and represents 
the range of channels selected to compute the integrated intensity map of each 
recombination line shown in Fig.~\ref{f:map_halpha_n2}. The setup and spectral 
window are indicated in the left panel of each row along with the 
corresponding HPBW.}
\label{f:spec_halpha_n2}
\end{figure*}

In this work, we focus our analysis on four positions located in Sgr~B2(N2), 
the secondary hot core of Sgr~B2(N). The first three positions correspond to
the continuum sources AN06, AN03, and AN02 extracted by \citet{SanchezMonge17} 
from their observations of Sgr~B2(N) performed with ALMA in the 1.2~mm 
atmospheric window with an angular resolution of 0.4$\arcsec$. These three
sources represent all the continuum sources within Sgr~B2(N2) in the list of 
\citet{SanchezMonge17}. Their spectral index between 211 and 275~GHz was found
to be 2.4, 3.2, and 3.1, respectively, suggesting that their 1.2~mm ALMA 
continuum emission is dominated by dust emission. In addition to these three
sources, we also analyzed the position called Sgr~B2(N2b) where we recently 
detected normal-propanol and iso-propanol \citep[][]{Belloche22}. The spectral 
index of this position in the 1.2~mm spectral window is approximately 3 
according to Fig.~7 of \citet{SanchezMonge17}. This suggests that the 1.2~mm 
ALMA continuum emission of N2b is dominated by dust as well. The coordinates
and 1.2~mm spectral indices of all selected positions are listed in 
Table~\ref{t:sources}. We describe below the additional information that we 
derived from the ReMoCA survey on the physical structure or stage of evolution 
of these four sources.

\subsubsection{AN03, a HC\ion{H}{II} region}
\label{sss:an03}

\begin{figure*}[!ht]
\centerline{\resizebox{\hsize}{!}{\includegraphics[angle=0]{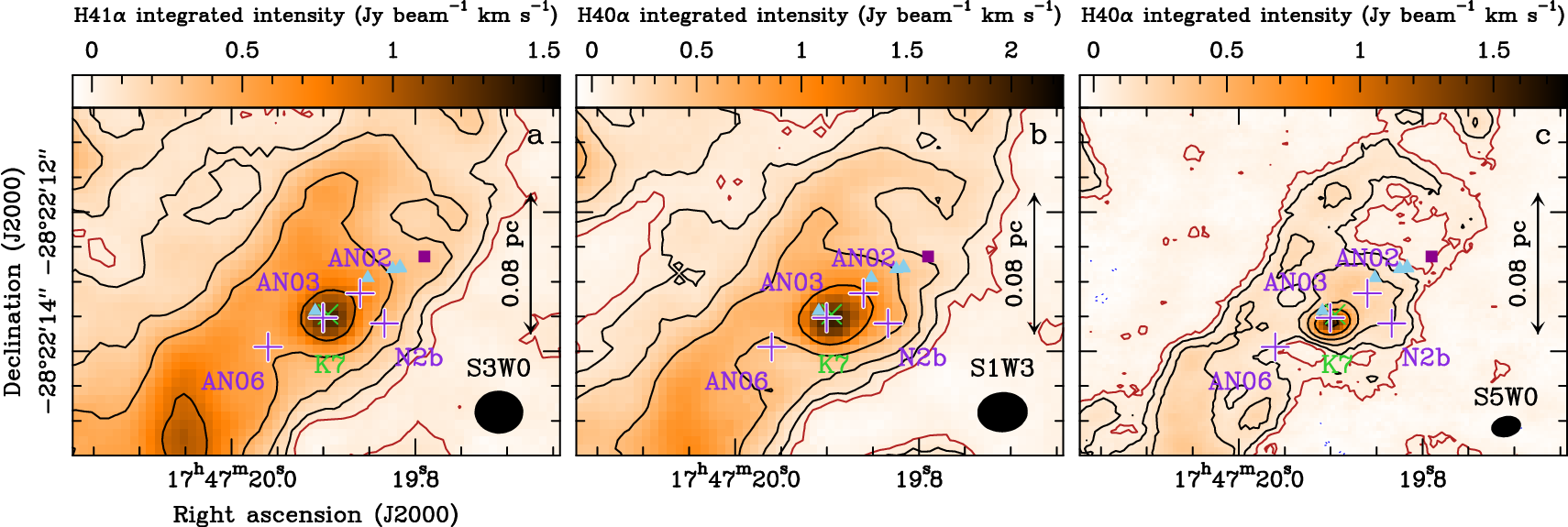}}}
\caption{ALMA integrated intensity maps of the hydrogen recombination lines
H41$\alpha$ (panel a) and H40$\alpha$ (panels b and c, for two setups with
different angular resolutions). The intensity was integrated over the velocity
range highlighted in dark gray in Fig.~\ref{f:spec_halpha_n2} in order to 
avoid contamination by molecular lines. The violet plus symbols mark the 
hot core positions AN06, AN03, AN02, and N2b. The green cross indicates the
VLA position of the HC\ion{H}{ii} region K7 at 7~mm from \citet{DePree15}. The
blue triangles and violet square indicate the VLA positions of the water and
Class II methanol masers reported by \citet{McGrath04} and \citet{Lu19}, 
respectively. The setup and spectral window numbers are given in the bottom 
right 
corner of each panel along with the associated beam size (HPBW). The values of
the noise level, $\sigma$, are 18, 21, and 15~mJy~beam$^{-1}$~km~s$^{-1}$, 
respectively. The contours start at $3\sigma$ (brown contour) and then 
increase by a factor of two at each step (black contours). The dotted blue 
contour, when present, shows the $-3\sigma$ level.}
\label{f:map_halpha_n2}
\end{figure*}

AN03 coincides with the hypercompact (HC) \ion{H}{II} region K7 that was 
identified by \citet{DePree15} in Very Large Array (VLA) continuum 
observations at 7~mm.
K7 is located at a J2000 equatorial position of $17^{\rm h}47^{\rm m}19{\fs}895, 
-28^\circ22'13{\farcs}47$, with uncertainties of 0.01s and 0.1$\arcsec$, and 
has a deconvolved full-width-at-half-maximum (FWHM) size of 
$\sim$0.08$\arcsec$ at 7~mm (660 au at the distance of Sgr~B2).
Several recombination lines of hydrogen were covered by the ReMoCA survey and 
detected toward Sgr~B2(N2). Figure~\ref{f:spec_halpha_n2} shows the ReMoCA 
spectrum at the frequencies of the H40$\alpha$ and H41$\alpha$ recombination 
lines toward the four positions of Sgr~B2(N2) analyzed in this article. We 
selected these lines because they suffer little from contamination by 
rotational emission of molecules (see the molecular contribution predicted
by our LTE model overlaid in 
blue in Fig.~\ref{f:spec_halpha_n2}). In order to investigate the spatial 
morphology of the emission of these recombination lines in Sgr~B2(N2), we 
integrated each line over a fixed velocity range, as highlighted in dark gray 
in Fig.~\ref{f:spec_halpha_n2}. The resulting integrated intensity maps are 
shown in Fig.~\ref{f:map_halpha_n2}. The H40$\alpha$ line was independently 
covered by two setups with differing angular resolutions (S1W3 and S5W0). 
Figures~\ref{f:spec_halpha_n2} and  \ref{f:map_halpha_n2} show the spectra and 
maps for both setups in order to illustrate the impact of the angular 
resolution. 

The maps of Fig.~\ref{f:map_halpha_n2} reveal a clear peak of H$\alpha$ 
emission at a J2000 equatorial position of $17^{\rm h}47^{\rm m}19{\fs}899, 
-28^\circ22'13{\farcs}59$ with an uncertainty of $\sim$0.1$\arcsec$, as 
measured in the S5W0 map which has the highest angular resolution (HPBW 
$0.43\arcsec \times 0.30\arcsec$). This H$\alpha$ emission peak coincides 
within the uncertainties with the positions of the 1.2~mm continuum source 
AN03 and the 7~mm HC\ion{H}{II} region K7. Figure~\ref{f:map_halpha_n2} also 
displays the positions of the water and methanol masers observed with the VLA 
by \citet{McGrath04} and \citet{Lu19}, respectively. AN03 seems to be 
associated with the easternmost water maser which has a systemic velocity of 
75.3~km~s$^{-1}$. Their angular separation is on the order of 0.15$\arcsec$ 
($\sim$1200~au).

The recombination line maps of Fig.~\ref{f:map_halpha_n2}
also show an arc of extended emission that corresponds to K5, a large shell of 
ionized gas seen in the VLA continuum maps of \citet{Gaume95} at 1.3~cm. The
fact that the morphology of this extended structure is very similar in all
three maps of Fig.~\ref{f:map_halpha_n2} gives us confidence that the maps 
suffer little from contamination by molecular emission.

\subsubsection{AN06, an isolated dust continuum source}
\label{sss:an06}

\begin{figure*}[!ht]
\centerline{\resizebox{\hsize}{!}{\includegraphics[angle=0]{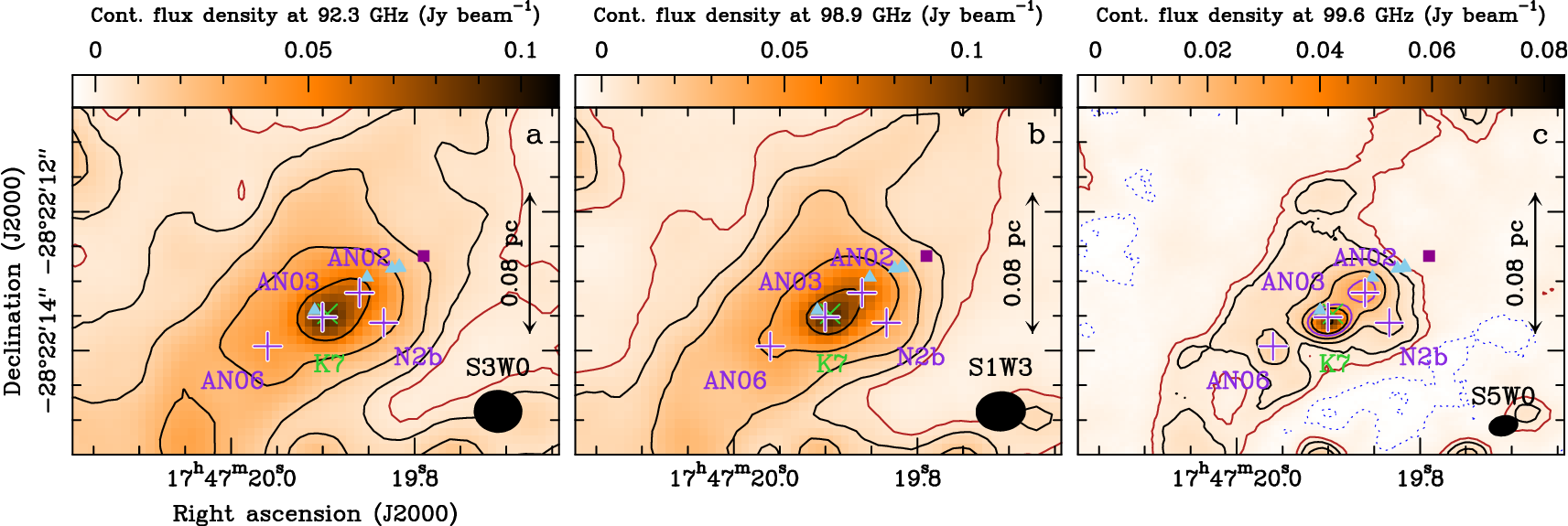}}}
\caption{ALMA continuum emission maps at 92.3, 98.9, and 99.6~GHz. The setup 
and spectral window numbers are given in the bottom right corner of each panel 
along with the associated beam size (HPBW). The symbols are the same as in 
Fig.~\ref{f:map_halpha_n2}. The values of the noise level, $\sigma$, are 0.73, 
0.97, and 0.42~mJy~beam$^{-1}$, respectively. The contours start at 
$5\sigma$ (brown contour) and then increase by a factor of two at each step 
(black contours). The additional violet contour in panel c is at $60\sigma$. 
It was added to emphasize the emission peak on AN02. Dotted blue contours,
when present, indicate the $-5\sigma$ and $-10\sigma$ levels.}
\label{f:map_3mmcont_n2}
\end{figure*}

Figure~\ref{f:map_3mmcont_n2} shows the maps of continuum emission measured
in the same ReMoCA spectral windows (S3W0, S1W3, and S5W0) as the 
recombination lines displayed in Fig.~\ref{f:map_halpha_n2}. These continuum 
emission maps trace a similar extended structure as the H$\alpha$ maps of 
Fig.~\ref{f:map_halpha_n2}, indicating that the extended continuum emission 
detected at 3~mm with ReMoCA contains a significant contribution from thermal 
free-free emission of ionized gas. However, the contours of the highest 
angular resolution map (S5W0, Fig.~\ref{f:map_3mmcont_n2}c) clearly reveal 
3~mm continuum peaks associated with the 1.2~mm continuum sources AN02 and 
AN06. This indicates that the compact 3~mm continuum emission around each 
source has a significant contribution from dust emission. AN06 is thus a 
bona-fide dust continuum source that is not associated with any HC\ion{H}{II} 
region because it has no compact counterpart in the ReMoCA maps of H$\alpha$ 
emission and in the VLA maps of thermal free-free emission of \citet{DePree15}.

\begin{figure*}[!ht]
\centerline{\resizebox{\hsize}{!}{\includegraphics[angle=0]{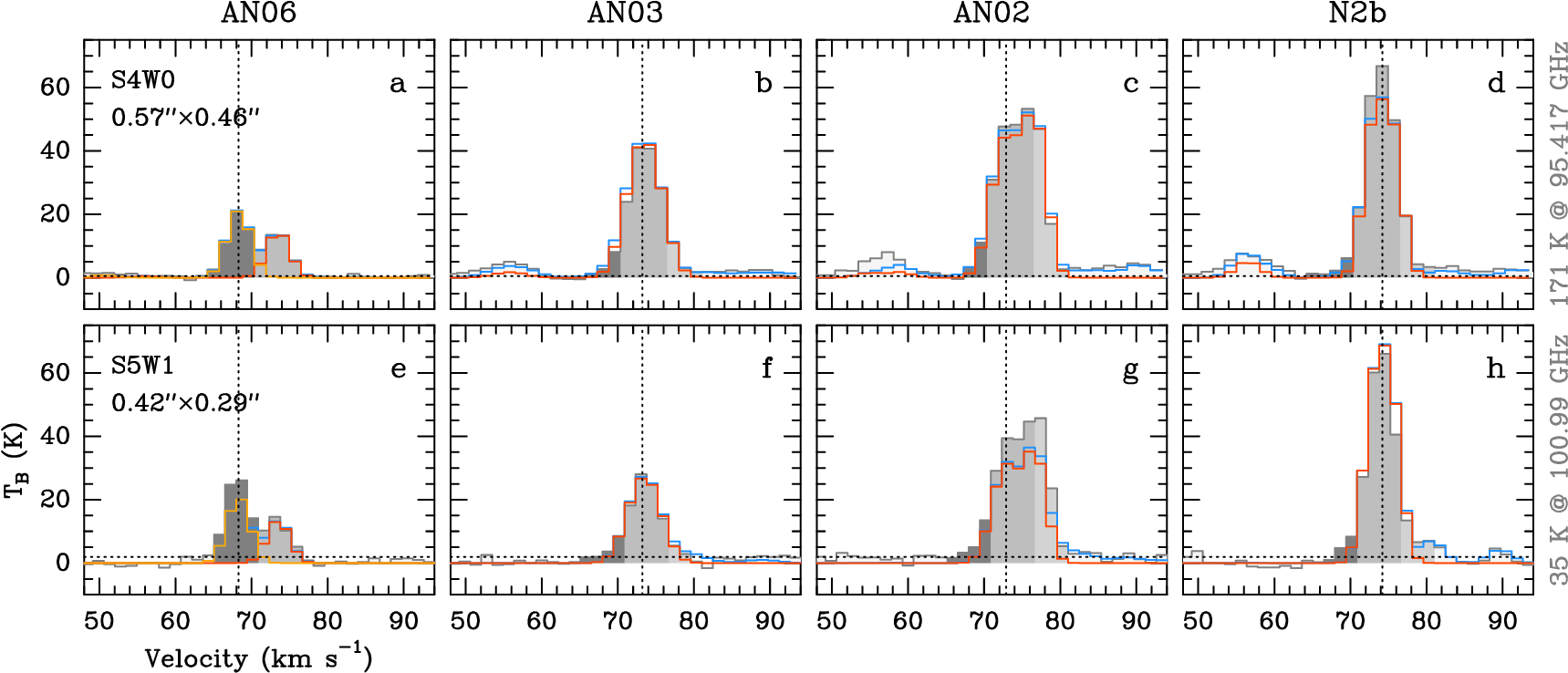}}}
\caption{ALMA continuum-subtracted spectra toward the hot core positions AN06, 
AN03, AN02, and N2b (from left to right) at the frequencies of two 
contamination-free rotational transitions of ethanol indicated on the right 
along with the energy of the upper level in temperature unit. In each panel, 
the vertical dashed line marks the systemic velocity of the source adopted for 
the LTE modeling of the molecular emission. The velocity axis refers to the 
rest frequency of the ethanol transition. The horizontal dashed line indicates 
the 3$\sigma$ noise level. The blue spectrum represents the LTE model that 
includes the contribution of all molecules identified so far. The red spectrum 
represents the LTE model of ethanol only. In the case of AN06, the synthetic 
ethanol spectra of the first and second velocity components are displayed in 
orange and red, respectively. The velocity ranges highlighted in the three 
darkest shades of gray are specific to each ethanol transition (but common to 
all sources, albeit with slight differences due to the finite spectral 
sampling) and represent the ranges of channels selected to compute the 
integrated intensity maps of each transition shown in 
Fig.~\ref{f:map_c2h5oh_n2}. The setup and spectral window are indicated in the 
left panel of each row along with the corresponding HPBW.}
\label{f:spec_c2h5oh_n2}
\end{figure*}

\begin{figure*}[!ht]
\centerline{\resizebox{\hsize}{!}{\includegraphics[angle=0]{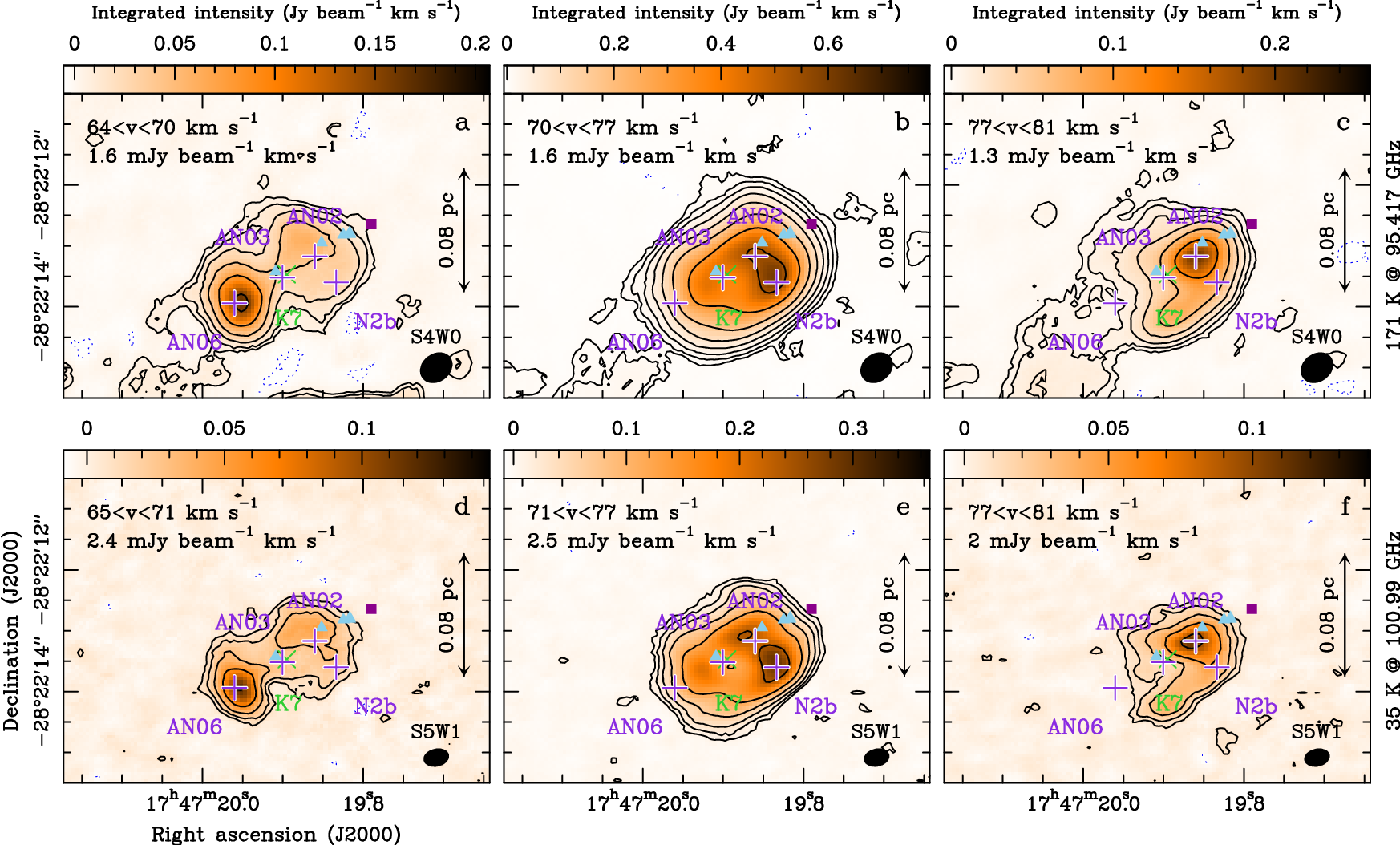}}}
\caption{ALMA integrated intensity maps of two contamination-free rotational 
transitions of ethanol with rest frequency and upper level energy indicated on 
the right of each row. The intensities were integrated over the velocity 
ranges highlighted in the three darkest shades of gray in 
Fig.~\ref{f:spec_c2h5oh_n2}. The integration range is indicated in the top 
left corner of each panel. The setup and spectral window numbers are given in 
the bottom right corner along with the associated beam size (HPBW). The symbols 
are the same as in Fig.~\ref{f:map_halpha_n2}. 
The noise level, $\sigma$, is indicated below the velocity range. The black 
contours start at $3\sigma$ and then increase by a factor of two at each step, 
except for the last contour of panel e which is at 320$\sigma$. The dotted 
blue contour, when present, indicates the $-3\sigma$ level.}
\label{f:map_c2h5oh_n2}
\end{figure*}

The ReMoCA spectra of AN06 reveal for most molecules that we analyzed two 
velocity components that are clearly distinct. This is illustrated in 
Fig.~\ref{f:spec_c2h5oh_n2} with two rotational transitions of ethanol that 
were carefully selected for their absence of contamination by other molecules. 
We divided the ethanol emission into three velocity ranges highlighted with 
the three darkest shades of gray in Fig.~\ref{f:spec_c2h5oh_n2}. The 
integrated intensity maps corresponding to these three velocity ranges are 
shown in Fig.~\ref{f:map_c2h5oh_n2}. While the maps in the velocity ranges 
71--77~km~s$^{-1}$ and 77--81~km~s$^{-1}$ do not show any particular structure 
toward AN06 (Figs.~\ref{f:map_c2h5oh_n2}b, c, e, and f), this source is 
clearly associated with a compact emission peak in the ethanol maps integrated 
from 65 to 71~km~s$^{-1}$ (Figs.~\ref{f:map_c2h5oh_n2}a and d). Therefore, we 
think that the dust continuum emission of AN06 is associated with the 
lower-velocity molecular component and we adopted a systemic velocity of 
68.3~km~s$^{-1}$ for this source. The higher velocity component detected 
toward AN06, at a systemic velocity of about 73.5~km~s$^{-1}$ and with lower
peak intensities than the lower velocity component, traces the edge 
of the Sgr~B2(N2) dense core and is likely not related to AN06 itself 
(Figs.~\ref{f:map_c2h5oh_n2}b and e). Both components have intrinsic line 
widths of about 3.5~km~s$^{-1}$. We conclude that AN06 is a bona-fide 
continuum source that is likely not embedded in Sgr~B2(N2) but rather lies in 
either the foreground or background of Sgr~B2(N2).

\subsubsection{AN02, a dust continuum source with two velocity components, 
possibly driving an outflow}
\label{sss:an02}

AN02 is associated with a 3~mm continuum peak in Fig.~\ref{f:map_3mmcont_n2}c 
and has no compact counterpart in H$\alpha$ emission at 3~mm
(Fig.~\ref{f:map_halpha_n2}) or thermal free-free emission at 7~mm
\citep[][]{DePree15}. It thus represents a bona-fide dust continuum source. 
Like for AN06, our LTE modeling of the molecular emission of AN02 required 
two velocity components to fit the asymmetric shape of the lines (see, e.g., 
Figs.~\ref{f:spec_c2h5oh_n2}c and g). However, with a velocity difference of 
only $\sim$4~km~s$^{-1}$ and intrinsic line widths of about 4 and 
3~km~s$^{-1}$, these two components are not as well separated as in the case 
of AN06. They have similar peak temperatures and both have a local 
maximum at the position of AN02 in the integrated intensity maps shown in 
Fig.~\ref{f:map_c2h5oh_n2}. Therefore, the data do not tell us whether the 
continuum source AN02 is associated with both components or only one of them. 
We assumed a systemic velocity of 72.9~km~s$^{-1}$ for AN02, which corresponds 
to the lower velocity component, but modeled both components together with 
velocity offsets of $\sim$0 and $\sim$4~km~s$^{-1}$.

Four water masers lie in the vicinity of AN02 toward the north-west, at 
angular distances of about 0.25$\arcsec$ and 0.6--0.7$\arcsec$. They are 
roughly aligned with AN02, suggesting that AN02 may drive an outflow that 
produces these masers. The easternmost maser, which we associated to AN03 in 
Sect.~\ref{sss:an03}, is located at a position that is roughly symmetric to the 
position of the westernmost masers with respect to AN02. Therefore, the 
easternmost maser may equally well be associated with AN02 rather than AN03. 
The three westernmost masers have velocities of 75.3, 74.0, and 
21.2~km~s$^{-1}$ \citep[][]{McGrath04}. The former velocities are both in 
between the velocities of the two molecular components associated with AN02
while the latter velocity is blueshifted by about $-50$~km~s$^{-1}$. The 
closest maser to AN02 has a velocity of 56.5 km~s$^{-1}$ that is blueshifted 
by about $-16$~km~s$^{-1}$. 

A Class II methanol maser was detected by \citet{Lu19} with the VLA in the 
vicinity of Sgr~B2(N2) (see violet square in Fig.~\ref{f:map_3mmcont_n2}). It 
is located about 1$\arcsec$ to the north-west of AN02. \citet{Mills18} reported
an astrometry offset affecting their VLA A-configuration data due to an error
in the position of their phase calibrator in the VLA catalog. Such
an offset may also affect the data set of \citet{Lu19} which was calibrated
with the same phase calibrator (X. Lu, priv. comm.), but the offset does not 
seem to be along the north-west direction (A. Ginsburg, priv. comm.). 
Therefore the source associated with the methanol maser is currently unknown.

\begin{figure*}
\centerline{\resizebox{0.7\hsize}{!}{\includegraphics[angle=0]{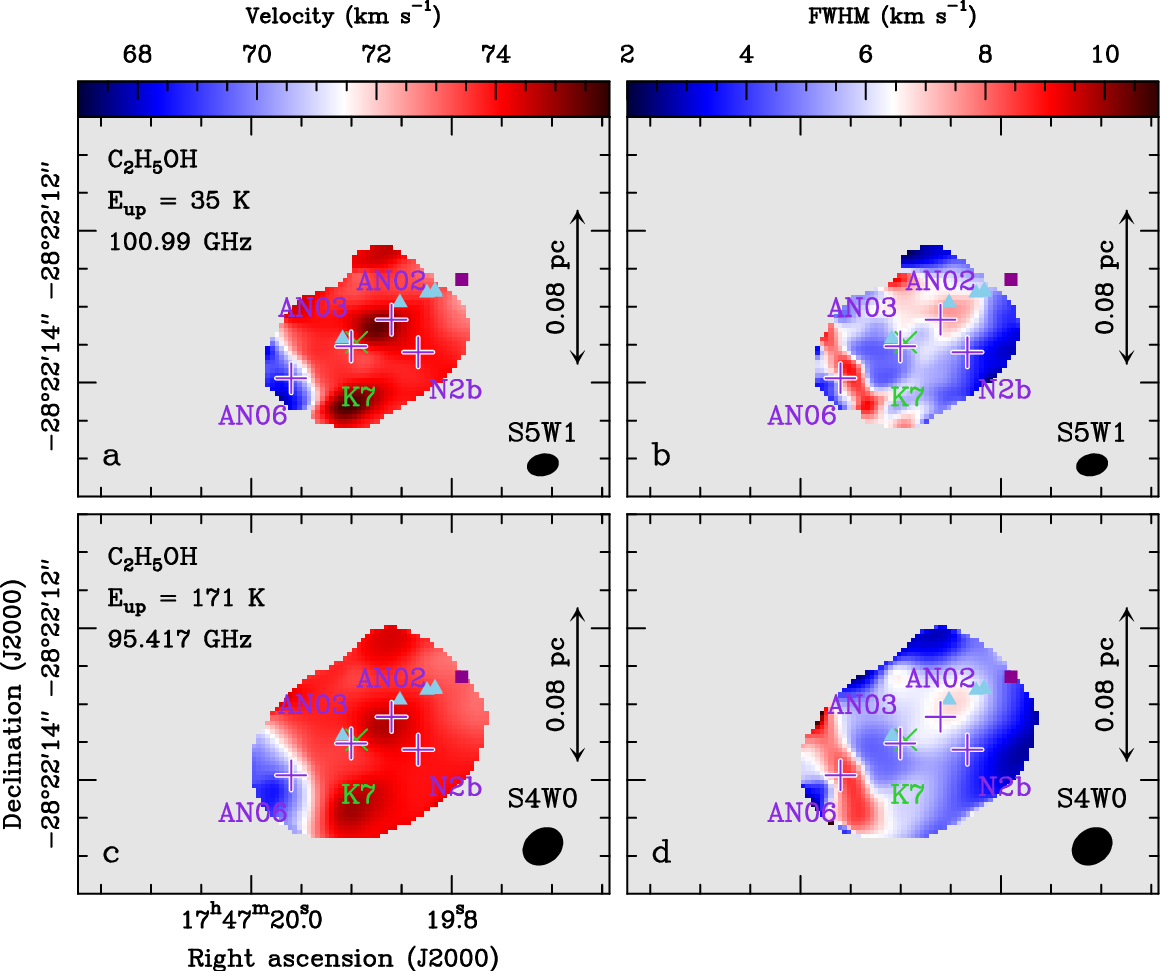}}}
\caption{Maps of centroid velocity (left column) and line width (right column) 
of two contamination-free rotational transitions of ethanol with rest 
frequency and upper level energy indicated in the top left corner of each row. 
The kinematic information is plotted only for pixels with a signal-to-noise 
ratio in peak intensity higher than 10. The setup and window numbers are given 
in the bottom right corner of each panel along with the associated beam size 
(HPBW). The symbols are the same as in Fig.~\ref{f:map_halpha_n2}. These two
transitions of ethanol have opacities on the order of 2 in N2b, which means
that the intrinsic line widths around N2b are somewhat smaller than the plotted
values.}
\label{f:map_fitgauss_n2}
\end{figure*}

We conclude that AN02 is a dust continuum source with two velocity components 
of molecular emission, which is possibly driving an outflow along the 
north-west/south-east direction.

\subsubsection{N2b, a dust continuum source with narrow lines}
\label{sss:n2b}

Figure~\ref{f:map_fitgauss_n2} shows maps of centroid velocity and line width
obtained by fitting a Gaussian function to the spectra of both transitions of 
ethanol used in Figs.~\ref{f:spec_c2h5oh_n2} and \ref{f:map_c2h5oh_n2}.
Figures~\ref{f:map_fitgauss_n2}b and d reveal that the line width decreases
toward the south-west of Sgr~B2(N2), reaching values as low as 
$\sim$2.7~km~s$^{-1}$. Such narrow lines are advantageous for the search for 
new molecules but the narrowest line widths are reached at the edge of the 
detected molecular emission where the emission becomes faint (see 
Fig.~\ref{f:map_c2h5oh_n2}). As a compromise between having narrow line widths 
and keeping high intensities, we selected position N2b for our previous work 
on propanol \citep[][]{Belloche22}. This position has intrinsic line widths on 
the order of 3.5~km~s$^{-1}$ and we adopted a systemic velocity of 
74.2~km~s$^{-1}$. Figures~\ref{f:map_c2h5oh_n2}b and e show that N2b is 
associated with a peak of integrated intensity of ethanol. It also corresponds
to an extension of 3~mm continuum emission in Fig.~\ref{f:map_3mmcont_n2}c,
which shows the map at highest angular resolution, while no such clear 
extension is visible in the H40$\alpha$ map of Fig.~\ref{f:map_halpha_n2}.
This suggests that N2b may be associated with a compact dust core like the other
three sources, even if it was not extracted as such by 
\citet{SanchezMonge17}. Their ALMA dust continuum map at 242~GHz also shows an 
extension at the position of N2b (see their Fig.~4).

Figure~\ref{f:map_fitgauss_n2} shows other features that are consistent with 
the source descriptions provided in the previous sections. The blue area 
toward AN06 in panels a and c reveals the low velocity component that we 
exclusively assigned to AN06 and that dominates its emission. The overlap 
along the line of sight between this velocity component and the dense core of 
Sgr~B2(N2) results in the sharp line width increase ($\sim$8~km~s$^{-1}$) at 
the location of the red band that touches AN06 in panels b and d. The 
existence of two velocity components toward AN02 also results in a local 
increase in the line width ($\sim$7~km~s$^{-1}$) and a slightly higher 
average centroid velocity ($\sim$75~km~s$^{-1}$) than the rest of Sgr B2(N2), 
which has a systemic velocity of $\sim$74~km~s$^{-1}$. Finally, an unresolved 
pocket of gas with a systemic velocity of $\sim$75~km~s$^{-1}$ stands out at a 
bit less than 1$\arcsec$ to the south of AN03. This region is not associated 
with any specific structure in the H$\alpha$ and continuum maps shown in
Figs.~\ref{f:map_halpha_n2} and \ref{f:map_3mmcont_n2}, therefore it is 
difficult to assess its nature, but it seems to be associated with a 
filamentary structure that is connected to AN02 in the ethanol maps integrated 
over the 77--81~km~s$^{-1}$ velocity range (Figs.~\ref{f:map_c2h5oh_n2}c and f).

\subsection{Chemical composition}
\label{ss:chemcomp}

\begin{table*}[!t]
 \begin{center}
\caption{
 List of molecules detected toward N2b, AN02, AN03, AN06, and AN06c2 with the ReMoCA survey.
 }
 \label{t:detections}
\vspace*{-1.2ex}
 \begin{tabular}{lccccclccccc}
 \hline\hline
\multicolumn{1}{c}{Molecule} & \multicolumn{1}{c}{N2b} & \multicolumn{1}{c}{AN02} & \multicolumn{1}{c}{AN03}  & \multicolumn{1}{c}{AN06}  & \multicolumn{1}{c}{AN06c2} & \multicolumn{1}{c}{Molecule} & \multicolumn{1}{c}{N2b} & \multicolumn{1}{c}{AN02} & \multicolumn{1}{c}{AN03}  & \multicolumn{1}{c}{AN06}  & \multicolumn{1}{c}{AN06c2} \\ 
 \hline
C$_2$H$_3$CN & d & d & d & d & d & CH$_3$SH & d & d & d & d & t \\ 
c-C$_2$H$_4$O & d & d & d & d & d & CN & a & a & a & a$^\star$ & a \\ 
syn-C$_2$H$_5$CHO & -- & -- & t & -- & -- & $^{13}$CO & d & d & d & d & d \\ 
C$_2$H$_5$CN & d & d & d & d & d & CS & a & a & a & a$^\star$ & a \\ 
C$_2$H$_5$OCHO & t & -- & -- & d & -- & H$_2$CO & d & d & d & d & d \\ 
C$_2$H$_5$OH & d & d & d & d & d & H$_2$CS & d & d & d & d & d \\ 
c-C$_3$H$_2$ & d & a & a & a$^\star$ & a & HC$_3$N & d & d & d & d & d \\ 
i-C$_3$H$_7$CN & d & d & d & -- & -- & HC$_5$N & t & -- & d & -- & -- \\ 
n-C$_3$H$_7$CN & d & d & d & d & d & HCCNC & t & -- & t & -- & -- \\ 
i-C$_3$H$_7$OH & d & -- & -- & -- & -- & HCN & a & d & d & a$^\star$ & a \\ 
n-C$_3$H$_7$OH & d & -- & -- & -- & -- & t-HCOOH & -- & t & d & -- & -- \\ 
CCH & a & a & a & a$^\star$ & a & HCO$^+$ & a & a & a & a$^\star$ & a \\ 
CCS & -- & a & a & a$^\star$ & a & HNC & a & a & a & a$^\star$ & a \\ 
CH$_2$CO & d & d & d & d & d & E-HNCHCN & t & -- & t & -- & -- \\ 
CH$_2$NH & d & d & d & d & t & HNCO & d & d & d & t & d \\ 
(CH$_2$OH)$_2$ & d & d & d & -- & -- & HNCS & d & t & d & t & t \\ 
CH$_2$(OH)CHO & d & d & d & -- & -- & HOCO$^+$ & -- & a & a & a$^\star$ & a \\ 
CH$_3$C$_3$N & t & t & d & -- & -- & HOC$^+$ & a & a & a & a$^\star$ & a \\ 
CH$_3$CCH & d & d & d & d & d & HSCN & t & -- & t & -- & -- \\ 
CH$_3$CHO & d & d & d & d & d & N$_2$H$^+$ & a & a & a & a$^\star$ & a \\ 
CH$_3$CN & d & d & d & d & d & NH$_2$CH$_2$CN & d & d & d & -- & -- \\ 
CH$_3$C(O)CH$_3$ & d & d & d & d & d & NH$_2$CHO & d & d & d & -- & d \\ 
CH$_3$C(O)NH$_2$ & t & d & d & -- & -- & NH$_2$CN & t & t & d & -- & -- \\ 
CH$_3$COOH & -- & t & t & -- & -- & NH$_2$D & t & t & d & d & d \\ 
CH$_3$NC & d & t & t & d & -- & NS$^+$ & a & a & a & a$^\star$ & a \\ 
CH$_3$NCO & d & d & d & d & d & OCS & d & d & d & d & d \\ 
CH$_3$NH2 & t & d & d & d & t & PN & a & a & a & a$^\star$ & a \\ 
CH$_3$NHCHO & d & d & d & -- & -- & SiO & a & a & a & a$^\star$ & a \\ 
CH$_3$OCH$_3$ & d & d & d & d & d & SO & d & d & d & d & d \\ 
CH$_3$OCHO & d & d & d & d & d & SO$_2$ & d & d & d & t & t \\ 
CH$_3$OH & d & d & d & d & d
 & & & & & & \\ 
 \hline
$N_{\rm em}$ & 35 & 33 & 39 & 26 & 23 &  &  &  &  &  & \\ 
$N_{\rm ten}$ & 10 & 7 & 6 & 3 & 5 &  &  &  &  &  & \\ 
$N_{\rm abs}$ & 11 & 13 & 13 & 14 & 14 &  &  &  &  &  & \\ 
$N_{\rm COM}$ & 23 & 22 & 24 & 17 & 14 &  &  &  &  &  & \\ 
$N_{\rm all}$ & 56 & 53 & 58 & 43 & 42 &  &  &  &  &  & \\ 
\hline
 \end{tabular}
 \end{center}
 \vspace*{-2.5ex}
\tablefoot{
 This table reports only one isotopolog per molecule. The full lists of molecules included in our Weeds models are provided in Tables~\ref{t:coldens_n2b}--\ref{t:coldens_an06}. Detections are labeled with "d", tentative detections with "t", detections only in absorption with "a", and nondetection with "--". We list at the bottom of the table the number of molecules detected in emission, $N_{\rm em}$, the number of molecules tentatively detected, $N_{\rm ten}$, the number of molecules detected only in absorption, $N_{\rm abs}$, the number of detected COMs, $N_{\rm COM}$, and the number of detected molecules including the tentative detections ($N_{\rm all}$ = $N_{\rm em}$+$N_{\rm ten}$+$N_{\rm abs}$). $^{(\star)}$ These molecules are detected in absorption but were technically assigned to AN06c2 for the calculation of the synthetic spectra.
 }
 \end{table*}

The LTE parameters of all the molecules included in the model of at least 
one of the four positions presented in Sect.~\ref{ss:sources}, obtained 
following the method described in Sect.~\ref{ss:weeds}, are listed in
Tables~\ref{t:coldens_n2b}--\ref{t:coldens_an06}. For a given catalog entry, 
each table indicates the number of spectral lines that we considered as 
detected, namely lines with a peak temperature above the 3$\sigma$ noise level 
and for which the LTE model of this entry accounts for at least about two 
thirds of the detected signal. Exceptions to the latter criterion occur for 
the setups with the largest beams (e.g., S1) for which the LTE models, which 
were optimized for the setups with the highest angular resolution (S4 and S5), 
sometimes clearly underestimate the emission in the larger beams due to the 
non-uniformity of the molecular emission across the Sgr~B2(N2) region. 
The excellent match between the LTE models and the ReMoCA spectra of N2b, 
AN02, AN03, and AN06 is illustrated in Fig.~\ref{f:spec_remoca_n2} that also
displays the ReMoCA spectrum of the ultracompact \ion{H}{ii} region K4 for 
comparison. K4 is located about 9$\arcsec$ to the North of Sgr~B2(N2).

Species that are considered as securely detected are labeled with a ``d'' in 
the third column of Tables~\ref{t:coldens_n2b}--\ref{t:coldens_an06}. The 
number of lines required to qualify a catalog entry as detected depends on the 
signal-to-noise ratio of the lines and whether or not other related entries 
(e.g., other isotopologs or vibrational states) are detected with consistent 
parameters. Entries that are considered as tentatively detected with only a 
few lines are labeled with a ``t''. Entries that are included in the LTE model 
because their predicted lines have peak temperatures close to or above the 
3$\sigma$ level but are heavily contaminated by emission or absorption of 
other species, and therefore cannot be considered as securely or tentatively 
identified, are labeled with a ``c'' to indicate that they merely contribute 
to the detected signal. They correspond typically to higher-energy vibrational 
states that are modeled with the same parameters as the lower-energy ones, or 
rarer isotopologs that are modeled with the parameters of the main species 
assuming a typical isotopic ratio. Species that are not detected are labeled 
with a ``n'' and the tables provide upper limits to their column densities.

\begin{figure}
\centerline{\resizebox{\hsize}{!}{\includegraphics[angle=0]{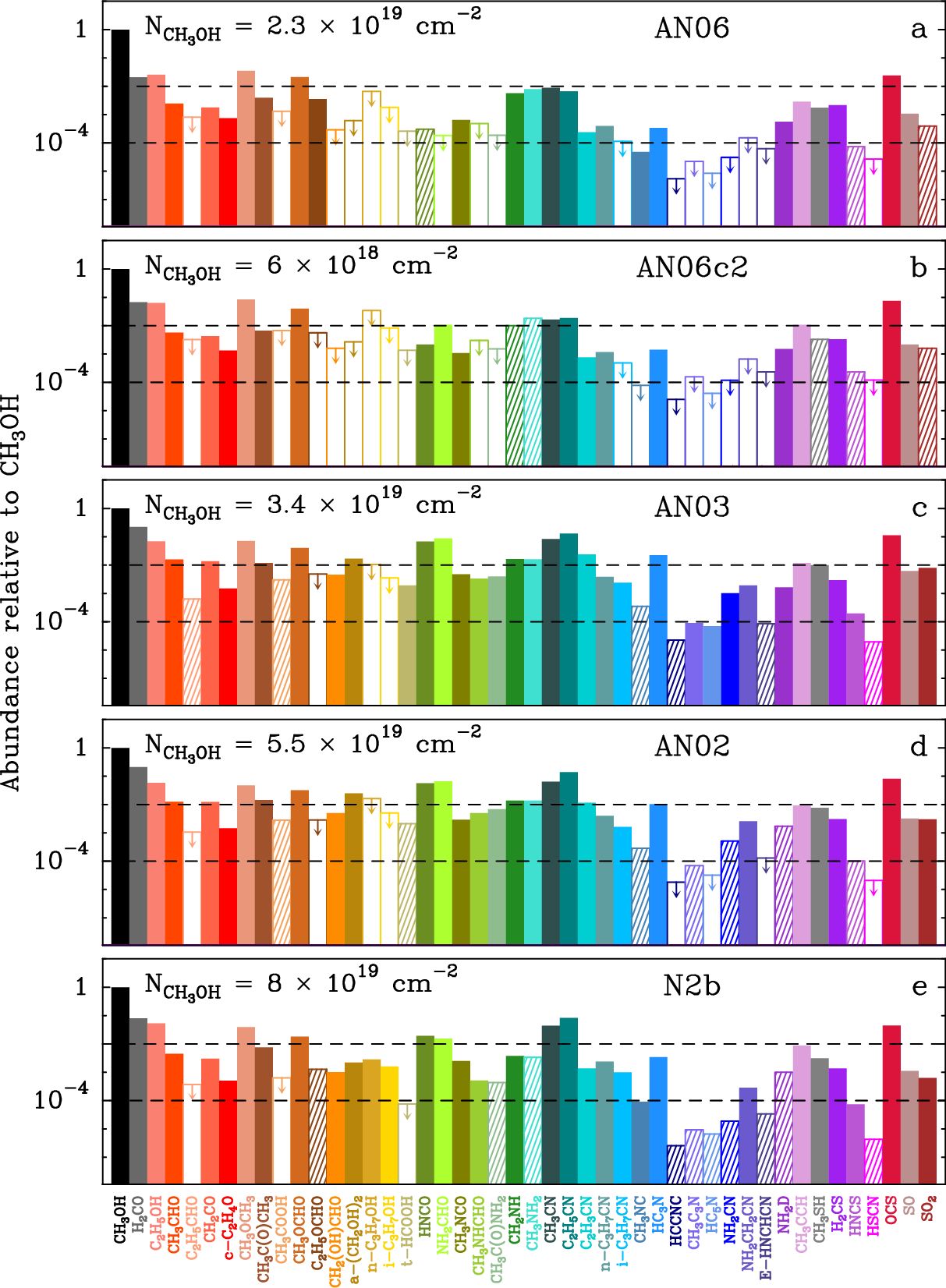}}}
\caption{Column densities derived toward N2b, AN02, AN03, and AN06 with our 
LTE modeling, normalized to the column density of methanol (see 
Tables~\ref{t:coldens_n2b}--\ref{t:coldens_an06}). The column density of 
methanol is indicated in the top left corner of each panel. The chemical 
compositions of the two velocity components detected toward AN06 are displayed 
separately (AN06 in the top panel and the second velocity component in the 
panel labeled AN06c2). The panel of AN02 reports the sum of the column 
densities of its two velocity components. Hatched bars show tentative 
detections while empty bars with downward arrows indicate upper limits. 
Molecules with status ``c'' in 
Tables~\ref{t:coldens_n2b}--\ref{t:coldens_an06} are here represented 
as upper limits. The dashed lines indicate levels of 1\% and 0.01\% with 
respect to methanol.}
\label{f:chemcomp_ch3oh}
\end{figure}

Table~\ref{t:detections} provides a concise overview of the detections 
obtained with the ReMoCA survey toward N2b, AN02, AN03, AN06, and AN06c2. This
table contains 61 molecules.
Thanks to the sensitivity of ALMA and the fact that the spectral confusion at
3~mm is not too severe, we were able to identify between 42 and 58 molecules 
(counting only the main isotopologs). Among these molecules, between 3 and 10 
are detected only tentatively and between 11 and 14 are seen only in 
absorption. Furthermore, many COMs are detected: between 22 and 24 toward N2b, 
AN02, and AN03, and about 15 toward AN06 and AN06c2.

In addition to the molecules included in the LTE model of at least one of the 
four positions, we provide upper limits to the column density of 146
relevant molecules that were searched for in the ReMoCA survey but not detected
(Tables~\ref{t:uplim_n2b}--\ref{t:uplim_an06}). The upper 
limits were obtained assuming the same LTE parameters as related molecules 
that were detected or typical LTE parameters of the respective source, keeping 
only the column density as a free parameter. The molecules are grouped by 
classes (O-bearing, O+N-bearing, N-bearing, S-bearing, S+O-bearing,
S+N-bearing, Cl-bearing, P-bearing, and hydrocarbons) and, within each 
class, by increasing entry index (tag) in our local Weeds database.

The column densities of 46 molecules extracted from 
Tables~\ref{t:coldens_n2b}--\ref{t:coldens_an06} and normalized to the column
density of methanol are displayed in Fig.~\ref{f:chemcomp_ch3oh}. We normalized
to methanol because it is the most abundant detected COM and because deriving 
reliable H$_2$ column densities is challenging. In order to compare the 
chemical compositions of the selected positions more easily, we show their 
abundances (relative to methanol) normalized to the abundances of N2b in 
Fig.~\ref{f:chemcomp_ch3oh_normn2b}. 

\subsection{Correlations of chemical composition between positions}
\label{ss:correl_pos}

\begin{figure*}
\centerline{\resizebox{0.82\hsize}{!}{\includegraphics[angle=0]{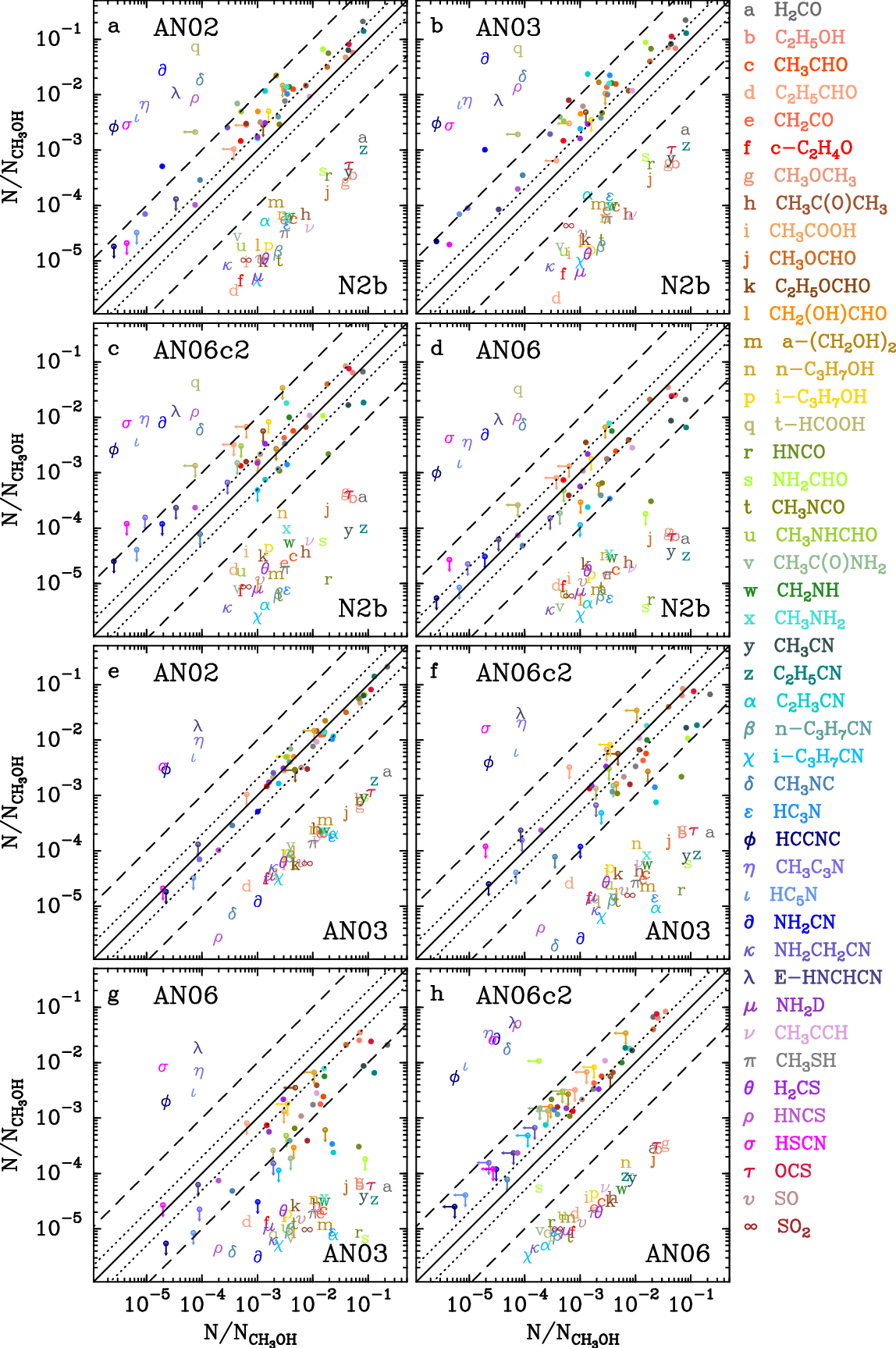}}}
\caption{Correlation plots of column densities normalized to methanol for 
various pairs of positions. The x- and y-axes of each panel correspond to the 
positions written in the bottom right and top left corners, respectively. The 
color coding of the molecules is the same as in 
Figs.~\ref{f:chemcomp_ch3oh} and 
\ref{f:chemcomp_ch3oh_normn2b} and is indicated on 
the right. Latine and Greek letters, reported at the same abscissa as the
corresponding molecule, were added to facilitate the identification of the 
data points. Filled data points represent firm and tentative detections while 
empty circles with arrows indicate upper limits toward at least one position.
The dashed and dotted lines indicate deviations by a factor 10 and 2, 
respectively.}
\label{f:correl_normch3oh}
\end{figure*}

Figure~\ref{f:correl_normch3oh} shows correlation plots of the column densities 
normalized to methanol for various pairs of positions.
The best correlation occurs between the chemical compositions of AN02 and AN03 
(Fig.~\ref{f:correl_normch3oh}e). Most data points deviate by 
much less than a factor of 2 from the 1:1 relation. The only data points that 
deviate by slightly more than a factor of 2 are HC$_3$N, HC$_5$N (upper limit 
toward AN02), and SO$_2$. C$_2$H$_3$CN, SO, and HNCS deviate by slightly less 
than a factor of 2. All six molecules are less abundant toward AN02 than toward
AN03. The upper limits derived for HCCNC, HSCN, and E-HNCHCN toward AN02 and 
for n- and i-C$_3$H$_7$OH, and C$_2$H$_5$OCHO, toward both AN02 and 
AN03 are not stringent enough to tell whether or not these molecules deviate 
from the overall correlation.

The next pair of positions that show a high degree of correlation are the two
velocity components of AN06 (Fig.~\ref{f:correl_normch3oh}h). In this case, 
however, the abundances (relative to methanol) of the second velocity 
component (AN06c2), which traces the gas at the edge of the dense core that 
contains AN03 and AN02, are systematically higher than the abundances of the 
main component (AN06), by roughly a factor of 2. There is one notable 
exception to the overall correlation: NH$_2$CHO is remarkably underabundant 
toward AN06, with an abundance upper limit that is nearly two orders of 
magnitude below the abundance measured toward AN06c2. HNCO is offset on the 
same side of the overall correlation, but only by a factor of about 4, while 
HC$_3$N, SO$_2$, and CH$_3$CCH are offset by a factor slightly larger than 2.

In contrast to the previous two pairs, the chemical composition of N2b 
correlates poorly with that of the other positions 
(Figs.~\ref{f:chemcomp_ch3oh_normn2b} and \ref{f:correl_normch3oh}a--d). 
Most molecules have abundances relative to methanol distributed between the 
1:1 and 10:1 lines in Figs.~\ref{f:correl_normch3oh}a and b, the abundances 
toward AN02 and AN03 being higher than toward N2b. The most extreme deviations 
from the 1:1 relation, by more than a factor of 20, occur for t-HCOOH (upper 
limit toward N2b) and NH$_2$CN. 

The correlation between N2b and AN06 is even poorer, with a spread of nearly 
two orders of magnitude (Figs.~\ref{f:chemcomp_ch3oh_normn2b}a and
\ref{f:correl_normch3oh}d). Most species are less abundant 
(relative to methanol) in AN06 than in N2b. The most extreme cases are HNCO and 
NH$_2$CHO which are almost two orders of magnitude (or more) less abundant, 
followed by the cyanides CH$_3$CN, C$_2$H$_5$CN, C$_2$H$_3$CN, n- and 
i-C$_3$H$_7$CN, and HC$_3$N (about one order of magnitude). The only molecules 
that are more abundant in AN06 than in N2b by at least a factor of two are 
C$_2$H$_5$OCHO and CH$_3$NH$_2$.

\begin{figure}
\centerline{\resizebox{\hsize}{!}{\includegraphics[angle=0]{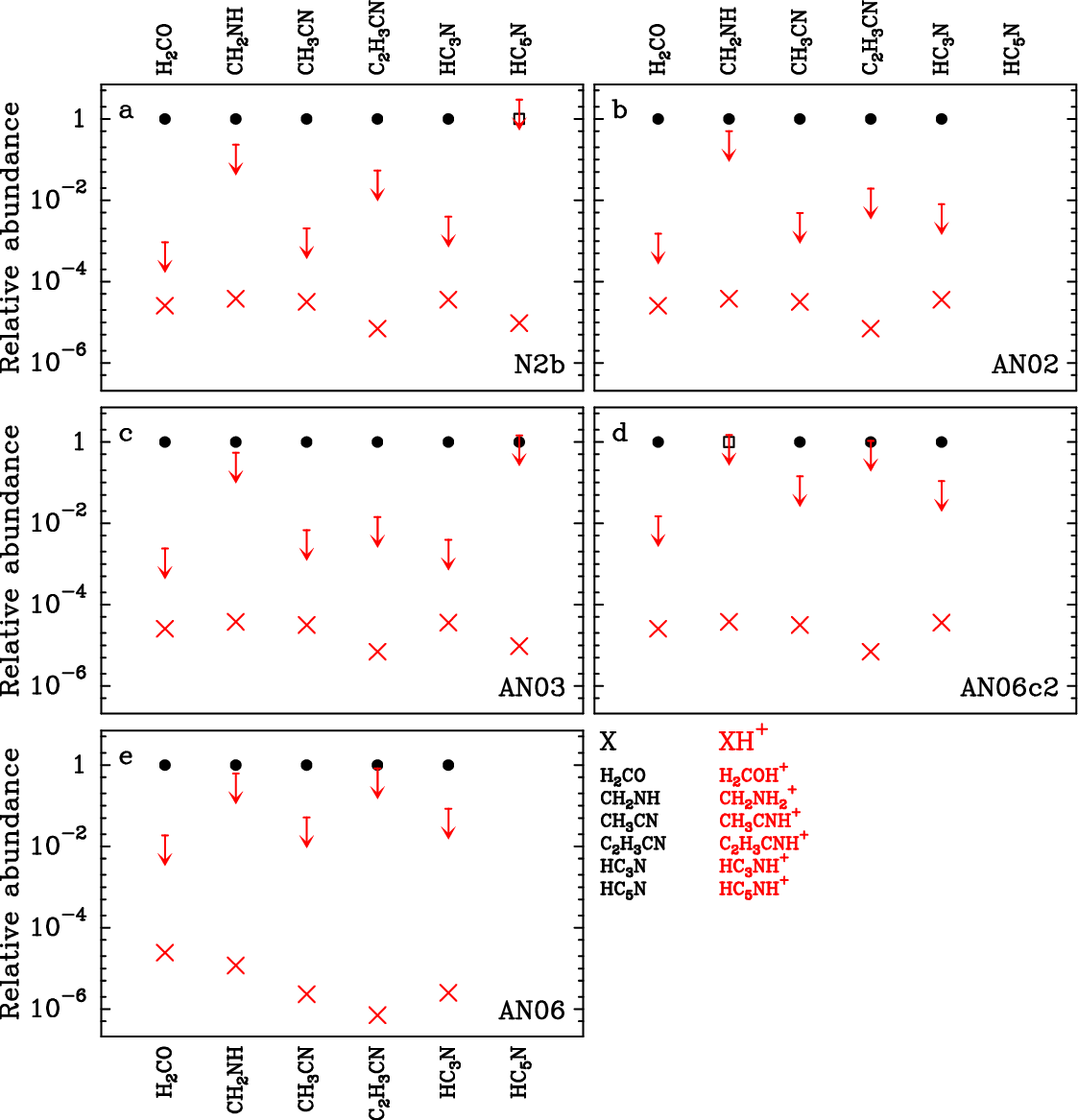}}}
\caption{Abundance of protonated molecules (red) relative to their neutral 
form (black). Each panel corresponds to one position as labeled in the bottom 
right corner. Empty squares and downward arrows represent 
tentative detections and upper limits, respectively. Crosses of the same color 
represent predictions of the chemical model described in 
Sect.~\ref{s:chemical_models}. The molecules are listed 
along the x-axis below and above the bottom and top panels, respectively. The
list of protonated molecules is indicated at the bottom right.}
\label{f:chemgroup_protonated}
\end{figure}

AN06c2 correlates a bit better with N2b than AN06 
(Figs.~\ref{f:chemcomp_ch3oh_normn2b}b and \ref{f:correl_normch3oh}c). The 
only species that deviate from the 1:1 relation by significantly more than a 
factor of 2 are CH$_3$NH$_2$ (more abundant in AN06c2) as well as HNCO and 
C$_2$H$_5$CN (more abundant in N2b).

Finally the chemical composition of both components of AN06 is poorly 
correlated with that of AN03 (Figs.~\ref{f:correl_normch3oh}f and g). All 
molecules are less abundant in AN06 than in AN03, by up to 2--3 orders of 
magnitude for HC$_3$N, C$_2$H$_3$CN, HNCO, and NH$_2$CHO. The situation is 
less extreme for AN06c2, with a handful of molecules close to the 1:1 
relation, most molecules between the 1:1 and 1:10 relations, and only HC$_3$N, 
C$_2$H$_3$CN, and HNCO being 15--30 times less abundant than in AN03.

In order to investigate if the lack of correlation between certain sources is
related to systematic differences between classes of molecules, we show in 
Fig.~\ref{f:correl_normch3oh_type} the same correlation plots as in 
Fig.~\ref{f:correl_normch3oh} but with colors coding for the atomic 
composition of the molecules: O-bearing in black, O+N-bearing in blue, 
N-bearing in red, pure hydrocarbon in green, and S-bearing in yellow. We 
notice that the correlations of the pairs N2b/AN02, N2b/AN06c2, N2b/AN06, and 
AN06/AN06c2 (Figs.~\ref{f:correl_normch3oh_type}a, c, d, and h) are much 
tighter for the S-bearing molecules than for the full sample of molecules. 
This is not the case for the pairs of sources that involve AN03 
(Figs.~\ref{f:correl_normch3oh_type}b, e, f, and g). Furthermore, the class of 
O+N-bearing molecules stands out in Figs.~\ref{f:correl_normch3oh_type}d 
and g: these molecules are located in the lower part of the correlation plot 
in both cases, revealing that they are underabundant in AN06 compared to N2b 
and AN03 (and AN02, given that it correlates well with AN03). In contrast to 
the S-bearing and O+N-bearing species, the O-bearing and N-bearing 
molecules do not show any obvious systematic differences between the five 
sources.

\begin{figure}
\centerline{\resizebox{\hsize}{!}{\includegraphics[angle=0]{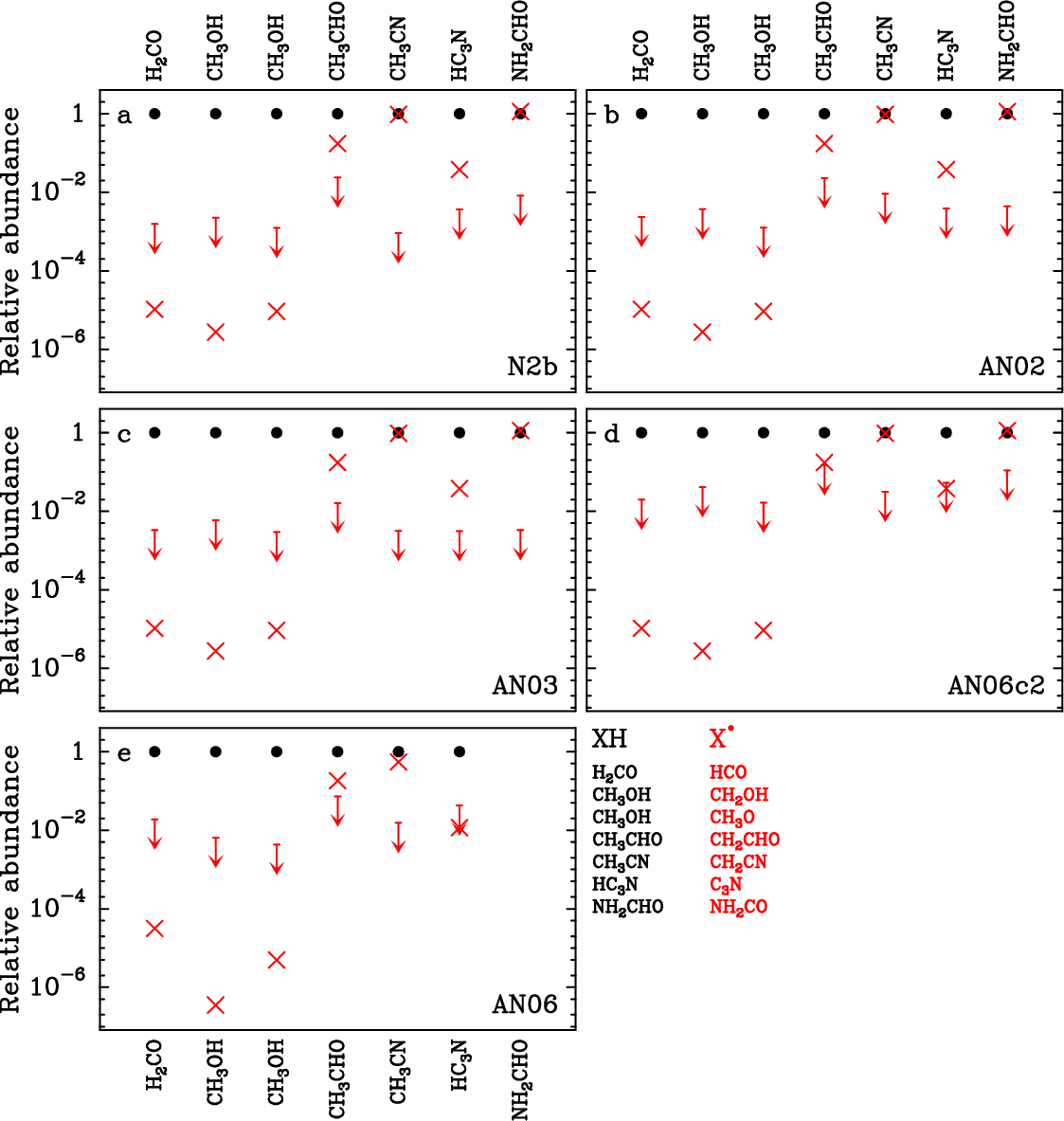}}}
\caption{Same as Fig.~\ref{f:chemgroup_protonated}, but for radicals (red) with 
respect to the hydrogenated form (black).}
\label{f:chemgroup_radical}
\end{figure}

To summarize, the pairs of positions in Sgr~B2(N2) that show a good 
correlation of their chemical composition (relative to methanol) are 
AN02/AN03 and AN06/AN06c2. The molecules that stand out in at least one panel 
of Fig.~\ref{f:correl_normch3oh} are HC$_3$N, HC$_5$N, C$_2$H$_3$CN, 
C$_2$H$_5$CN, NH$_2$CN, CH$_3$NH$_2$, NH$_2$CHO, HNCO, t-HCOOH, C$_2$H$_5$OCHO,
and SO$_2$. Finally, AN03 stands out for its S-bearing molecular 
content, which poorly correlates with the other sources, and AN06 for its 
underabundant O+N-bearing species compared to N2b, AN03, and AN02 
(Fig.~\ref{f:correl_normch3oh_type}).

\subsection{Correlations between families of molecules}
\label{ss:correl_fam}

Figures~\ref{f:chemgroup_protonated}--\ref{f:isomers} explore the
chemical composition of each selected position by plotting abundance ratios of 
specific pairs of species for different families of molecules. 

\subsubsection{Protonated molecules}
\label{sss:protonated}

We compare in Fig.~\ref{f:chemgroup_protonated} six protonated molecules 
to their neutral form. None of these cations are detected toward any position. 
The most stringent constraint is obtained for formaldehyde toward N2b, with 
[H$_2$COH$^+$]/[H$_2$CO] $< 10^{-3}$. 

\subsubsection{Radicals}
\label{sss:radicals}

Radicals related to formaldehyde, methanol, acetaldehyde, methyl cyanide, 
cyanoacetylene, and formamide are not detected either 
(Fig.~\ref{f:chemgroup_radical}). The most stringent constraint is obtained 
for methyl cyanide with [CH$_2$CN]/[CH$_3$CN] $< 10^{-3}$, again toward N2b.

\subsubsection{Degree of bond saturation}

\begin{figure}
\centerline{\resizebox{\hsize}{!}{\includegraphics[angle=0]{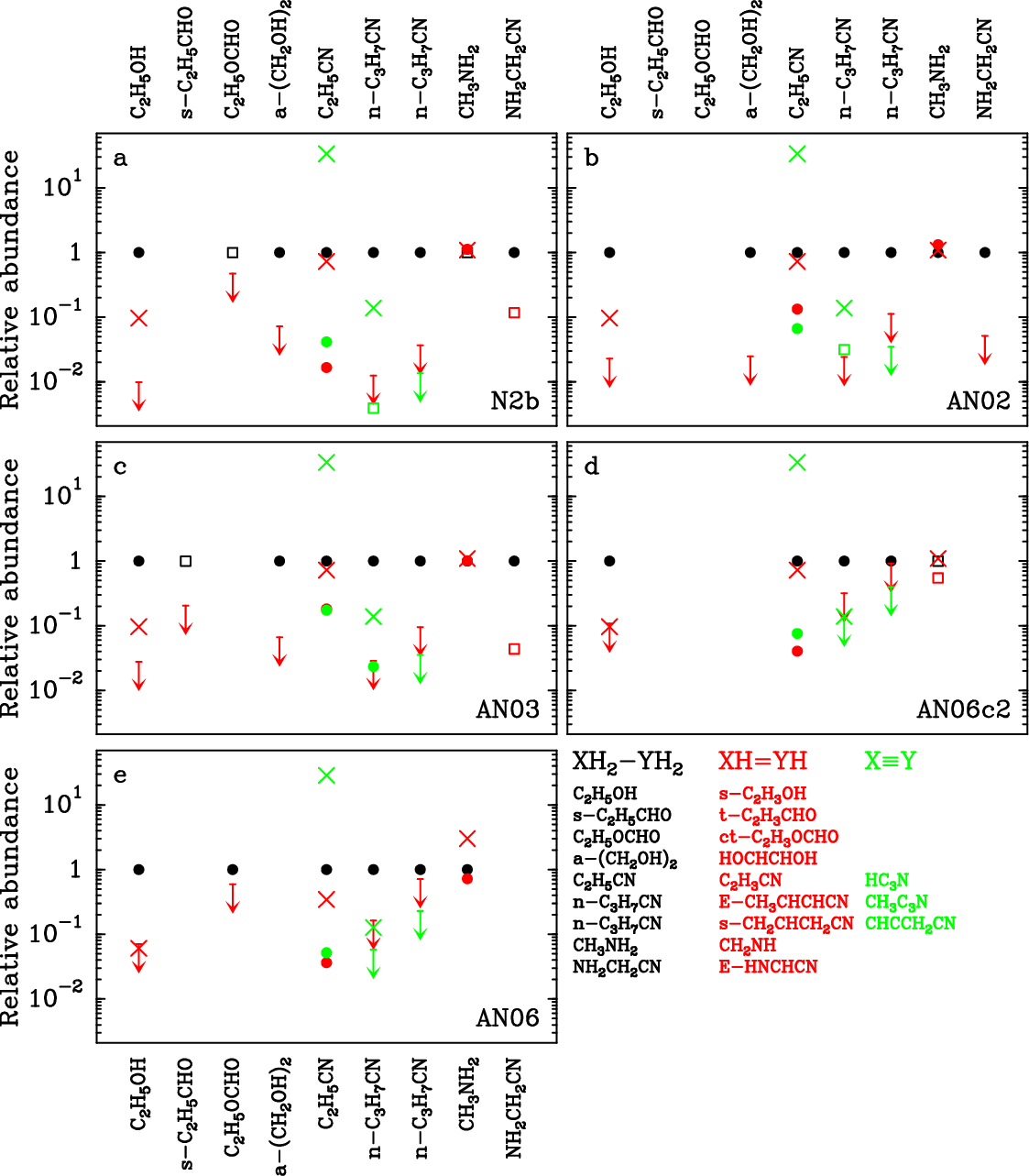}}}
\caption{Same as Fig.~\ref{f:chemgroup_protonated}, but for molecules with a
double (red) or triple (green) bond with respect to the saturated form 
with a single bond (black).}
\label{f:chemgroup_unsaturated}
\end{figure}

Figure~\ref{f:chemgroup_unsaturated} shows the abundance ratios of saturated 
(single bond) and unsaturated (double or triple bond) molecules for nine 
families of molecules. In nearly all cases, the double-bond form is much less 
abundant than the single-bond one, by up to at least two orders of magnitude 
for the pairs C$_2$H$_5$OH/C$_2$H$_3$OH and 
\hbox{n-C$_3$H$_7$CN}/E-CH$_3$CHCHCN in N2b. The only clear exception 
is the pair CH$_3$NH$_2$/CH$_2$NH which has an abundance ratio of about 1 
toward all positions, the unsaturated form even slightly dominating over the 
saturated one in AN02 and, marginally, N2b. 

The three molecules C$_2$H$_5$CN/C$_2$H$_3$CN/HC$_3$N are detected toward all 
five positions, with the saturated form being the most abundant one by factors
of 5 (in AN03) to 50 (in N2b). The relative ratios of the double-bond and 
triple-bond molecules depend on the position: they have a similar abundance in 
AN03 but the triple-bond one dominates over the double-bond one by a factor 
$\sim$2 in N2b, AN06, and AN06c2, while the opposite is true in AN02, with 
C$_2$H$_3$CN dominating over HC$_3$N by a factor of 2. At the next stage in 
complexity in the alkyl cyanide family, the unsaturated forms of 
n-C$_3$H$_7$CN are less abundant than the saturated one by at least one order 
of magnitude in N2b, AN02, and AN03. The triple-bond species CH$_3$C$_3$N is 
(tentatively) more abundant than the double-bond one, E-CH$_3$CHCHCN, in AN02, 
which is opposite to what we found for the pair HC$_3$N/C$_2$H$_3$CN in the 
same source.

\subsubsection{CN, CHO, and COOH functional groups}

\begin{figure}
\centerline{\resizebox{\hsize}{!}{\includegraphics[angle=0]{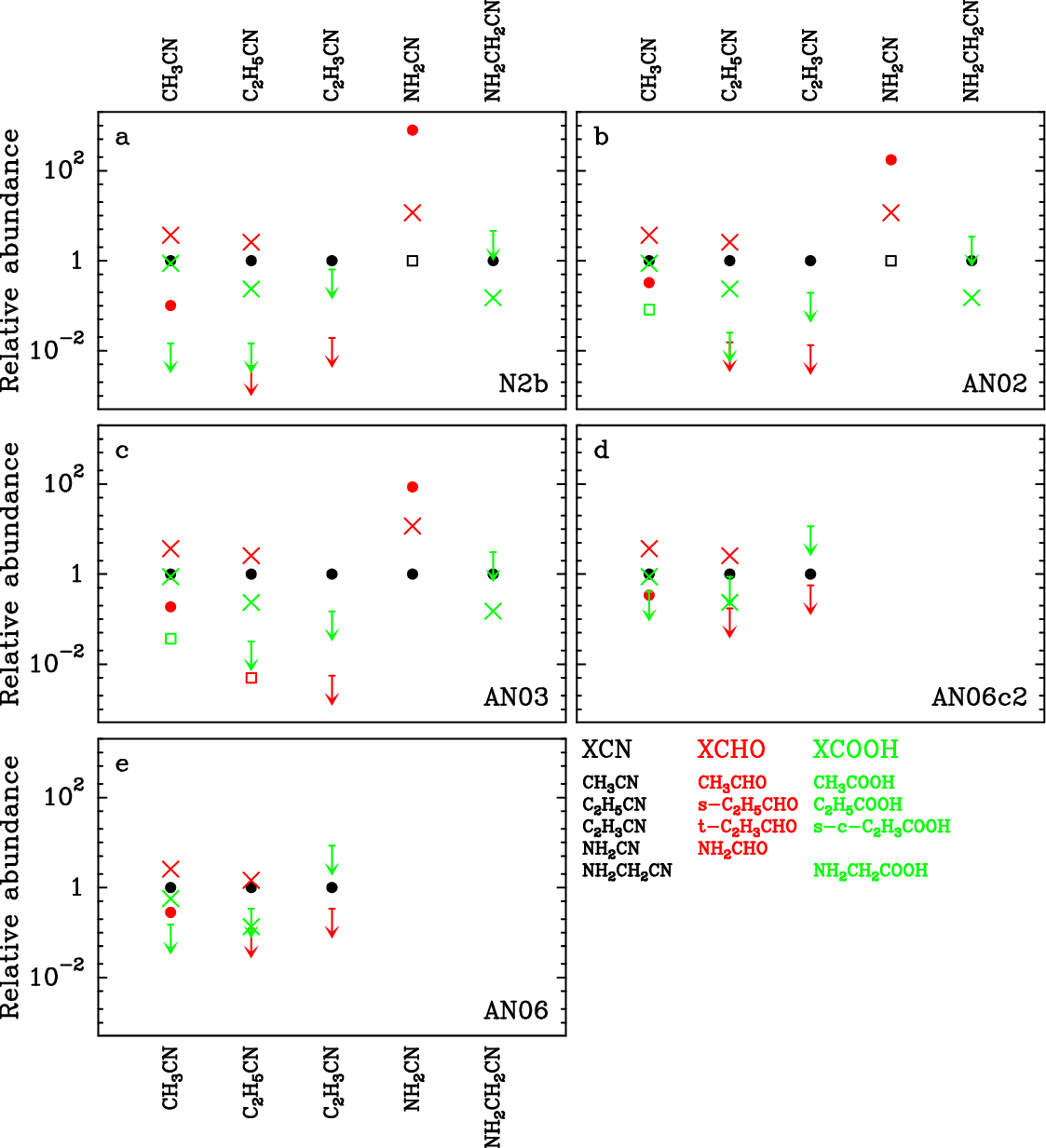}}}
\caption{Same as Fig.~\ref{f:chemgroup_protonated}, but for molecules with 
O-bearing functional groups -CHO and -COOH with respect to molecules with a 
-CN functional group.
}
\label{f:chemgroup_hydrolyzed}
\end{figure}

We examine in Fig.~\ref{f:chemgroup_hydrolyzed} how the abundances of five 
groups of molecules behave for molecules that share the same backbone but 
terminate with a nitrile functional 
group (-CN), an aldehyde group (-CHO), or a carboxylic group (-COOH).
No obvious pattern is visible. Overall, the CN-bearing molecules are more
abundant than their CHO- or COOH-bearing counterparts, with the notable 
exception of NH$_2$CN which is about two orders of magnitude less abundant 
than NH$_2$CHO in all three positions where it is detected (N2b, AN02, AN03). 
While acetaldehyde is at most a factor of 10 less abundant than methyl cyanide,
we find that propanal is at least two orders of magnitude less abundant than
ethyl cyanide in nearly all positions (the upper limit is inconclusive for 
AN06c2). In addition to the simple carboxylic acid HCOOH, only one 
complex carboxylic acid, CH$_3$COOH, is 
(tentatively) detected toward AN02 and AN03, with an abundance about one order
of magnitude lower than CH$_3$CN. If this ratio also holds for the pair 
NH$_2$CH$_2$CN/NH$_2$CH$_2$COOH, then glycine would have an abundance at least
one order of magnitude lower than its current upper limit.

\subsubsection{Oxydized and reduced forms}

\begin{figure}
\centerline{\resizebox{\hsize}{!}{\includegraphics[angle=0]{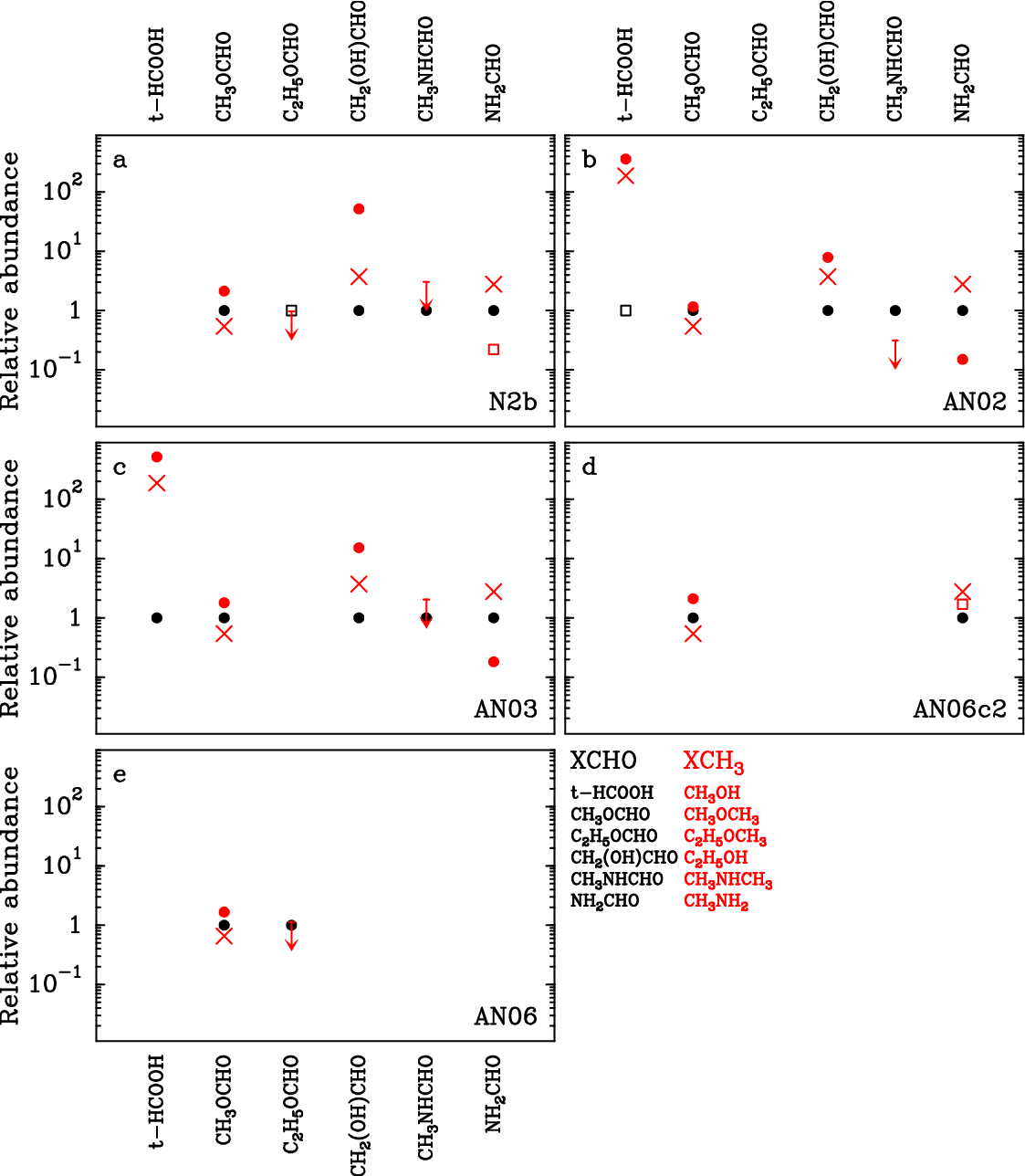}}}
\caption{Same as Fig.~\ref{f:chemgroup_protonated}, but for the reduced form 
(-CH$_3$ functional group) with respect to the aldehyde form (-CHO functional 
group).}
\label{f:chemgroup_reduced}
\end{figure}

We show in Fig.~\ref{f:chemgroup_reduced} how the abundances behave when we
go from an oxydized form (aldehyde functional group) to a reduced form (CH$_3$
group). Here again, no obvious pattern is seen. While CH$_3$OH and 
C$_2$H$_5$OH largely dominate over t-HCOOH and CH$_2$(OH)CHO, respectively, 
in the sources where t-HCOOH or CH$_2$(OH)CHO is detected, it is the opposite 
for CH$_3$NH$_2$ which is less abundant than NH$_2$CHO in N2b, AN02, and AN03 
(but maybe not in AN06c2). The pair CH$_3$OCH$_3$/CH$_3$OCHO lies in between
these two cases, with an abundance ratio close to 1 in all five sources. The 
upper limits obtained for CH$_3$NHCH$_3$ and C$_2$H$_5$OCH$_3$ suggest 
that they behave like one of the latter two cases.

\subsubsection{Size of carbon backbone}
 
\begin{figure}
\centerline{\resizebox{\hsize}{!}{\includegraphics[angle=0]{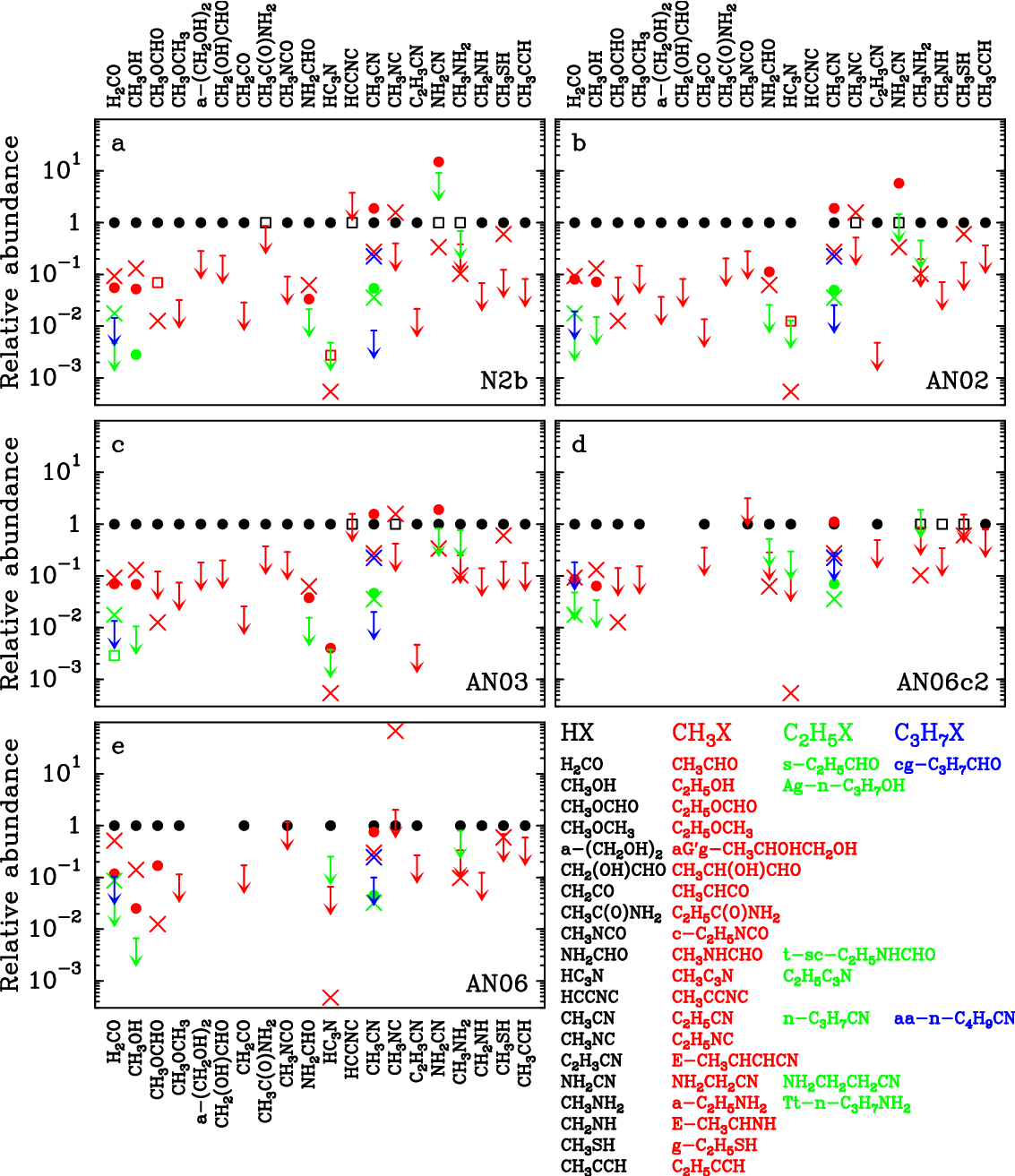}}}
\caption{Same as Fig.~\ref{f:chemgroup_protonated}, but for series of 
molecules with an additional CH$_2$ group in their backbone at each
further step in complexity.}
\label{f:chemgroup_homolog}
\end{figure}

We investigate in Fig.~\ref{f:chemgroup_homolog} how the abundances vary with 
the size of the carbon backbone for 20 series of homologous molecules. H$_2$CO, 
CH$_3$OH, and NH$_2$CHO show a similar pattern: an abundance drop of about one 
order of magnitude when a molecule has an additional CH$_2$ group at 
one end of its heavy-atom backbone (CH$_3$CHO, C$_2$H$_5$OH, and CH$_3$NHCHO), 
and a further drop of about one order of magnitude for a second additional
CH$_2$ group (s-C$_2$H$_5$CHO, tentatively detected in AN03,
n-C$_3$H$_7$OH, detected in N2b, and C$_2$H$_5$NHCHO, with a stringent
upper limit in AN02). This behavior is also seen for CH$_3$OCHO for the case of
one additional CH$_2$ group, either with (tentative) detections or 
constraining upper limits of C$_2$H$_5$OCHO. We do not have results for the 
next degree of complexity in this case. 

The same behavior is observed for the series of alkyl cyanides C$_2$H$_5$CN, 
n-C$_3$H$_7$CN (detection), and n-C$_4$H$_9$CN (upper limit), at least in N2b, 
but CH$_3$CN does not fit into this pattern because it has a somewhat lower 
abundance than C$_2$H$_5$CN in nearly all five sources. Toward N2b, in 
addition to the previous cases, the derived upper limits shown in red also 
indicate an abundance drop of about one order of magnitude (at least) for one
additional CH$_2$ group for the following species: CH$_3$OCH$_3$, 
a-(CH$_2$OH)$_2$, CH$_2$(OH)CHO, CH$_2$CO, CH$_3$NCO, CH$_3$NC, C$_2$H$_3$CN, 
CH$_2$NH, CH$_3$SH, and CH$_3$CCH. The 
upper limits are inconclusive for CH$_3$C(O)NH$_2$ and CH$_3$NH$_2$ in N2b, 
but they do imply an abundance drop by nearly an order of magnitude (at least) 
in AN02. The case of HC$_3$N is more extreme, with an abundance drop of 2--3 
orders of magnitude for CH$_3$C$_3$N in N2b, AN02, and AN03. For the 
two positions were HCCNC is tentatively detected, the upper limits derived for 
CH$_3$CCNC are inconclusive. 

The only clear exception to the general behavior noted in the previous
paragraph is NH$_2$CN, for which the next stage in backbone length, 
NH$_2$CH$_2$CN, is one order of magnitude more abundant in N2b. However, the 
latter molecule corresponds to an additional CH$_2$ 
group between the nitrogen and carbon atoms of NH$_2$CN. We do not have 
results for the molecule with an additional CH$_2$ group at the end of the 
backbone (CH$_3$NHCN).

\subsubsection{Structural isomers}
\label{sss:isomers}

\begin{figure}
\centerline{\resizebox{\hsize}{!}{\includegraphics[angle=0]{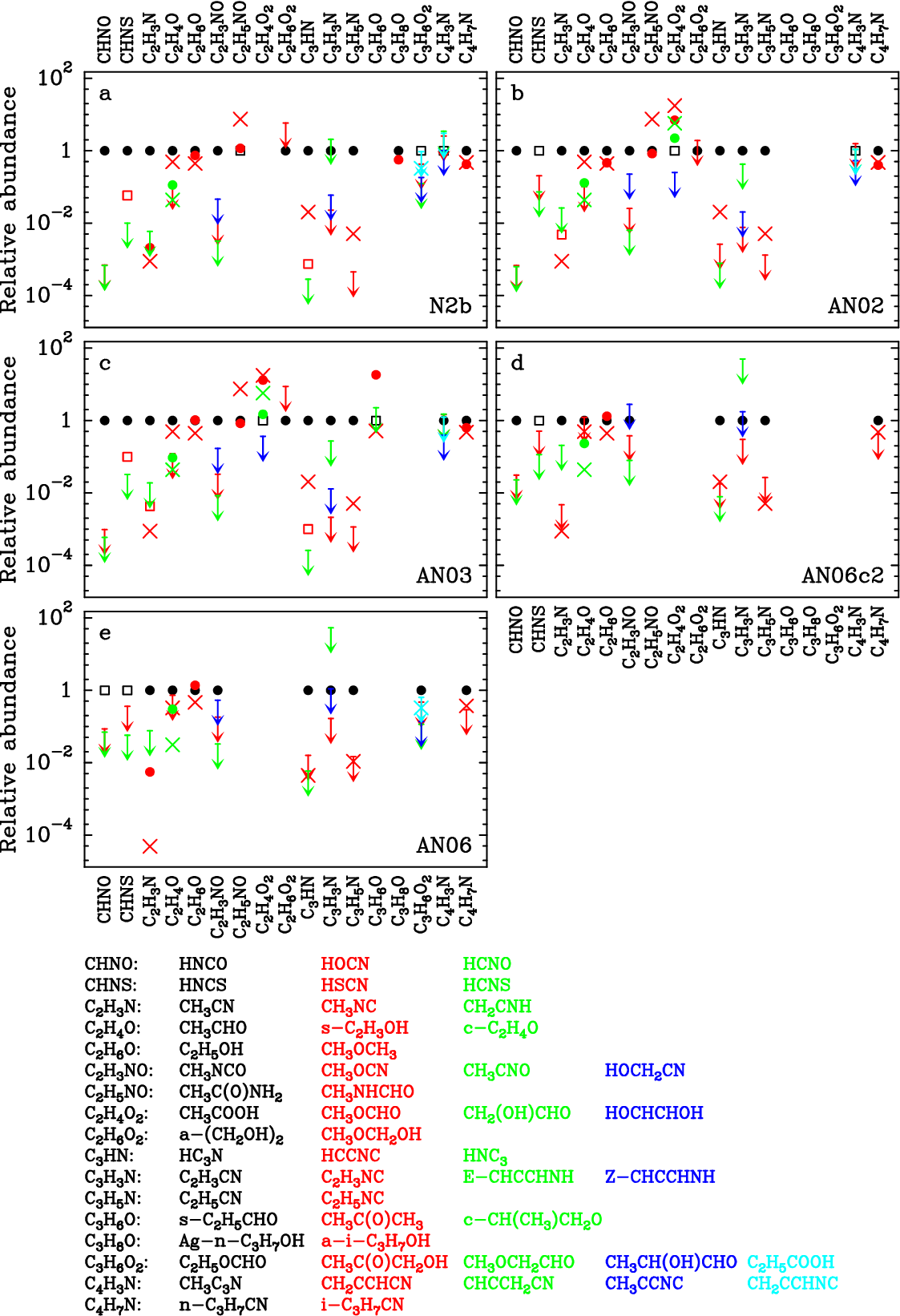}}}
\caption{Same as Fig.~\ref{f:chemgroup_protonated}, but for groups of 
structural isomers. The elemental composition of the groups is labeled along 
the x-axis. The list of molecules belonging to each group is provided at the 
bottom of the figure with their respective colors as used in the plots.
}
\label{f:isomers}
\end{figure}

Finally, we examine the relative abundances of structural isomers, that is
molecules with the same elemental composition but a different arrangement of 
their constituent atoms. The results are displayed in Fig.~\ref{f:isomers} for 
17 groups of species with a molecular size ranging from 4 to 12 atoms.
Several pairs of isomers stand out for having similar abundances:
C$_2$H$_5$OH/CH$_3$OCH$_3$, CH$_3$C(O)NH$_2$/CH$_3$NHCHO, 
CH$_3$COOH/CH$_2$(OH)CHO (tentatively), n/i-C$_3$H$_7$OH, and 
n/i-C$_3$H$_7$CN. On the contrary, one isomer (given in parentheses) 
dominates by more than two orders of magnitude in the following groups: CHNO 
(HNCO), C$_2$H$_3$N (CH$_3$CN), C$_3$HN (HC$_3$N), and C$_3$H$_5$N 
(C$_2$H$_5$CN). We note that methyl acetate (CH$_3$C(O)OCH$_3$),
which was detected in Orion~KL \citep[][]{Tercero13,Tercero18}, is 
missing in the C$_3$H$_6$O$_2$ family shown in Fig.~\ref{f:isomers}. We have a
preliminary spectroscopic entry for this molecule but it suffers from several 
issues that currently prevent us from deriving reliable column densities or 
upper limits for the Sgr~B2(N2) sources.

Focusing only on the detections and tentative detections displayed in 
Fig.~\ref{f:isomers}, we do not find any significant difference in the isomer
ratios between the five positions. These isomer ratios thus appear to be 
relatively robust with respect to the potential range of physical conditions 
or age spanned by this small sample of sources.

\subsection{Comparison of chemical composition of Sgr~B2(N2) to other sources}
\label{ss:comp_lit}

\begin{figure}
\centerline{\resizebox{\hsize}{!}{\includegraphics[angle=0]{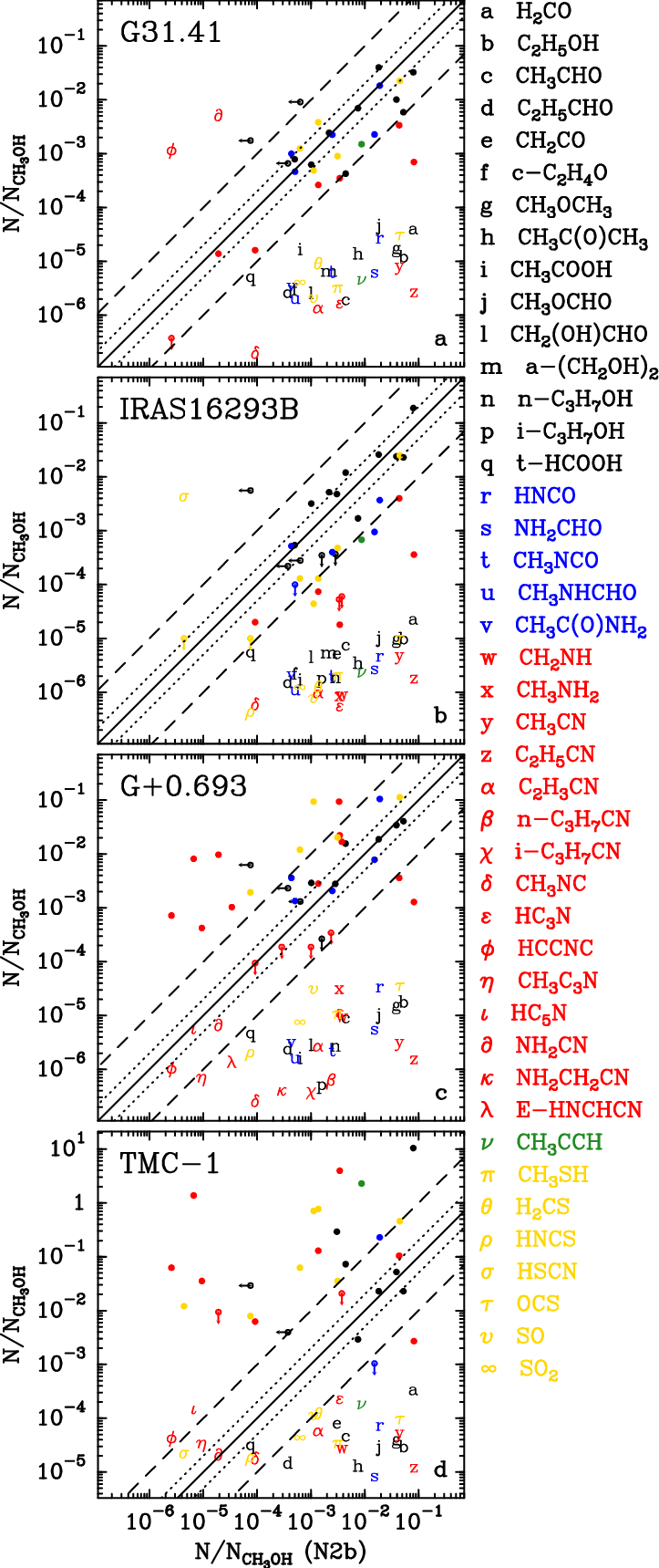}}}
\caption{Same as Fig.~\ref{f:correl_normch3oh_type} but comparing sources from 
the literature to N2b.}
\label{f:correl_normch3oh_type_n2b}
\end{figure}

\begin{figure}
\centerline{\resizebox{\hsize}{!}{\includegraphics[angle=0]{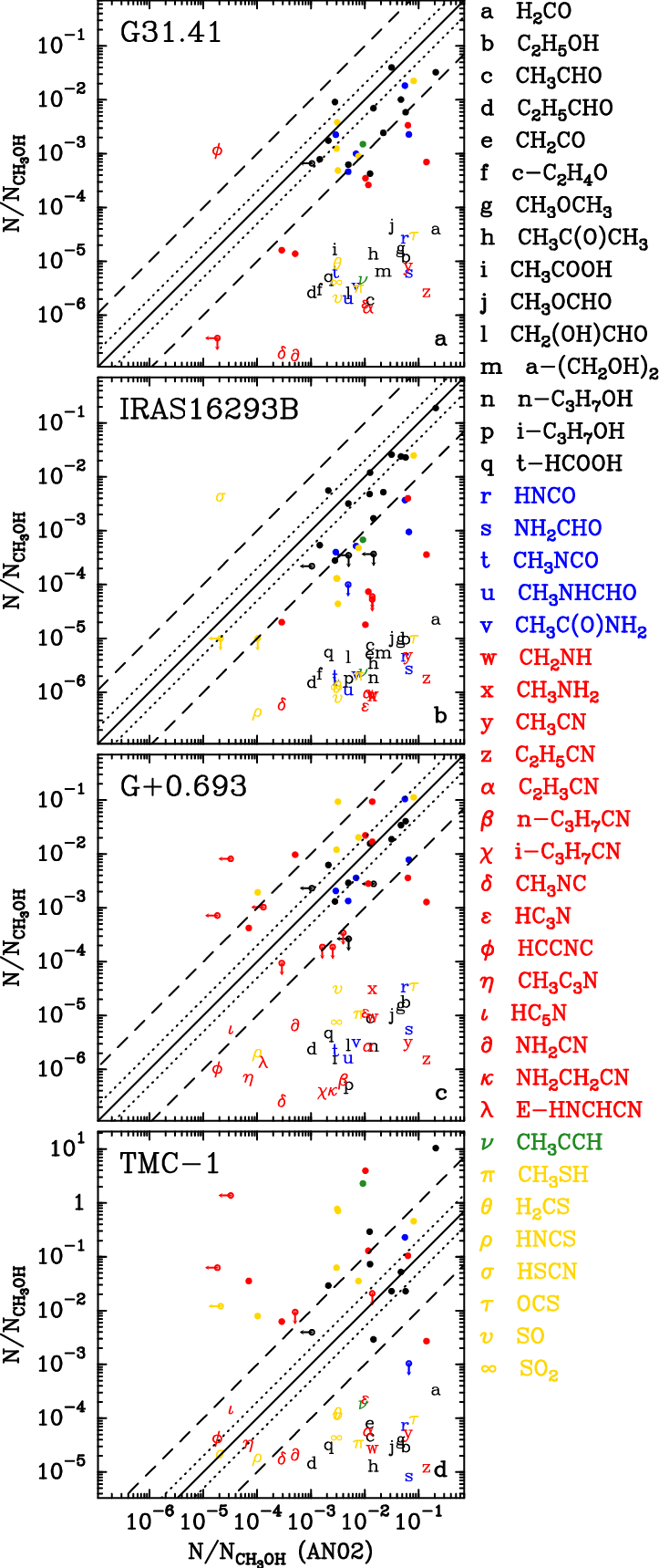}}}
\caption{Same as Fig.~\ref{f:correl_normch3oh_type} but comparing sources from 
the literature to AN02.}
\label{f:correl_normch3oh_type_an02}
\end{figure}

\begin{figure}
\centerline{\resizebox{\hsize}{!}{\includegraphics[angle=0]{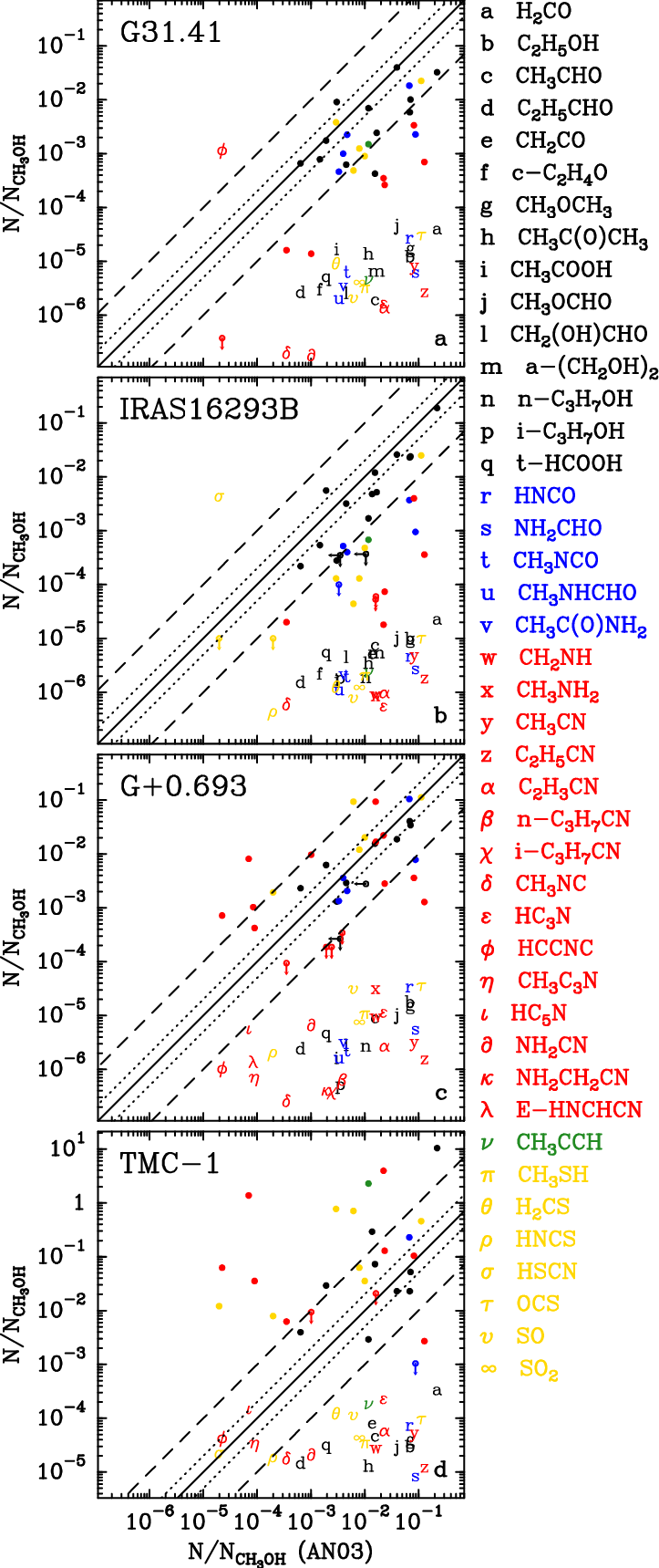}}}
\caption{Same as Fig.~\ref{f:correl_normch3oh_type} but comparing sources from 
the literature to AN03.}
\label{f:correl_normch3oh_type_an03}
\end{figure}

\begin{figure}
\centerline{\resizebox{\hsize}{!}{\includegraphics[angle=0]{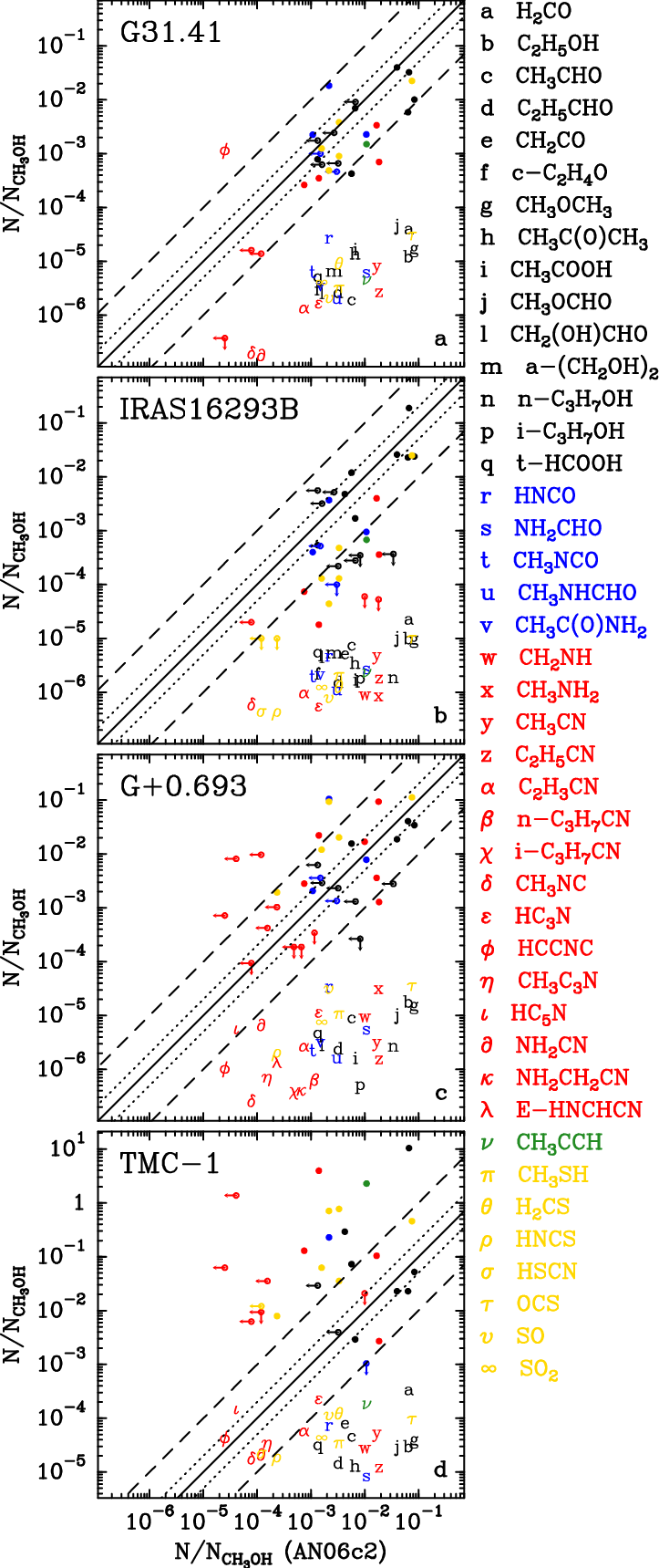}}}
\caption{Same as Fig.~\ref{f:correl_normch3oh_type} but comparing sources from 
the literature to AN06c2.}
\label{f:correl_normch3oh_type_an06c2}
\end{figure}

\begin{figure}
\centerline{\resizebox{\hsize}{!}{\includegraphics[angle=0]{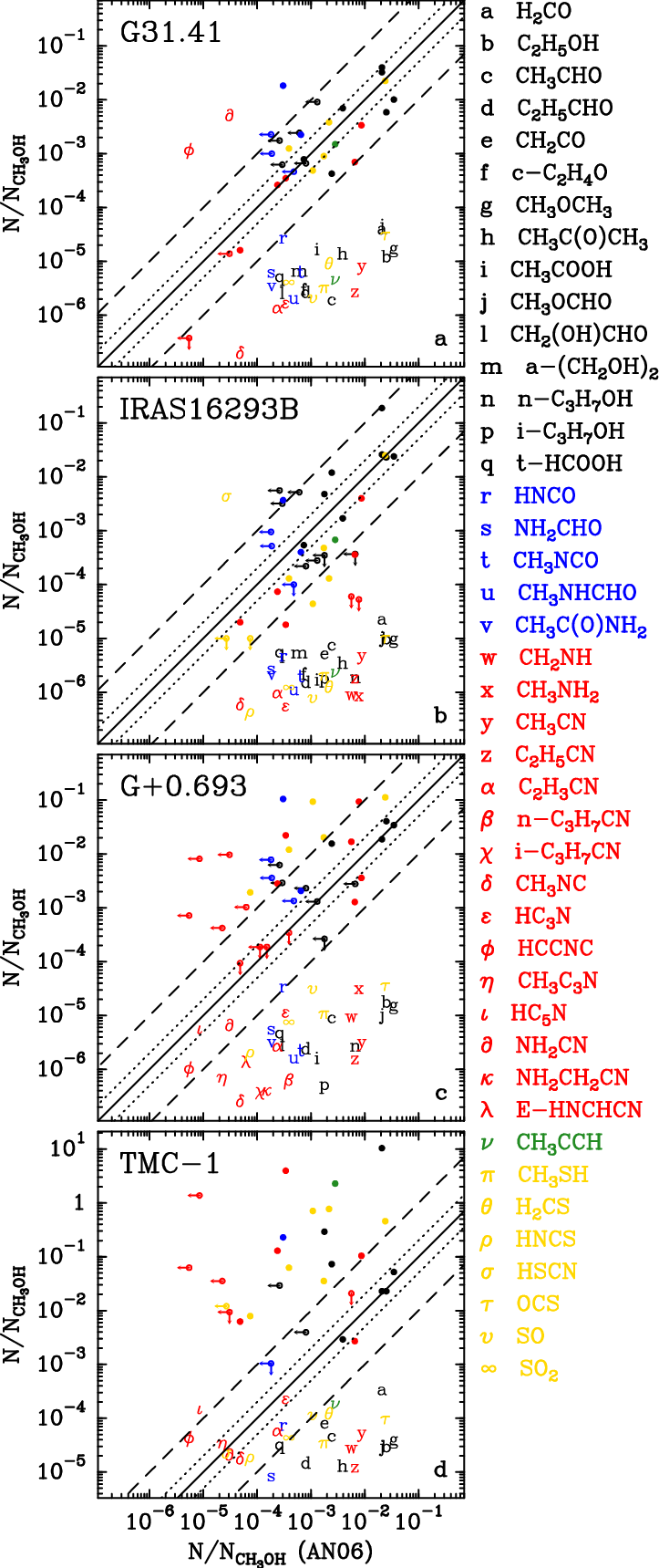}}}
\caption{Same as Fig.~\ref{f:correl_normch3oh_type} but comparing sources from 
the literature to AN06.}
\label{f:correl_normch3oh_type_an06}
\end{figure}

Figures~\ref{f:correl_normch3oh_type_n2b}--\ref{f:correl_normch3oh_type_an06} 
compare the column densities relative to methanol derived in 
Sect.~\ref{ss:chemcomp} toward the sources of Sgr~B2(N2) to the 
composition of 
sources that have been studied in detail in the literature: the hot core 
G31.41+0.31, hereafter G31.41 
\citep[GUAPOS interferometric survey, e.g.,][]{Mininni20,LopezGallifa24}, the 
hot corino IRAS~16293--2422~B, hereafter IRAS16293B
\citep[PILS interferometric survey, e.g.,][]{Jorgensen16,Drozdovskaya19}, 
the (possibly shocked) molecular cloud region G+0.693--0.027, hereafter G+0.693
\citep[Yebes and IRAM 30\,m single-dish survey, e.g.,][]{Rivilla22c,JimenezSerra22}\footnote{A comparison between G+0.693 and Sgr~B(N2) on the basis of our previous survey EMoCA, which had a lower angular resolution (1.6$\arcsec$) and could not disentangle the four sources embedded in Sgr~B2(N2), was presented by \citet{JimenezSerra25}.},
and the starless core TMC-1 in the Taurus molecular cloud 
\citep[mainly the QUIJOTE single-dish survey, e.g.,][]{Cernicharo21,Agundez25}. 
The column densities of these sources compiled from numerous articles 
published by the teams of these surveys, as well as a few others in the
case of TMC-1 \citep[e.g.,][]{Gratier16,Tennis23}, are listed in 
Table~\ref{t:coldenslit}. In the cases where this had 
not been done, we applied a posteriori vibrational or conformational 
corrections to the published column densities. We report in 
Table~\ref{t:correl_lit} the Pearson correlation coefficients of the
abundance distributions shown in 
Figs.~\ref{f:correl_normch3oh_type_n2b}--\ref{f:correl_normch3oh_type_an06},
for all molecules and per class of molecules. The calculation does not take the 
upper limits into account and it was done only for samples with at least four 
items.

\begin{table*}[!t]
 \begin{center}
 \caption{
 Correlations between the chemical composition of Sgr~B2(N2) sources and sources from the literature.
}
 \label{t:correl_lit}
 \vspace*{-1.2ex}
 \begin{tabular}{lccccccccccccccc}
 \hline\hline
 \multicolumn{1}{c}{Source} & \multicolumn{3}{c}{All\tablefootmark{(a)}} & \multicolumn{3}{c}{O\tablefootmark{(b)}} & \multicolumn{3}{c}{N\tablefootmark{(c)}} & \multicolumn{3}{c}{O+N\tablefootmark{(d)}} & \multicolumn{3}{c}{S\tablefootmark{(e)}} \\ 
  & \multicolumn{1}{c}{\hspace*{-1.4ex} $n$\tablefootmark{(f)}} & \multicolumn{1}{c}{\hspace*{-2ex} $r$\tablefootmark{(g)}} & \multicolumn{1}{c}{\hspace*{-1.5ex} $P$\tablefootmark{(h)}} & \multicolumn{1}{c}{\hspace*{-1.4ex} $n$\tablefootmark{(f)}} & \multicolumn{1}{c}{\hspace*{-2ex} $r$\tablefootmark{(g)}} & \multicolumn{1}{c}{\hspace*{-1.5ex} $P$\tablefootmark{(h)}} & \multicolumn{1}{c}{\hspace*{-1.4ex} $n$\tablefootmark{(f)}} & \multicolumn{1}{c}{\hspace*{-2ex} $r$\tablefootmark{(g)}} & \multicolumn{1}{c}{\hspace*{-1.5ex} $P$\tablefootmark{(h)}} & \multicolumn{1}{c}{\hspace*{-1.4ex} $n$\tablefootmark{(f)}} & \multicolumn{1}{c}{\hspace*{-2ex} $r$\tablefootmark{(g)}} & \multicolumn{1}{c}{\hspace*{-1.5ex} $P$\tablefootmark{(h)}} & \multicolumn{1}{c}{\hspace*{-1.4ex} $n$\tablefootmark{(f)}} & \multicolumn{1}{c}{\hspace*{-2ex} $r$\tablefootmark{(g)}} & \multicolumn{1}{c}{\hspace*{-1.5ex} $P$\tablefootmark{(h)}} \\ 
 \hline
\multicolumn{16}{c}{N2b} \\ 
G31.41 & \hspace*{-1.4ex} 26 & \hspace*{-2ex} $0.79$$^{+0.11}_{-0.21}$ & \hspace*{-1.5ex} 1($-6$) & \hspace*{-1.4ex} 9 & \hspace*{-2ex} $0.80$$^{+0.16}_{-0.52}$ & \hspace*{-1.5ex} 1($-2$) & \hspace*{-1.4ex} 6 & \hspace*{-2ex} $0.94$$^{+0.05}_{-0.39}$ & \hspace*{-1.5ex} 5($-3$) & \hspace*{-1.4ex} 5 & \hspace*{-2ex} $0.83$$^{+0.15}_{-1.02}$ & \hspace*{-1.5ex} 8($-2$) & \hspace*{-1.4ex} 5 & \hspace*{-2ex} $0.81$$^{+0.18}_{-1.07}$ & \hspace*{-1.5ex} 1($-1$) \\ 
I16293B & \hspace*{-1.4ex} 25 & \hspace*{-2ex} $0.68$$^{+0.17}_{-0.29}$ & \hspace*{-1.5ex} 2($-4$) & \hspace*{-1.4ex} 10 & \hspace*{-2ex} $0.87$$^{+0.10}_{-0.34}$ & \hspace*{-1.5ex} 1($-3$) & \hspace*{-1.4ex} 5 & \hspace*{-2ex} $0.77$$^{+0.21}_{-1.12}$ & \hspace*{-1.5ex} 1($-1$) & \hspace*{-1.4ex} 4 & \hspace*{-2ex} $0.74$$^{+0.26}_{-1.51}$ & \hspace*{-1.5ex} 3($-1$) & \hspace*{-1.4ex} 5 & \hspace*{-2ex} $0.96$$^{+0.04}_{-0.45}$ & \hspace*{-1.5ex} 9($-3$) \\ 
G+0.693 & \hspace*{-1.4ex} 27 & \hspace*{-2ex} $0.53$$^{+0.23}_{-0.34}$ & \hspace*{-1.5ex} 5($-3$) & \hspace*{-1.4ex} 6 & \hspace*{-2ex} $0.92$$^{+0.07}_{-0.47}$ & \hspace*{-1.5ex} 8($-3$) & \hspace*{-1.4ex} 11 & \hspace*{-2ex} $0.35$$^{+0.44}_{-0.67}$ & \hspace*{-1.5ex} 3($-1$) & \hspace*{-1.4ex} 5 & \hspace*{-2ex} $0.77$$^{+0.21}_{-1.12}$ & \hspace*{-1.5ex} 1($-1$) & \hspace*{-1.4ex} 5 & \hspace*{-2ex} $0.83$$^{+0.16}_{-1.03}$ & \hspace*{-1.5ex} 8($-2$) \\ 
TMC-1 & \hspace*{-1.4ex} 24 & \hspace*{-2ex} $0.14$$^{+0.37}_{-0.42}$ & \hspace*{-1.5ex} 5($-1$) & \hspace*{-1.4ex} 7 & \hspace*{-2ex} $0.32$$^{+0.55}_{-0.89}$ & \hspace*{-1.5ex} 5($-1$) & \hspace*{-1.4ex} 8 & \hspace*{-2ex} $-0.16$$^{+0.77}_{-0.62}$ & \hspace*{-1.5ex} 7($-1$) & 1 & \ldots & \ldots  & \hspace*{-1.4ex} 7 & \hspace*{-2ex} $0.69$$^{+0.26}_{-0.82}$ & \hspace*{-1.5ex} 8($-2$) \\ 
 \hline
\multicolumn{16}{c}{AN02} \\ 
G31.41 & \hspace*{-1.4ex} 28 & \hspace*{-2ex} $0.69$$^{+0.16}_{-0.27}$ & \hspace*{-1.5ex} 6($-5$) & \hspace*{-1.4ex} 11 & \hspace*{-2ex} $0.63$$^{+0.26}_{-0.58}$ & \hspace*{-1.5ex} 4($-2$) & \hspace*{-1.4ex} 6 & \hspace*{-2ex} $0.94$$^{+0.06}_{-0.41}$ & \hspace*{-1.5ex} 6($-3$) & \hspace*{-1.4ex} 5 & \hspace*{-2ex} $0.65$$^{+0.33}_{-1.20}$ & \hspace*{-1.5ex} 2($-1$) & \hspace*{-1.4ex} 5 & \hspace*{-2ex} $0.80$$^{+0.19}_{-1.08}$ & \hspace*{-1.5ex} 1($-1$) \\ 
I16293B & \hspace*{-1.4ex} 27 & \hspace*{-2ex} $0.65$$^{+0.18}_{-0.29}$ & \hspace*{-1.5ex} 3($-4$) & \hspace*{-1.4ex} 12 & \hspace*{-2ex} $0.86$$^{+0.10}_{-0.29}$ & \hspace*{-1.5ex} 3($-4$) & \hspace*{-1.4ex} 5 & \hspace*{-2ex} $0.72$$^{+0.26}_{-1.17}$ & \hspace*{-1.5ex} 2($-1$) & \hspace*{-1.4ex} 4 & \hspace*{-2ex} $0.80$$^{+0.20}_{-1.50}$ & \hspace*{-1.5ex} 2($-1$) & \hspace*{-1.4ex} 5 & \hspace*{-2ex} $0.98$$^{+0.02}_{-0.24}$ & \hspace*{-1.5ex} 3($-3$) \\ 
G+0.693 & \hspace*{-1.4ex} 25 & \hspace*{-2ex} $0.47$$^{+0.26}_{-0.38}$ & \hspace*{-1.5ex} 2($-2$) & \hspace*{-1.4ex} 7 & \hspace*{-2ex} $0.89$$^{+0.09}_{-0.48}$ & \hspace*{-1.5ex} 7($-3$) & \hspace*{-1.4ex} 8 & \hspace*{-2ex} $0.26$$^{+0.55}_{-0.81}$ & \hspace*{-1.5ex} 5($-1$) & \hspace*{-1.4ex} 5 & \hspace*{-2ex} $0.80$$^{+0.19}_{-1.08}$ & \hspace*{-1.5ex} 1($-1$) & \hspace*{-1.4ex} 5 & \hspace*{-2ex} $0.85$$^{+0.14}_{-0.97}$ & \hspace*{-1.5ex} 7($-2$) \\ 
TMC-1 & \hspace*{-1.4ex} 22 & \hspace*{-2ex} $0.27$$^{+0.35}_{-0.44}$ & \hspace*{-1.5ex} 2($-1$) & \hspace*{-1.4ex} 8 & \hspace*{-2ex} $0.53$$^{+0.37}_{-0.81}$ & \hspace*{-1.5ex} 2($-1$) & \hspace*{-1.4ex} 6 & \hspace*{-2ex} $0.09$$^{+0.75}_{-0.87}$ & \hspace*{-1.5ex} 9($-1$) & 1 & \ldots & \ldots  & \hspace*{-1.4ex} 6 & \hspace*{-2ex} $0.62$$^{+0.33}_{-1.00}$ & \hspace*{-1.5ex} 2($-1$) \\ 
 \hline
\multicolumn{16}{c}{AN03} \\ 
G31.41 & \hspace*{-1.4ex} 29 & \hspace*{-2ex} $0.63$$^{+0.18}_{-0.29}$ & \hspace*{-1.5ex} 2($-4$) & \hspace*{-1.4ex} 12 & \hspace*{-2ex} $0.69$$^{+0.22}_{-0.50}$ & \hspace*{-1.5ex} 1($-2$) & \hspace*{-1.4ex} 6 & \hspace*{-2ex} $0.94$$^{+0.05}_{-0.39}$ & \hspace*{-1.5ex} 5($-3$) & \hspace*{-1.4ex} 5 & \hspace*{-2ex} $0.74$$^{+0.25}_{-1.15}$ & \hspace*{-1.5ex} 2($-1$) & \hspace*{-1.4ex} 5 & \hspace*{-2ex} $0.70$$^{+0.28}_{-1.18}$ & \hspace*{-1.5ex} 2($-1$) \\ 
I16293B & \hspace*{-1.4ex} 28 & \hspace*{-2ex} $0.58$$^{+0.21}_{-0.32}$ & \hspace*{-1.5ex} 1($-3$) & \hspace*{-1.4ex} 13 & \hspace*{-2ex} $0.90$$^{+0.07}_{-0.21}$ & \hspace*{-1.5ex} 3($-5$) & \hspace*{-1.4ex} 5 & \hspace*{-2ex} $0.66$$^{+0.31}_{-1.19}$ & \hspace*{-1.5ex} 2($-1$) & \hspace*{-1.4ex} 4 & \hspace*{-2ex} $0.78$$^{+0.22}_{-1.50}$ & \hspace*{-1.5ex} 2($-1$) & \hspace*{-1.4ex} 5 & \hspace*{-2ex} $0.93$$^{+0.06}_{-0.65}$ & \hspace*{-1.5ex} 2($-2$) \\ 
G+0.693 & \hspace*{-1.4ex} 29 & \hspace*{-2ex} $0.59$$^{+0.20}_{-0.31}$ & \hspace*{-1.5ex} 8($-4$) & \hspace*{-1.4ex} 8 & \hspace*{-2ex} $0.89$$^{+0.09}_{-0.40}$ & \hspace*{-1.5ex} 3($-3$) & \hspace*{-1.4ex} 11 & \hspace*{-2ex} $0.44$$^{+0.38}_{-0.66}$ & \hspace*{-1.5ex} 2($-1$) & \hspace*{-1.4ex} 5 & \hspace*{-2ex} $0.79$$^{+0.19}_{-1.09}$ & \hspace*{-1.5ex} 1($-1$) & \hspace*{-1.4ex} 5 & \hspace*{-2ex} $0.86$$^{+0.13}_{-0.96}$ & \hspace*{-1.5ex} 6($-2$) \\ 
TMC-1 & \hspace*{-1.4ex} 26 & \hspace*{-2ex} $0.25$$^{+0.33}_{-0.40}$ & \hspace*{-1.5ex} 2($-1$) & \hspace*{-1.4ex} 9 & \hspace*{-2ex} $0.62$$^{+0.29}_{-0.70}$ & \hspace*{-1.5ex} 8($-2$) & \hspace*{-1.4ex} 8 & \hspace*{-2ex} $-0.07$$^{+0.74}_{-0.67}$ & \hspace*{-1.5ex} 9($-1$) & 1 & \ldots & \ldots  & \hspace*{-1.4ex} 7 & \hspace*{-2ex} $0.68$$^{+0.26}_{-0.83}$ & \hspace*{-1.5ex} 9($-2$) \\ 
 \hline
\multicolumn{16}{c}{AN06c2} \\ 
G31.41 & \hspace*{-1.4ex} 20 & \hspace*{-2ex} $0.68$$^{+0.18}_{-0.34}$ & \hspace*{-1.5ex} 1($-3$) & \hspace*{-1.4ex} 7 & \hspace*{-2ex} $0.77$$^{+0.20}_{-0.74}$ & \hspace*{-1.5ex} 5($-2$) & \hspace*{-1.4ex} 4 & \hspace*{-2ex} $0.81$$^{+0.18}_{-1.49}$ & \hspace*{-1.5ex} 2($-1$) & 3 & \ldots & \ldots  & \hspace*{-1.4ex} 5 & \hspace*{-2ex} $0.89$$^{+0.10}_{-0.86}$ & \hspace*{-1.5ex} 4($-2$) \\ 
I16293B & \hspace*{-1.4ex} 21 & \hspace*{-2ex} $0.80$$^{+0.11}_{-0.24}$ & \hspace*{-1.5ex} 1($-5$) & \hspace*{-1.4ex} 8 & \hspace*{-2ex} $0.86$$^{+0.11}_{-0.46}$ & \hspace*{-1.5ex} 6($-3$) & \hspace*{-1.4ex} 4 & \hspace*{-2ex} $0.81$$^{+0.19}_{-1.49}$ & \hspace*{-1.5ex} 2($-1$) & 3 & \ldots & \ldots  & \hspace*{-1.4ex} 5 & \hspace*{-2ex} $0.95$$^{+0.04}_{-0.49}$ & \hspace*{-1.5ex} 1($-2$) \\ 
G+0.693 & \hspace*{-1.4ex} 18 & \hspace*{-2ex} $0.39$$^{+0.33}_{-0.48}$ & \hspace*{-1.5ex} 1($-1$) & \hspace*{-1.4ex} 4 & \hspace*{-2ex} $0.82$$^{+0.17}_{-1.48}$ & \hspace*{-1.5ex} 2($-1$) & \hspace*{-1.4ex} 6 & \hspace*{-2ex} $0.08$$^{+0.76}_{-0.86}$ & \hspace*{-1.5ex} 9($-1$) & 3 & \ldots & \ldots  & \hspace*{-1.4ex} 5 & \hspace*{-2ex} $0.82$$^{+0.17}_{-1.05}$ & \hspace*{-1.5ex} 9($-2$) \\ 
TMC-1 & \hspace*{-1.4ex} 19 & \hspace*{-2ex} $0.04$$^{+0.44}_{-0.46}$ & \hspace*{-1.5ex} 9($-1$) & \hspace*{-1.4ex} 7 & \hspace*{-2ex} $0.23$$^{+0.61}_{-0.86}$ & \hspace*{-1.5ex} 6($-1$) & \hspace*{-1.4ex} 4 & \hspace*{-2ex} $-0.66$$^{+1.48}_{-0.33}$ & \hspace*{-1.5ex} 3($-1$) & 1 & \ldots & \ldots  & \hspace*{-1.4ex} 6 & \hspace*{-2ex} $0.64$$^{+0.32}_{-1.00}$ & \hspace*{-1.5ex} 2($-1$) \\ 
 \hline
\multicolumn{16}{c}{AN06} \\ 
G31.41 & \hspace*{-1.4ex} 20 & \hspace*{-2ex} $0.74$$^{+0.15}_{-0.30}$ & \hspace*{-1.5ex} 2($-4$) & \hspace*{-1.4ex} 7 & \hspace*{-2ex} $0.79$$^{+0.17}_{-0.69}$ & \hspace*{-1.5ex} 3($-2$) & \hspace*{-1.4ex} 5 & \hspace*{-2ex} $0.91$$^{+0.09}_{-0.79}$ & \hspace*{-1.5ex} 3($-2$) & 2 & \ldots & \ldots  & \hspace*{-1.4ex} 5 & \hspace*{-2ex} $0.85$$^{+0.14}_{-0.98}$ & \hspace*{-1.5ex} 7($-2$) \\ 
I16293B & \hspace*{-1.4ex} 21 & \hspace*{-2ex} $0.81$$^{+0.11}_{-0.23}$ & \hspace*{-1.5ex} 1($-5$) & \hspace*{-1.4ex} 8 & \hspace*{-2ex} $0.81$$^{+0.15}_{-0.56}$ & \hspace*{-1.5ex} 1($-2$) & \hspace*{-1.4ex} 5 & \hspace*{-2ex} $0.88$$^{+0.11}_{-0.88}$ & \hspace*{-1.5ex} 5($-2$) & 2 & \ldots & \ldots  & \hspace*{-1.4ex} 5 & \hspace*{-2ex} $0.89$$^{+0.11}_{-0.87}$ & \hspace*{-1.5ex} 4($-2$) \\ 
G+0.693 & \hspace*{-1.4ex} 17 & \hspace*{-2ex} $0.33$$^{+0.37}_{-0.51}$ & \hspace*{-1.5ex} 2($-1$) & \hspace*{-1.4ex} 4 & \hspace*{-2ex} $0.76$$^{+0.23}_{-1.51}$ & \hspace*{-1.5ex} 2($-1$) & \hspace*{-1.4ex} 6 & \hspace*{-2ex} $0.08$$^{+0.76}_{-0.86}$ & \hspace*{-1.5ex} 9($-1$) & 2 & \ldots & \ldots  & \hspace*{-1.4ex} 5 & \hspace*{-2ex} $0.86$$^{+0.13}_{-0.95}$ & \hspace*{-1.5ex} 6($-2$) \\ 
TMC-1 & \hspace*{-1.4ex} 20 & \hspace*{-2ex} $0.11$$^{+0.42}_{-0.46}$ & \hspace*{-1.5ex} 6($-1$) & \hspace*{-1.4ex} 7 & \hspace*{-2ex} $0.10$$^{+0.69}_{-0.81}$ & \hspace*{-1.5ex} 8($-1$) & \hspace*{-1.4ex} 5 & \hspace*{-2ex} $-0.13$$^{+0.98}_{-0.78}$ & \hspace*{-1.5ex} 8($-1$) & 1 & \ldots & \ldots  & \hspace*{-1.4ex} 6 & \hspace*{-2ex} $0.71$$^{+0.26}_{-0.95}$ & \hspace*{-1.5ex} 1($-1$) \\ 
 \hline
 \end{tabular}
 \end{center}
 \vspace*{-2.5ex}
 \tablefoot{
 I16293B stands for IRAS16293B.
 \tablefoottext{a}{All molecules.}
 \tablefoottext{b}{O-bearing molecules.}
 \tablefoottext{c}{N-bearing molecules.}
 \tablefoottext{d}{O+N-bearing molecules.}
 \tablefoottext{e}{S-bearing molecules.}
 \tablefoottext{f}{Number of data points used to compute the correlation coefficient.}
 \tablefoottext{g}{Pearson correlation coefficient and its 95\% confidence interval.}
 \tablefoottext{h}{$P$-value. $x$($y$) means $x \times 10^{y}$.}
 }
 \end{table*}

\citet{Jorgensen20} reported a good correlation between the chemical 
compositions of IRAS16293B (derived from the PILS survey) and Sgr~B2(N2) 
(obtained from the EMoCA survey at an angular resolution of 
$\sim$1.6$\arcsec$) after normalizing the column densities of the O-bearing 
and N-bearing molecules to CH$_3$OH and HNCO (or CH$_3$CN), respectively. 
Given that Sgr~B2(N2) consists of several sources that were not resolved by the
EMoCA survey, we revisit this correlation by taking advantage of the higher 
angular resolution of the ReMoCA survey.
Figures~\ref{f:correl_normch3oh_type_n2b}b--\ref{f:correl_normch3oh_type_an03}b
show a clear differentiation of the four main classes of molecules (O-bearing,
O+N-bearing, N-bearing, and S-bearing) between IRAS16293B and
N2b/AN02/AN03. For each class of molecules taken separately, there is a good
correlation between IRAS16293B and N2b/AN02/AN03, but the four groups are 
shifted with respect to each other: the O-bearing molecules occupy the top 
part of the distribution, the N-bearing species the bottom part, and the 
O+N-bearing and S-bearing molecules lie in between. The correlation 
coefficients listed in Table~\ref{t:correl_lit} confirm this visual impression:
the coefficients of IRAS16293B for the individual classes of molecules are 
systematically larger than for the whole sample of molecules. The tightest 
correlation is obtained for the S-bearing molecules, followed by the O-bearing 
ones. The N-bearing molecules are the ones that correlate 
the least, with correlation coefficients between 0.66 and 0.77. The O+N-bearing 
species have a coefficient that lies in between, except for N2b where it is 
slightly smaller than for the N-bearing molecules. With a 
coefficient larger than 0.8, the overall correlation seems to be stronger for
AN06c2 and AN06 than for N2b/AN02/AN03 but this may be biased by the smaller 
sample of molecules (21 versus 25--28) and the upper limits of several 
O-bearing, N-bearing, and O+N-bearing species in  
Figs.~\ref{f:correl_normch3oh_type_an06c2}b and
\ref{f:correl_normch3oh_type_an06}b suggest significant deviations from the 
overall correlation.

Among the four sources from the literature displayed in 
Figs.~\ref{f:correl_normch3oh_type_n2b}--\ref{f:correl_normch3oh_type_an03}, 
the one that shows the best overall correlation with N2b, AN02, and AN03 is 
the hot core G31.41 (Table~\ref{t:correl_lit}). One reason for this better 
overall correlation compared to IRAS16293B is that the differentiation between 
the different classes of molecules is less pronounced. Still, like in the case 
of IRAS16293B, the N-bearing species occupy the lower part of the distribution 
in 
Figs.~\ref{f:correl_normch3oh_type_n2b}a--\ref{f:correl_normch3oh_type_an03}a: 
they are underabundant by about one order of magnitude in G31.41, especially 
compared to AN02 and AN03. However, in contrast to IRAS16293B, the O-bearing
molecules are well mixed with the O+N-bearing species and the S-bearing
molecules in the overall distribution. Among the four classes of molecules,
the tightest correlation occurs for the N-bearing molecules with a coefficient 
larger than 0.9. The O-bearing molecules are the ones that correlate the 
least. This is even more true for N2b when we consider the upper limits 
obtained for t-HCOOH and CH$_3$COOH which are not taken into account in the 
calculation of the correlation coefficients (labels q and i in 
Fig.~\ref{f:correl_normch3oh_type_n2b}a). However, these two molecules 
do not stand out in the O-bearing class in the cases of AN02 and AN03 where 
they are detected. There is also a strong overall correlation between G31.41
and AN06 (Fig.~\ref{f:correl_normch3oh_type_an06}a), except for the class of
O+N-bearing molecules, with HNCO, NH$_2$CHO, and CH$_3$C(O)NH$_2$ being 
overabundant in G31.41 by about or more than one order of magnitude.

In contrast to G31.41 and IRAS16293B, there is no overall correlation between 
G+0.693 and the Sgr~B2(N2) sources, although they are all located in the same 
molecular cloud complex. The lack of correlation results mainly from the 
distribution of the N-bearing species which have correlation coefficients 
between 0.08 and 0.44 (Table~\ref{t:correl_lit}). For each other class of 
molecules, we find a much higher degree of correlation, with all correlation 
coefficients being above 0.75. The S-bearing species are located above 
the O+N-bearing molecules and O-bearing species in 
Figs.~\ref{f:correl_normch3oh_type_n2b}c--\ref{f:correl_normch3oh_type_an06}c, 
meaning that, relative to O+N-bearing molecules and O-bearing species, 
the S-bearing molecules are more prominent in G+0.693 than in the Sgr~B2(N2) 
sources.

Finally, the situation is even more extreme for the starless core TMC-1. There 
is absolutely no overall correlation between this source and the Sgr~B2(N2) 
sources, even for the O- or N-bearing molecules taken separately. 
The only class of molecules that shows some (low) degree of correlation is the 
S-bearing class but, as indicated by the smaller correlation coefficients 
(0.6$-$0.7), the dispersion of the data points in 
Figs.~\ref{f:correl_normch3oh_type_n2b}d--\ref{f:correl_normch3oh_type_an06}d
is larger than for the other three sources from the literature.

To summarize, the class of N-bearing species is the one that reveals the 
largest variance compared to the other classes of molecules: with respect to 
the Sgr~B2(N2) sources, it shows the tightest correlation for G31.41, a poorer 
correlation for IRAS16293B, and no correlation at all for G+0.693 and TMC-1. 
The N-bearing species are underabundant by 1--2 orders of magnitude in G31.41 
and IRAS16293B with respect to N2b, AN02, and AN03. This is also the case, 
albeit in a less pronounced manner, for the S-bearing and O+N-bearing 
molecules. In contrast to the N-bearing species, the class of S-bearing 
molecules has the smallest variance: it shows a high degree of correlation for 
G31.41, IRAS16293B, and G+0.693 with respect to the Sgr~B2(N2) sources (and a 
moderate degree of correlation for TMC-1). This is also the case, albeit to a 
lesser degree, for the class of O+N-bearing molecules. Finally, the O-bearing 
species stand out in G31.41, with a poorer correlation to the Sgr~B2(N2) 
sources. Overall, the O-bearing species are the ones that are the closest to 
the 1:1 relation but this is biased by the fact that we analyzed the abundances 
relative to methanol.

\section{Comparison with chemical models}
\label{s:chemical_models}

To aid in our interpretation of the varied molecular column densities 
determined toward each source, we make use of a pre-existing grid of gas-grain 
chemical models. \citet{Shope24} used the three-phase (gas/surface/bulk-ice) 
chemical kinetics model MAGICKAL to simulate hot-core chemistry under a range 
of physical parameters. Their chemical model and network were identical to 
the ``{\texttt{final}'' model used by \citet{Garrod22}. Their physical 
treatment, consisting of a cold, isothermal collapse (stage~1) followed by a 
dynamically static warm-up phase (stage~2), was the same as that used by 
\cite{Garrod13}, and similar to various related implementations. The 
\citeauthor{Shope24}~grid varied several key parameters within their generic 
model: (i) the cosmic-ray ionization rate (CRIR) used throughout stages~1 and 
2 of the model; (ii) the final gas density, $n_{\mathrm{H}}$, reached during 
stage~1, which carries through to stage~2; (iii) the warm-up timescale, 
$t_{\mathrm{wu}}$, parameterized as the time taken to reach 200~K during 
stage~2; and (iv) the visual extinction, $A_{\mathrm{V}}$, used at the 
beginning of stage~1, which in turn scales the evolving $A_{\mathrm{V}}$ value 
as collapse occurs. Parameters (i)--(iii) were varied logarithmically 
\citep[$n_{\mathrm{H}} = 2 \times 10^6 - 2 \times 10^{10}$~cm$^{-3}$; 
$\zeta = 1.30 \times 10^{-18} - 1.30 \times 10^{-15}$~s$^{-1}$; 
$t_{\mathrm{wu}} = 2 \times 10^4 - 2 \times 10^6$~yr; see Table~1 
of][]{Shope24}. The initial visual extinction was tested with two values
(2~mag and 3~mag). Due to the warm-up timescale only being relevant to 
stage~2, this resulted in a total of 84 stage-1 model runs, and 756 stage-2 
runs, i.e. 9 stage-2 models for each preceding stage-1 run. Each 
stage-1/stage-2 combination is a single-point (0-D) representation of a hot 
core through its chemical evolution from cold, diffuse conditions to a dense, 
hot core that reaches a final temperature of 400~K.

Although the main focus of the \citet{Shope24} models was a comparison with 
NGC~6334I, they also compared their model grid with the EMoCA-derived column 
densities for Sgr~B2(N2) presented by \citet{Jorgensen20} 
\citep[see][]{Belloche16}. This comparison involved finding the grid model 
with the best collective match to the observational column densities, each 
taken as a ratio with the column density of methanol, i.e.
$R_{\textrm{obs},i} = N$(i)/$N$(CH$_3$OH). Since the models produce only 
single-point fractional abundances, the peak value for each molecule (during 
stage~2) was taken in ratio with the peak methanol abundance, i.e.
$R_{\textrm{mod},i} = X_{\mathrm{peak}}$(i)/$X_{\mathrm{peak}}$(CH$_3$OH), to 
represent the column density ratio \citep[see 
\citeauthor{Shope24}~\citeyear{Shope24} and][ for more technical discussions 
of the nuances of such comparisons]{Belloche19}. The logarithm of the 
quotient of the two ratios, i.e.
$\log(R_{\textrm{mod},i}/R_{\textrm{obs},i}) = m_i$, indicates the number of orders 
of magnitude that a modeled species diverges from its observed value. For 
species for which only an upper limit is determined, a modeled value that 
exceeds the observational value is treated in the same way, while a model 
value below the observed upper limit is treated as a perfect match, i.e.
$m_i = 0$. To determine the overall quality of match for a particular model, 
the root mean square of $m_i$ over all species $i$ is calculated, with the 
best matching model having the lowest value. We note that this choice of 
matching parameter tends to favor models in which multiple species show a 
moderately good match, over models in which the match is very good for some 
species and very poor for others. In other words, species for which the model 
results are very divergent from the observations have a strong negative impact 
on the overall matching parameter.

Particularly noteworthy in the analysis of \citet{Shope24} is their 
determination of the best-matching model for Sgr B2(N2) to have a warm-up 
timescale $2 \times 10^6$~yr and a cosmic-ray ionization rate of 
$1.3 \times 10^{-17}$~s$^{-1}$. The characteristic warm-up timescale should be 
considered simply a parameter, which is most 
important in determining how rapidly an already hot core becomes yet hotter; 
however, this timescale value is the longest of those tested in the grid. 
Meanwhile, the best-matching CRIR corresponds to the canonical interstellar 
value, $\zeta_0$, which is rather lower than might be expected for Sgr B2(N2), 
based on other recent studies \citep[e.g.,][]{Bonfand17}. However, the 
determination of this match was based on the column densities of 11 chemical 
species, plus two upper limits, with several species omitted. Our present 
ReMoCA observations provide both a larger selection of molecules with which to 
compare, all of which are included in the model chemical network, as well as a 
sample of cores that the higher angular resolution of ReMoCA allowed us to
distinguish. Collective fitting of this dataset may allow a more 
accurate determination of likely physical conditions in each core.

Because of these differences in the observational dataset, our method diverges 
slightly from that used by \citet{Shope24}. Crucially, we assume that all five 
cores must experience the same CRIR. To impose this constraint, we determine 
firstly the best-matching model for each core within each fixed CRIR value; 
this produces a total of five match parameters for each CRIR value. We then 
take the root mean square of these five, so that each CRIR has an associated 
matching parameter that identifies the best collective value for the entire 
dataset (these values are shown in Fig.~\ref{f:model_zeta}). We consider the 
best model for each source to be the one with the lowest matching parameter 
within the subset for which the CRIR has been collectively determined as the 
best match.

Some of the detected species are not included in our match calculations, as 
they are not present in the chemical network used in the model grid. However, 
following \citet{Shope24}, we also elect not to include certain species in 
the calculations even though they are present in the models. This includes 
species such as acetone (CH$_3$C(O)CH$_3$) and propanal (C$_2$H$_5$CHO), for 
which the network is sparse 
\citep[although it has since been improved by][]{Belloche22}. Others, like 
CH$_3$NCO, have an uncertain chemistry that is not well reproduced by the 
models. Some omitted species, such as HC$_3$N and HC$_5$N, are not uniquely 
associated with the hot stage of the core, and therefore could be unduly 
influenced by chemistry in the lower density gas (either in the observations 
or during the early stages of the models). We also exclude the sulfur-bearing 
species OCS, H$_2$CS, and CH$_3$SH, as the dominant form of sulfur in dense 
clouds is uncertain and the chemistry is thus poorly defined. We retain SO and 
SO$_2$ in the fit, as they are mainly produced in the gas phase in hot cores, 
following ice desorption, and they have been well reproduced by chemical 
models for a long time, given an appropriate initial sulfur abundance (here, 
$X_{\mathrm{init}}(S) = 8 \times 10^{-8}$). We note that the initial elemental 
abundances used in the models were not varied in the grid; \citet{Shope24} 
used the values reported in Table~1 of \citet{Garrod13}. 

There is a total of 37 species that could in principle be included in the 
matching routine. Twelve are omitted (as described above), leaving 25, 
of which between 0 and 8 are observational upper limits.
In contrast to \citet{Shope24}, for species bearing a nitrile (-CN) group the 
matching parameter is based on a ratio with CH$_3$CN instead of CH$_3$OH, in 
keeping with the comparison methods of \citet{Jorgensen20}. For methyl 
cyanide (CH$_3$CN) itself, the ratio with methanol is retained.

Using this updated matching parameter method, we compare the observational 
data with the model grid in two ways. Firstly, we determine the best matching 
models based on an unrestricted comparison with all models; secondly, we 
restrict the comparison to a narrower range of gas densities that are 
reflective of observational values or estimates for each source (see 
Table~\ref{t:restricted}). Values for AN02, AN03, and AN06 are based on 
determinations of $n$(H$_2$) by \cite{SanchezMonge17}. The ReMoCA 
continuum emission is a factor of $\sim$3 fainter toward N2b than toward AN02 
(see Fig.~\ref{f:map_3mmcont_n2}). Therefore we extend the restricted density 
range of this source to lower values. As we have seen in Sect.~\ref{sss:an06}, 
two velocity components overlap along the line of sight to AN06. The secondary 
component, AN06c2, traces the edge of the dense core that contains AN02, AN03, 
and N2b, while the main component, AN06, is likely in the foreground or 
background. The continuum emission analyzed by \citet{SanchezMonge17} contains 
the contribution of both components, yet we think that it is dominated by AN06.
This motivates our decision to restrict the density range explored for AN06c2
to lower values.

\begin{table}
    \caption[]{Range of final (i.e.~stage-2) gas densities from the model grid
 used in the comparison with each source, in the density-restricted setup.}
        \label{t:restricted}
        \centering
        \begin{tabular}{ll}
        \hline
        \noalign{\smallskip}
        Source &  $n_{\mathrm{H,final}}$ (cm$^{-3}$) \\
        \noalign{\smallskip}
        \hline
        \noalign{\smallskip}
        N2b    &  $2 \times 10^7 - 2 \times 10^9$ \\
        AN02   &  $2 \times 10^8 - 2 \times 10^9$ \\
        AN03   &  $2 \times 10^8 - 2 \times 10^9$ \\
        AN06c2 &  $2 \times 10^6 - 2 \times 10^8$ \\
        AN06   &  $2 \times 10^7 - 2 \times 10^9$ \\
        \noalign{\smallskip}
        \hline
        \end{tabular}
\end{table}

\begin{figure}[!t]
\resizebox{0.93\hsize}{!}{\includegraphics[angle=0]{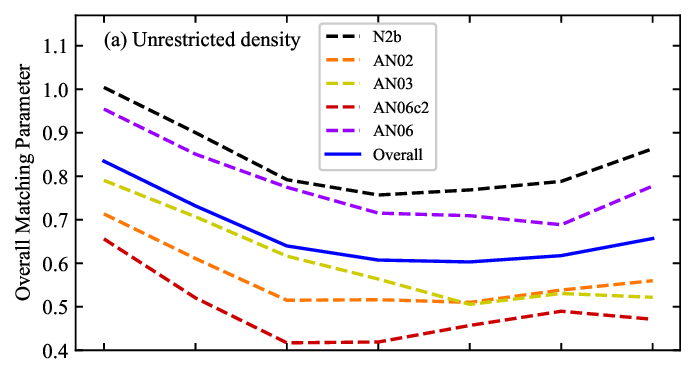}}\\
\resizebox{\hsize}{!}{\includegraphics[angle=0]{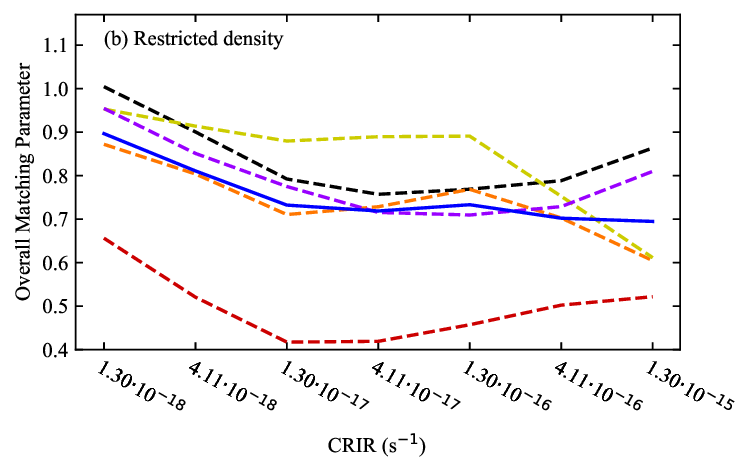}}
\vspace*{-2ex}
\caption{Overall matching parameter obtained for the comparison of the 
grid of chemical model results with observational abundances of all species 
across all sources, assuming a uniform CRIR. A lower matching parameter 
indicates a closer match between models and observations. Panel (a): 
unrestricted density. Panel (b): density restricted to the ranges 
indicated in Table~\ref{t:restricted}.}
\label{f:model_zeta}
\end{figure}

Figure~\ref{f:model_zeta} indicates the collective matching parameter over all 
five sources, along with the individual subvalues for each source, at each 
CRIR value; panel (a) shows the results for the unrestricted setup, while 
panel (b) shows those of the density-restricted comparison. 
In both setups, CRIR values below the canonical interstellar value, $\zeta_0$, 
are disfavored. In the unrestricted setup, the best overall match occurs with 
$\zeta = 10 \zeta_0$. In the density-restricted setup, the best match is 
obtained at the maximum CRIR value tested, 
$\zeta = 100 \zeta_0 = 1.3 \times 10^{-15}$~s$^{-1}$.

It is notable that the preference for the most extreme CRIR value is strongly 
associated with the variation in the AN02 and AN03 matching parameters; 
indeed, the overall matching parameter is worse for all CRIR values in the 
restricted setup as a result of this dependence. In the unrestricted case, the 
best-matching models for both of these sources (with $\zeta = 10 \zeta_0$) 
take \hbox{stage-2} gas densities of $n_{\mathrm{H}} = 2 \times 10^{6}$~cm$^{-3}$, 
which are much lower than the observational estimates. Values of 
$2 \times 10^{8}$~cm$^{-3}$ (with $\zeta = 100 \zeta_0$) provide the best match 
in the restricted setup. Thus, at these higher densities for AN02 and 
AN03, a further elevated CRIR is required for the best match with observations, even though the other sources are provided a somewhat worse match under those 
conditions. 

Our analysis continues with the exclusive use of the density-restricted 
comparisons. Figure~\ref{f:model_match} plots the 
matching parameter of each model for each observed source, ordered in bands 
corresponding to a fixed CRIR value. The bands are ordered from the best to 
the worst overall matching parameter. Within each band, and for each 
individual source, the matching parameter is plotted from best to worst. We 
note that there is no correspondence between model numbers among curves in the 
same band, as the results for each source-model combination are individually 
sorted. Furthermore, due to the density restrictions shown in 
Table~\ref{t:restricted}, different sources have different numbers of grid 
models available to them. The plots indicate simply the degree of variation in 
matching parameter between different models for each source. Source AN06c2 
universally shows the best matching parameter, comparing the curves at fixed 
sorted-model values. Matching the other sources is generally harder, 
i.e.~there are fewer models that reproduce them well. This may be related to 
the fact that AN06c2 has the most upper limits, making a match easier. 
However, AN06 has the same number of upper limits, while being a worse match 
in all cases, when comparing best with best. The matching parameters for AN02 
and AN03 generally appear to become the most rapidly divergent with increasing 
sorted-model value, i.e.~they have a narrow range of physical conditions that 
produce a ``good'' match with observations.

\begin{figure*}[!t]
\resizebox{0.52\hsize}{!}{\includegraphics[angle=0]{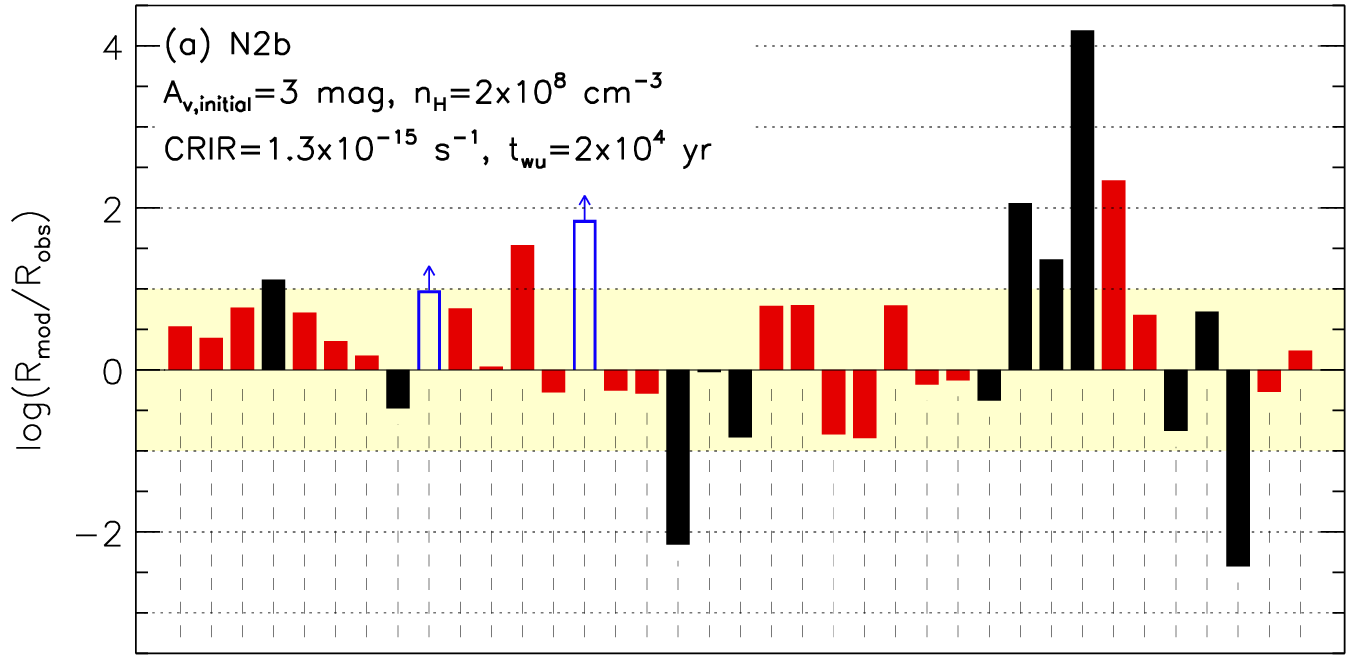}}\\[0.4ex]
\resizebox{0.52\hsize}{!}{\includegraphics[angle=0]{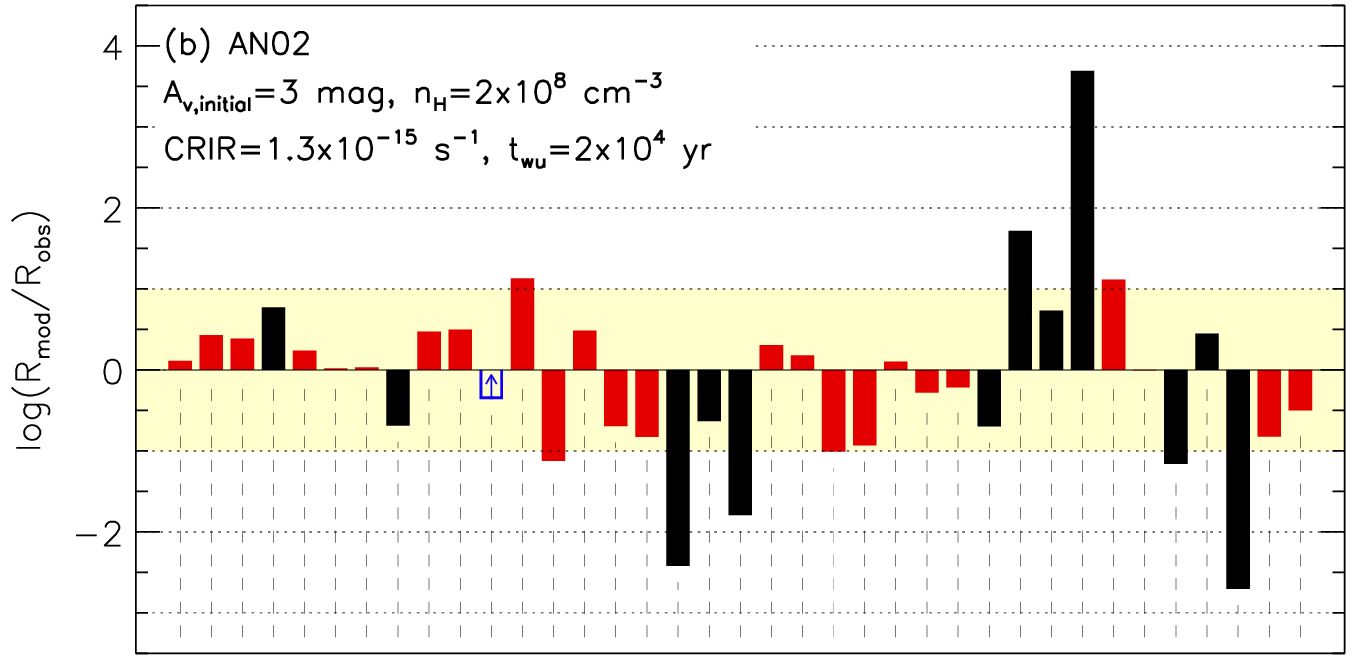}}\resizebox{0.472\hsize}{!}{\includegraphics[angle=0]{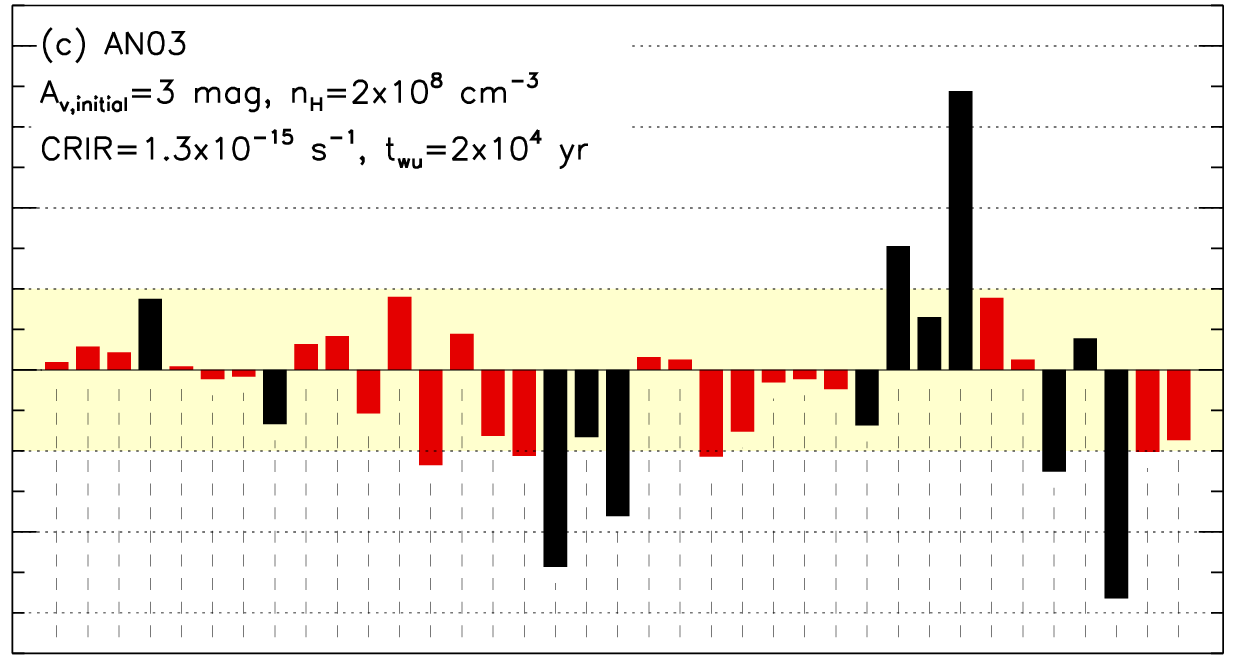}}\\[0.4ex]
\resizebox{0.52\hsize}{!}{\includegraphics[angle=0]{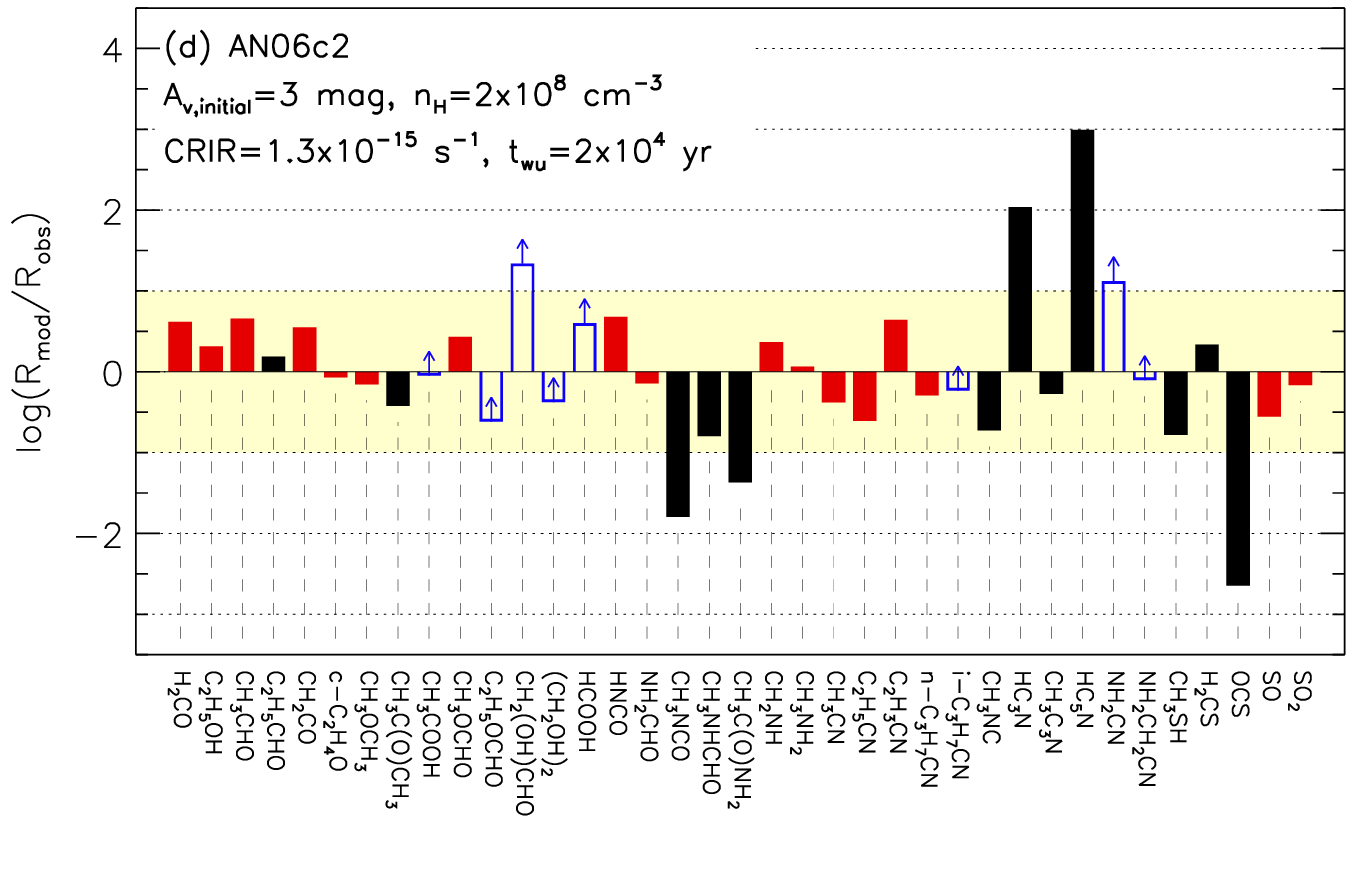}}\resizebox{0.472\hsize}{!}{\includegraphics[angle=0]{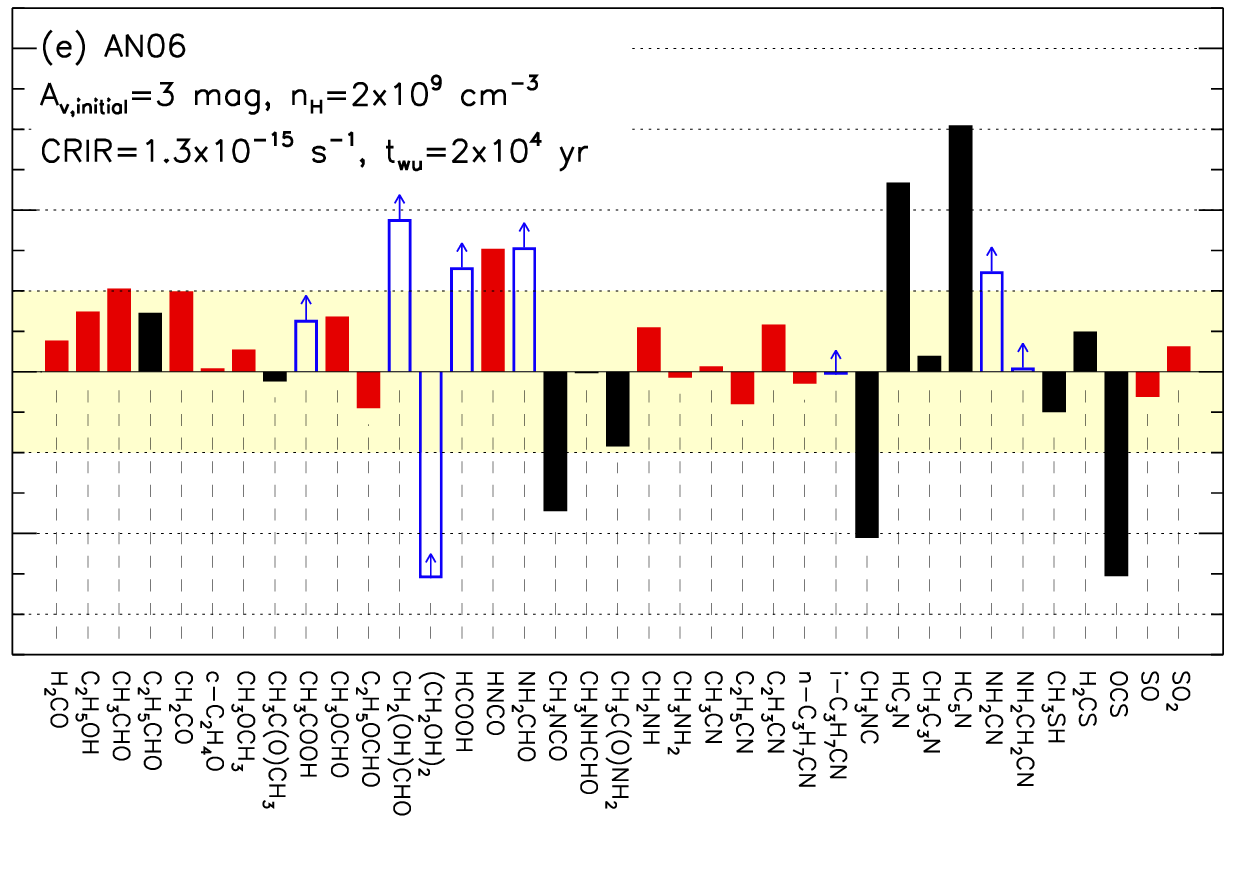}}
\vspace*{-3ex}
\caption{Comparison of best-fit model abundances to observational results, 
using the density-restricted setup. 
Bars indicate the number of orders of magnitude by which each abundance ratio 
in the best-matching model exceeds (or otherwise) the observational value. 
Data correspond to the best-matching model for each individual source, within 
the subset of grid models with $\zeta=100 \zeta_0$, which is found to produce 
the best overall match across all sources. The shaded area represents values 
where the models and observations vary by 1 oom or less. 
Unfilled blue bars with an arrow indicate that the comparison is based on an 
observational upper limit. Black bars indicate observed species that were not 
included in the matching parameter analysis.}
\label{f:model_bars}
\end{figure*}

Figure~\ref{f:model_bars} shows the comparison between models and observations 
for each source; the best-matching model is shown in each case, for the 
best-matching CRIR of $\zeta = 1.3 \times 10^{-15}$~s$^{-1}$. Unfilled blue 
bars with an arrow indicate observational upper limits, while black bars 
indicate species that are omitted from the matching procedure, but which 
nevertheless exist in the models.
With the CRIR value fixed, the key remaining parameters distinguishing the 
best-match models for each source are the gas density and the warm-up 
timescale. Particularly notable is the fact that, with the exception of AN06, 
the best match for each source is provided by the same model, taking values 
$\zeta=100 \zeta_0$, 
$n_{\mathrm{H,final}} = 2 \times 10^8$~cm$^{-3}$, 
$t_{\mathrm{wu}} = 2 \times 10^4$~yr, and 
$A_{\mathrm{V,init}} = 3$~mag.
The best-matching model for AN06 has a factor 10 higher gas density, 
but is 
otherwise the same. We note that the best warm-up timescale for all sources is 
the shortest of those tested, and $2.5\times$ shorter than the ``fast'' 
warm-up timescale used in various past models by \citet{Garrod13} and others.

Many of the species omitted from the matching procedure are, unsurprisingly, 
not well reproduced. But in general, most common hot-core species included in 
the matching routine are reasonably well reproduced by the models, i.e.~are 
within one order of magnitude (oom) of the observed values (as marked by the 
shaded region). For example, the structural isomers methyl formate and 
glycolaldehyde are both generally somewhat overproduced, although they are 
mostly within 1~oom of the observed values, and they also tend to scale with 
each other among the plots shown. However, due to the fact that four of the 
sources share a single best-matching model, any apparent variations in the 
quality of match between that (or any) pair of species among those four 
sources is solely due to variations in the observed values. In the two 
best-match models representative of either AN06 or collectively of AN02, AN03, 
AN06c2, and N2b, glycolaldehyde is overabundant compared with methyl formate. 

H$_2$CO, C$_2$H$_5$OH, CH$_3$CHO, and CH$_2$CO are consistently, if only 
modestly, overproduced for all five source-model matches. Species that appear 
consistently underproduced include CH$_3$CN and C$_2$H$_5$CN, although this is 
not the case for the AN06 model. HNCO and NH$_2$CHO are in some cases 
underproduced and in some cases overproduced, although they are mostly 
discrepant in the same sense; again, this apparent variation is essentially 
caused by matching a single model to a selection of sources that behave 
somewhat differently.

Figures.~\ref{f:chemgroup_protonated}--\ref{f:isomers} show not only the 
observational ratios but the equivalent values from the best-matching chemical 
models for each source. Chemical model results are indicated by crosses. 
Although in our matching-parameter analysis we have only used molecules that 
are detected in at least one of our source datasets, it is valuable 
to compare the best-matching models with the observations for a much broader 
range of species, including those that have been omitted from the
matching parameter.

The model abundances for protonated molecules with respect to their 
unprotonated forms (Fig.~\ref{f:chemgroup_protonated}) are comfortably below 
all of the observational upper limits. Ion abundances in the models (during 
the hot stage, and under otherwise fixed CRIR conditions) tend to scale 
inversely with gas density, which influences collisional lifetimes. Within the 
range of densities tested in the restricted setup, the models would be 
unlikely to reach the observed upper limits. The model-source comparisons are 
therefore not a strong constraint on ion abundances.

Gas-phase destruction rates for radicals also tend to scale with gas density. 
The modeled values for radicals HCO, CH$_3$O, and CH$_2$OH with respect to 
H$_2$CO and CH$_3$OH (shown in Fig.~\ref{f:chemgroup_radical}}) are 
substantially below the observational upper limits. The other four radicals 
plotted (CH$_2$CHO, CH$_2$CN, C$_3$N, and NH$_2$CO) are in some cases 
far above the upper limits. This may be caused by 
a lack of appropriate destruction mechanisms for these radicals; in 
particular, reactions with (abundant) atomic hydrogen could lead to their 
rapid destruction, as is the case for HCO, CH$_3$O, and CH$_2$OH 
\citep[see, e.g.,][]{Tsang86}, leading to the production of H$_2$ and another 
stable species. Other disproportionation and radical-pair production reactions
are also possible, where energetically 
favorable. In the models, reactions with atomic H are frequently the most 
important destruction mechanisms for radicals that have them; more important 
even than sequential protonation and recombination, or CR-driven 
photodissociation. Atomic-H reactions for CH$_2$CHO, CH$_2$CN, NH$_2$CO, and 
C$_3$N are not currently present in the network (although see Bonfand et 
al.,~in prep.). In the latter case, however, such a pathway is unlikely, due 
to the absence of energetically favorable products. As such, the fact that the 
modeled C$_3$N/HC$_3$N ratio exceeds the observational upper limit for sources 
N2b, AN02, and AN03 indicates that a model with higher gas density could 
remedy this. However, the fact that HC$_3$N was not included in the matching 
parameter could also affect the ratio.

Similarly, the ratios between low- and high-saturation species, 
C$_2$H$_5$CN/HC$_3$CN and the equivalent larger 
nitriles (see Fig.~\ref{f:chemgroup_unsaturated}) may also be affected
by the difficulties of including those species in the matching 
parameters. The abundances of C$_2$H$_3$CN and C$_2$H$_3$OH are somewhat 
dependent on the branching ratio of the recombination of protonated ethyl 
cyanide and protonated ethanol, respectively, which could explain their 
overabundances for most model-source combinations. In 
Fig.\ref{f:chemgroup_hydrolyzed}, \hbox{-CHO} and \hbox{-COOH} group-bearing species 
are somewhat overproduced relative to -CN, although relative to each other they 
are much closer to the observations. The NH$_2$CHO/NH$_2$CN ratio is higher 
than expected.

Modeled ratios of -CHO versus -CH$_3$ group-bearing species shown in 
Fig.~\ref{f:chemgroup_reduced} are generally not far off the observations,
although the CH$_2$(OH)CHO/C$_2$H$_5$OH ratio is elevated in the case of N2b. 
As described by \citet{Shope24}, this may indicate that a higher gas density 
model is required.

As may be seen in Fig.~\ref{f:chemgroup_homolog}, modeled ratios between the 
different nitrile homologs (C$_n$H$_{2n+1}$CN, $n=1-4$) show a somewhat varied 
match with observations, with C$_2$H$_5$CN somewhat underproduced in all cases, 
while normal butyl cyanide is generally overproduced. The general 
underproduction of 
C$_2$H$_5$CN may be related to an overconversion to the larger cyanides in the 
ice mantles. For the CH$_3$OCHO series, in cases where C$_2$H$_5$OCHO is 
detected, the ethyl to methyl formate abundance appears low in the models. As 
noted by \cite{Belloche09}, production of ethyl formate in the models is
dominated by a methyl-radical addition to another radical whose precursor is 
methyl formate. In the present model grid, this occurs largely within the bulk 
ice, driven by CR-induced UV photoproduction of radicals. A number of 
different factors determining the availability of the CH$_3$ radical (such as 
CH$_4$ abundance and photodissociation efficiency) could be at play in 
determining the eventual modeled abundance of ethyl formate. Again, four of 
the sources are being matched with the same single model, so when they vary 
observationally, disagreements should become apparent. A higher resolution 
grid, with differentiated source-model matches for all sources, might lead to 
better individual matches for the molecules mentioned here.

Among the structural isomers shown in Fig.~\ref{f:isomers}, CH$_3$CN/CH$_3$NC 
is not a good match for source AN06, although it is not bad for the other 
sources. In the present chemical network, which includes reactions used by 
\cite{Willis20}, the abundance of CH$_3$NC is particularly sensitive to the 
reaction H + CH$_3$NC $\rightarrow$ HCN + CH$_3$, whose activation-energy 
barrier is highly uncertain. At the higher density experienced in the AN06 
best-match model, this reaction would have a more pronounced effect. CH$_3$NC 
was already omitted from the matching parameter due to the uncertainties in 
its chemistry. The poor match for AN06 could indicate that the barrier is too 
low in the network.

Figure~\ref{f:isomers} also shows the ratios of C$_2$H$_4$O$_2$ species. 
CH$_3$OCHO (methyl formate, MF) and CH$_2$(OH)CHO (glycolaldehyde, GA) 
generally show roughly appropriate ratios with each other, although they are 
both very moderately overproduced in comparison with CH$_3$COOH (acetic acid, 
AA) for sources where it is detected (AN02 and AN03). The production of acetic 
acid in the models occurs through somewhat different chemical pathways that 
involve acetaldehyde, CH$_3$CHO \citep[see][]{Garrod08,Garrod22}. In the model 
representing N2b, AN02, AN03, and AN06c2, the MF/GA/AA ratio is 18/5.8/1.
In the model representing AN06, a similar ratio 17/4.0/1 is achieved. These 
may be compared with the observed ratios of 29/1.6/$<$1 (N2b), 12/1.8/1 
(AN02), and 14/1.5/1 (AN03). The modeled abundance of glycolaldehyde in 
particular, especially in comparison with source N2b, could be considered 
notably in excess of the observational value, when considering MF:GA ratios. 
\citet{Shope24} considered the abundance of glycolaldehyde in particular and 
the possible explanations for extremely low ratios with methyl formate toward 
some sources. Those authors noted that a combination of high gas density and 
long timescale in the models was capable of reducing GA abundances during the 
warm-up period. Chemically, this occurred through the brief adsorption of 
gas-phase atomic H onto the grain surfaces, followed by reaction with a 
glycolaldehyde molecule that had recently become exposed on the surface but 
had not yet desorbed. This process was found to be effective under conditions 
where H adsorption would remain rapid for a long period of time.

The best-match models for the five sources studied in the present work achieve 
fairly high gas densities, but do not experience long enough periods for the 
effect to take hold strongly. In a less constrained physical model, or a more 
precise treatment of the warm-up process, GA destruction might be more 
efficient. As noted by \citeauthor{Shope24}, the key period for GA destruction 
is not the entire warm-up period from $\sim$10--400~K, but the much shorter 
period between the start and end of major water-ice desorption. A somewhat 
longer period in this more limited temperature range would still be consistent 
with hot-core lifetimes on the order of a few 10$^4$~yr. More dynamically 
accurate simulations of hot-core chemical evolution are necessary to test 
these possibilities.

\section{Discussion}
\label{s:discussion}

\subsection{Chemical evolution within Sgr~B2(N2)}
\label{ss:discussion_sgrb2n2}

The ReMoCA survey has allowed us to derive the detailed chemical composition of
four hot cores embedded in Sgr~B2(N2) and a position, AN06c2, located at the 
edge of the dense core (Sect.~\ref{ss:chemcomp}). As demonstrated in 
Sect.~\ref{ss:correl_pos}, there is a tight correlation between the chemical 
compositions of AN02 and AN03, despite the fact that AN03 contains a 
HC\ion{H}{ii} region while AN02 does not (Sects.~\ref{sss:an03} and 
\ref{sss:an02}). While we could worry that the small angular separation 
between AN02 and AN03 ($\sim$0.64$\arcsec$) may have resulted in a mutual 
contamination of their observationally derived chemical compositions, we can 
exclude this interpretation because the correlation between N2b and AN02 is 
looser (Fig.~\ref{f:correl_normch3oh}a) despite their even smaller separation 
($\sim$0.55$\arcsec$). The correlation between AN02 and AN03 is thus genuine. 
The presence of the HC\ion{H}{ii} region indicates that AN03 is more 
evolved than AN02 or more massive (hence more luminous). The similarity of the 
masses derived from dust emission by \citet{SanchezMonge17} for AN02 and AN03 
tends to discard the latter interpretation. According to the time sequence 
derived by \citet{Nony24} for the high-mass star forming protocluster W49N, 
the lifetime of hot cores is $6 \times 10^4$~yr, including the phase when a 
hypercompact or ultracompact (UC) \ion{H}{ii} region coexists with the hot 
core emission. W49N contains a bit less than half as many hot cores with 
H/UC\ion{H}{ii} regions as hot cores without, which suggests a statistical
lifetime of $\sim$$2 \times 10^4$~yr for the former. Hot core lifetimes are 
similar in Sgr~B2(N) and W49N \citep[][]{Bonfand17} and so we conclude that 
the chemistry has not evolved significantly during a timescale of
$\sim$$2 \times 10^4$~yr around AN03 on a scale of 0.7$\arcsec$ 
($\sim$5700~au). With a size of $\sim$0.08$\arcsec$, the HC\ion{H}{ii} region 
is one order of magnitude smaller than the molecular emission probed with 
ReMoCA toward AN03 ($\sim$0.7$\arcsec$--1$\arcsec$). This is likely the reason 
why it has not yet had time to alter the chemistry of the hot core.
As a matter of fact, there is no significant difference between AN02 
and AN03 in terms of rotational temperature of their molecular emission
(see Fig.~\ref{f:correl_temp}e).

The hot core N2b is most likely in an earlier evolutionary stage than AN02 and 
AN03 because, unlike these sources, it is neither associated with a 
HC\ion{H}{ii} region nor with masers. The rotational temperatures of its 
molecular emission are on average a bit lower than the temperatures of AN02 
and AN03 by about 20~K (Fig.~\ref{f:correl_temp}a--b), which may support an
earlier evolutionary stage. As reported in Sect.~\ref{sss:n2b}, N2b is 
associated with a compact dust core and its elevated rotational temperatures of 
$\sim$150~K suggest that it has already formed a protostar, unless its 
temperature structure is imposed by its neighbor AN02. If N2b contains a 
nascent protostar, then it is at most a few $10^4$~yr younger than AN02. 
Therefore, it is remarkable that its chemical composition shows significant 
differences compared to that of AN02 and AN03. Overall, N2b's molecular 
abundances 
relative to methanol are a factor $\sim$2--4 lower than those of AN02 and AN03, 
and their dispersion is larger than in the correlation plot of AN02 versus 
AN03 (compare panels a--b and e of Fig.~\ref{f:correl_normch3oh}). Two
organic molecules stand out with abundances that are even lower by more than 
one order of magnitude in N2b: NH$_2$CN and t-HCOOH. It is tempting to 
conclude that the chemistry evolves significantly during the initial phase (a 
few 10$^4$~yr) of thermal desorption at the hot core stage, which may suggest 
that gas-phase processes play an important role in reshaping the chemical
composition after desorption of the molecules from the grains. In this context,
it would be interesting to understand why NH$_2$CN and \hbox{t-HCOOH} in 
particular appear to be so sensitive to this phase.

We proposed in Sect.~\ref{sss:an06} that AN06 is a hot core that is not 
embedded in the dense region that contains AN02, AN03, and N2b. Like N2b, it is
not associated to a HC\ion{H}{ii} region or any masers. AN06 may thus
be in an early evolutionary stage, maybe even earlier than N2b given its 
slightly lower rotational temperatures (see Fig.~\ref{f:correl_temp}d). Its
chemical composition correlates with that of N2b, albeit with a much larger 
dispersion than AN02 versus AN03 (compare panels d and e of
Fig.~\ref{f:correl_normch3oh}). If AN06 and N2b are at a similar stage of 
evolution, then the differences in their chemical compositions suggest that 
they reside in different environments. Alternatively, AN06 may be less massive
than the other hot cores of Sgr~B2(N2) and the chemical differences may result
from a difference in evolutionary stage. \citet{SanchezMonge17} derived a 
mass of AN06 that is 30\% smaller than the mass of AN03. However, their dust
continuum observations cannot disentangle the two velocity components that 
overlap along the line of sight. The mass calculated for AN06 may thus be 
largely overestimated. The most striking chemical difference between AN06 and 
N2b is that HNCO and NH$_2$CHO are nearly two orders of magnitude less 
abundant in the former. Understanding the cause of this huge deviation from 
the overall correlation may tell us what controls the larger dispersion of the 
correlation between AN06 and N2b. Finally, while the chemical compositions of 
AN06 and AN06c2 correlate well apart from an overall shift by a factor $\sim$2 
(Fig.~\ref{f:correl_normch3oh}h), the abundance of formamide stands out by 
nearly two orders of magnitude. In this respect, the composition of AN06c2 is
much more similar to that of the dense region that contains AN02, AN03, and 
N2b. This supports the idea that AN06c2 traces the edge of the region and AN06 
overlaps along the line of sight but is physically unrelated to this 
region.

\subsection{Tight correlation between Sgr~B2(N2) and G31.41}
\label{ss:discussion_g31p41}

Located in the Scutum spiral arm at a distance of 3.75~kpc \citep[][]{Immer19},
the hot core G31.41 is a small protocluster of luminosity 
$\sim$$4.5 \times 10^4$~$L_\odot$ that consists of four high-mass young 
stellar objects (YSOs) with gas masses of 15--26~$M_\odot$ within a radius of 
400--500~au \citep[][]{Beltran21}. Each of them drives an outflow 
\citep[][]{Beltran22} but none of them harbors a 
HC\ion{H}{ii} region \citep[][]{Cesaroni10,Beltran21}. AN03 and AN02 have
gas masses of about 500~$M_\odot$ within a radius of about $2800$~au 
\citep[][]{SanchezMonge17}. Assuming power-law density profiles with a 
power-law index in the range from $-1.5$ to $-2$, then the mass enclosed in
a radius of 500~au is 40--90~$M_\odot$, implying that AN03 and AN02 may form 
higher-mass stars than G31.41's YSOs by a factor 2--3, or even higher given
that AN03 already contains a HC\ion{H}{ii} region. Among the four sources 
from the literature investigated in Sect.~\ref{ss:comp_lit}, G31.41 is the one
with the tightest correlation of its overall chemical composition with that of 
N2b, AN02, and AN03. The overall chemical composition of hot cores thus seems 
to be relatively insensitive to the environmental features that distinguish the 
Galactic center region from the Galactic disk (e.g., level of turbulence, 
CRIR, gas temperature). The O-bearing species show a poorer correlation 
between G31.41 and N2b, AN02, and AN03 than the N-bearing ones. This suggests 
that the former may be more sensitive to the environmental conditions than the 
latter.

\subsection{Chemical segregation between classes of molecules}
\label{ss:discussion_classes}

Chemical segregation between O-bearing and N-bearing molecules has been 
reported in various star forming regions since the 1990s 
\citep[see, e.g., \citeauthor{Qin22}~\citeyear{Qin22} and short reviews on 
this topic in Sect.~4.3 of \citeauthor{Jorgensen20}~\citeyear{Jorgensen20} and 
Sect.~4.7 of][]{Busch24}. Among the four classes of molecules investigated in 
Sect.~\ref{s:results} (\hbox{O-, N-}, N+O-, and S-bearing), the N-bearing class 
stands out as the one with the largest variance across the sample of sources 
investigated in Sect.~\ref{ss:comp_lit}. For 
this class of molecules, the composition of Sgr~B2(N2)'s hot cores correlates 
well with that of the other two hot cores/corinos G31.41 and IRAS16293B but 
does not correlate at all with that of G+0.693 and TMC-1. This is in stark 
contrast to the class of S-bearing molecules which always shows a correlation 
with Sgr~B2(N2)'s hot cores, strong in the case of G31.41, IRAS16293B, and 
G+0.693, and moderate in the case of TMC-1. The O-bearing molecules lie in 
between, with a strong correlation in the case of G31.41, IRAS16293B, and 
G+0.693, but a poor one in the case of TMC-1.

TMC-1 is a cold ($\sim$10~K) starless core of moderate density 
\citep[$3-8 \times 10^4$~cm$^{-3}$, see, e.g.,][]{Pratap97,Lique06} located in 
the low-mass star forming molecular cloud Taurus. Part of its molecular 
composition may be controlled by gas-phase chemistry 
\citep[e.g.,][]{Cernicharo21}, but pure gas-phase chemical models have 
difficulties in reproducing the abundances of certain molecules like 
acetaldehyde \citep[e.g.,][]{Agundez25}, suggesting that non-thermal 
desorption of molecules formed on the grains, triggered by, e.g., cosmic rays 
or the release of chemical energy, likely contributes to the composition of the
gas phase as well \citep[see, e.g., the case of CH$_3$OH, C$_2$H$_5$OH, and 
C$_2$H$_5$CHO in][]{Agundez23}. 

G+0.693 is also a starless region of moderate density 
\citep[$10^4$--$10^5$~cm$^{-3}$,][]{Zeng20,Colzi24} and low dust temperature 
\citep[$\sim$20~K,][]{Etxaluze13}, but it is located in the Sgr~B2 molecular 
cloud complex, about 54$\arcsec$ away from Sgr~B2(N2), which means that they 
both have been subject to similar levels of turbulence, cosmic rays, and 
radiation (UV photons, X-rays) during their evolution. The chemistry of 
G+0.693 is thought to be 
dominated by low-velocity shocks that eject molecules from the grain mantles 
through sputtering \citep[e.g.,][]{RequenaTorres06,Zeng18}. If we consider 
that most molecules are formed on the grains and that the molecular 
composition of Sgr~B2(N2)'s hot cores was in large part inherited from the 
composition of the grain mantles formed during the prestellar phase, then 
differences in the compositions of G+0.693 and Sgr~B2(N2)'s hot cores could
result from the shocks that affect G+0.693. Alternatively, the
composition of the hot cores may have been altered by processes on the grains
during the warm-up phase. However, it is also likely that the large difference 
in gas density between G+0.693 and Sgr~B2(N2)'s hot cores has a major effect 
on the post-desorption 
behavior of gas-phase molecules. Lower gas densities in G+0.693 should result 
in both a slower rate of destruction and a different balance of destructive 
ions, depending on how complete is the loss of ice due to sputtering during 
the passage of a shock (Willis et al., in prep.). The amount of ammonia in the 
gas can assist in the destruction of some molecules (especially amine 
group-bearing species) while enhancing the survival of others 
\citep{Taquet16,Garrod23}.

The CRIR in the Galactic center region is higher than in the Galactic disk by 
1--2 orders of magnitude \citep[e.g.,][]{Indriolo15,LePetit16}. The fact that 
N-bearing species of Sgr~B2(N2)'s hot cores correlate well with hot 
cores/corinos in the Galactic disk (G31.41 and IRAS16293B) but neither with 
G+0.693 nor with TMC-1 implies that the CRIR is not the dominant factor 
producing the large variance of the N-bearing class. We conclude that the 
class of N-bearing molecules reacts more sensitively to shocks (G+0.693), 
low-temperature gas-phase chemistry subsequent to non-thermal desorption 
(TMC-1), or density (G+0.693 and TMC-1) than O-bearing and S-bearing 
species. This is in line with the findings
of \citet{Busch24} who reported an abundance enhancement of cyanopolyynes
relative to methanol in post-shock gas in the outflow of Sgr~B2(N)'s main hot 
core. In contrast, the strong correlation of the O-bearing content of 
Sgr~B2(N2)'s hot cores with G+0.693 and the poor one with TMC-1 suggest that, 
in comparison to the processes that control the O-bearing content of hot 
cores, shocks do not greatly modify the relative abundances of the class of 
O-bearing species while low-temperature gas-phase chemistry does to some 
extent.

We showed in Sect.~\ref{ss:comp_lit} that there is a tight correlation between
the chemical compositions of the low-mass protostar IRAS16293B and that of 
N2b, AN02, and AN03 after normalizing the column densities by class of 
molecules, but the classes are globally shifted in abundance with respect to 
each other: N-bearing molecules are underabundant by about 2~oom in 
IRAS16293B, O+N-bearing and S-bearing species by about 1 oom. Similar shifts 
between classes of molecules also exist in the case of G31.41, but they are 
less pronounced: 1--1.5~oom for the N-bearing and $<$1~oom for the 
O+N-bearing and S-bearing species. G31.41 seems to be forming stars of lower 
mass than Sgr~B2(N2)'s sources (see Sect.~\ref{ss:discussion_g31p41}), but the 
difference may only be a factor of a few while there is a much larger gap in 
mass between G31.41 and IRAS16293B ($>$1~oom). Therefore it seems unlikely 
that the abundance shifts between classes of molecules result primarily from 
physical differences between regions forming low-mass and high-mass stars, 
such as the temperature during the prestellar phase, the duration of the 
prestellar phase, or the level of UV radiation during the warm-up phase. 

Shocks may play a role in hot cores and corinos, via, e.g.,
accretion shocks of infalling material onto a circumstellar disk. This could 
in turn lead to somewhat different desorption behavior, including only 
partial desorption of ices, given a sufficiently weak shock. This could lead 
to only the outer layers being desorbed. These outer layers are often 
considered to be relatively water-poor, and have been
associated with observed apolar spectral signatures of CO and CO$_2$
\citep[][]{Garrod11}, while being richer in CO 
and its products such as methanol. It is likely that these layers would also 
be rich in O-bearing COMs (which are also ultimately related to CO), if they 
exist in the ice. On the other hand, the deeper ice layers are shown by models 
to have a much greater NH$_3$ fraction. One might expect that amine-group 
bearing molecules would also be more abundant in those deeper ice layers, 
while their binding energies with water also tend to be greater than many of 
the CO-related species. Thus if only weak accretion shocks are active, only 
the weakly bound upper layers of O-bearing molecule-rich ice might be released 
into the gas. In hot cores, the accretion shocks may be stronger (leading to 
more complete ice desorption, either by sputtering or thermal desorption), and 
the heating of the grains by the protostar is likely to lead to a more global, 
sustained, and vigorous heating of the grains than in a hot corino. Hot cores 
may therefore undergo a more complete desorption of the ices than hot corinos, 
in general, leading to greater N-bearing species being released.

Other mechanisms may also lead to sporadic heating of the ices, such as 
accretion outbursts. Low-mass protostars may flash-heat their surroundings 
multiple times \citep[e.g.][]{Vorobyov15}. This could again lead to the 
desorption of only the outer and more weakly bound ice layers, which would be 
O-rich and N-poor. It is also suggested, based on simulations \citep{Meyer21}, 
that outbursts are also common for high-mass protostars.

In principle, any of these processes that may lead to only partial ice 
desorption might well encourage the preferential release of O-rich material, 
while leaving deeper, N-rich material in place. Differences in molecular O/N 
ratios between sources would then be an indicator of how violently and 
completely the ice mantles are desorbed.

Unfortunately, while current chemical models can operationalize 
differences in binding energies between species, hot core models do not 
currently distinguish between ice layers beyond the typical bulk+surface-layer 
paradigm, and they are also not yet capable of taking local ice surface 
composition into account in the determination of binding energies. Testing the 
above idea will require substantial upgrades in model capabilities.

\subsection{Chemical model behavior}
\label{ss:model_discussion}

The chemical model grid is capable of reproducing the observed abundances to a 
reasonable degree of accuracy (within 1 oom) in most cases. However, the 
rather coarse resolution of the grid means that we are unable to determine 
clearly the possible variations in physical conditions that would lead to the 
observed chemical distinctions between the five sources. The method imposes 
the requirement that a single value of CRIR be used in the matching routine, 
under the assumption that the same rate should be applicable to the entire 
region. A narrow range of gas density values is also imposed, based on 
observational constraints. Under these conditions, a single set of model 
parameters is found to be the best match to four of the sources. It is 
plausible that with a much higher resolution grid, better individual matches 
could be obtained for each source that would still be consistent with the 
imposed restrictions.

The matching method is nevertheless valuable in identifying two key parameter 
values: firstly, that the CRIR must necessarily be ``high'' to best match all 
five sources; and secondly, that the warm-up timescale for all best-matching 
models is $2 \times 10^4$~yr, the shortest value tested. Although the warm-up 
timescale in these models is a parameterized value that need not necessarily 
match the true dynamical timescale, the best-match value is both consistent 
across sources and in keeping with expected lifetimes for hot cores, on the 
order $6 \times 10^4$~yr \citep[][]{Nony24,Bonfand17}. We note that our grid 
also includes values $3.56$ and $6.32 \times 10^4$~yr, which did not produce 
the best match for any of the sources. 

Again, with four out of five sources being reproduced by the same chemical 
model, the modeling comparison itself has limited value in determining the 
origins of the chemical differences between sources. However, the models can 
help to indicate what conditions could be adjusted to produce a closer match.

For example, the lower observed values of HNCO and NH$_2$CHO toward AN06, as 
opposed to N2b, might indicate that different CRIR values are required between 
sources. The models indicate that the extreme $\zeta = 100 \zeta_0$ value 
suppresses the abundances of HNCO and NH$_2$CHO in the ices during the cold 
collapse stage, due to the enhanced UV field. Imposition of such a strict 
agreement in CRIR between Sgr B2(N2) sources might therefore be inappropriate; 
however, to account for a factor of nearly two orders of magnitude would 
require substantial variation, if the differences in these molecular abundances 
were caused by this alone.

Alternatively, other UV ionization sources could be active, requiring a 
slightly lower overall CRIR for all sources, and allowing more influence for 
variations in visual extinction between sources. Moreover, a more careful 
treatment of the \hbox{2-D/3-D} structure during stages~1 and 2 would help to 
determine how important external UV may be to the eventual detected abundances 
of certain COMs, by allowing regions of the core with different initial visual 
extinctions to contribute to the molecular abundances that are ultimately 
observed. Using the simple physical model that we have employed here, it is 
not possible to determine which regions of a natal cold core may later 
contribute most strongly to the release of icy COMs into the gas phase. Also, 
as noted above, shocks or luminosity outbursts could be important to the 
release of the ice mantles, determining both the location of COM desorption 
and the degree of ice loss, which could in turn lead to variations in which 
molecules are released. Testing such ideas requires dedicated coupled 
chemical-dynamical modeling that cannot currently be carried out in high 
enough volumes to allow a model-grid comparison similar to what is used here.

\subsection{Comparison to ALMA 1.2~mm survey of Sgr~B2(N)}
\label{comp_1mm}

\citet{Moeller25} derived the chemical composition of several dozen hot 
cores identified by \citet{SanchezMonge17} on the basis of an imaging spectral 
line survey performed with ALMA between 211 and 275~GHz with a spectral 
resolution of 490~kHz (0.7--0.5~km~s$^{-1}$) and an angular resolution ranging 
from 0.4$\arcsec$ to 0.7$\arcsec$. They used super-resolution data 
cubes restored with a beam of 0.4$\arcsec$. For each source, they modeled
a spectrum averaged over a polygon (see the shaded blue polygons in their 
Fig.~2). Following \citet{SanchezMonge17}, they divided Sgr~B2(N2) into three 
sources (AN02, AN03, and AN06) with polygon sizes on the order of 2$\arcsec$ 
$\times$ 1$\arcsec$. Their polygon around AN02 also includes the position that 
we named N2b. 

Figure~\ref{f:correl_chemcomp_1mm} compares the column densities that we
derived toward AN02, AN03, AN06, and AN06c2 to the column densities of the 
``Core Components'' reported for AN02, AN03, and AN06 in Tables~L.57, L.59, 
and L.65 of \citet{Moeller25}, respectively. In the case of AN02, most 
molecules were modeled by these authors with only one velocity component with 
a larger line width and thus we compared their 1.2~mm column densities to the 
sum of the column densities that we derived for the two velocity components 
that we identified in this source with ReMoCA. For both C$_2$H$_5$CN and 
CH$_3$CN, we used only the main component reported by \citeauthor{Moeller25} 
and ignored the other component that has a six orders of magnitude lower column 
density\footnote{On top of its extremely low column density, the weak 
component of C$_2$H$_5$CN has a puzzling velocity offset of $-$159~km~s$^{-1}$ 
in Table~L.57 of \citet{Moeller25}. We do not know what this component 
represents.}. For CH$_3$OH, we added the column densities of the two 
components modeled by \citeauthor{Moeller25}. In the case of AN03 (Table~L.59 
of \citeauthor{Moeller25}), we selected only the high-column-density 
components derived for CH$_3$CN and CH$_3$OH, and we ignored the HC$_3$N 
component that has a line width smaller than 0.5~km~s$^{-1}$. Finally, 
\citeauthor{Moeller25} modeled the spectrum of AN06 with two velocity 
components 
similar to ours only for C$_2$H$_5$CN and OCS. Most other molecules were 
modeled with a single velocity component with a larger line width. Therefore, 
apart from C$_2$H$_5$CN and OCS, Figs.~\ref{f:correl_chemcomp_1mm}c and d 
display the same 1.2~mm column densities.

Given that \citeauthor{Moeller25} used spectra averaged over large polygons 
while we 
analyzed spectra from a single pixel, we do not expect the ReMoCA column 
densities to match exactly the 1.2~mm column densities. While there is a 
decent overall correlation between the column densities of both studies, 
Fig.~\ref{f:correl_chemcomp_1mm}a shows that the ReMoCA column densities of 
AN02 are generally higher (by a factor of $\sim$2) than the 1.2~mm ones. This 
must result in part from the fact that the 1.2~mm column densities were derived
from spectra that were averaged over typical sizes of 2$\arcsec$ $\times$ 
1$\arcsec$ while the ReMoCA ones are peak column densities. In addition to 
this, the 1.2~mm column densities of AN02 were derived assuming an emission 
size of 1.7$\arcsec$ while, guided by their integrated intensity maps, we 
modeled most molecules with emission sizes (FWHM) of 0.7--1.0$\arcsec$. 
Furthermore, we do not know if the column densities reported by 
\citeauthor{Moeller25} account for vibrational and conformational corrections. 
These corrections can be as high as a factor 1.5--2 for some of the molecules 
displayed in Fig.~\ref{f:correl_chemcomp_1mm} (for instance C$_2$H$_3$CN, 
C$_2$H$_5$CN, HC$_3$N, CH$_3$OCHO). The situation is similar for AN03, which 
was modeled with an emission size of 1.6$\arcsec$ by \citeauthor{Moeller25} 
versus 0.7--1.0$\arcsec$ in our study, with a slightly better match between the 
two studies compared to AN02 (Fig.~\ref{f:correl_chemcomp_1mm}b). 

The case of AN06 looks different (Figs.~\ref{f:correl_chemcomp_1mm}c and d). 
The 1.2~mm column densities are in many cases higher than the ReMoCA ones (by 
more than a factor of two). We think that this is due to the fact that the 
aperture used by \citeauthor{Moeller25} includes more emission from the 
Sgr~B2(N2) dense core, meaning more contamination from the AN03 region, than 
our single-pixel spectra. As a matter of fact, our Weeds models for AN06 were 
as far as possible optimized based on setups S4 and S5, which have the smallest 
beams in our survey (0.3--0.5$\arcsec$). The line emission of AN06 in the 
ReMoCA setups with larger beams (S1--S3) is generally underestimated by our 
Weeds models, due to contamination by the AN03 environment, while it is well 
fitted for S4 and S5.

One molecule stands out in Fig.~\ref{f:correl_chemcomp_1mm}. The column 
densities reported for NH$_2$D by \citeauthor{Moeller25} are much higher (up to
two orders of magnitude for AN03) than the column densities derived from the 
ReMoCA survey. The ReMoCA column density toward AN03 was derived from four 
detected transitions while \citeauthor{Moeller25} detected only one transition 
in some of their sources. In their Fig.~F.20, they show the spectra toward AN05 
and AN08 only, and so it is unclear to us if the NH$_2$D column densities 
reported for AN02, AN03, and AN06 are reliable. Their Fig.~F.20 indicates that 
there is significant contamination by other molecules at the frequency of the 
NH$_2$D transition toward AN08. Therefore, it could well be that the emission
detected toward AN02, AN03, and AN06 at the frequency of this NH$_2$D 
transition is dominated by these contaminating molecules and does not trace 
NH$_2$D.

The comparison presented in this section strengthens our confidence in the 
reliability of the column densities derived from the ReMoCA survey. They 
faithfully represent the genuine chemical composition of the four hot cores 
AN02, AN03, AN06, and N2b. Thanks to the lower degree of spectral confusion at 
3~mm compared to the 1.2~mm range\footnote{As explained in 
Sect.~\ref{s:introduction}, our previous single-dish survey of Sgr~B2(N) at 3, 
2, and 1.3~mm already showed that spectral confusion and optical depth of the 
continuum emission are severe issues at 1.3~mm for this source 
\citep[][]{Belloche13}.}, 
the ReMoCA survey has allowed us to detect many more molecules (up to 58 toward 
AN03), in particular many more COMs: while \citeauthor{Moeller25} reported the 
detection of ten COMs toward AN02, AN03, and AN06, we managed to identify 22, 
24, and 17 COMs toward these sources, respectively (see
Table~\ref{t:detections}).

\subsection{Unidentified lines}
\label{ss:ulines}

There are still many unidentified lines in the ReMoCA spectra of the sources 
embedded in Sgr~B2(N2) (see Fig.~\ref{f:spec_remoca_n2}). Table~\ref{t:ulines} 
provides 
the list of unidentified lines that are brighter than 10~K (signal-to-noise 
ratio higher than 10--30 depending on the setup) in the spectrum of N2b. There 
are 255 such lines, which translates into about eight bright unidentified lines 
per GHz on average. As was already the case for our earlier single-dish survey
of Sgr~B2(N) \citep[][]{Belloche13}, we think that most of these bright
unidentified lines correspond to rotational transitions from within 
vibrationally excited states of molecules which have their vibrational ground
state and in some cases a few vibrational states already included in our model 
but for which spectroscopic predictions of higher vibrational states are still 
missing in our database, for instance ethyl cyanide or ethanol.

\section{Conclusions}
\label{s:conclusions}
We used the imaging spectral line survey ReMoCA performed with ALMA to probe 
the chemical composition of Sgr~B2(N2), the secondary hot molecular core of 
the high-mass star forming protocluster Sgr~B2(N). At the angular resolution 
of the survey, Sgr~B2(N2) consists of four hot cores (N2b, AN02, AN03, and 
AN06). Two velocity components were detected toward AN06. The main component 
traces a hot core associated with a compact dust continuum source. It is 
likely in the foreground or background of the dense core Sgr~B2(N2). The 
second velocity component, which we called AN06c2, traces the edge of 
Sgr~B2(N2). AN02 is associated with a dust continuum source and possibly 
drives an outflow. It is likely younger than AN03 which is associated with a 
HC\ion{H}{II} region. N2b is associated with a fainter dust continuum source, 
has narrow line widths, and is possibly younger than AN02 and AN03. All four 
sources have line-rich spectra. We identified up to 58 molecules, including up 
to 24 COMs. In addition to these molecules, many less abundant isotopologs 
were also detected. We derived the molecular composition of the four hot 
cores as well as AN06c2 under the assumption of LTE. The main results of our 
analysis are the following:
\begin{enumerate}
\item The pairs of sources that show the best correlations of their chemical
compositions are AN02/AN03 and AN06/AN06c2. AN03 stands out for its S-bearing
molecular content that poorly correlates with that of the other sources, and 
AN06 for its underabundant O+N-bearing molecules.
\item AN06 may be in a similar evolutionary stage as N2b. Their compositions
correlate with each other, albeit with a larger dispersion than AN02 versus 
AN03. The most striking difference between AN06 and N2b is that HNCO and 
NH$_2$CHO are almost two orders of magnitude less abundant in the former.
\item Protonated molecules XH$^+$ and radicals X$^\bullet$ were not detected. 
The most stringent constraint was obtained in both cases for N2b, with 
abundance ratios XH$^+$/X and X$^\bullet$/XH lower than 10$^{-3}$.
\item Molecules with a double bond are in nearly all cases much less abundant
than the corresponding single-bond species, by up to at least two orders of
magnitude. The only clear exception is the pair CH$_2$NH/CH$_3$NH$_2$ with a 
ratio close to unity. 
\item Only one complex carboxylic acid was (tentatively) detected. If 
the abundance ratio NH$_2$CH$_2$COOH/NH$_2$CH$_2$CN is similar to the ratio
CH$_3$COOH/CH$_3$CN then the abundance of glycine must be at least one order of 
magnitude below its current ReMoCA upper limit.
\item The abundances of series of homologous molecules drop by about one order 
of magnitude at each further step in complexity, except for NH$_2$CN, 
which is less abundant than NH$_2$CH$_2$CN by one order of magnitude,
and CH$_3$CN, which has a similar abundance as C$_2$H$_5$CN.
\setcounter{myenum}{\value{enumi}}
\end{enumerate} 

To gain more insights into the workings of interstellar chemistry, we compared 
the chemical composition of Sgr~B2(N2)'s hot cores to that of other sources 
that have been studied in detail in the literature: the lower-density Sgr~B2 
source G+0.693, thought to be affected by shocks, the hot core G31.41 in the 
Galactic disk, the hot corino IRAS16293B, and the cold starless core TMC-1, 
both located in the Solar neighborhood. We focused in particular on the 
behavior of four classes of molecules (O-bearing, N-bearing, O+N-bearing, and
S-bearing). The main results of this comparison are the following:
\begin{enumerate}
\setcounter{enumi}{\themyenum}
\item The source that shows the tightest correlation with N2b, AN02, and AN03 
is G31.41. This implies that the overall chemical composition of hot cores is
relatively insensitive to the environmental features that distinguish the 
Galactic center region from the Galactic disk.
\item For each class of molecules taken separately, there is also
a good correlation between IRAS16293B and Sgr~B2(N2)'s hot cores, but the
four classes are shifted with respect to each other. Such a segregation
also exists for G31.41 relative to Sgr~B2(N2)'s sources but it is less 
pronounced.
\item There is no overall correlation between G+0.693 and Sgr B2(N2)'s hot 
cores. The lack of correlation results mainly from the N-bearing species,
while the other classes correlate better. The S-bearing species are more
prominent in G+0.693.
\item There is absolutely no correlation between the compositions of TMC-1 and 
Sgr~B2(N2)'s hot cores.
\item Among the four classes of molecules, the class of N-bearing species is 
the one that reveals the largest variance. Firstly, its abundance 
distribution with respect to Sgr~B2(N2)'s hot cores goes from a tight 
correlation (G31.41) to no correlation (G+0.693 and TMC-1). Secondly, this
class shows overall shifts with respect to the other classes that can be 
large, up to two orders of magnitude (IRAS16293B).
\item In contrast, the class of S-bearing molecules
has the smallest variance, with a high degree of correlation for G31.41,
IRAS16293B, and G+0.693 with respect to Sgr~B2(N2).
\setcounter{myenum}{\value{enumi}}
\end{enumerate}
We also compared the molecular composition of Sgr~B2(N2)'s sources to a grid
of generic hot-core models computed with the chemical kinetics code 
\hbox{MAGICKAL}. The main results of this comparison are the following:
\begin{enumerate}
\setcounter{enumi}{\themyenum}
\item We confirm previous evidence for a strongly elevated CRIR for 
Sgr~B2(N2). The models produce their best collective match with observational 
abundances for the five sources adopting a CRIR value 
$\zeta = 1.3 \times 10^{-15}$~s$^{-1}$, which is the highest value tested. The 
shortest warm-up timescale ($2 \times 10^4$~yr) is found to produce the 
optimum match with all sources.
\item The chemical model grid is not able to distinguish adequately the 
chemical differences between sources. Four of the sources are best matched by 
a single model, under conditions where the gas density is constrained to an 
observational range of values. The models therefore provide little direct 
indication of the origins of chemical differences between the sources.
\end{enumerate}
The model grid might provide more information given higher resolution in
physical conditions. However, the match with observations may be much further 
improved by the use of an explicit treatment of gas dynamics in tandem with 
the chemistry.

We conclude from the observed behavior of the class of N-bearing molecules 
that this
class reacts more sensitively to shocks (G+0.693), low-temperature gas phase 
chemistry after non-thermal desorption (TMC-1), or density (G+0.693 and TMC-1) 
than the classes of O-bearing and S-bearing species. A possible interpretation 
of the segregation between N-bearing and O-bearing molecules is that only
partial ice desorption might encourage the preferential release of 
the outer ice layers that are rich in CO and related O-bearing species,
while leaving deeper, N-rich material in place. The overall abundance shifts 
observed in the 
gas phase between the classes of N-bearing and O-bearing molecules may thus 
indicate how violently and completely the ice mantles are desorbed. Testing 
this idea will require substantial improvements of chemical models that do not 
currently distinguish between ice layers beyond the typical bulk+surface-layer 
paradigm.

\begin{acknowledgements}
AB thanks all the physicists who sent him spectroscopic predictions over the
past two decades, in particular E. Alonso, J.~L. Alonso, B. Arenas, 
L. Bizzocchi, L. Bonah, S. Br\"unken, C. Cabezas, Ning Chen, L. Coudert, 
C. Endres, Z. Fried, E. Gougoula, S. Gruet, B. Heyne, V. Ilyushin, Z. Kisiel, 
I. Kleiner, K. Kobayashi, L. Kolesnikov\'a, J. Kouck\'y, L. Margul\`es, 
A. Maris, M.-A. Martin-Drumel, B. McGuire, C. Medcraft, I. Medvedev, 
M. Melosso, P. Misra, R. Motiyenko, M. Ordu, M. Sanz-Novo, M. Schnell, 
P. Stahl, O. Zingsheim, and Luyao Zou. We value the long-term efforts of the
spectroscopic community that crucially feed the databases used by 
astrophysicists.
RTG thanks the National Science Foundation for funding through the Astronomy 
\& Astrophysics program (grant number 2206516).
We thank B. Shope for advice on the chemical model grid.
This paper makes use of the following ALMA data: ADS/JAO.ALMA\#2016.1.00074.S. 
ALMA is a partnership of ESO (representing its member states), NSF (USA), and 
NINS (Japan), together with NRC (Canada), NSC and ASIAA (Taiwan), and KASI 
(Republic of Korea), in cooperation with the Republic of Chile. The Joint ALMA 
Observatory is operated by ESO, AUI/NRAO, and NAOJ. The interferometric data 
are available in the ALMA archive at https://almascience.eso.org/aq/.
Part of this work has been carried out within the Collaborative
Research Center 956, sub-project B3, funded by the Deutsche
Forschungsgemeinschaft (DFG, the German research foundation) -- project 
ID 184018867. HSPM acknowledges support by the DFG through the Collaborative 
Research Center 1601, sub-projects A4 and Inf -- project ID 500700252.
\end{acknowledgements}

\bibliographystyle{aa}
\bibliography{aa54411-25}

\begin{appendix}
\label{appendix}

\onecolumn
\section{Column densities from the literature}
\label{a:coldenslit}

Table~\ref{t:coldenslit} reports column densities and rotational temperatures 
of G31.41, IRAS16293B, G+0.693, and TMC-1 that we collected from the 
literature. These column density values were used in 
Figs.~\ref{f:correl_normch3oh_type_n2b}--\ref{f:correl_normch3oh_type_an06}.

\begin{table*}[h!]
 \begin{center}
 \caption{
 Column densities of G31.41, IRAS16293B, G+0.693, and TMC-1 collected from the literature.
}
 \label{t:coldenslit}
 \vspace*{-1.2ex}
 \begin{tabular}{lcrccrcrrcrrc}
 \hline\hline
 \multicolumn{1}{c}{Molecule} & \multicolumn{3}{c}{G31.41} & \multicolumn{3}{c}{IRAS16293B} & \multicolumn{3}{c}{G+0.693}  & \multicolumn{3}{c}{TMC-1} \\ 
  & \multicolumn{1}{c}{$N$\tablefootmark{(a)}} & \multicolumn{1}{c}{$T$\tablefootmark{(b)}} & \multicolumn{1}{c}{Ref\tablefootmark{(c)}} & \multicolumn{1}{c}{$N$\tablefootmark{(a)}} & \multicolumn{1}{c}{$T$\tablefootmark{(b)}} & \multicolumn{1}{c}{Ref\tablefootmark{(c)}} & \multicolumn{1}{c}{$N$\tablefootmark{(a)}} & \multicolumn{1}{c}{$T$\tablefootmark{(b)}} & \multicolumn{1}{c}{Ref\tablefootmark{(c)}} & \multicolumn{1}{c}{$N$\tablefootmark{(a)}} & \multicolumn{1}{c}{\hspace*{-1ex}$T$\tablefootmark{(b)}} & \multicolumn{1}{c}{\hspace*{-1ex}Ref\tablefootmark{(c)}}\\ 
  & \multicolumn{1}{c}{(cm$^{-2}$)} & \multicolumn{1}{c}{(K)} & & \multicolumn{1}{c}{(cm$^{-2}$)} & \multicolumn{1}{c}{(K)} & & \multicolumn{1}{c}{(cm$^{-2}$)} & \multicolumn{1}{c}{(K)} & & \multicolumn{1}{c}{(cm$^{-2}$)} & \multicolumn{1}{c}{\hspace*{-1ex}(K)} \\ 
 \hline
CH$_3$OH &8.0(19) & 153 &1 &1.0(19) & 300 &6,7,8 &3.2(16) &  10 &21 &4.8(13) &  10 &35 \\ 
H$_2$CO &2.6(18) &  50 &2 &1.9(18) & 106 &6,9 &\ldots & \ldots & \ldots &5.0(14) &   5 &35 \\ 
C$_2$H$_5$OH &4.7(17) & 119 &1 &2.3(17) & 300 &6,8 &1.3(15) &  10 &21 &1.1(12) &   6 &35 \\ 
CH$_3$CHO &3.4(16) &  82 &1 &1.2(17) & 125 &6,8 &5.0(14) &   9 &22 &3.5(12) &  10 &35 \\ 
C$_2$H$_5$CHO &5.3(16)$^\star$ & 150 &2 &2.2(15) & 125 &6,10 &7.4(13) &  12 &22 &1.9(11) &   6 &35 \\ 
CH$_2$CO &\ldots & \ldots & \ldots &4.8(16) & 125 &8 &\ldots & \ldots & \ldots &1.4(13) &  10 &35 \\ 
c-C$_2$H$_4$O &6.3(16) & 132 &2 &5.4(15) & 125 &6,10 &\ldots & \ldots & \ldots &\ldots & \ldots & \ldots \\ 
CH$_3$OCH$_3$ &8.1(17) &  98 &1 &2.4(17) & 125 &6,8 &1.1(15) &  10 &21 &2.5(12) &   4 &35 \\ 
CH$_3$C(O)CH$_3$ &5.6(17) & 170 &1 &1.7(16) & 125 &6,10 &\ldots & \ldots & \ldots &1.4(11) &   6 &35 \\ 
CH$_3$COOH &7.3(17)$^\star$ & 250 &3 &2.8(15) & 300 &6,7 &4.2(13) &  17 &23 &\ldots & \ldots & \ldots \\ 
CH$_3$OCHO &3.2(18)$^\star$ & 221 &3 &2.6(17) & 300 &6,8 &6.0(14) &  13 &24 &1.1(12) &   5 &35 \\ 
CH$_2$(OH)CHO &5.0(16) & 128 &3 &3.2(16) & 300 &6,7 &9.3(13) &  22 &25 &\ldots & \ldots & \ldots \\ 
a-(CH$_2$OH)$_2$ &1.9(17)$^\star$ & 120 &1 &5.2(16) & 300 &6,7 &\ldots & \ldots & \ldots &\ldots & \ldots & \ldots \\ 
n-C$_3$H$_7$OH &\ldots & \ldots & \ldots &$<$3.7(15)$^\star$ & 100 &11 &8.9(13) &  14 &26 &\ldots & \ldots & \ldots \\ 
i-C$_3$H$_7$OH &\ldots & \ldots & \ldots &$<$3.5(15)$^\star$ & 100 &11 &$<$8.5(12) &  12 &26 &\ldots & \ldots & \ldots \\ 
t-HCOOH &1.4(17) & 152 &4 &5.6(16) & 300 &6,8 &2.0(14) &  10 &27 &1.4(12) &  10 &35 \\ 
HNCO &1.5(18)$^\star$ & 217 &5 &3.7(16) & 100 &6,12 &3.4(15) &  17 &28 &1.1(13) &   8 &35 \\ 
NH$_2$CHO &1.8(17)$^\star$ & 150 &5 &9.5(15) & 300 &6,13 &2.5(14) &   5 &29 &$<$5.0(10) &  10 &36 \\ 
CH$_3$NCO &1.8(17)$^\star$ & 122 &5 &4.0(15) & 100 &12 &6.6(13) &   8 &28 &\ldots & \ldots & \ldots \\ 
CH$_3$NHCHO &3.7(16) & 285 &5 &$<$1.0(15) & 300 &14 &4.3(13) &   7 &29 &\ldots & \ldots & \ldots \\ 
CH$_3$C(O)NH$_2$ &8.0(16) & 285 &5 &5.2(15)$^\star$ & 300 &15,14 &1.1(14) &   8 &29 &\ldots & \ldots & \ldots \\ 
CH$_2$NH &\ldots & \ldots & \ldots &$<$6.0(14) & 120 &16 &5.4(14) &  10 &28 &$<$1.0(12) &   7 &37 \\ 
CH$_3$NH$_2$ &\ldots & \ldots & \ldots &$<$5.3(14) & 100 &16 &3.0(15) &  16 &28 &\ldots & \ldots & \ldots \\ 
CH$_3$CN &2.7(17) & 111 &1 &4.0(16) & 110 &6,17 &1.1(14) &  14 &28 &5.0(12) &   7 &39 \\ 
C$_2$H$_5$CN &5.6(16) &  83 &1 &3.6(15) & 110 &17 &4.1(13) &  18 &28 &1.3(11) &   6 &38 \\ 
C$_2$H$_3$CN &2.1(16) & 104 &1 &7.4(14) & 110 &17 &9.0(13) &  11 &28 &6.2(12) &   4 &38 \\ 
n-C$_3$H$_7$CN &\ldots & \ldots & \ldots &\ldots & \ldots & \ldots &$<$1.1(13) &  15 &28 &\ldots & \ldots & \ldots \\ 
i-C$_3$H$_7$CN &\ldots & \ldots & \ldots &\ldots & \ldots & \ldots &$<$6.0(12) &  15 &28 &\ldots & \ldots & \ldots \\ 
CH$_3$NC &1.3(15)$^\star$ & 150 &2 &2.0(14) & 150 &6,18 &$<$3.0(12) &  15 &28 &3.0(11) &   7 &39 \\ 
HC$_3$N &2.8(16) &  50 &2 &1.8(14) & 100 &6,17 &7.1(14) &  12 &28 &1.9(14) &   5 &40 \\ 
HCCNC &$<$3.0(13) &  50 &2 &\ldots & \ldots & \ldots &2.3(13) &   7 &28 &3.0(12) &  10 &41 \\ 
CH$_3$C$_3$N &\ldots & \ldots & \ldots &\ldots & \ldots & \ldots &1.4(13) &  19 &30 &1.7(12) &   7 &42 \\ 
HC$_5$N &\ldots & \ldots & \ldots &\ldots & \ldots & \ldots &2.6(14) &  16 &28 &6.6(13) &   8 &43 \\ 
NH$_2$CN &1.1(15) &  50 &2 &\ldots & \ldots & \ldots &3.1(14) &   7 &28 &$<$4.5(11) &  10 &44 \\ 
NH$_2$CH$_2$CN &\ldots & \ldots & \ldots &\ldots & \ldots & \ldots &$<$6.0(12) &  15 &28 &\ldots & \ldots & \ldots \\ 
E-HNCHCN &\ldots & \ldots & \ldots &\ldots & \ldots & \ldots &3.3(13) &   8 &31 &\ldots & \ldots & \ldots \\ 
CH$_3$CCH &1.2(17) & 150 &2 &6.8(15) & 100 &19 &\ldots & \ldots & \ldots &1.1(14) &   5 &45 \\ 
CH$_3$SH &7.2(16) & 150 &2 &4.8(15) & 125 &6,20 &6.5(14) &   8 &27 &1.7(12) &   9 &35 \\ 
H$_2$CS &3.1(17) &  50 &2 &1.3(15) & 125 &6,20 &\ldots & \ldots & \ldots &3.7(13) & ? &35 \\ 
HNCS &\ldots & \ldots & \ldots &$<$1.0(14) & 125 &20 &6.2(13) &  20 &32 &3.8(11) &   5 &35 \\ 
HSCN &\ldots & \ldots & \ldots &$<$1.0(14) & 125 &20 &\ldots & \ldots & \ldots &5.8(11) &   5 &35 \\ 
OCS &1.8(18) &  50 &2 &2.5(17) & 125 &6,20 &3.6(15) &  23 &32 &2.2(13) &   6 &35 \\ 
SO &3.9(16) &  50 &2 &4.4(14) & 125 &6,20 &3.0(15) &   7 &33 &3.4(13) &   4 &35 \\ 
SO$_2$ &1.0(17) &  56 &2 &1.3(15) & 125 &6,20 &3.8(14) &  19 &34 &3.0(12) & ? &35 \\ 
 \hline
 \end{tabular}
 \end{center}
 \vspace*{-3.5ex}
 \tablefoot{
 \tablefoottext{a}{$x(y)$ means $x \times 10^{y}$. The star symbol ($^\star$) indicates that we applied an a posteriori vibrational or conformational  correction to the column density.}
 \tablefoottext{b}{Rotational temperature.}
 \tablefoottext{c}{References: 1: \citet{Mininni23}; 2: \citet{LopezGallifa24}; 3: \citet{Mininni20}; 4: \citet{GarciaDeLaConcepcion22}; 5: \citet{Colzi21}; 6: \citet{Drozdovskaya19}; 7: \citet{Jorgensen16}; 8: \citet{Jorgensen18}; 9: \citet{Persson18}; 10: \citet{Lykke17}; 11: \citet{Manigand21}; 12: \citet{Ligterink17}; 13: \citet{Coutens16}; 14: \citet{Ligterink18b}; 15: \citet{Jorgensen20}; 16: \citet{Ligterink18a}; 17: \citet{Calcutt18a}; 18: \citet{Calcutt18b}; 19: \citet{Calcutt19}; 20: \citet{Drozdovskaya18}; 21: \citet{RequenaTorres06}; 22: \citet{SanzNovo22}; 23: \citet{SanzNovo23}; 24: \citet{SanzNovo25}; 25: \citet{Rivilla22a}; 26: \citet{JimenezSerra22}; 27: \citet{RodriguezAlmeida21}; 28: \citet{Zeng18}; 29: \citet{Zeng23}; 30: \citet{Rivilla22c}; 31: \citet{Rivilla19}; 32: \citet{SanzNovo24a}; 33: \citet{Rivilla22b}; 34: \citet{SanzNovo24b}; 35: \citet{Agundez25}; 36: \citet{Cernicharo20b}; 37: \citet{Margules22}; 38: \citet{Cernicharo24a}; 39: \citet{Tennis23}; 40: \citet{Tercero24}; 41: \citet{Cernicharo20a}; 42: \citet{Marcelino21}; 43: \citet{Cernicharo24b}; 44: \citet{Irvine84}; 45: \citet{Gratier16}.}
 }
 \end{table*}

\onecolumn
\section{Bright unidentified lines}
\label{a:ulines}

Table~\ref{t:ulines} provides the list of unidentified lines that are 
brighter than 10~K in the ReMoCA spectrum of N2b.

\begin{table*}[h!]
 \begin{center}
\caption{
 List of unidentified lines brighter than 10 K in the ReMoCA spectrum of N2b.
 }
 \label{t:ulines}
\vspace*{-1.2ex}
\begin{tabular}{rccrccrccrccrcc}
 \hline\hline
\multicolumn{1}{c}{Freq.\tablefootmark{(a)}} & \multicolumn{1}{c}{\hspace*{-1ex}$T$\tablefootmark{(b)}} & \multicolumn{1}{c}{\hspace*{-2.ex}S.W\tablefootmark{(c)}} & \multicolumn{1}{c}{Freq.\tablefootmark{(a)}} & \multicolumn{1}{c}{\hspace*{-1ex}$T$\tablefootmark{(b)}} & \multicolumn{1}{c}{\hspace*{-2.ex}S.W\tablefootmark{(c)}} & \multicolumn{1}{c}{Freq.\tablefootmark{(a)}} & \multicolumn{1}{c}{\hspace*{-1ex}$T$\tablefootmark{(b)}} & \multicolumn{1}{c}{\hspace*{-2.ex}S.W\tablefootmark{(c)}} & \multicolumn{1}{c}{Freq.\tablefootmark{(a)}} & \multicolumn{1}{c}{\hspace*{-1ex}$T$\tablefootmark{(b)}} & \multicolumn{1}{c}{\hspace*{-2.ex}S.W\tablefootmark{(c)}} & \multicolumn{1}{c}{Freq.\tablefootmark{(a)}} & \multicolumn{1}{c}{\hspace*{-1ex}$T$\tablefootmark{(b)}} & \multicolumn{1}{c}{\hspace*{-2.ex}S.W\tablefootmark{(c)}} \\ 
 \hline
 84141.3 & \hspace*{-1ex}31 & \hspace*{-2.ex}1.0 &  90308.4 & \hspace*{-1ex}18 & \hspace*{-2.ex}2.1 &  98730.0 & \hspace*{-1ex}32 & \hspace*{-2.ex}5.0 & 104953.3 & \hspace*{-1ex}16 & \hspace*{-2.ex}3.2 & 108933.6 & \hspace*{-1ex}36 & \hspace*{-2.ex}4.3\\ 
 84158.3 & \hspace*{-1ex}12 & \hspace*{-2.ex}1.0 &  90339.8 & \hspace*{-1ex}19 & \hspace*{-2.ex}2.1 &  98750.4 & \hspace*{-1ex}52 & \hspace*{-2.ex}5.0 & 105004.3 & \hspace*{-1ex}11 & \hspace*{-2.ex}3.2 & 108953.7 & \hspace*{-1ex}10 & \hspace*{-2.ex}4.3\\ 
 84517.1 & \hspace*{-1ex}10 & \hspace*{-2.ex}1.0 &  90544.2 & \hspace*{-1ex}19 & \hspace*{-2.ex}2.1 &  98769.8 & \hspace*{-1ex}14 & \hspace*{-2.ex}5.0 & 105167.3 & \hspace*{-1ex}12 & \hspace*{-2.ex}3.2 & 108966.5 & \hspace*{-1ex}15 & \hspace*{-2.ex}4.3\\ 
 84705.3 & \hspace*{-1ex}28 & \hspace*{-2.ex}1.0 &  90739.2 & \hspace*{-1ex}28 & \hspace*{-2.ex}2.1 &  98773.3 & \hspace*{-1ex}35 & \hspace*{-2.ex}5.0 & 105293.0 & \hspace*{-1ex}12 & \hspace*{-2.ex}3.3 & 109009.2 & \hspace*{-1ex}18 & \hspace*{-2.ex}4.3\\
\hline
 \end{tabular}
 \end{center}
 \vspace*{-2.5ex}
\tablefoot{
 \tablefoottext{a}{Rest frequency in MHz.}
 \tablefoottext{b}{Peak temperature in K.}
 \tablefoottext{c}{Setup (S) and spectral window (W).}
}
 \end{table*}

\clearpage
\onecolumn
\section{Spectroscopic references}
\label{a:references}

Table~\ref{t:spectrobib} provides the list of relevant references for the 448 
spectroscopic entries that were used in this work to model the observed 
spectra or derive upper limits to the column densities of nondetected 
molecules.

\LTcapwidth=\textwidth

\begin{longtable}{lrlcrl}
 \caption{\label{t:spectrobib}
 Spectroscopic references of molecules analyzed in this work.
} \\ 
 \hline\hline
 \multicolumn{1}{c}{Molecule and vib. state\tablefootmark{a}} & \multicolumn{1}{c}{Tag\tablefootmark{b}} & \multicolumn{1}{c}{DB\tablefootmark{c}} & \multicolumn{1}{c}{Version} & \multicolumn{2}{c}{References\tablefootmark{d}} \\ 
 \hline
 \endfirsthead
 \caption{continued.} \\ 
 \hline\hline
 \multicolumn{1}{c}{Molecule and vib. state\tablefootmark{a}} & \multicolumn{1}{c}{Tag\tablefootmark{b}} & \multicolumn{1}{c}{DB\tablefootmark{c}} & \multicolumn{1}{c}{Version} & \multicolumn{2}{c}{References\tablefootmark{d}} \\ 
 \hline
 \endhead
 \hline
 \endfoot
 & & & & & Creswell and Schwendemann (1977) \kill 
NH$_2$D & 18501 & CDMS & 1 & Spectro: &  \citet{DeLucia75} \\ 
 & & & & & \citet{Cohen82} \\ 
 & & & & & \citet{Fusina88} \\ 
 & & & & $C_{\rm vib}$: &  \citet{Snels06} \\ 
CCH & 25501 & CDMS & 3 & Spectro: &  \citet{Mueller00b} \\ 
 & & & & & \citet{Padovani09} \\ 
 & & & & & \citet{Sastry81} \\ 
 & & & & & \citet{Woon95} \\
\hline 
 \end{longtable}
 \tablefoot{
 \tablefoottext{a}{When no vibrational state is indicated, the entry generally contains the ground state only.}
 \tablefoottext{b}{Entry number in our Weeds local database. For entries from the CDMS, JPL, or LSD databases, it corresponds to the molecule tag in these databases, except for several CDMS entries with hyperfine structure downloaded from the CDMS documentation for which we had to assign specific tags, or for entries containing several vibrational states that we split into individual entries to which we assigned new tags.}
 \tablefoottext{c}{Database which the spectroscopic entry originates from. CDMS: Cologne Database for Molecular Spectrosocpy (https://cdms.astro.uni-koeln.de); LSD: Lille Spectroscopic Database (https://lsd.univ-lille.fr/); JPL: molecular spectroscopy database of the Jet Propulsion Laboratory (https://spec.jpl.nasa.gov/); private: entry obtained from collaborators or generated by us from published studies.}
 \tablefoottext{d}{Spectro: references reporting line lists or dipole moment measurements or calculation used to produce the spectroscopic entry. $C_{\rm vib}$: references used to compute the vibrational correction to the partition function. $C_{\rm conf}$: references used to compute the conformational correction to the partition function.}
 }

%\clearpage
%\onecolumn
\section{Spectra}
\label{a:spectra}

Figure~\ref{f:spec_remoca_n2} shows the complete ReMoCA spectral survey
toward five positions in Sgr~B2(N): the four hot-core positions N2b, AN02, 
AN03, and AN06, and the ultracompact (UC) \ion{H}{ii} region K4. The 
continuum-subtracted spectra were shifted vertically by 0, 200, 400, 600, and 
900 K, respectively. The continuum level is indicated in blue on the right 
below each spectrum. LTE synthetic spectra computed with Weeds and containing 
the contribution of all identified molecules are overlaid in red on the 
hot-core spectra. The parameters of these LTE models are listed in 
Tables~\ref{t:coldens_n2b}-\ref{t:coldens_an06}. The molecular identification 
of a selection of spectral lines is indicated in various colors: magenta for 
rotational lines in the vibrational ground state, teal for rotational lines in 
vibrationally excited states (with the suffix v added to the label), and 
orange for lines seen in absorption. Green labels above the K4 spectrum mark 
the frequencies of hydrogen and helium recombination lines. To facilitate the 
visual matching between the spectral lines and their magenta or teal labels, 
short vertical bars are displayed in the same color below each hot-core 
spectrum. Each panel corresponds to half a spectral window (W) of a single 
spectral setup (S). The setup index (from 1 to 5) and the spectral window 
index (from 0 to 3) of each panel are indicated in the bottom left corner 
along with the HPBW. The spectra are displayed in brightness temperature 
scale and were corrected for primary beam attenuation. The panels are ordered 
in increasing frequency of the spectral windows. There is a substantial
frequency overlap between the upper sideband of setups 1 and 2 and the lower
sideband of setups 4 and 5. The angular resolution of these pairs of setups
differs by a factor $\sim$1.5, which motivates our decision to display all
spectral windows in Fig.~\ref{f:spec_remoca_n2}, despite their frequency 
overlap.

\begin{figure*}
\centerline{\resizebox{0.85\hsize}{!}{\includegraphics[angle=270]{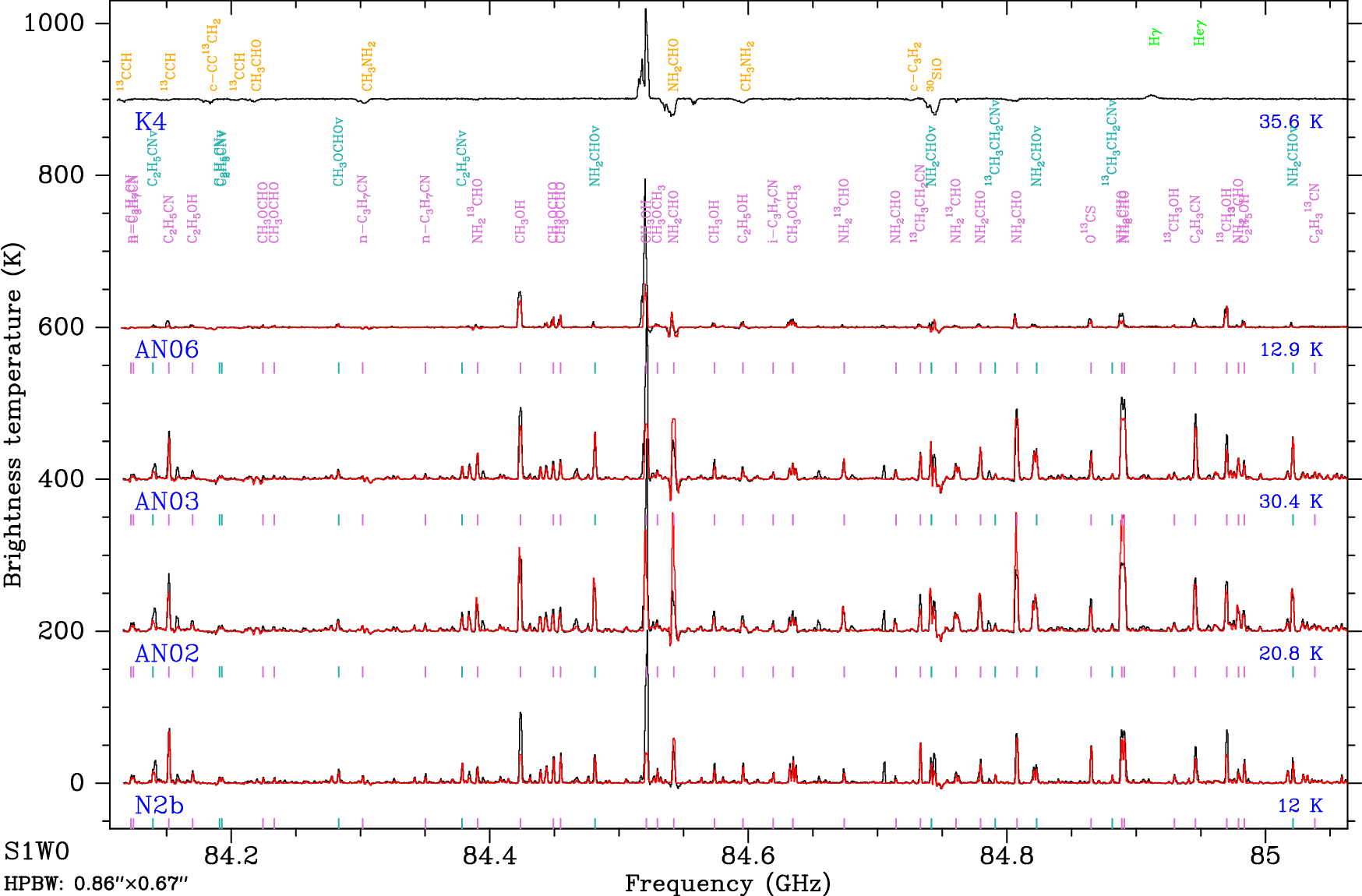}}}
\caption{ALMA continuum-subtracted spectra of the hot core positions N2b,
AN02, AN03, AN06 and the UC\ion{H}{ii} region K4 (black).
LTE synthetic spectra are overlaid in red. See Appendix~\ref{a:spectra} for a
detailed description of the figure.}
\label{f:spec_remoca_n2}
\end{figure*}

\clearpage
\begin{figure*}
\addtocounter{figure}{-1}
\centerline{\resizebox{0.85\hsize}{!}{\includegraphics[angle=270]{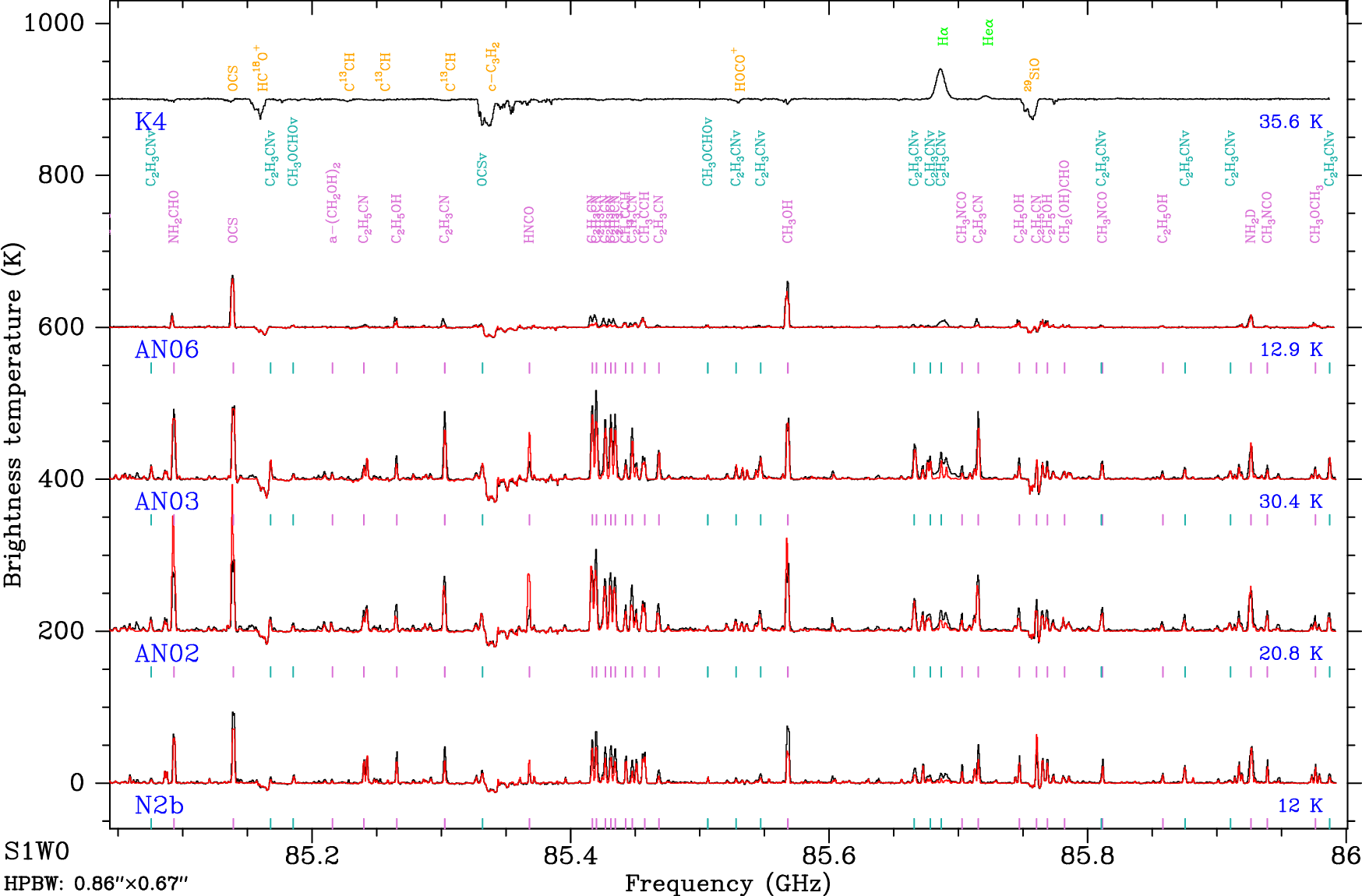}}}
\caption{continued.
}
\end{figure*}

\clearpage
\begin{figure*}
\addtocounter{figure}{-1}
\centerline{\resizebox{0.85\hsize}{!}{\includegraphics[angle=270]{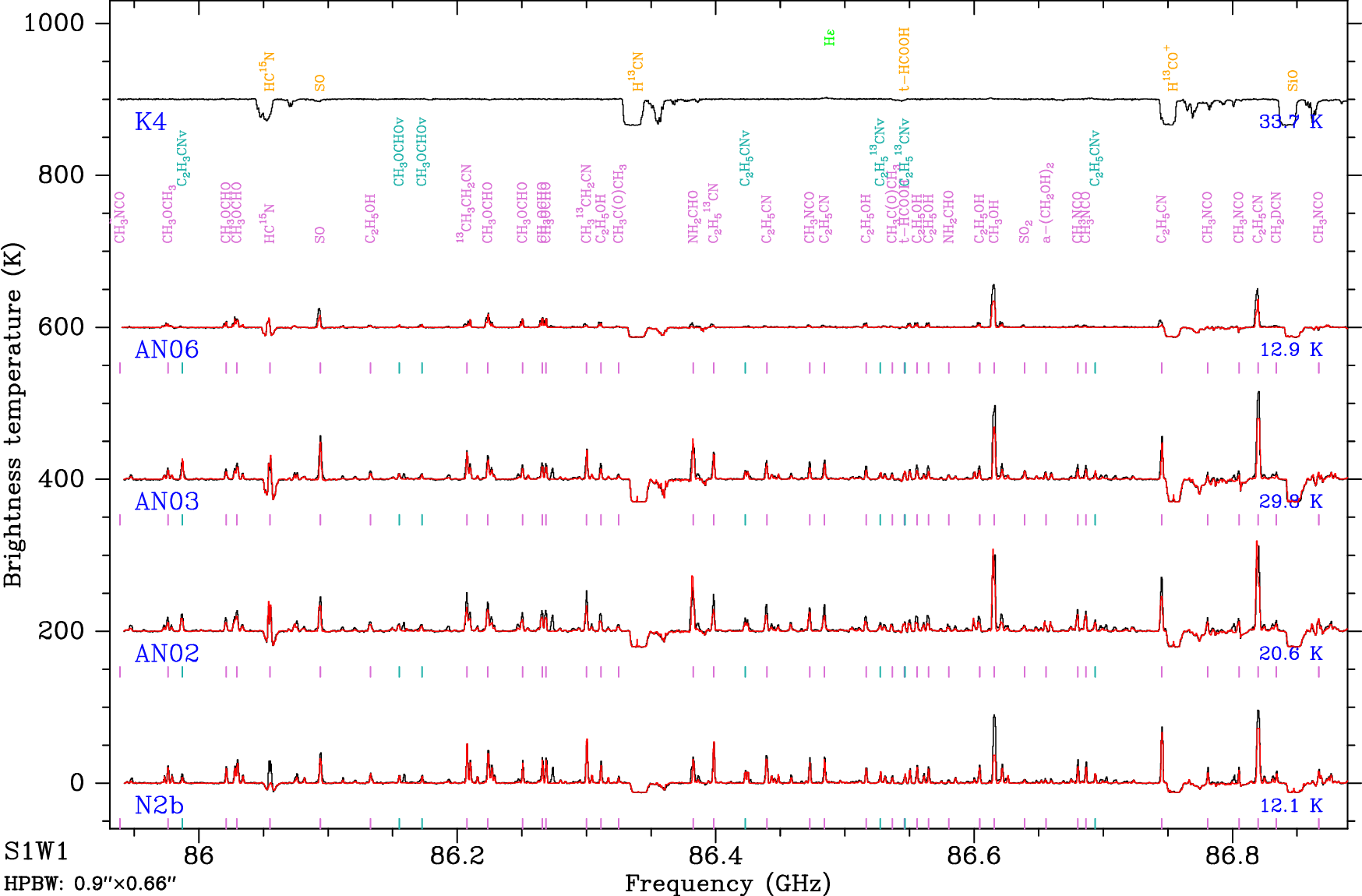}}}
\caption{continued.
}
\end{figure*}

\clearpage
\begin{figure*}
\addtocounter{figure}{-1}
\centerline{\resizebox{0.85\hsize}{!}{\includegraphics[angle=270]{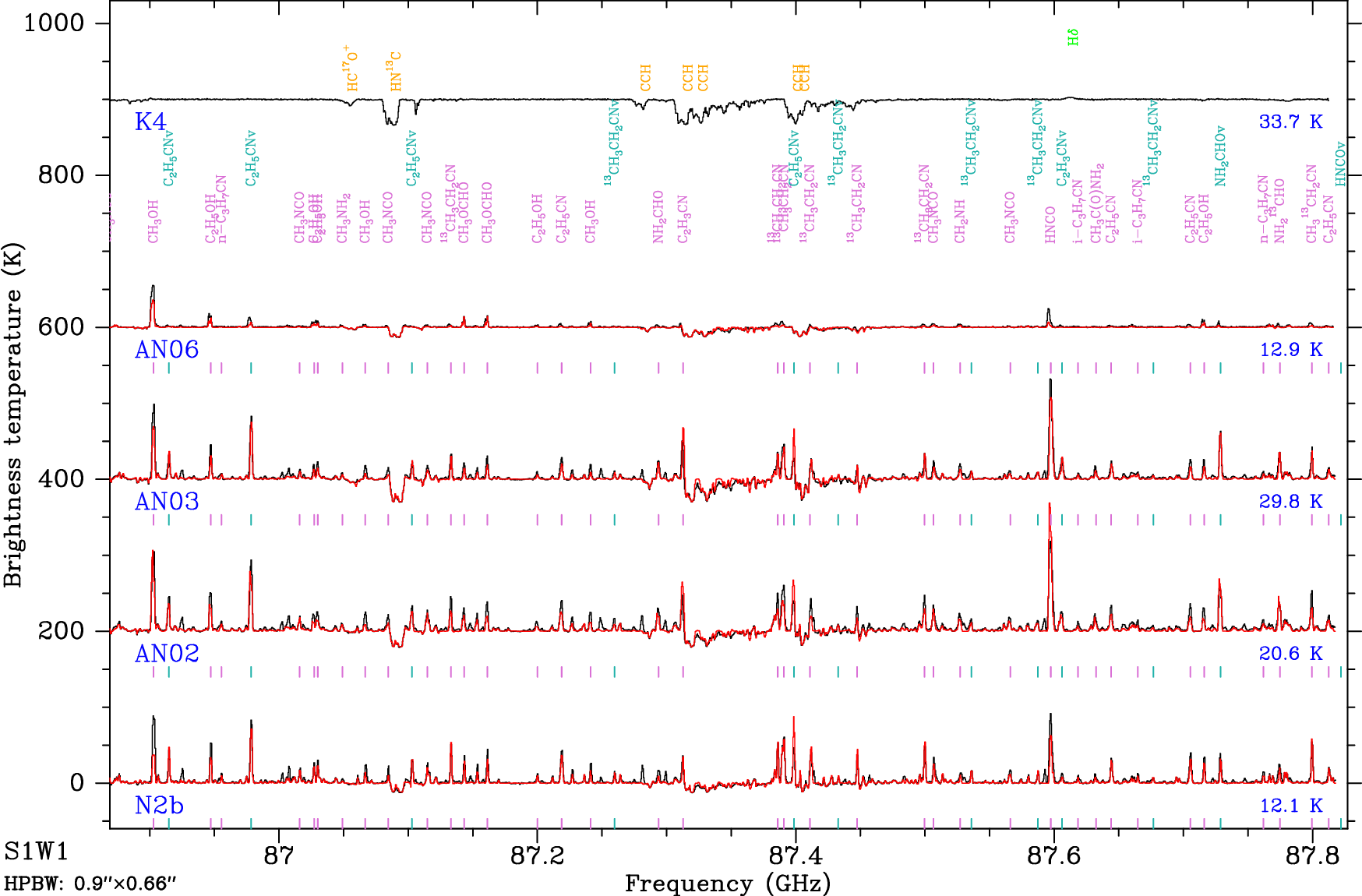}}}
\caption{continued.
}
\end{figure*}

\clearpage
\begin{figure*}
\addtocounter{figure}{-1}
\centerline{\resizebox{0.85\hsize}{!}{\includegraphics[angle=270]{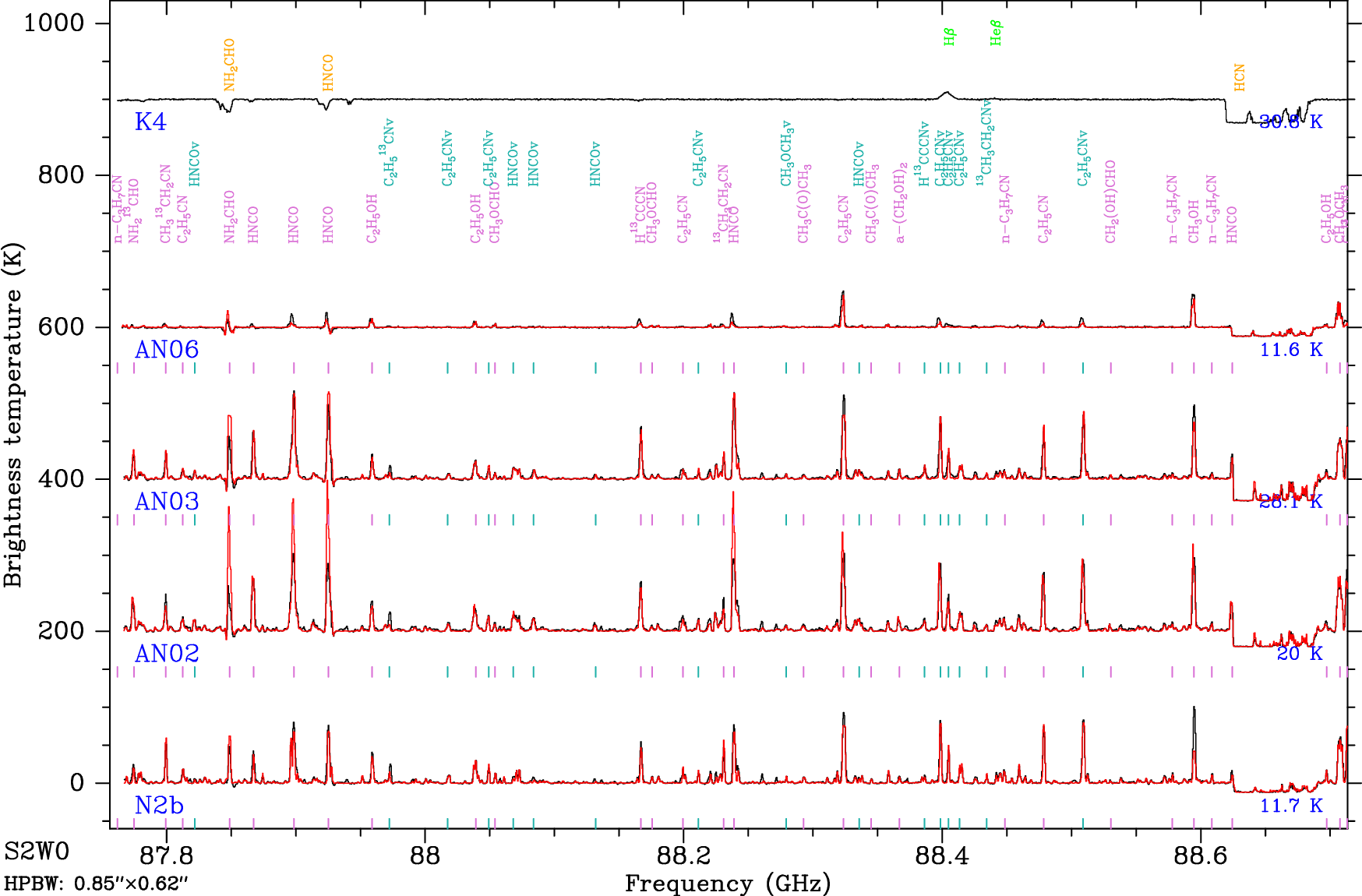}}}
\caption{continued.
}
\end{figure*}

\clearpage
\begin{figure*}
\addtocounter{figure}{-1}
\centerline{\resizebox{0.85\hsize}{!}{\includegraphics[angle=270]{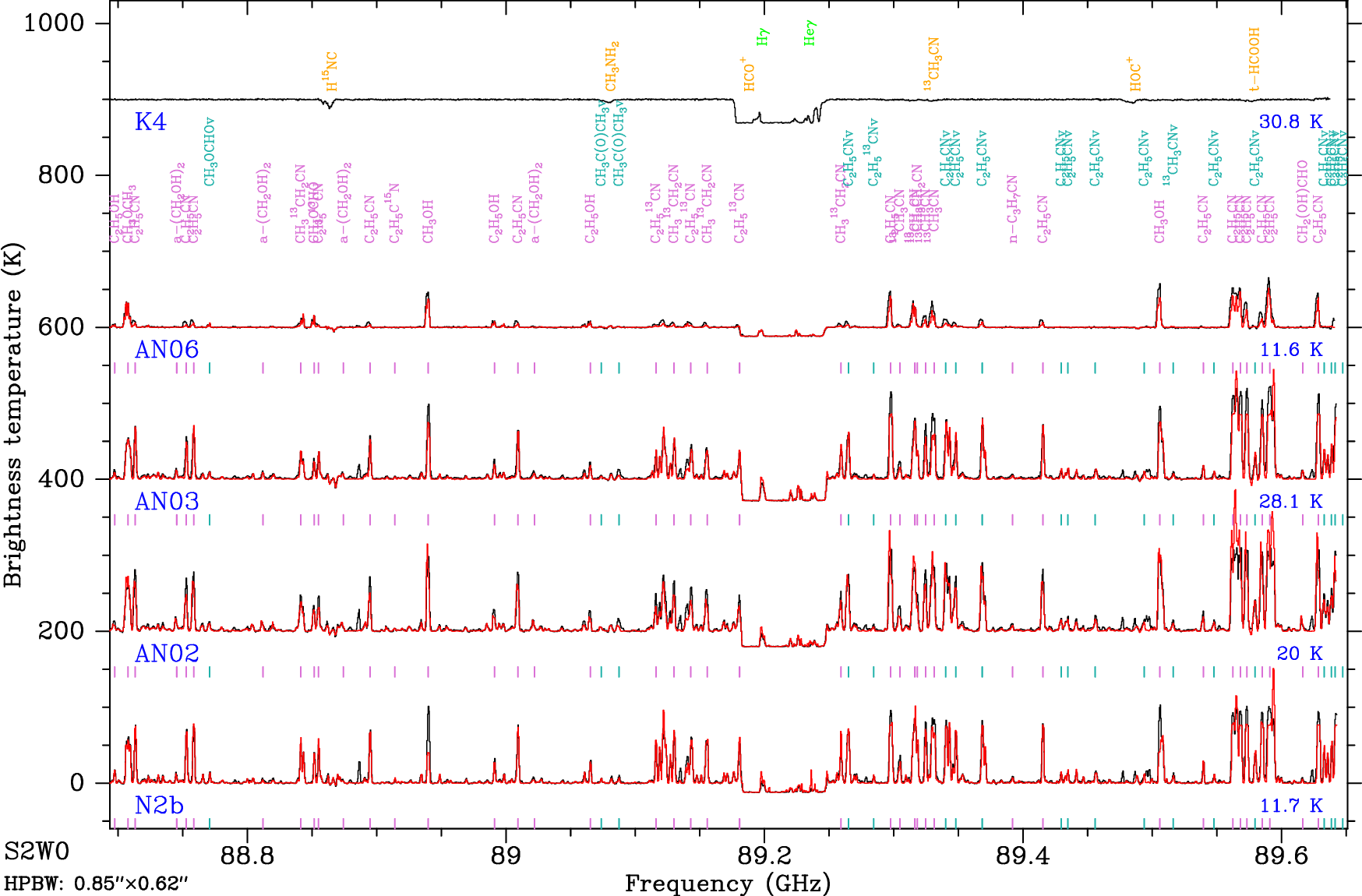}}}
\caption{continued.
}
\end{figure*}

\clearpage
\begin{figure*}
\addtocounter{figure}{-1}
\centerline{\resizebox{0.85\hsize}{!}{\includegraphics[angle=270]{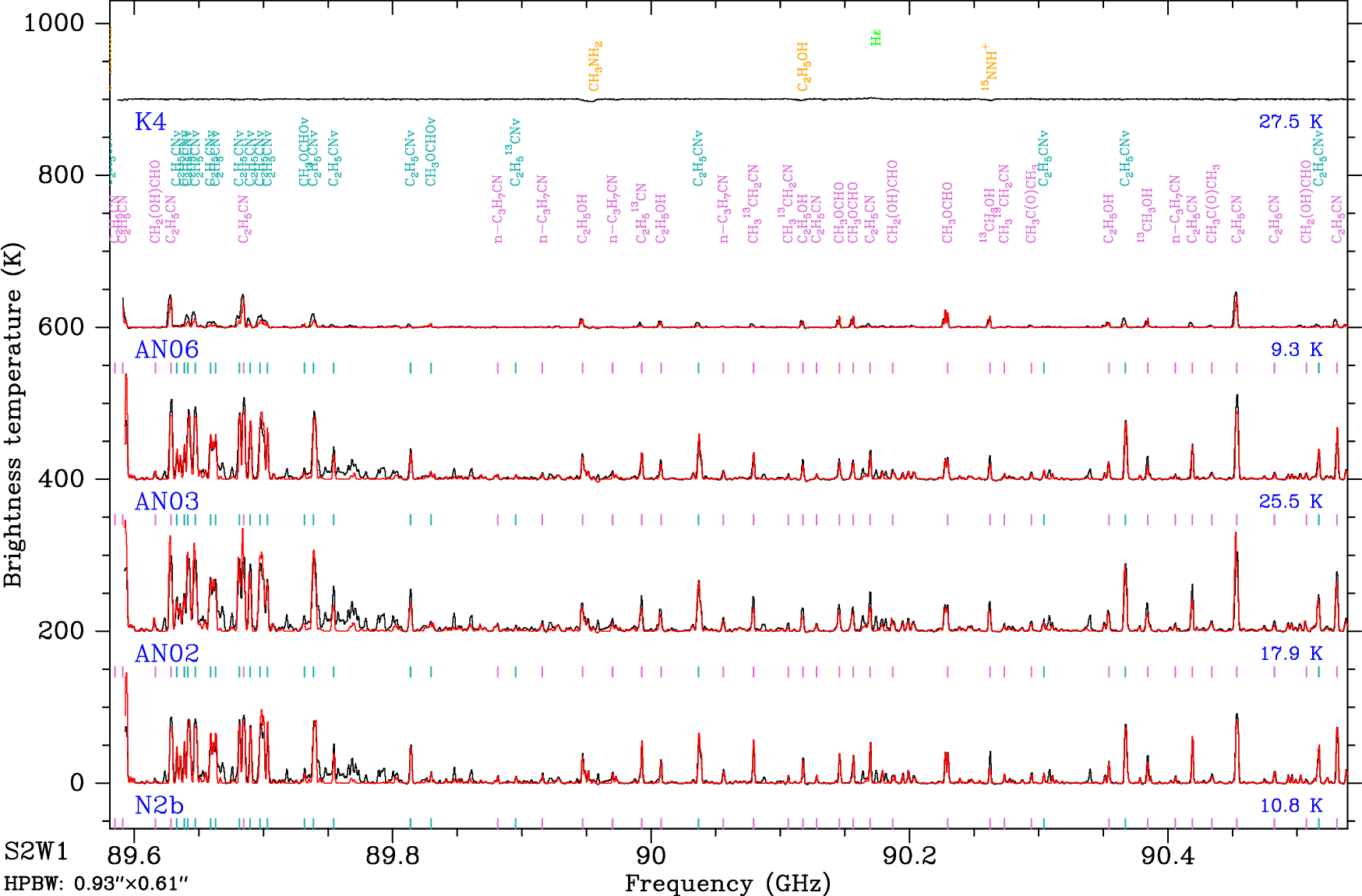}}}
\caption{continued.
}
\end{figure*}

\clearpage
\begin{figure*}
\addtocounter{figure}{-1}
\centerline{\resizebox{0.85\hsize}{!}{\includegraphics[angle=270]{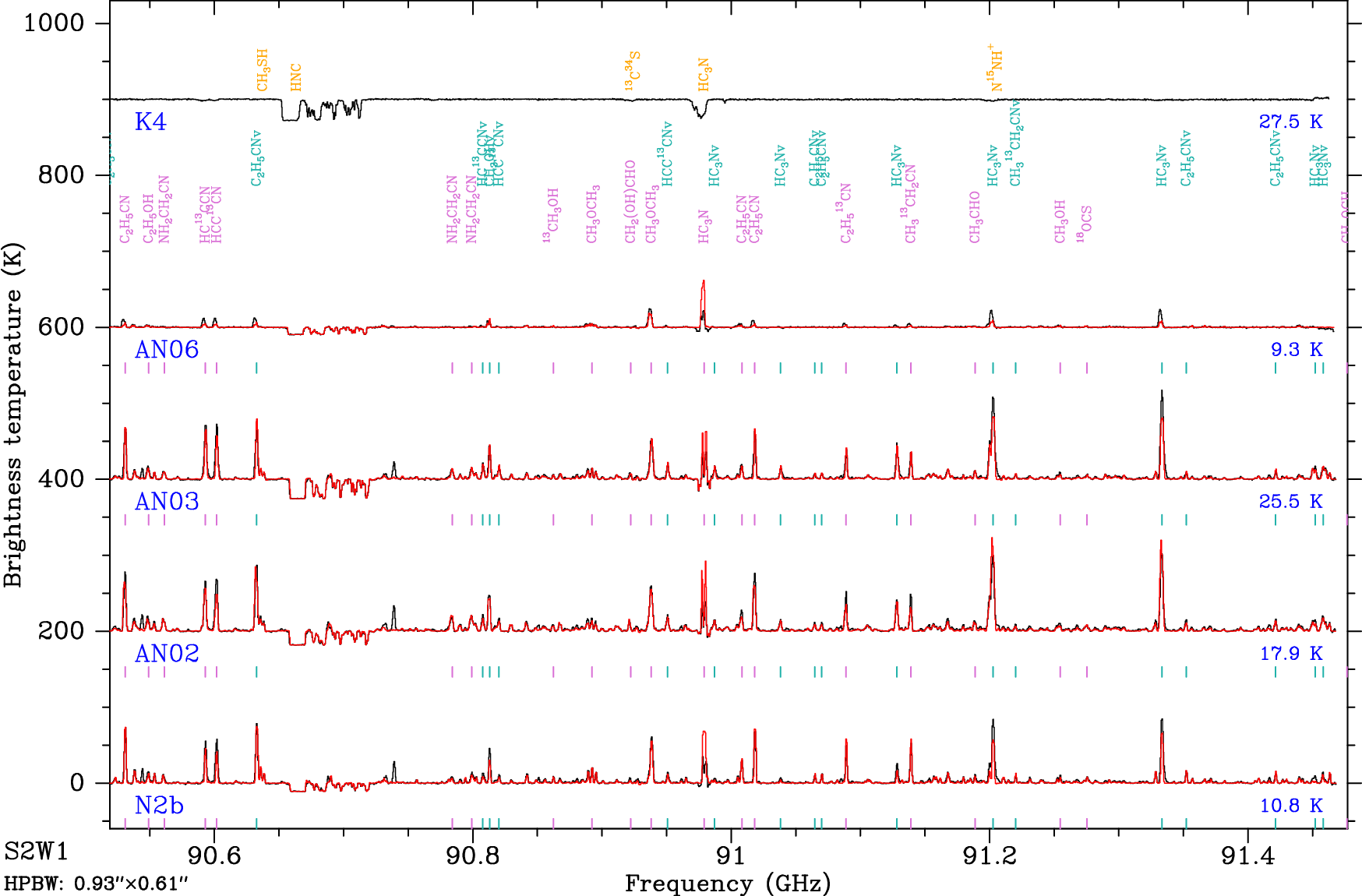}}}
\caption{continued.
}
\end{figure*}

\clearpage
\begin{figure*}
\addtocounter{figure}{-1}
\centerline{\resizebox{0.85\hsize}{!}{\includegraphics[angle=270]{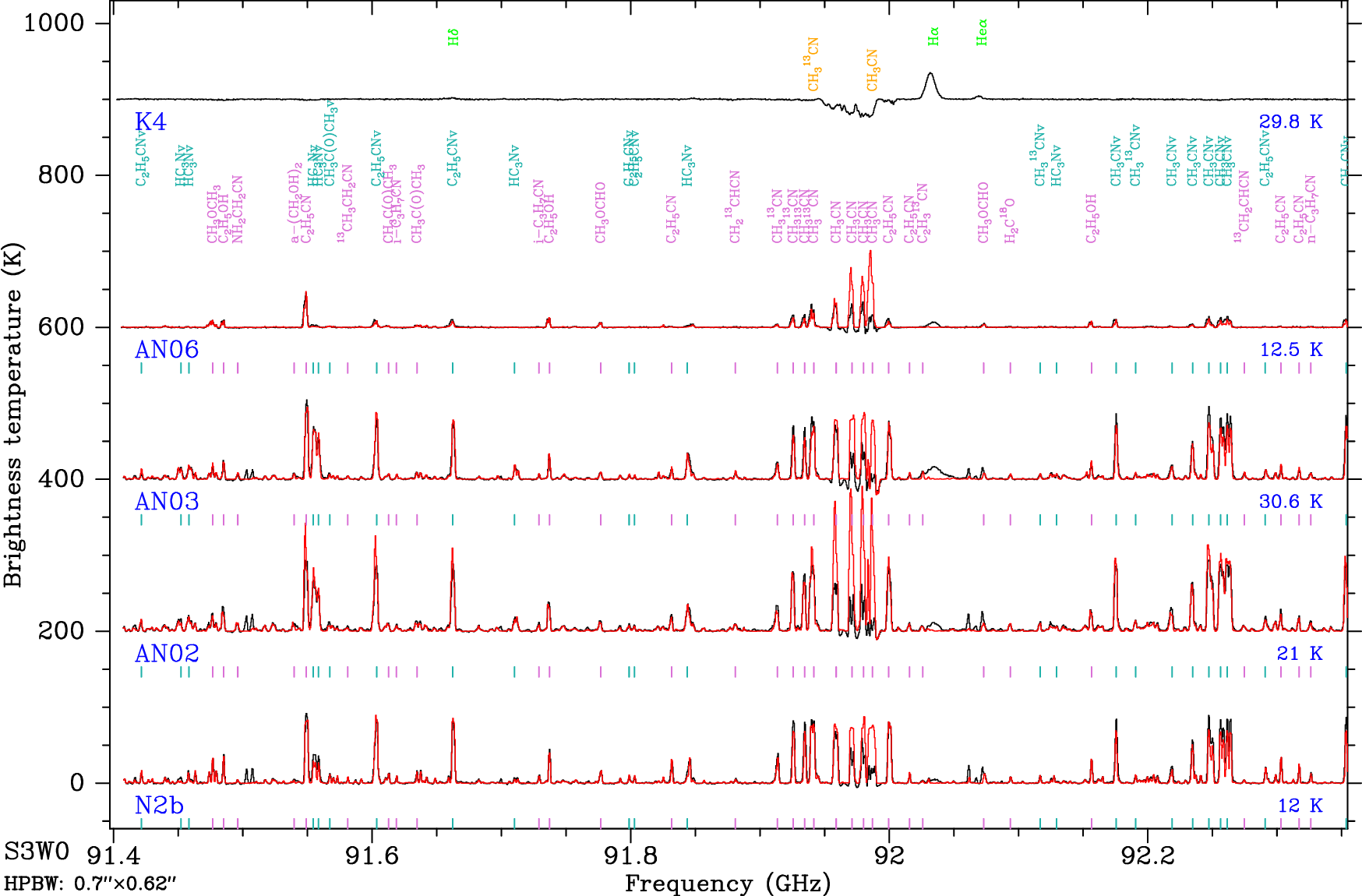}}}
\caption{continued.
}
\end{figure*}

\clearpage
\begin{figure*}
\addtocounter{figure}{-1}
\centerline{\resizebox{0.85\hsize}{!}{\includegraphics[angle=270]{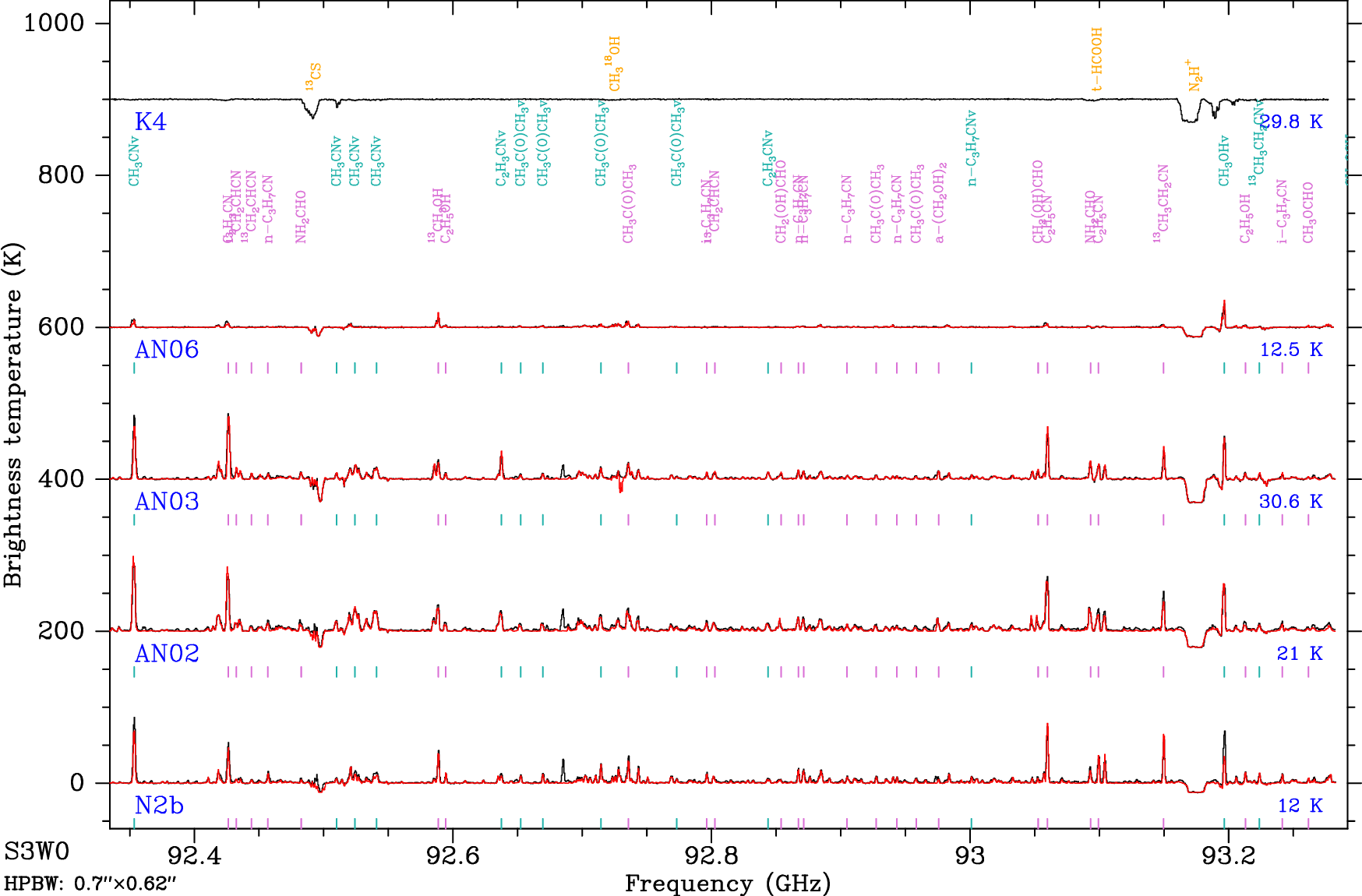}}}
\caption{continued.
}
\end{figure*}

\clearpage
\begin{figure*}
\addtocounter{figure}{-1}
\centerline{\resizebox{0.85\hsize}{!}{\includegraphics[angle=270]{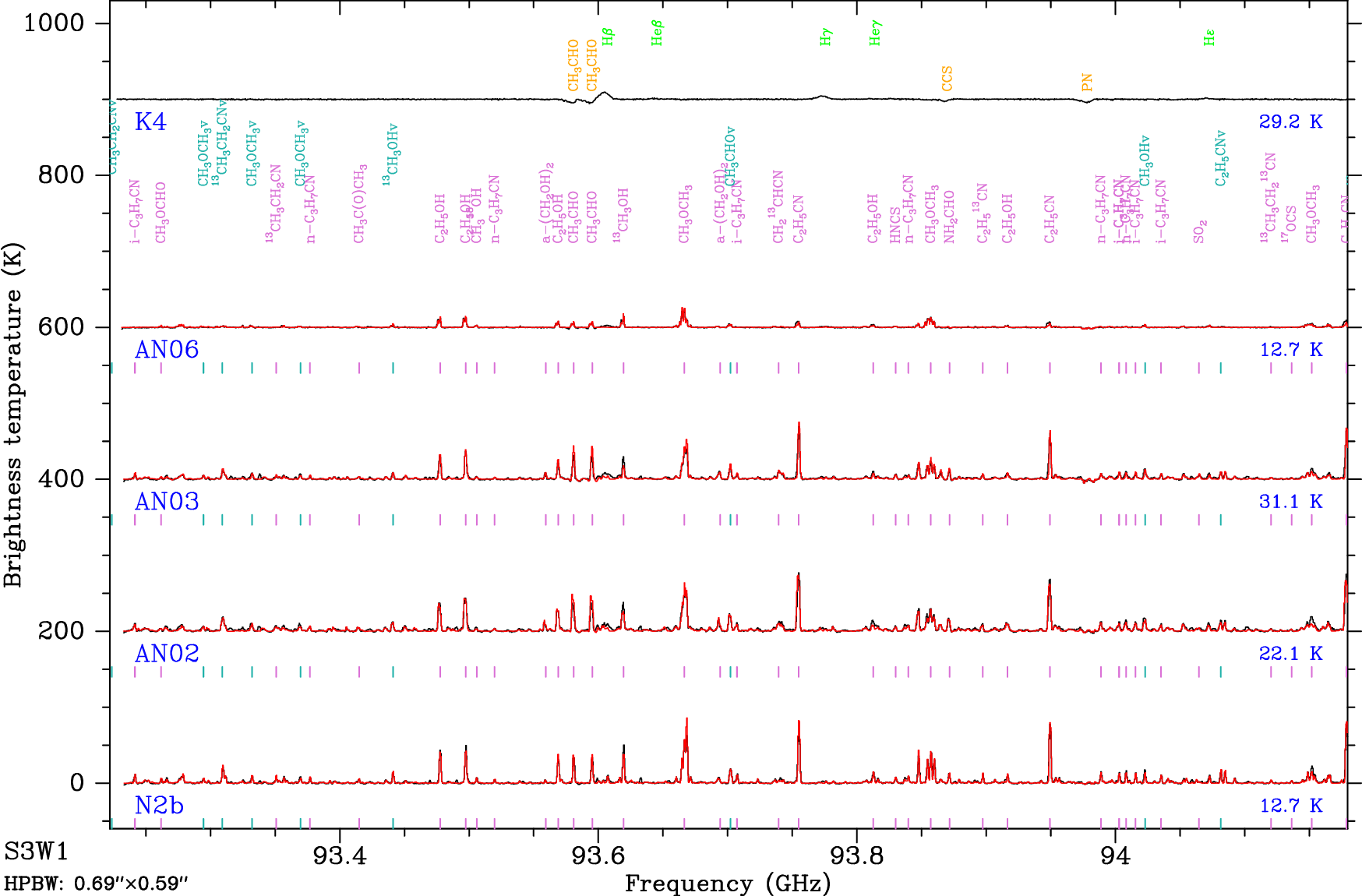}}}
\caption{continued.
}
\end{figure*}

\clearpage
\begin{figure*}
\addtocounter{figure}{-1}
\centerline{\resizebox{0.85\hsize}{!}{\includegraphics[angle=270]{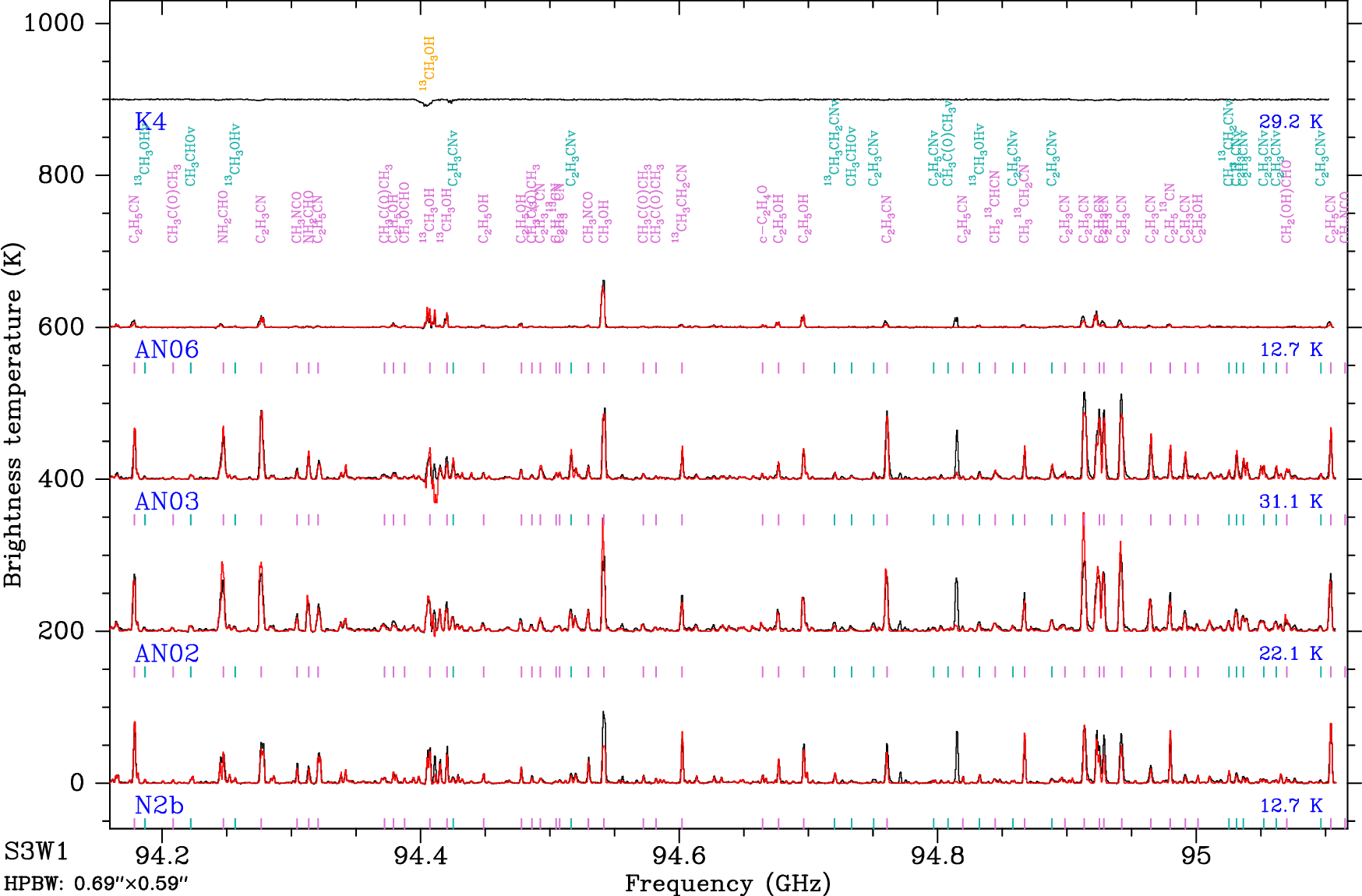}}}
\caption{continued.
}
\end{figure*}

\clearpage
\begin{figure*}
\addtocounter{figure}{-1}
\centerline{\resizebox{0.85\hsize}{!}{\includegraphics[angle=270]{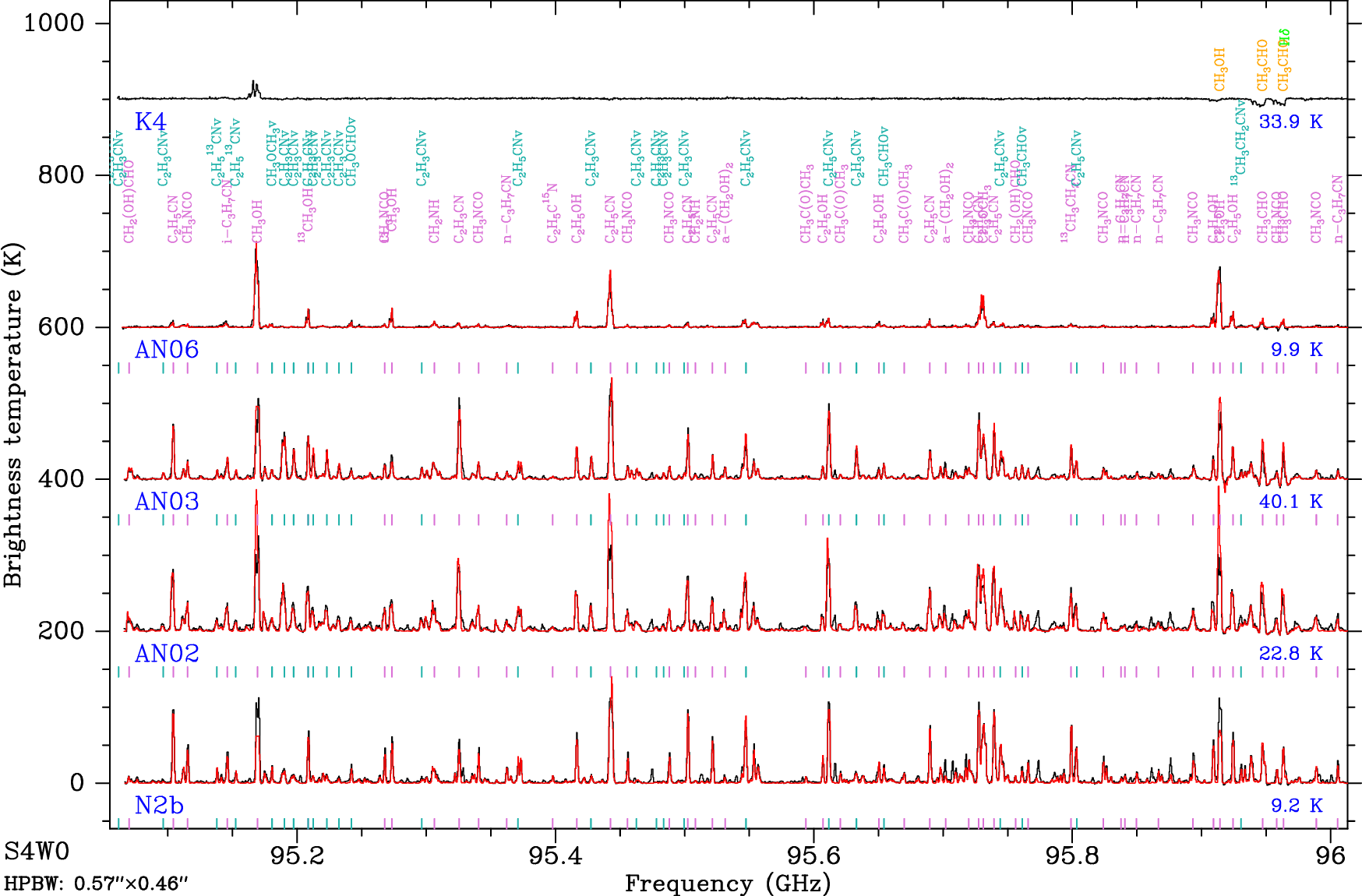}}}
\caption{continued.
}
\end{figure*}

\clearpage
\begin{figure*}
\addtocounter{figure}{-1}
\centerline{\resizebox{0.85\hsize}{!}{\includegraphics[angle=270]{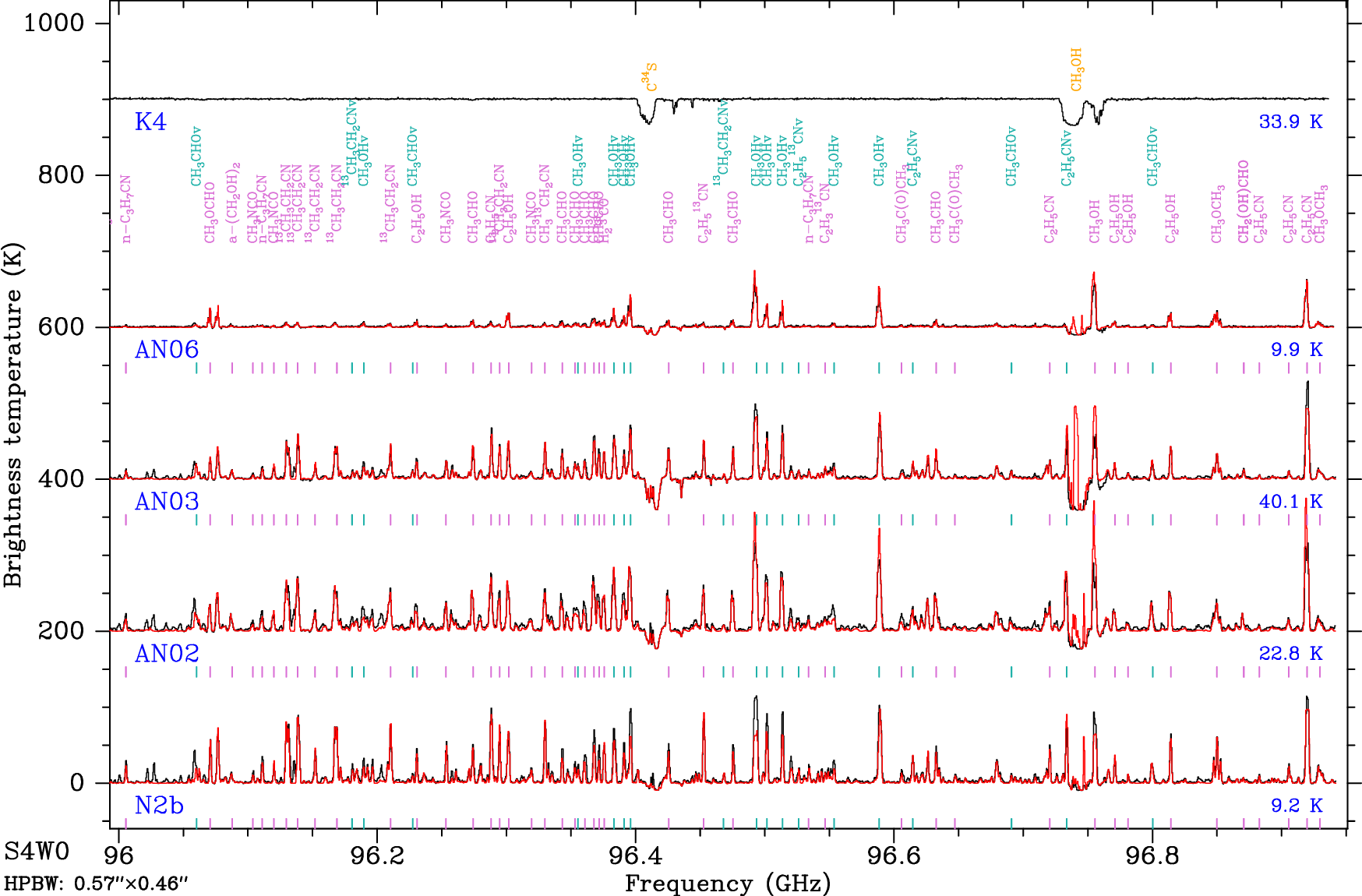}}}
\caption{continued.
}
\end{figure*}

\clearpage
\begin{figure*}
\addtocounter{figure}{-1}
\centerline{\resizebox{0.85\hsize}{!}{\includegraphics[angle=270]{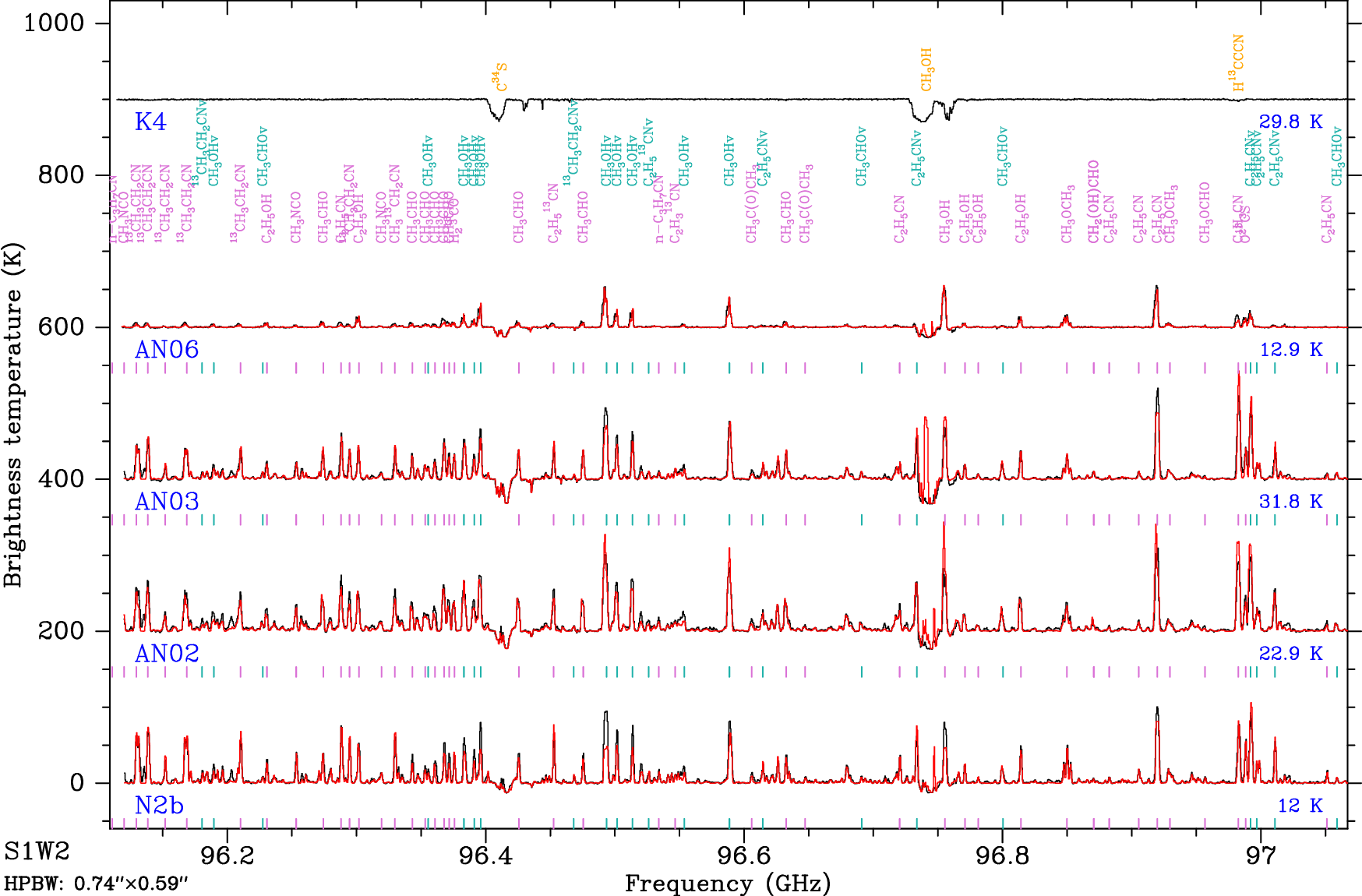}}}
\caption{continued.
}
\end{figure*}

\clearpage
\begin{figure*}
\addtocounter{figure}{-1}
\centerline{\resizebox{0.85\hsize}{!}{\includegraphics[angle=270]{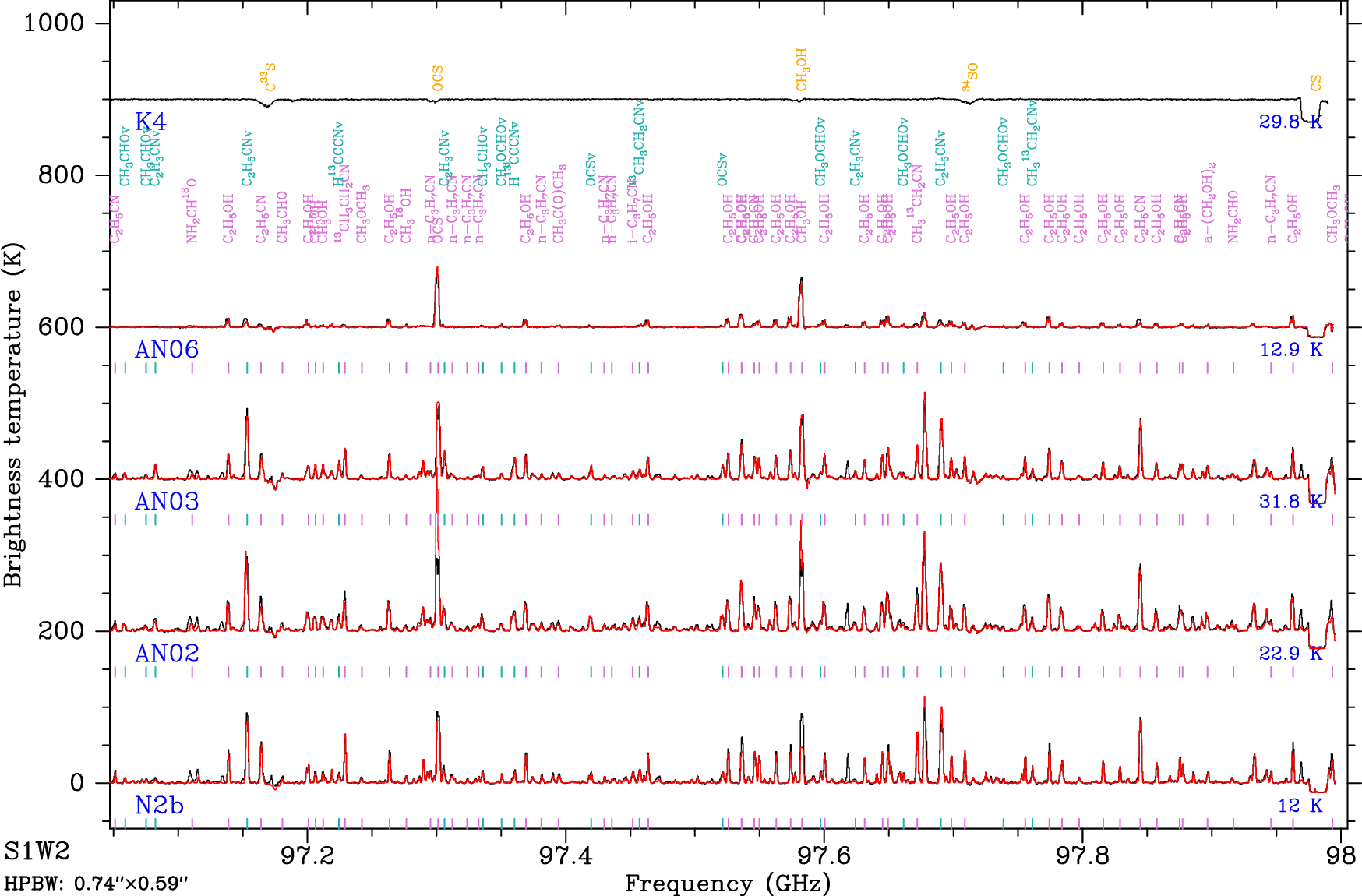}}}
\caption{continued.
}
\end{figure*}

\clearpage
\begin{figure*}
\addtocounter{figure}{-1}
\centerline{\resizebox{0.85\hsize}{!}{\includegraphics[angle=270]{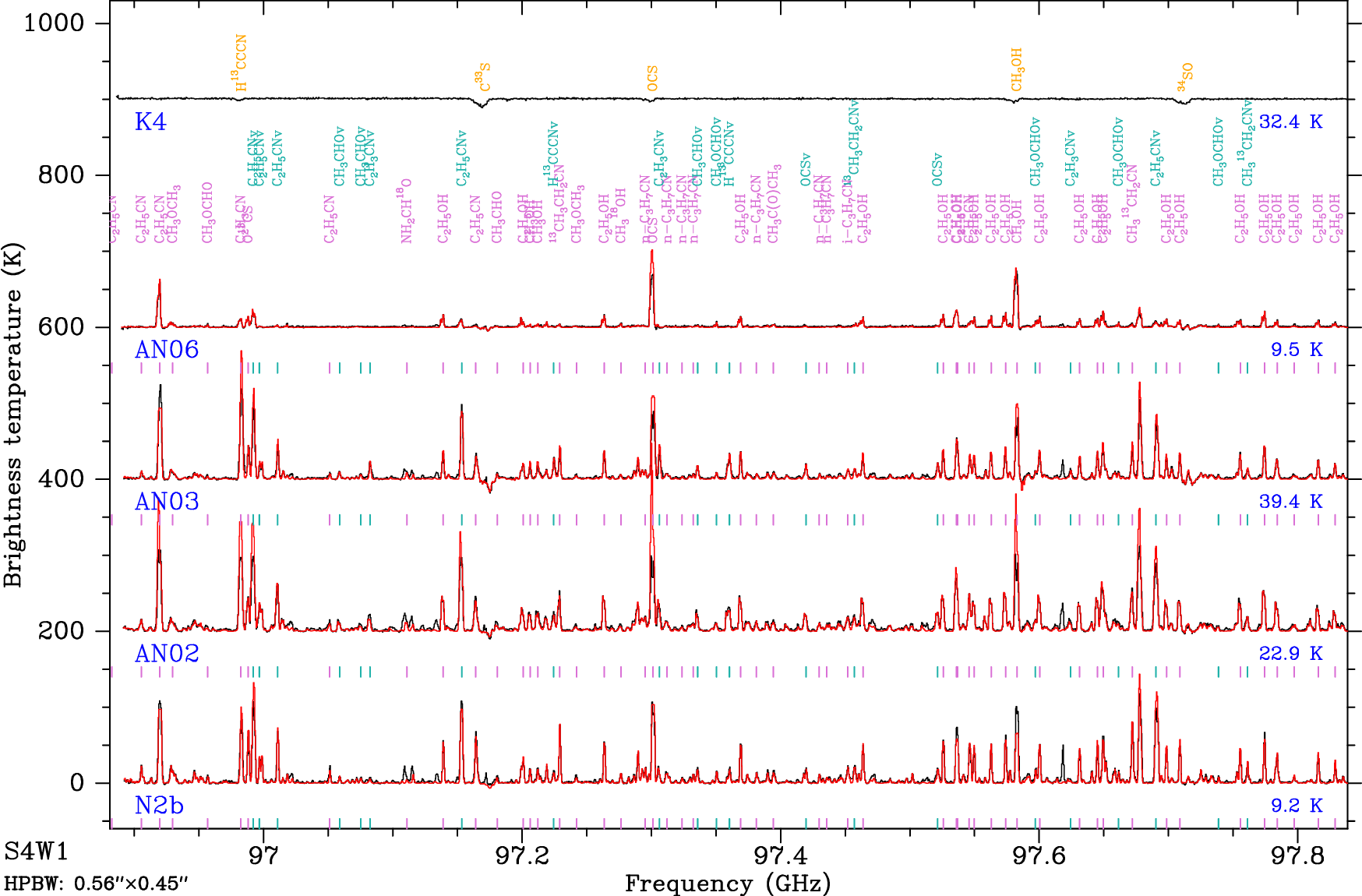}}}
\caption{continued.
}
\end{figure*}

\clearpage
\begin{figure*}
\addtocounter{figure}{-1}
\centerline{\resizebox{0.85\hsize}{!}{\includegraphics[angle=270]{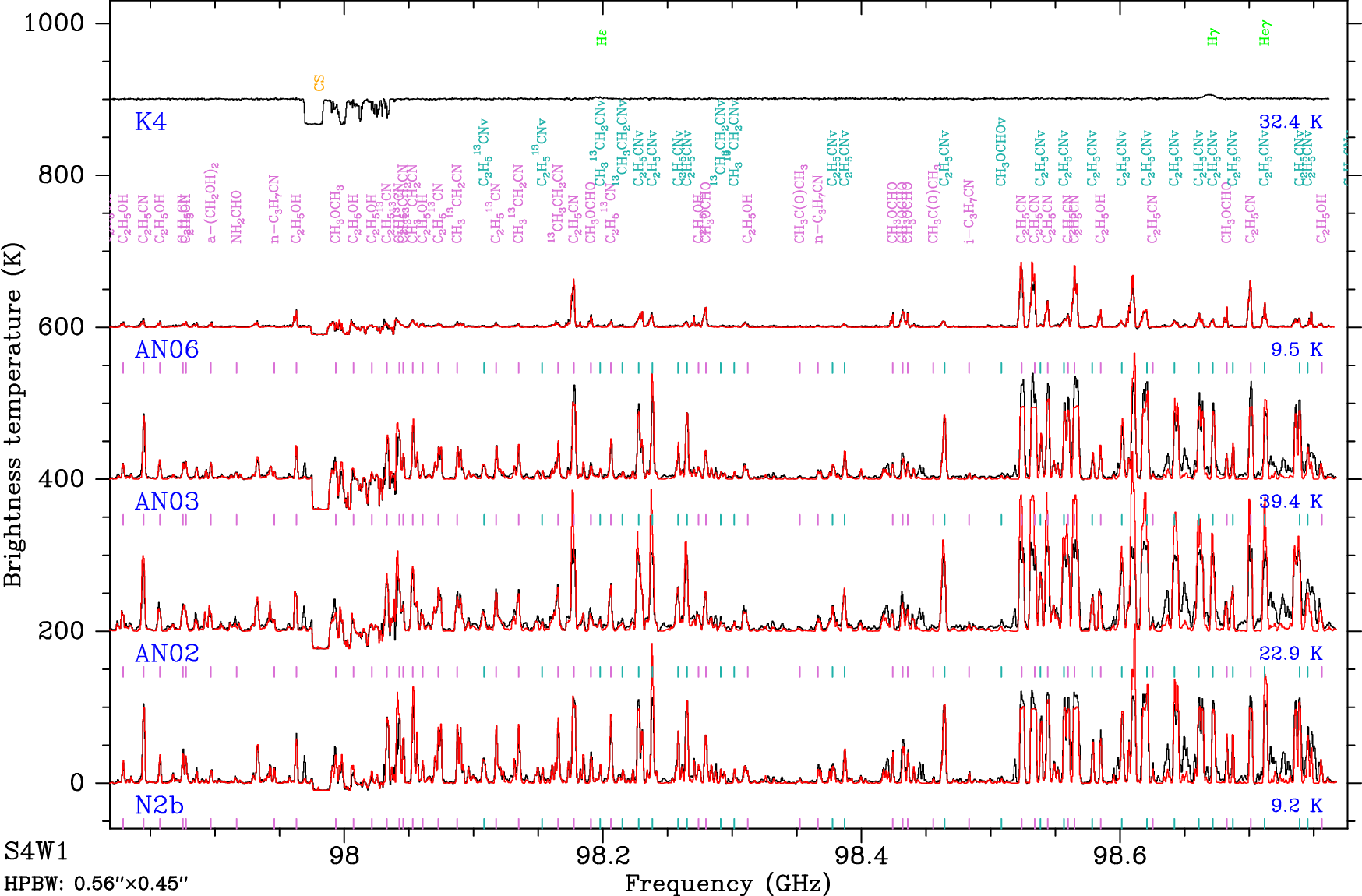}}}
\caption{continued.
}
\end{figure*}

\clearpage
\begin{figure*}
\addtocounter{figure}{-1}
\centerline{\resizebox{0.85\hsize}{!}{\includegraphics[angle=270]{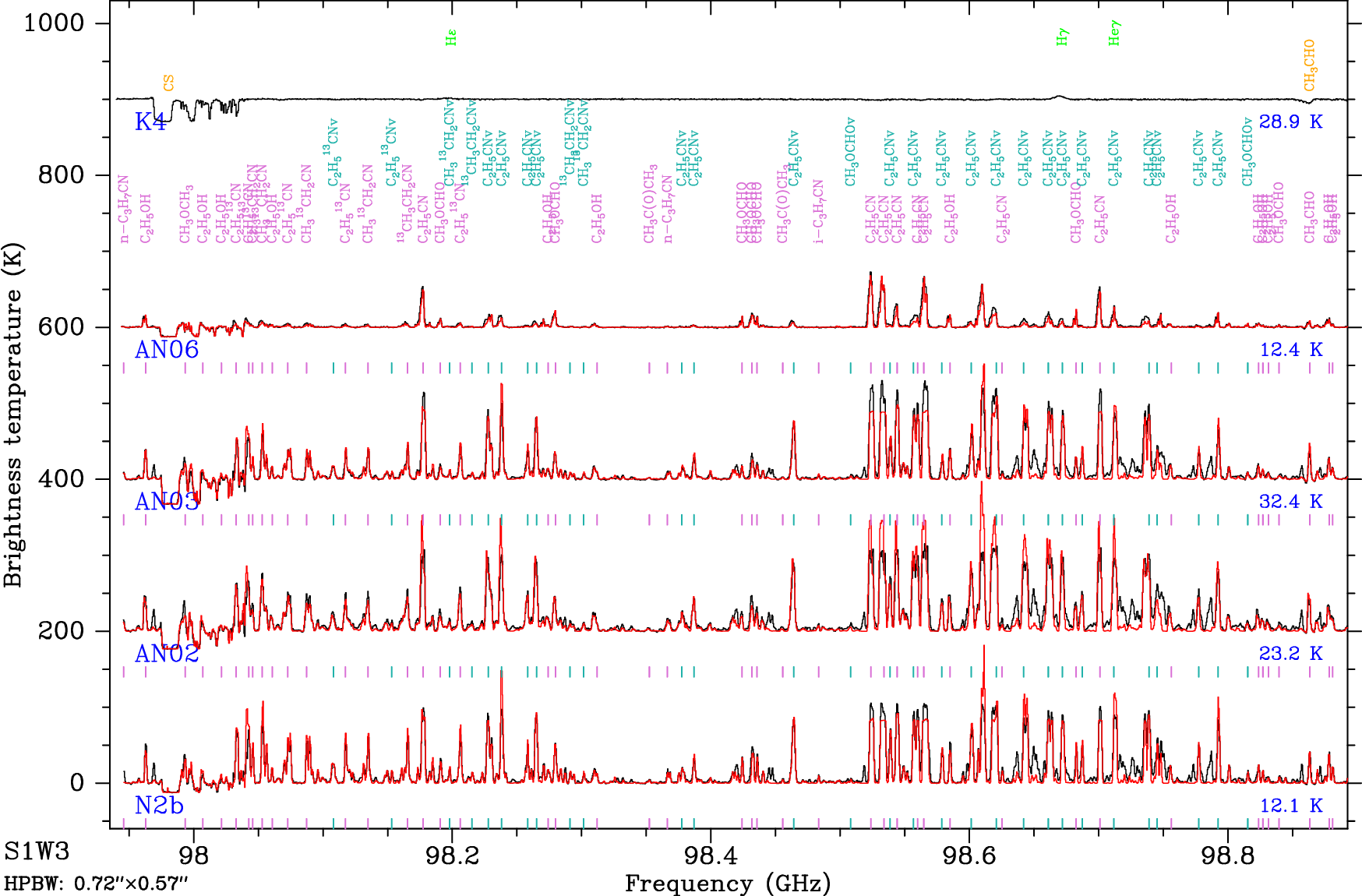}}}
\caption{continued.
}
\end{figure*}

\clearpage
\begin{figure*}
\addtocounter{figure}{-1}
\centerline{\resizebox{0.85\hsize}{!}{\includegraphics[angle=270]{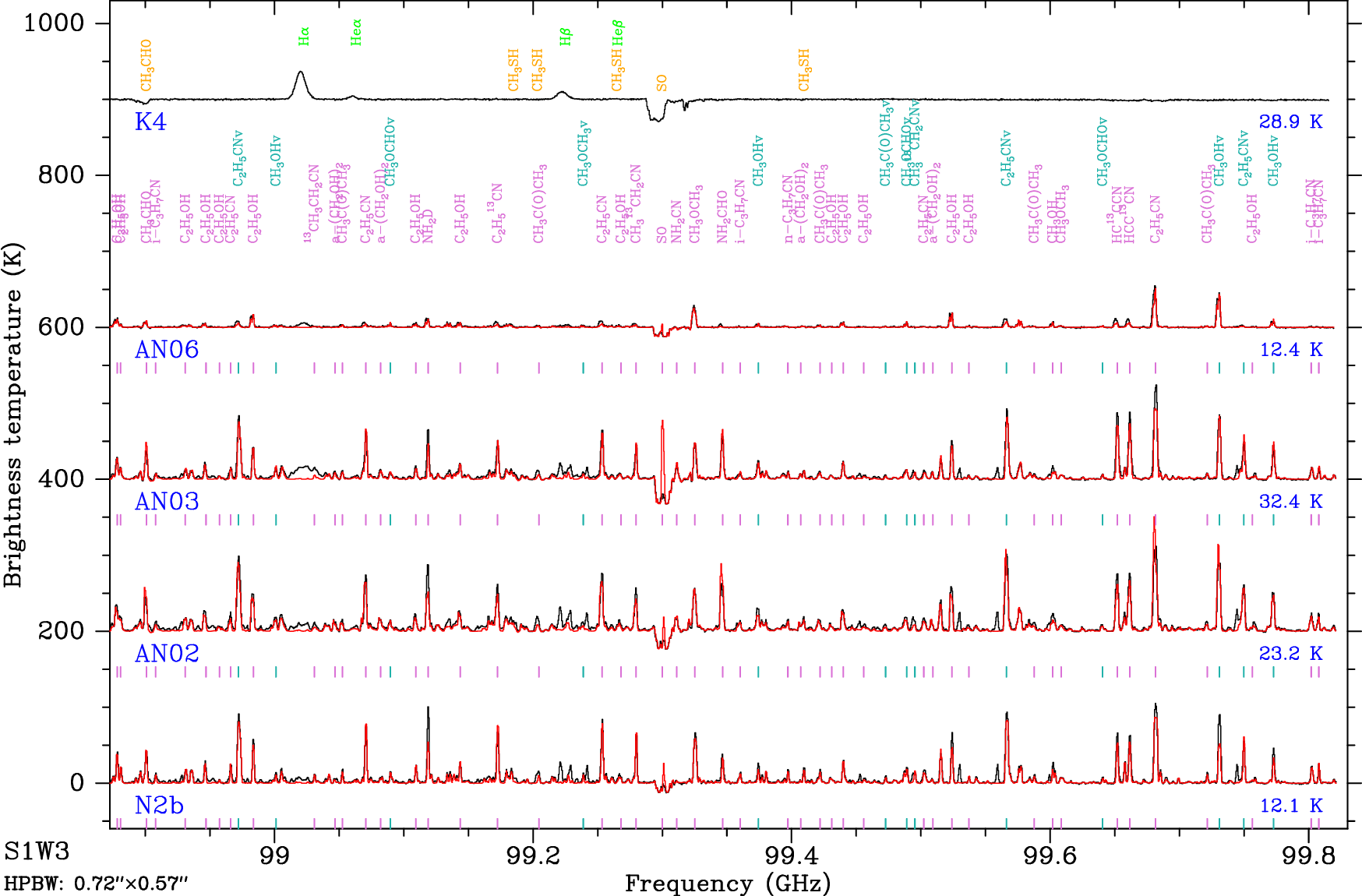}}}
\caption{continued.
}
\end{figure*}

\clearpage
\begin{figure*}
\addtocounter{figure}{-1}
\centerline{\resizebox{0.85\hsize}{!}{\includegraphics[angle=270]{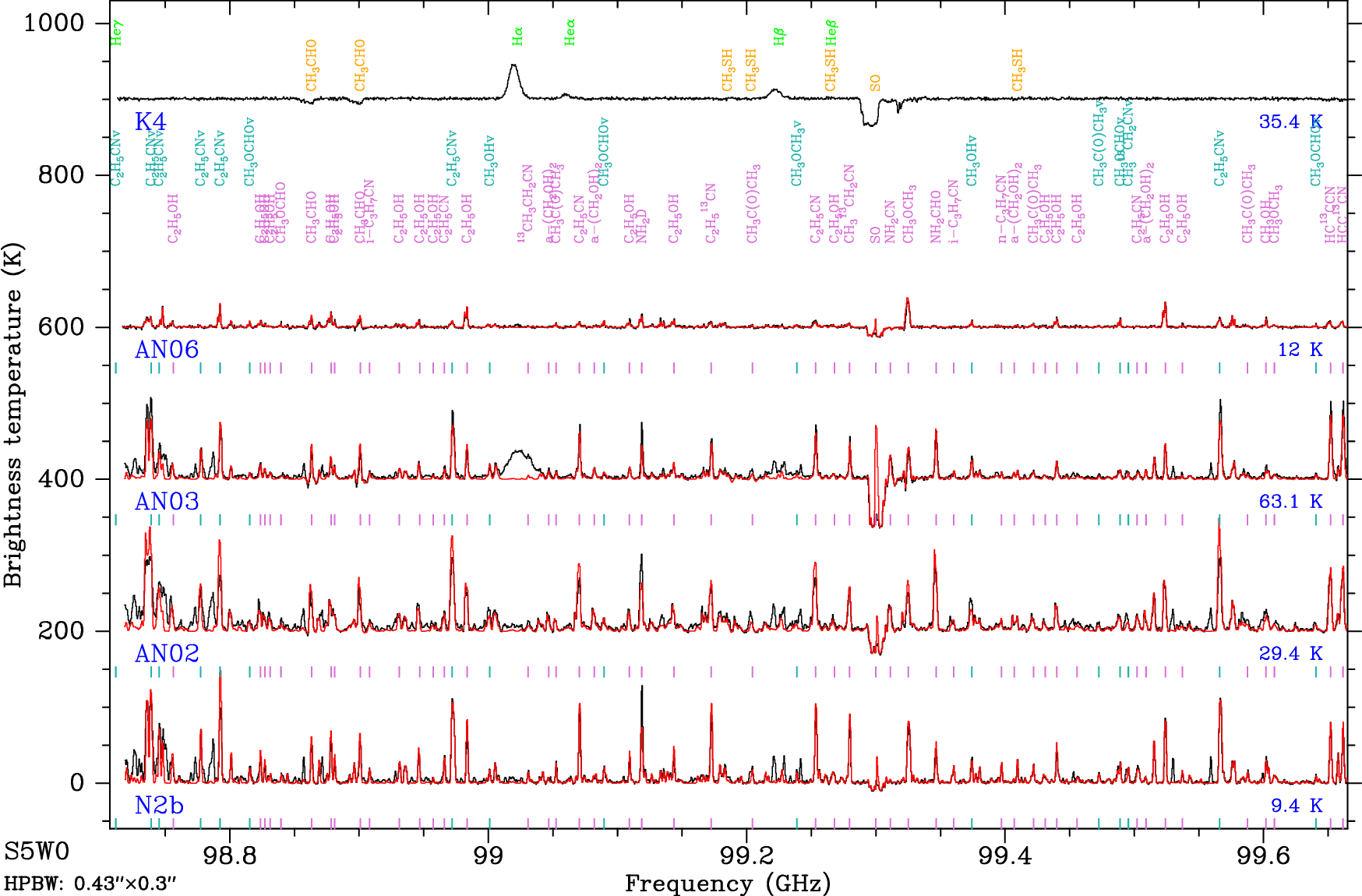}}}
\caption{continued.
}
\end{figure*}

\clearpage
\begin{figure*}
\addtocounter{figure}{-1}
\centerline{\resizebox{0.85\hsize}{!}{\includegraphics[angle=270]{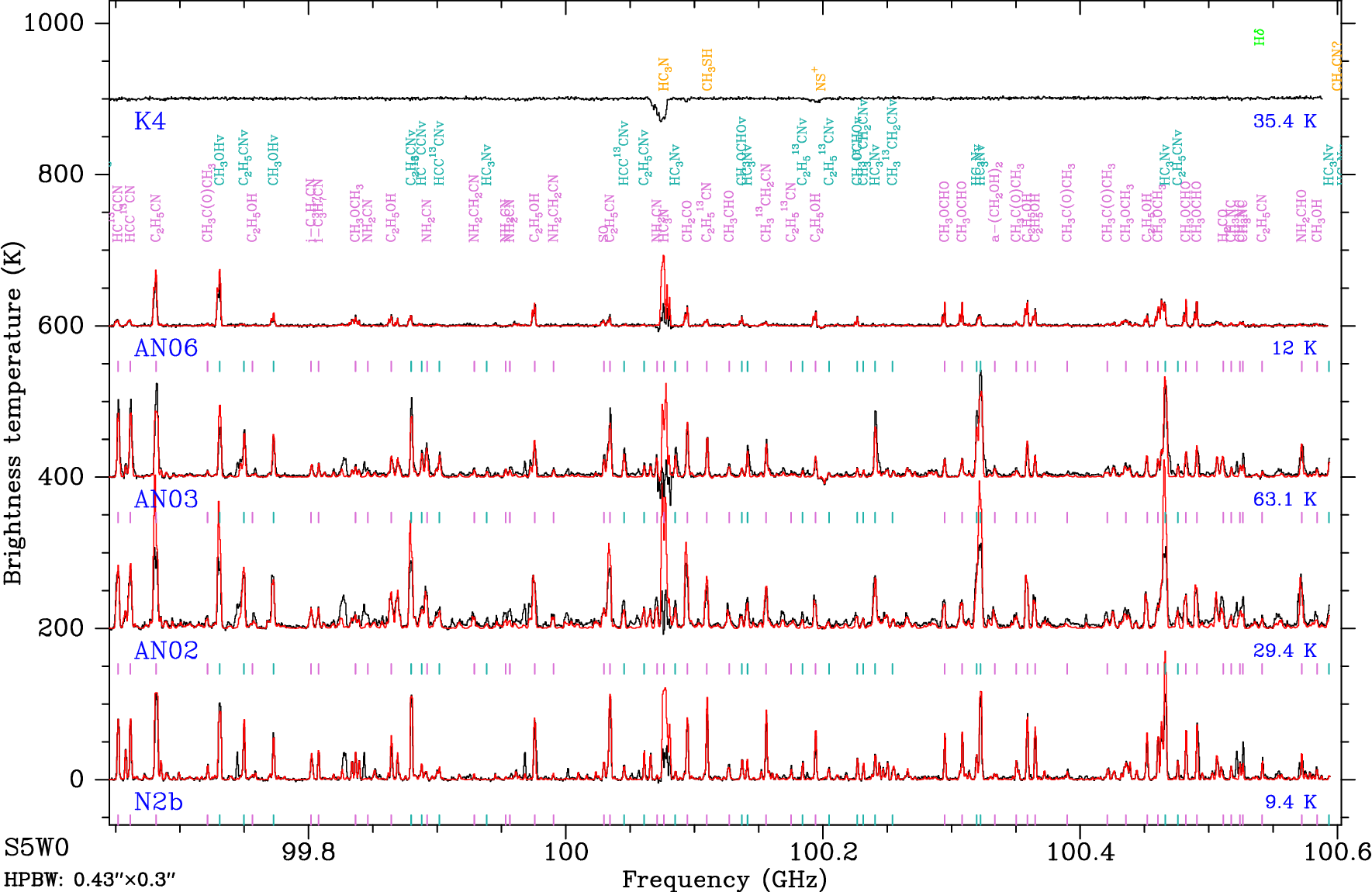}}}
\caption{continued.
}
\end{figure*}

\clearpage
\begin{figure*}
\addtocounter{figure}{-1}
\centerline{\resizebox{0.85\hsize}{!}{\includegraphics[angle=270]{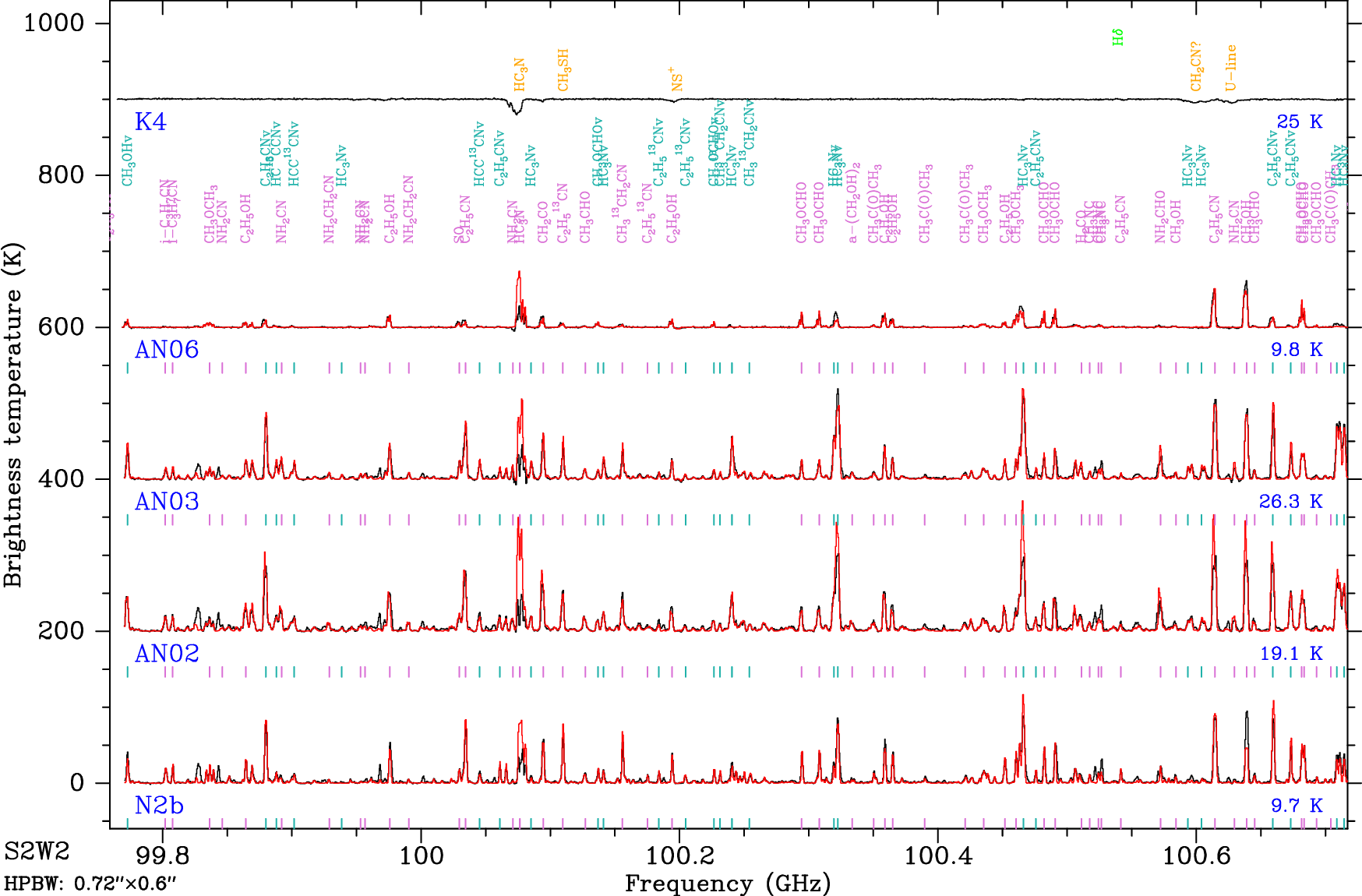}}}
\caption{continued.
}
\end{figure*}

\clearpage
\begin{figure*}
\addtocounter{figure}{-1}
\centerline{\resizebox{0.85\hsize}{!}{\includegraphics[angle=270]{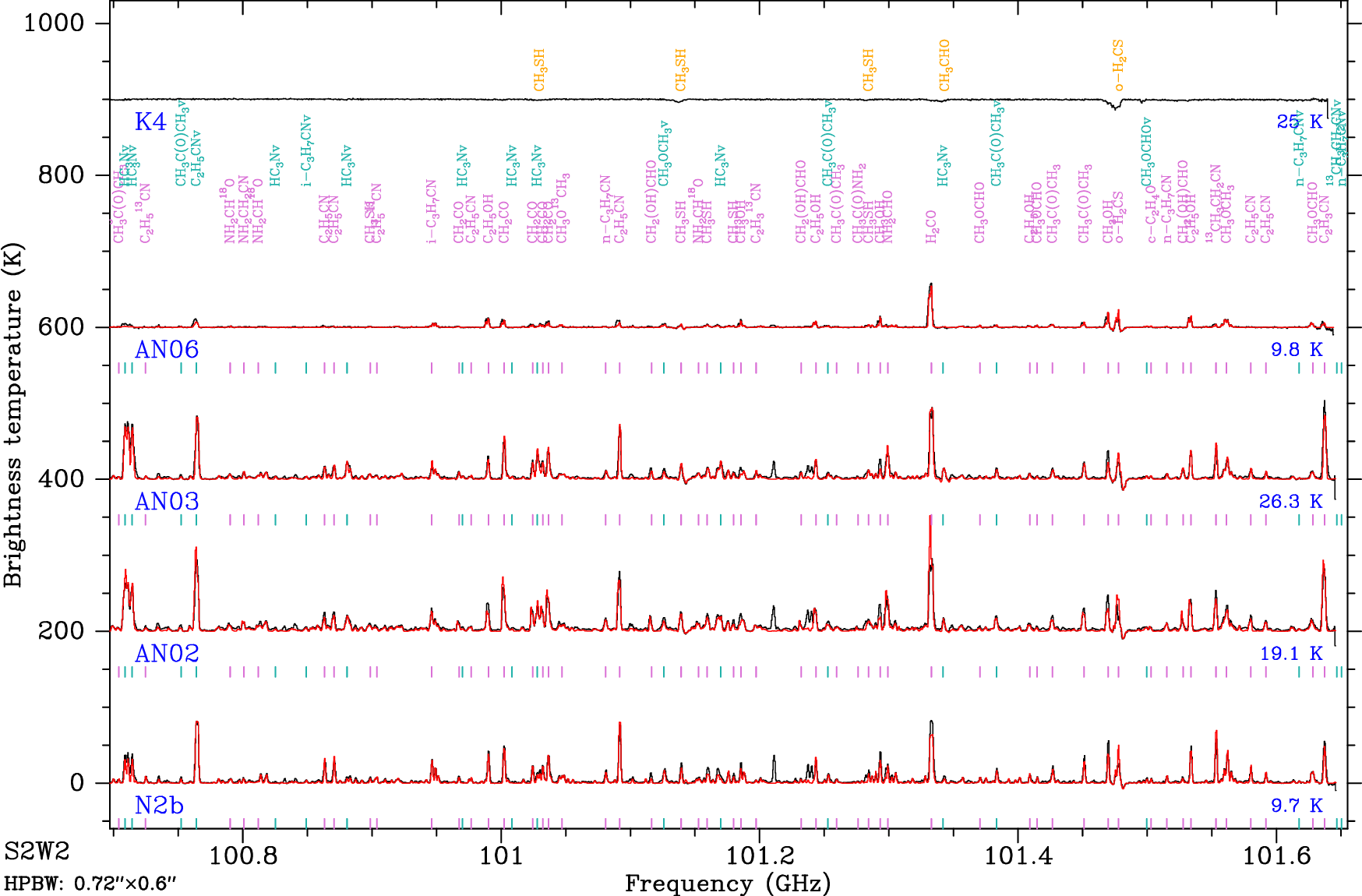}}}
\caption{continued.
}
\end{figure*}

\clearpage
\begin{figure*}
\addtocounter{figure}{-1}
\centerline{\resizebox{0.85\hsize}{!}{\includegraphics[angle=270]{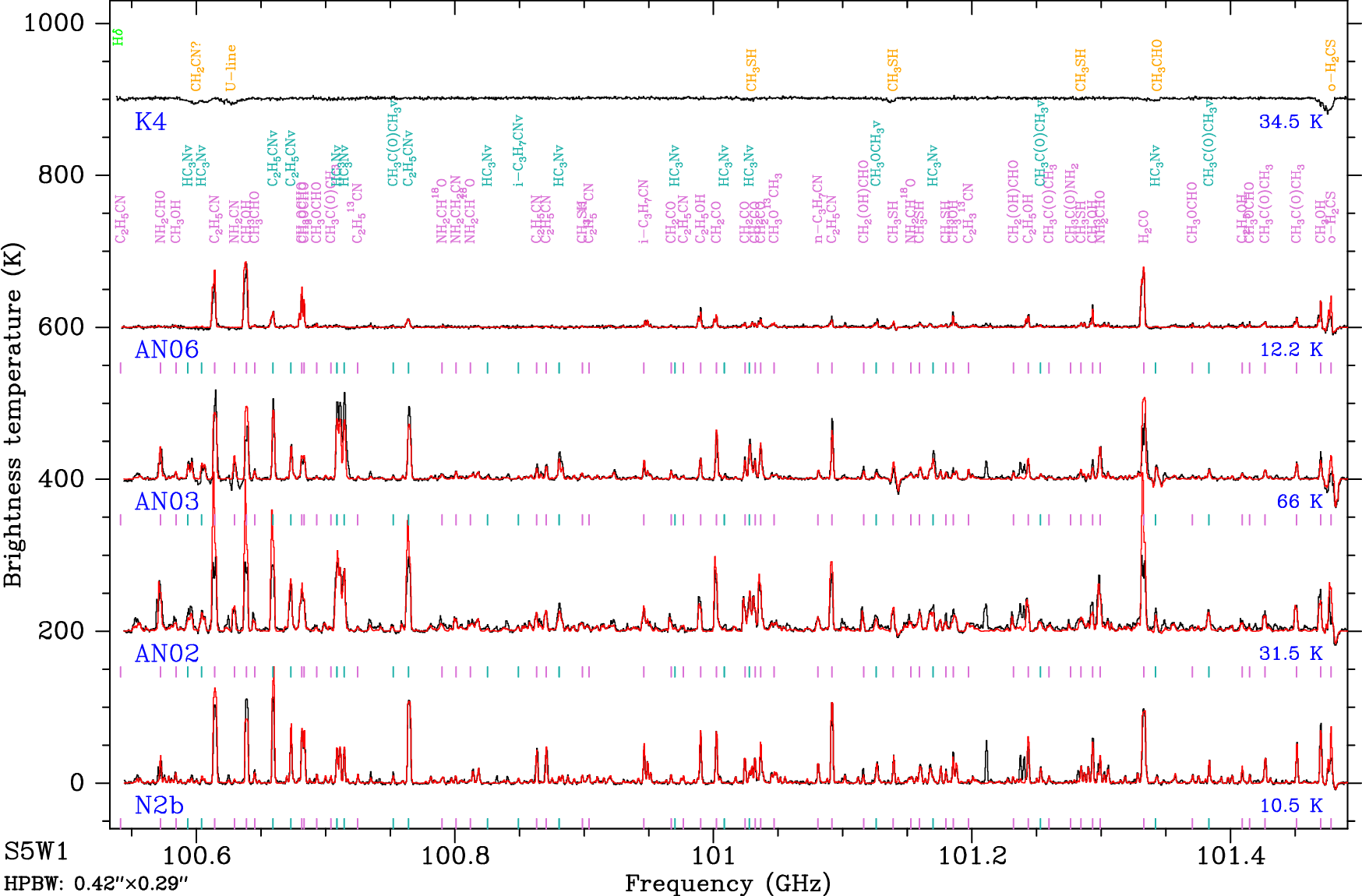}}}
\caption{continued.
}
\end{figure*}

\clearpage
\begin{figure*}
\addtocounter{figure}{-1}
\centerline{\resizebox{0.85\hsize}{!}{\includegraphics[angle=270]{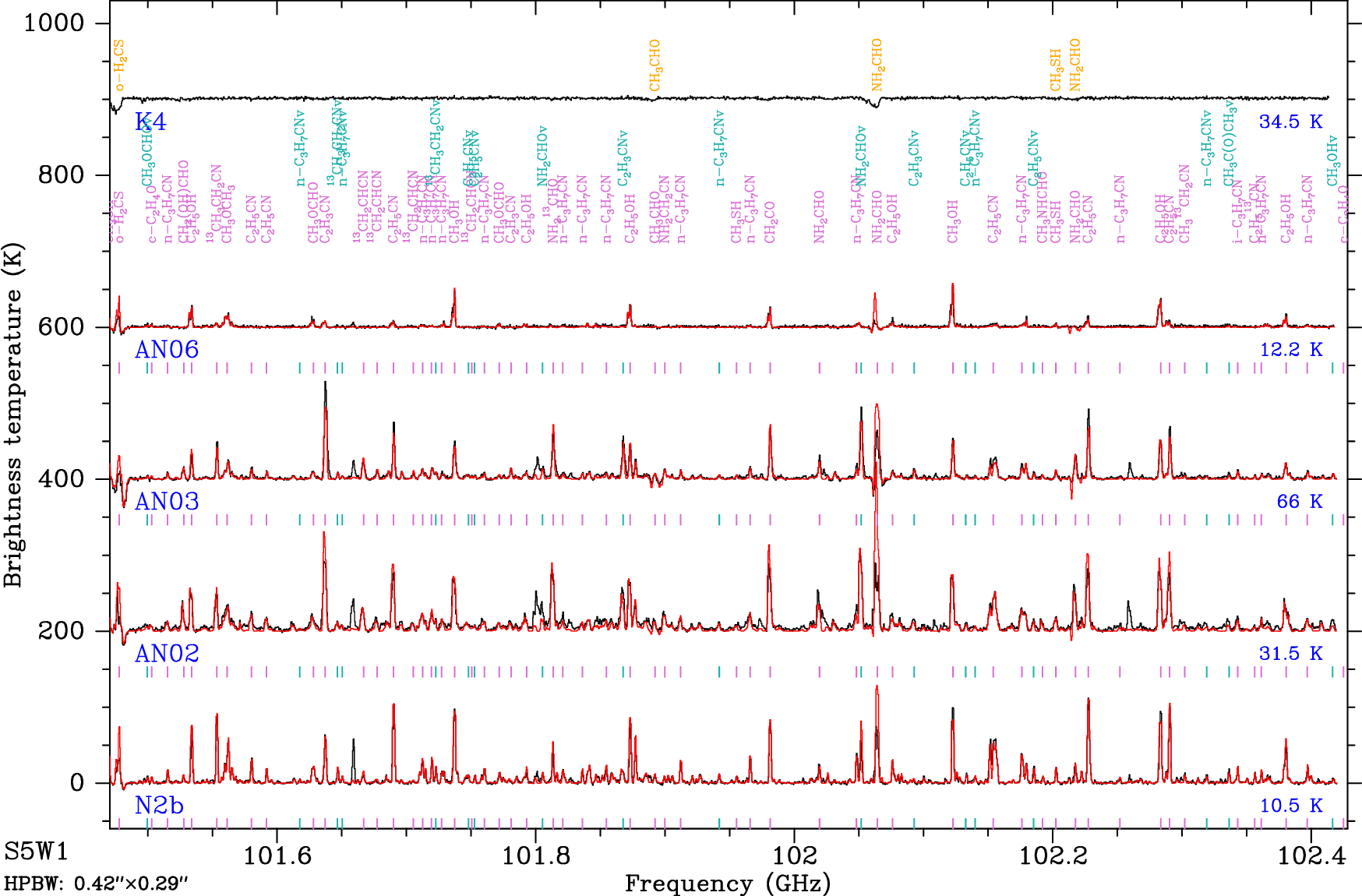}}}
\caption{continued.
}
\end{figure*}

\clearpage
\begin{figure*}
\addtocounter{figure}{-1}
\centerline{\resizebox{0.85\hsize}{!}{\includegraphics[angle=270]{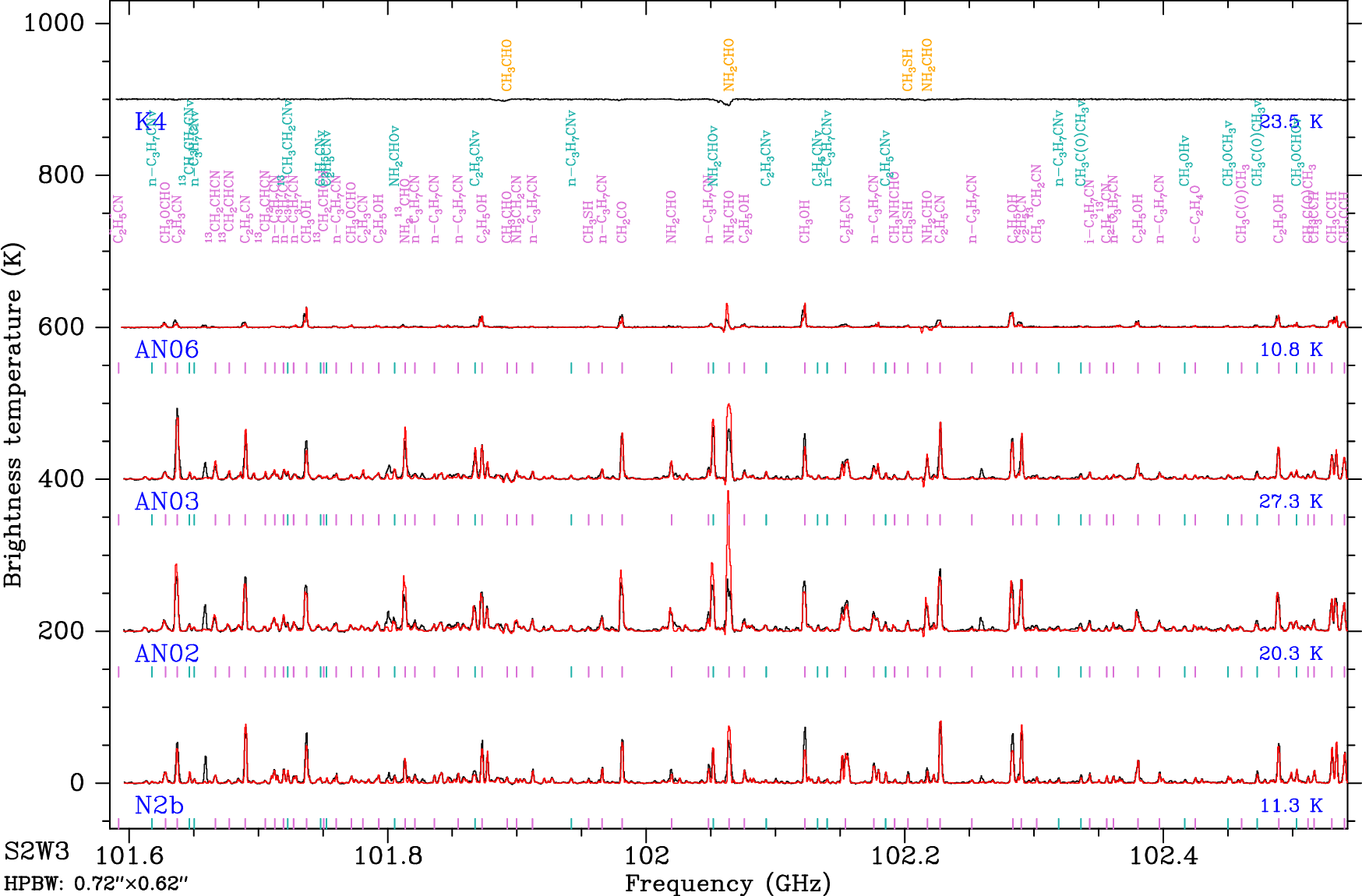}}}
\caption{continued.
}
\end{figure*}

\clearpage
\begin{figure*}
\addtocounter{figure}{-1}
\centerline{\resizebox{0.85\hsize}{!}{\includegraphics[angle=270]{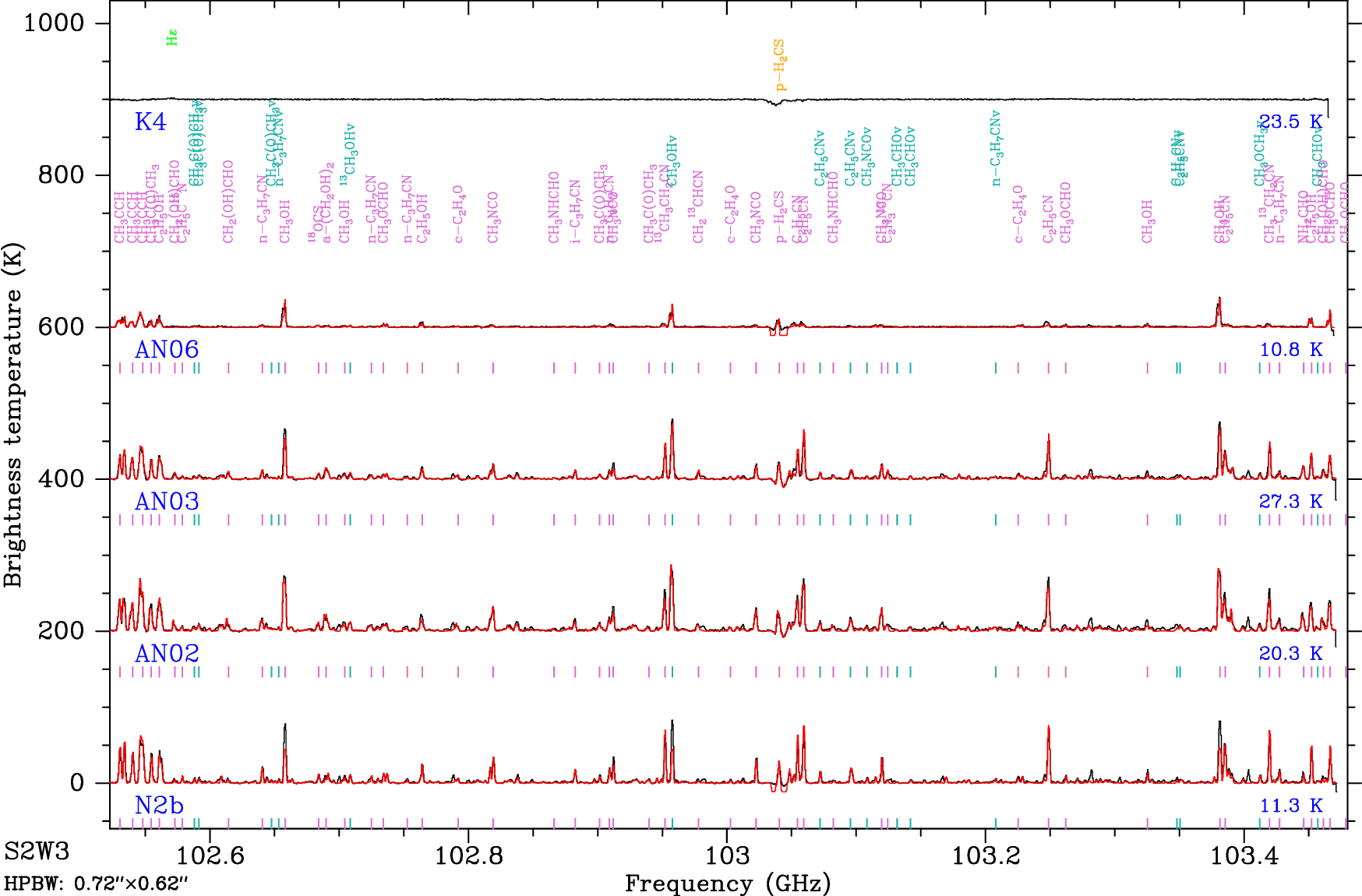}}}
\caption{continued.
}
\end{figure*}

\clearpage
\begin{figure*}
\addtocounter{figure}{-1}
\centerline{\resizebox{0.85\hsize}{!}{\includegraphics[angle=270]{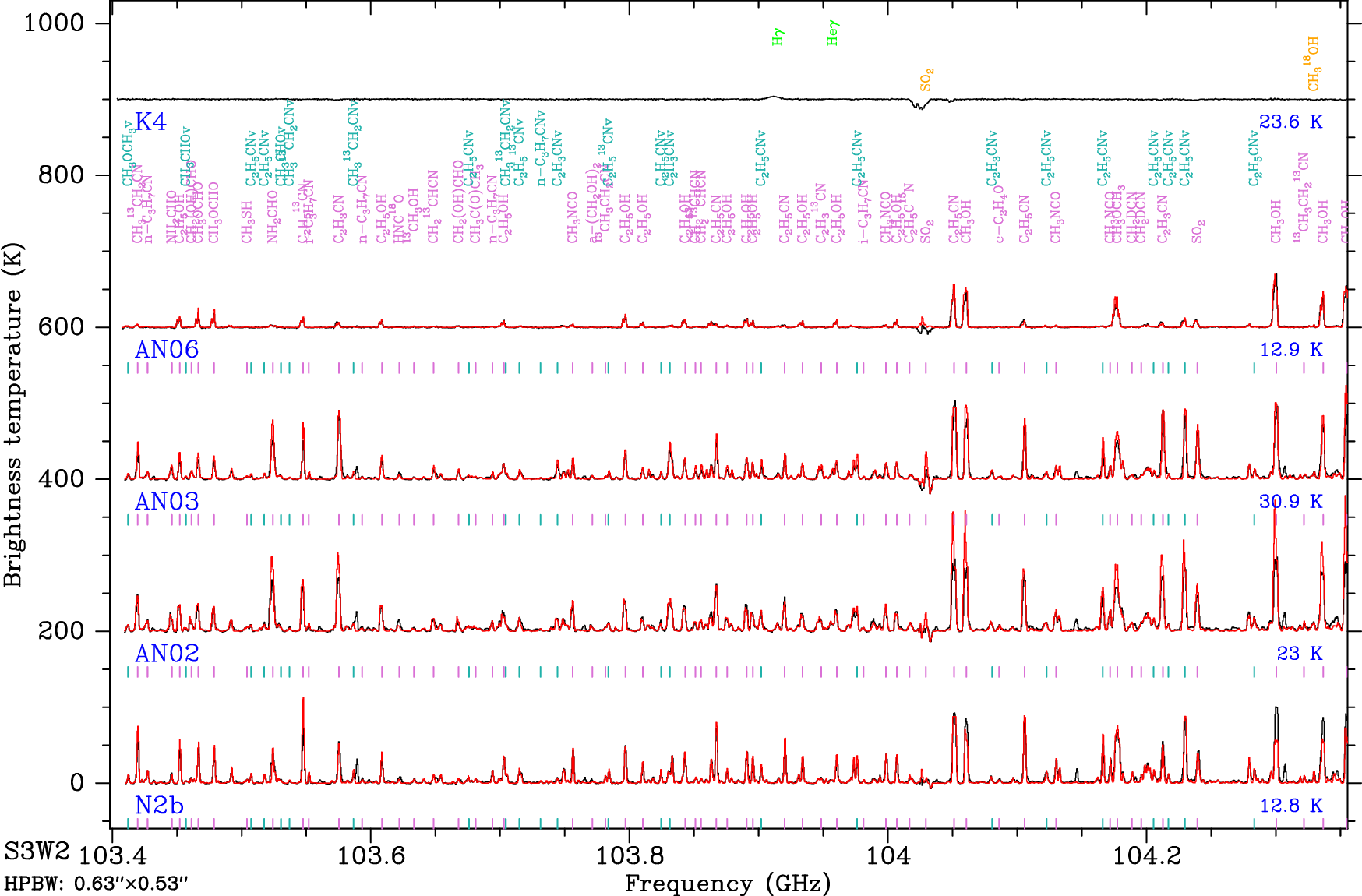}}}
\caption{continued.
}
\end{figure*}

\clearpage
\begin{figure*}
\addtocounter{figure}{-1}
\centerline{\resizebox{0.85\hsize}{!}{\includegraphics[angle=270]{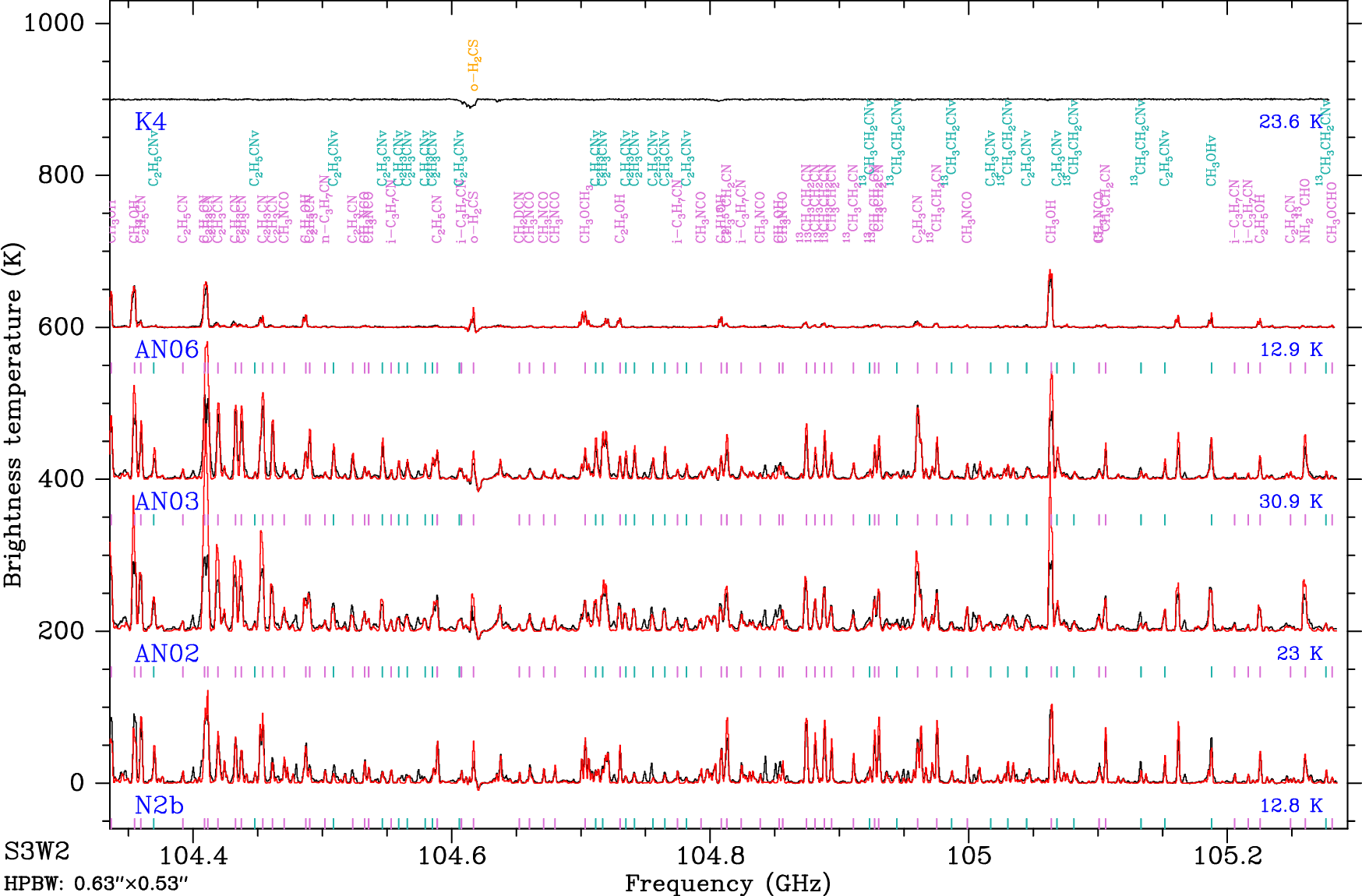}}}
\caption{continued.
}
\end{figure*}

\clearpage
\begin{figure*}
\addtocounter{figure}{-1}
\centerline{\resizebox{0.85\hsize}{!}{\includegraphics[angle=270]{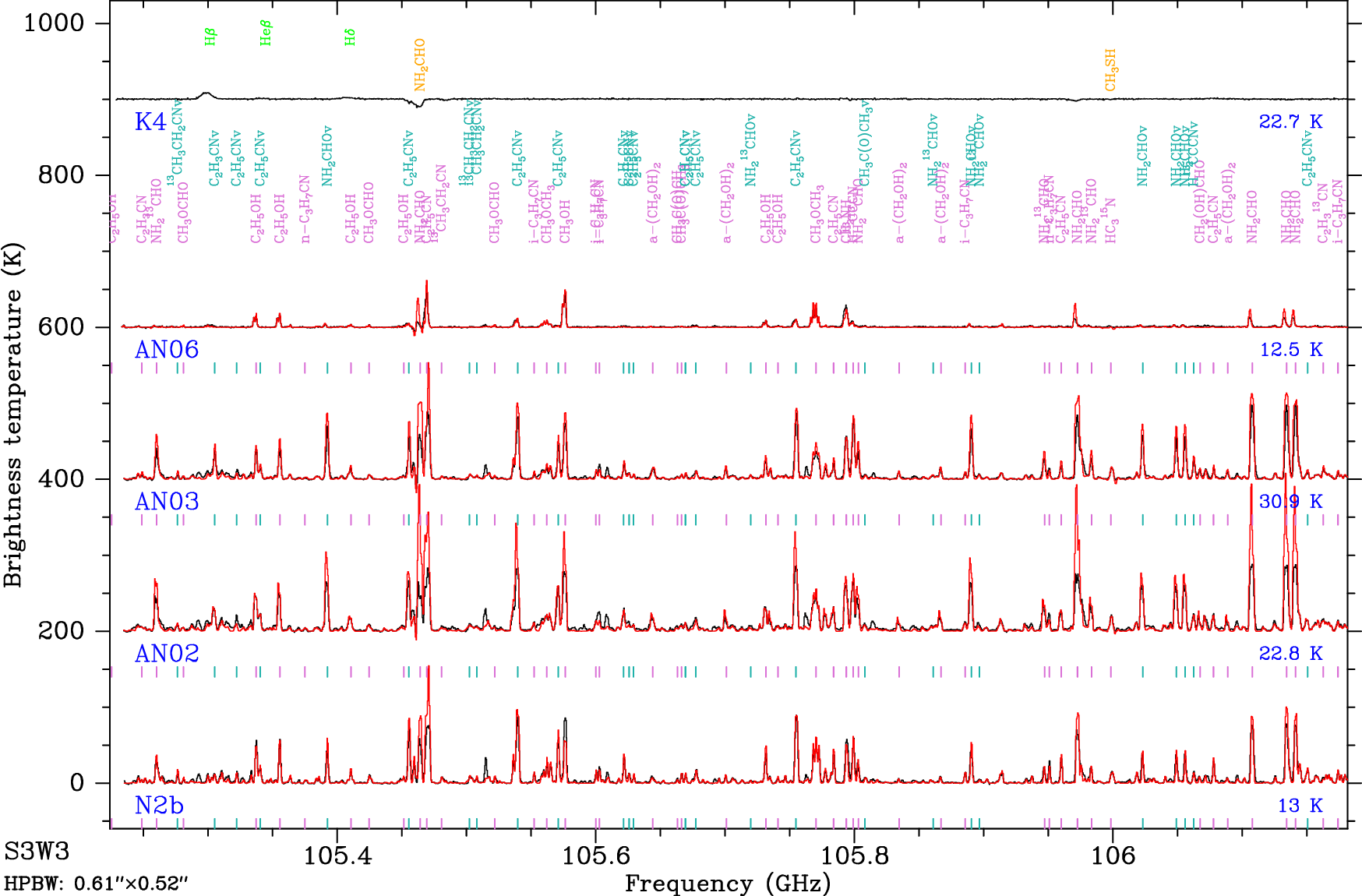}}}
\caption{continued.
}
\end{figure*}

\clearpage
\begin{figure*}
\addtocounter{figure}{-1}
\centerline{\resizebox{0.85\hsize}{!}{\includegraphics[angle=270]{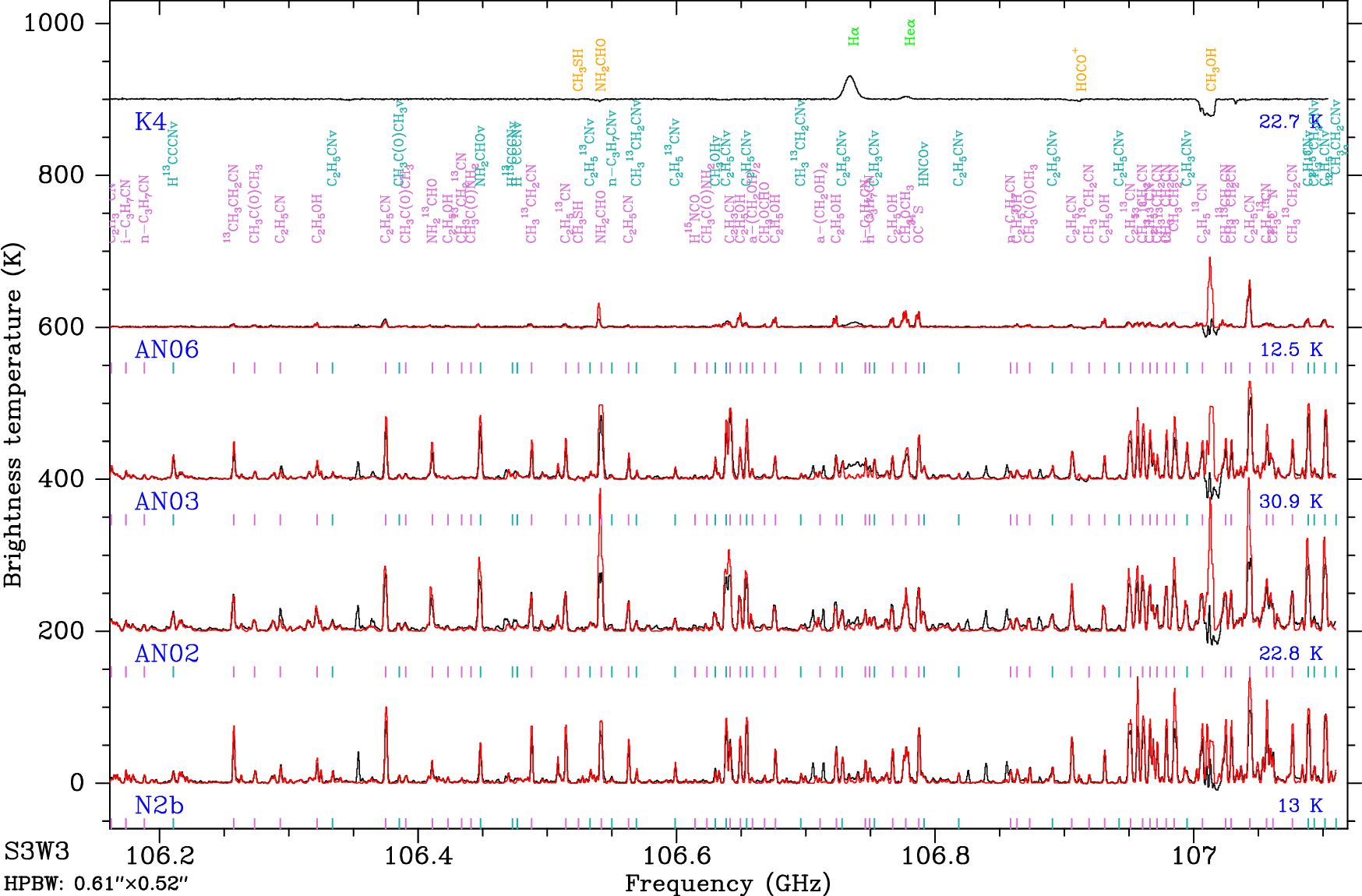}}}
\caption{continued.
}
\end{figure*}

\clearpage
\begin{figure*}
\addtocounter{figure}{-1}
\centerline{\resizebox{0.85\hsize}{!}{\includegraphics[angle=270]{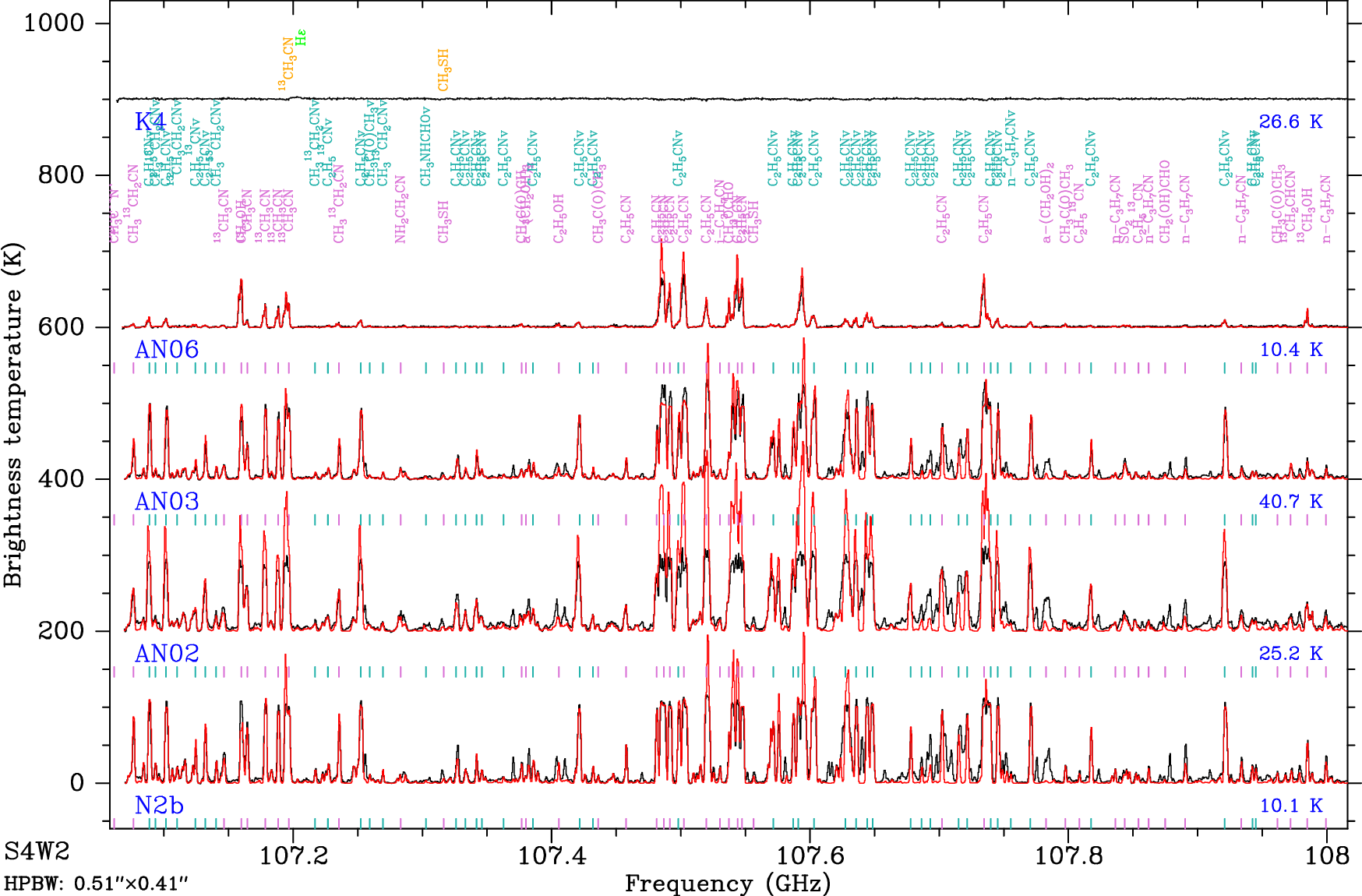}}}
\caption{continued.
}
\end{figure*}

\clearpage
\begin{figure*}
\addtocounter{figure}{-1}
\centerline{\resizebox{0.85\hsize}{!}{\includegraphics[angle=270]{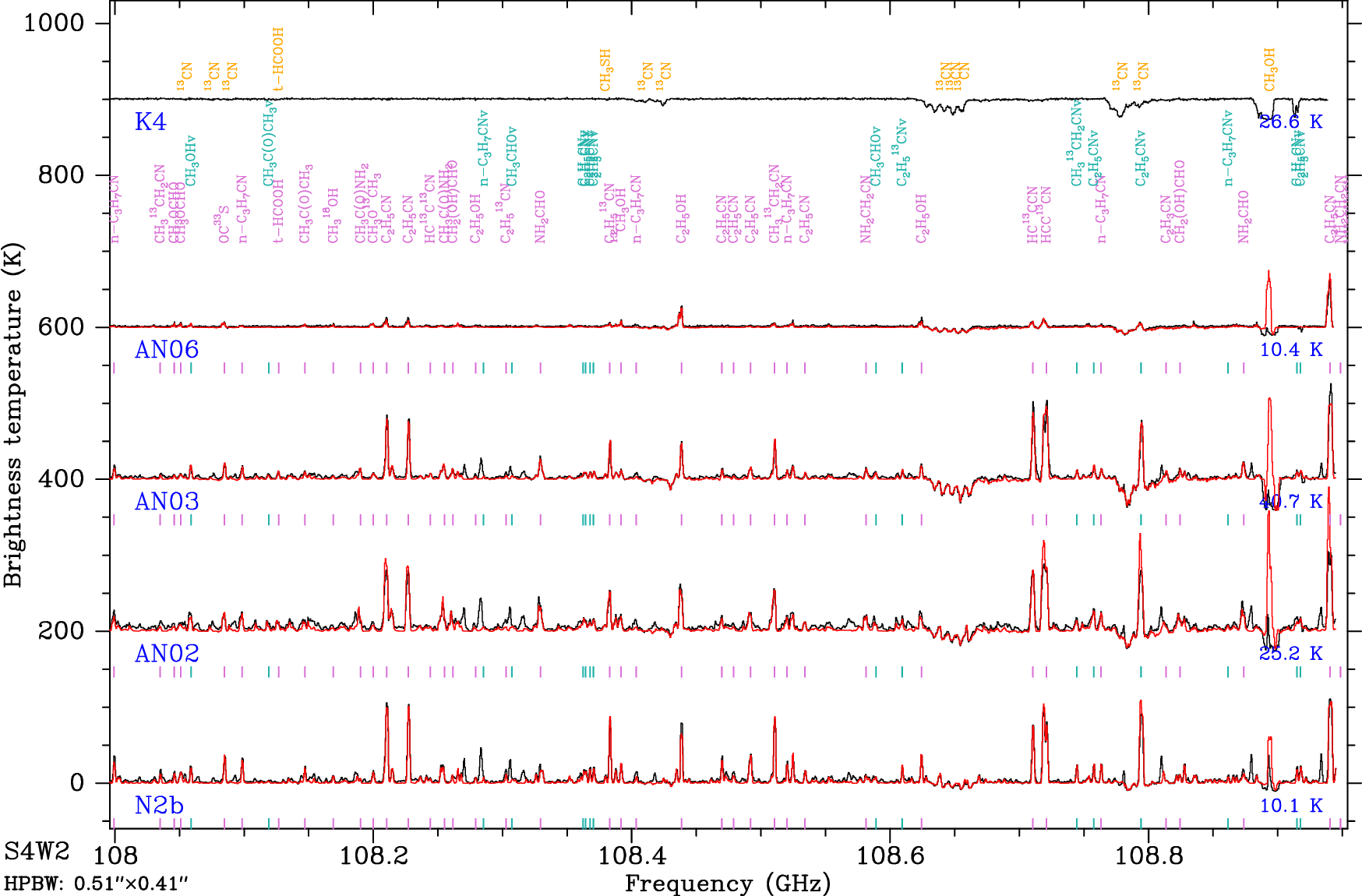}}}
\caption{continued.
}
\end{figure*}

\clearpage
\begin{figure*}
\addtocounter{figure}{-1}
\centerline{\resizebox{0.85\hsize}{!}{\includegraphics[angle=270]{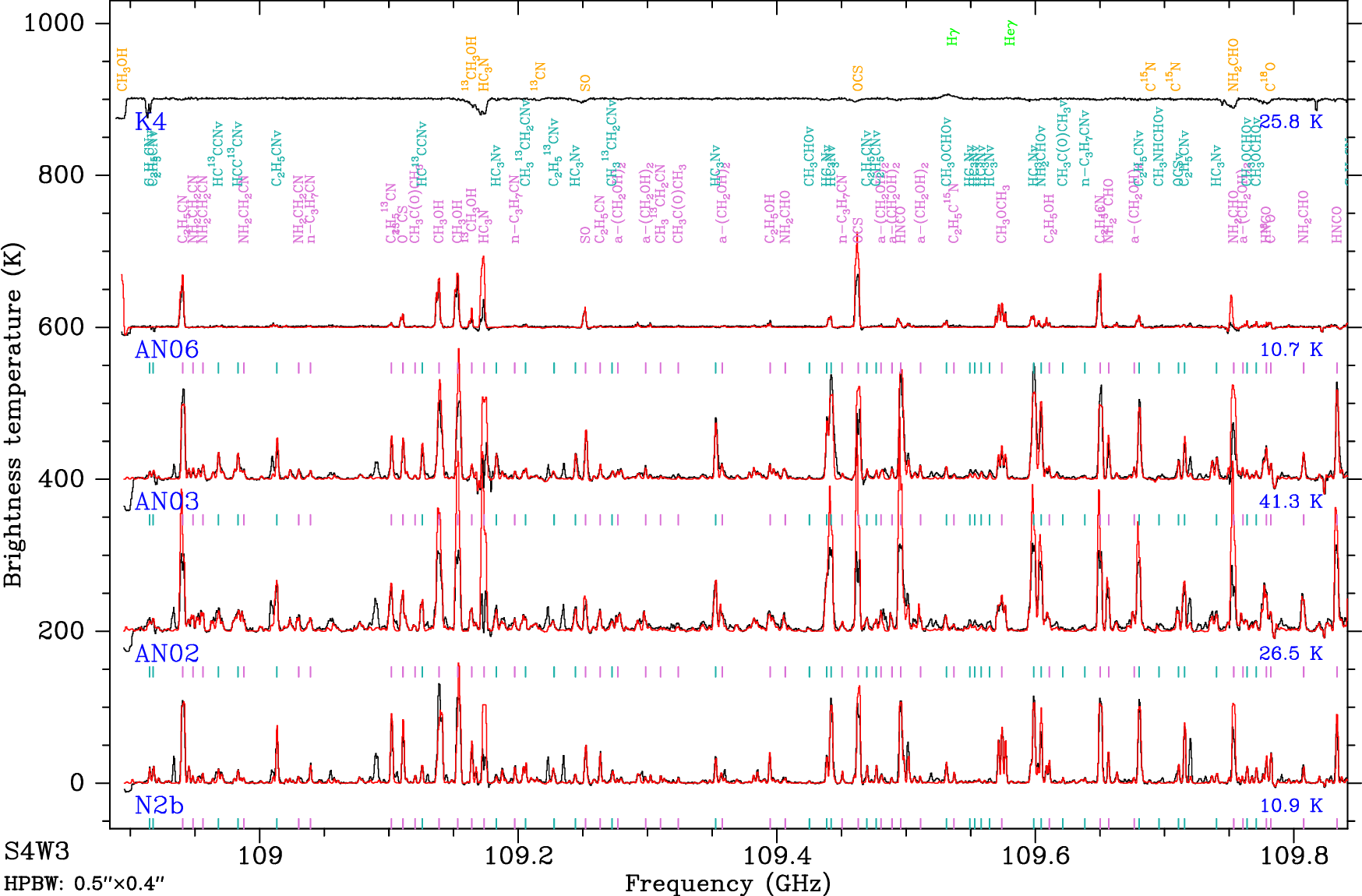}}}
\caption{continued.
}
\end{figure*}

\clearpage
\begin{figure*}
\addtocounter{figure}{-1}
\centerline{\resizebox{0.85\hsize}{!}{\includegraphics[angle=270]{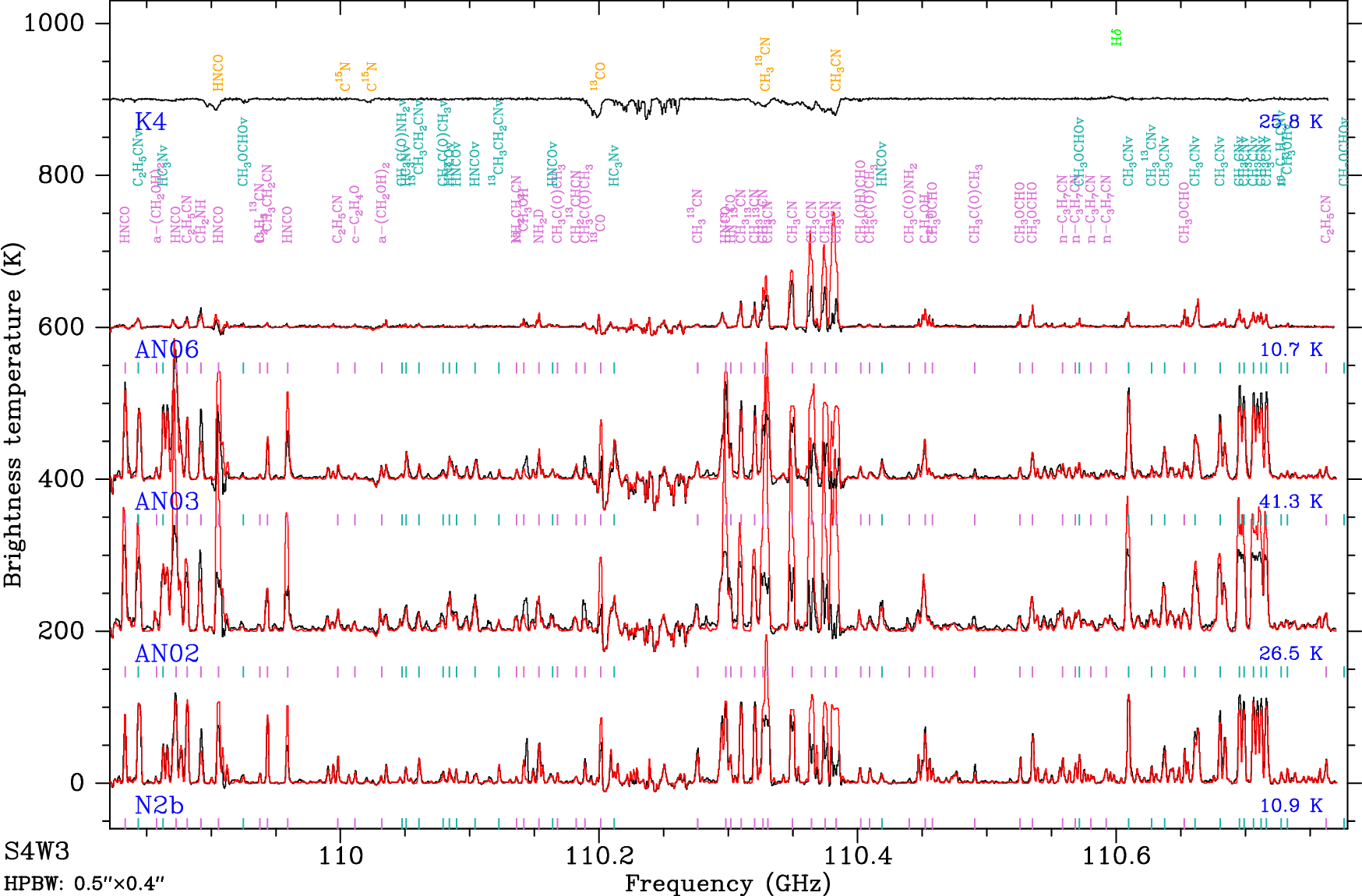}}}
\caption{continued.
}
\end{figure*}

\clearpage
\begin{figure*}
\addtocounter{figure}{-1}
\centerline{\resizebox{0.85\hsize}{!}{\includegraphics[angle=270]{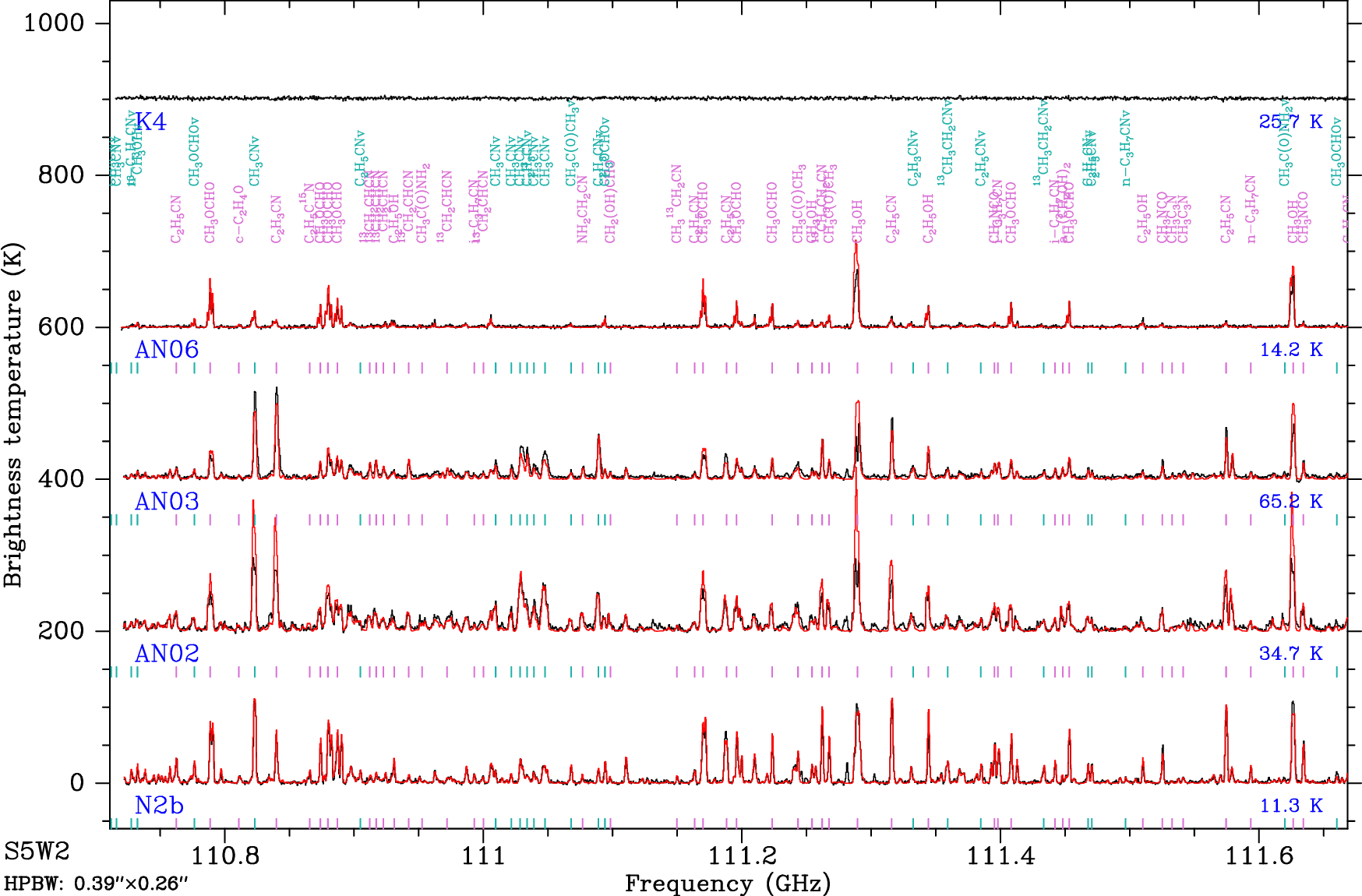}}}
\caption{continued.
}
\end{figure*}

\clearpage
\begin{figure*}
\addtocounter{figure}{-1}
\centerline{\resizebox{0.85\hsize}{!}{\includegraphics[angle=270]{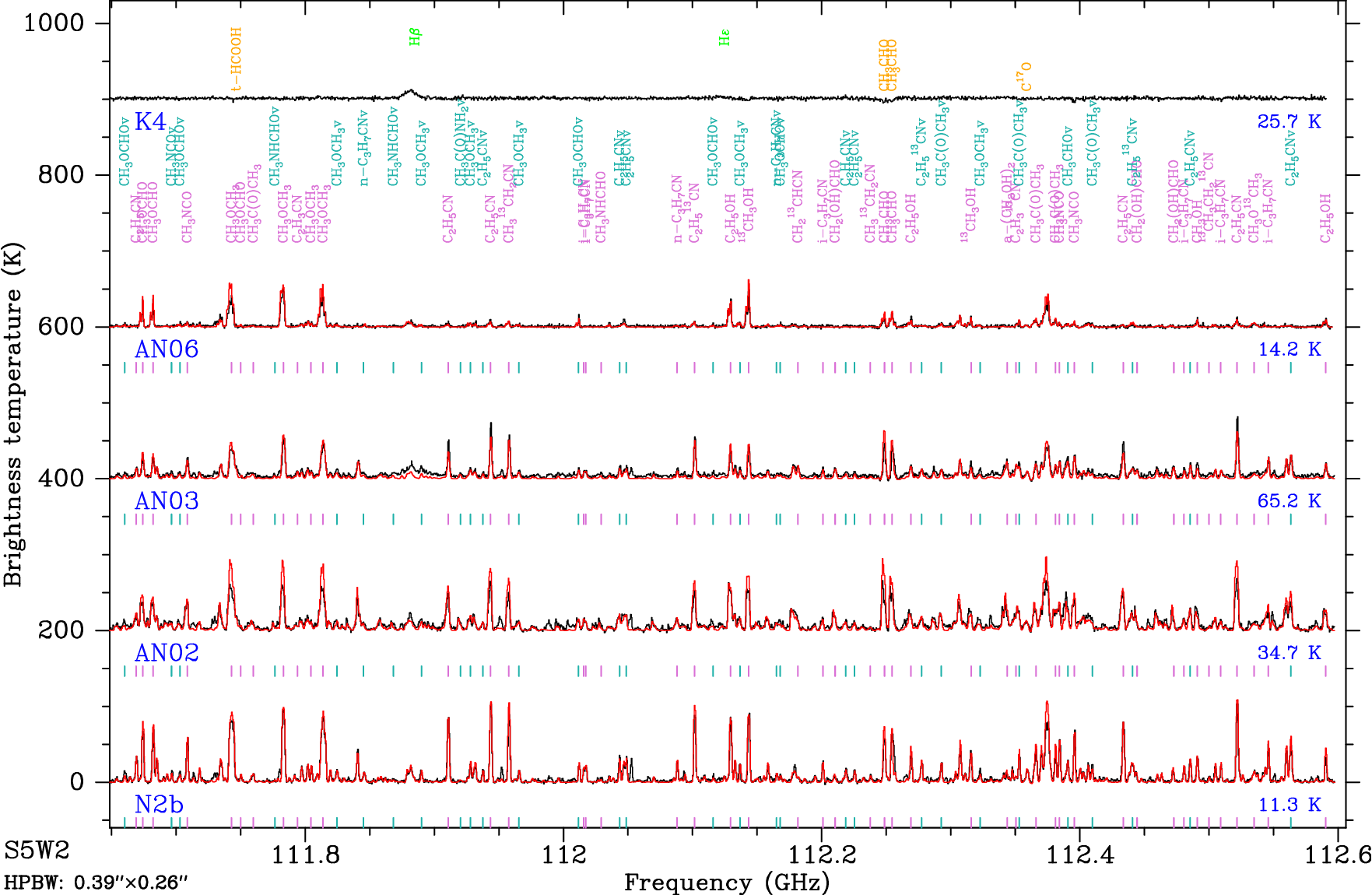}}}
\caption{continued.
}
\end{figure*}

\clearpage
\begin{figure*}
\addtocounter{figure}{-1}
\centerline{\resizebox{0.85\hsize}{!}{\includegraphics[angle=270]{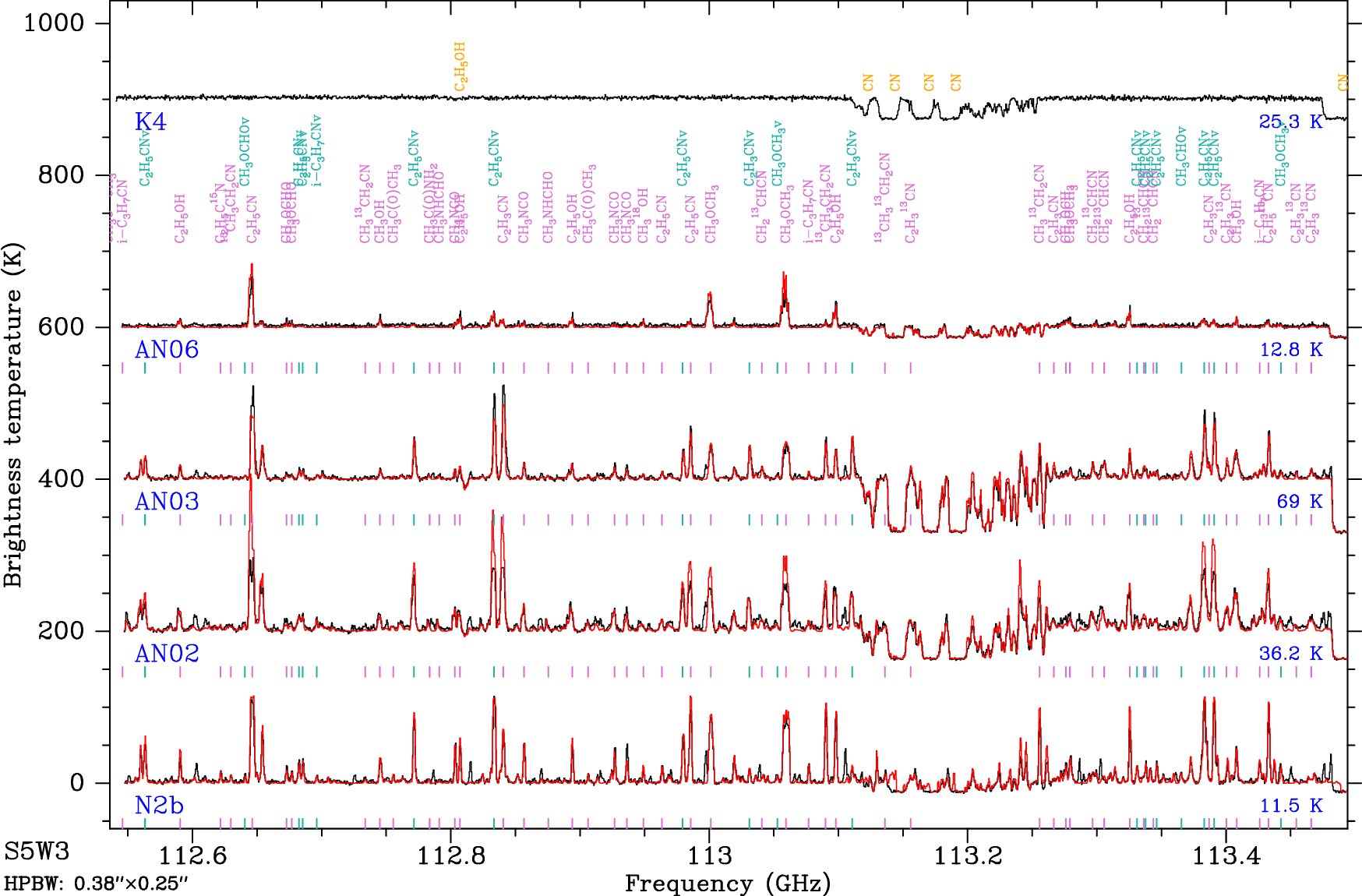}}}
\caption{continued.
}
\end{figure*}

\clearpage
\begin{figure*}
\addtocounter{figure}{-1}
\centerline{\resizebox{0.85\hsize}{!}{\includegraphics[angle=270]{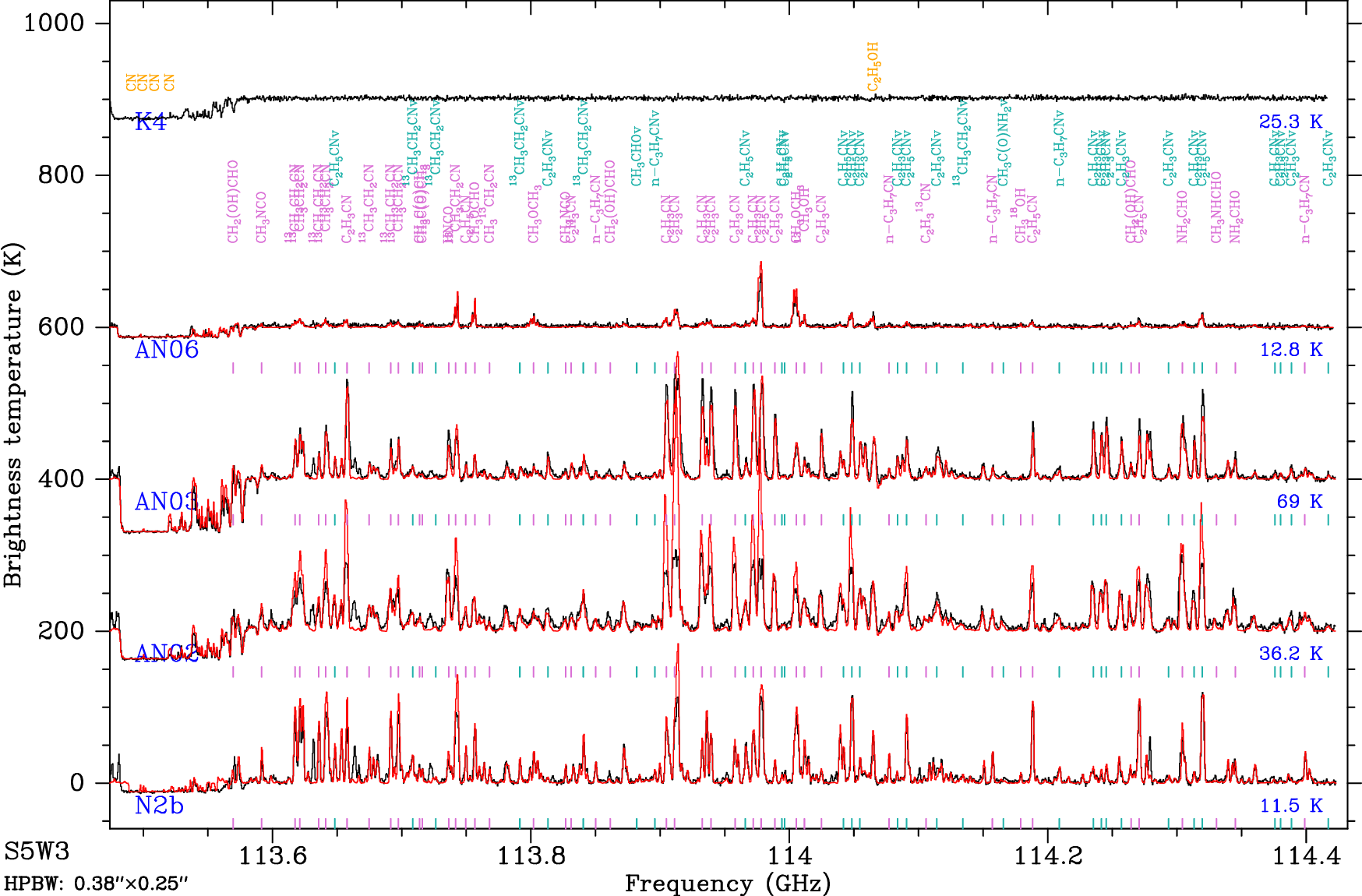}}}
\caption{continued.
}
\end{figure*}

\clearpage
\onecolumn
\section{LTE model parameters and column density upper limits}
\label{a:lteparam}

Tables~\ref{t:coldens_n2b}--\ref{t:coldens_an06} list the parameters of our
best-fit LTE models of N2b, AN02, AN03, the second velocity component toward
AN06 (called AN06c2), and the first velocity component toward AN06, 
respectively.
Tables~\ref{t:uplim_n2b}--\ref{t:uplim_an06} provide the column density upper 
limits derived for 146 molecules that were not detected in any of the
investigated positions.

\begin{longtable}{lccrccrcccrrr}
 \caption{\label{t:coldens_n2b}
 Parameters of our best-fit LTE model of N2b.
} \\ 
 \hline\hline
 \multicolumn{1}{c}{Molecule} & \multicolumn{1}{c}{\hspace*{-1ex}Tag\tablefootmark{a}} & \multicolumn{1}{c}{St.\tablefootmark{b}} & \multicolumn{1}{c}{\hspace*{-2ex}$N_{\rm det}$\tablefootmark{c}} & \multicolumn{1}{c}{Size\tablefootmark{d}} & \multicolumn{1}{c}{$T_{\mathrm{rot}}$\tablefootmark{e}} & \multicolumn{1}{c}{$N$\tablefootmark{f}} & \multicolumn{1}{c}{$C_{\rm{vib}}$\tablefootmark{g}} & \multicolumn{1}{c}{\hspace*{-1ex}$C_{\rm{conf}}$\tablefootmark{h}} & \multicolumn{1}{c}{\hspace*{-1ex}$\Delta V$\tablefootmark{i}} & \multicolumn{1}{c}{\hspace*{-1ex}$V_{\mathrm{off}}$\tablefootmark{j}} & \multicolumn{1}{c}{$\frac{N_{\rm ref}}{N}$\tablefootmark{k}} & \multicolumn{1}{c}{$\frac{N}{N_{\rm ref}}$\tablefootmark{k}} \\ 
  & & & & \multicolumn{1}{c}{$''$} & \multicolumn{1}{c}{K} & \multicolumn{1}{c}{cm$^{-2}$} & & & \multicolumn{2}{c}{\hspace*{-1.5ex} km~s$^{-1}$} & & \\ 
 \hline
 \endfirsthead
 \caption{continued.} \\ 
 \hline\hline
 \multicolumn{1}{c}{Molecule} & \multicolumn{1}{c}{\hspace*{-1ex}Tag\tablefootmark{a}} & \multicolumn{1}{c}{St.\tablefootmark{b}} & \multicolumn{1}{c}{\hspace*{-2ex}$N_{\rm det}$\tablefootmark{c}} & \multicolumn{1}{c}{Size\tablefootmark{d}} & \multicolumn{1}{c}{$T_{\mathrm{rot}}$\tablefootmark{e}} & \multicolumn{1}{c}{$N$\tablefootmark{f}} & \multicolumn{1}{c}{$C_{\rm{vib}}$\tablefootmark{g}} & \multicolumn{1}{c}{\hspace*{-1ex}$C_{\rm{conf}}$\tablefootmark{h}} & \multicolumn{1}{c}{\hspace*{-1ex}$\Delta V$\tablefootmark{i}} & \multicolumn{1}{c}{\hspace*{-1ex}$V_{\mathrm{off}}$\tablefootmark{j}} & \multicolumn{1}{c}{$\frac{N_{\rm ref}}{N}$\tablefootmark{k}} & \multicolumn{1}{c}{$\frac{N}{N_{\rm ref}}$\tablefootmark{k}} \\ 
 \hline
 \endhead
 \hline
 \endfoot
\hline 
 C$_2$H$_3$CN, $\varv=0$$^\star$ & \hspace*{-1ex} 53902 & d & \hspace*{-2ex} 43 & 1.00 &  200 & 1.1 (17) &  1.00 & \hspace*{-1ex}  1.00 & \hspace*{-1ex} 4.5 & \hspace*{-1ex} $0.0$ & \ldots & 1 \\ 
 \hspace*{4ex}$\varv_{11}=1$ & \hspace*{-1ex} 53903 & d & \hspace*{-2ex} 30 & 1.00 &  200 & 1.1 (17) &  1.00 & \hspace*{-1ex}  1.00 & \hspace*{-1ex} 4.5 & \hspace*{-1ex} $0.0$ & \ldots & 1 \\ 
 \hspace*{4ex}$\varv_{15}=1$ & \hspace*{-1ex} 53904 & d & \hspace*{-2ex} 20 & 1.00 &  200 & 1.1 (17) &  1.00 & \hspace*{-1ex}  1.00 & \hspace*{-1ex} 4.5 & \hspace*{-1ex} $0.0$ & \ldots & 1 \\ 
 \hspace*{4ex}$\varv_{11}=2$ & \hspace*{-1ex} 53905 & d & \hspace*{-2ex} 3 & 1.00 &  200 & 1.1 (17) &  1.00 & \hspace*{-1ex}  1.00 & \hspace*{-1ex} 4.5 & \hspace*{-1ex} $0.0$ & \ldots & 1 \\ 
 \hspace*{4ex}$\varv_{10}=1$ & \hspace*{-1ex} 53906 & c & \hspace*{-2ex} 0 & 1.00 &  200 & 1.1 (17) &  1.00 & \hspace*{-1ex}  1.00 & \hspace*{-1ex} 4.5 & \hspace*{-1ex} $0.0$ & \ldots & 1 \\ 
 \hspace*{4ex}$\varv_{11}=1/\varv_{15}=1$ & \hspace*{-1ex} 53907 & c & \hspace*{-2ex} 0 & 1.00 &  200 & 1.1 (17) &  1.00 & \hspace*{-1ex}  1.00 & \hspace*{-1ex} 4.5 & \hspace*{-1ex} $0.0$ & \ldots & 1 \\ 
 \hspace*{4ex}$\varv_{15}=2$ & \hspace*{-1ex} 53908 & c & \hspace*{-2ex} 0 & 1.00 &  200 & 1.1 (17) &  1.00 & \hspace*{-1ex}  1.00 & \hspace*{-1ex} 4.5 & \hspace*{-1ex} $0.0$ & \ldots & 1 \\ 
 \hspace*{4ex}$\varv_{14}=1$ & \hspace*{-1ex} 53909 & c & \hspace*{-2ex} 0 & 1.00 &  200 & 1.1 (17) &  1.00 & \hspace*{-1ex}  1.00 & \hspace*{-1ex} 4.5 & \hspace*{-1ex} $0.0$ & \ldots & 1 \\
\hline 
 \end{longtable}
 \tablefoot{
 \tablefoottext{a}{Entry number in our Weeds local database. See note b of Table~\ref{t:spectrobib}.}
 \tablefoottext{b}{Status: d: detection, t: tentative detection, c: contributes to detected signal close to or above 3$\sigma$ level and included in LTE model, n: nondetection.}
 \tablefoottext{c}{Number of detected lines. One line of a given species may mean a group of transitions of this species that are blended together. Transitions detected in absorption are not counted.}
 \tablefoottext{d}{Source diameter (\textit{FWHM}).}
 \tablefoottext{e}{Rotational temperature.}
 \tablefoottext{f}{Total column density of the molecule. $X$ ($Y$) means $X \times 10^Y$.}
 \tablefoottext{g}{Correction factor that was applied to the column density to account for the contribution of vibrationally or torsionally excited states, in the cases where this contribution was not included in the partition function of the spectroscopic predictions. In most cases, this factor was estimated in the harmonic approximation.}
 \tablefoottext{h}{Correction factor that was applied to the column density to account for the contribution of conformers, in the cases where this contribution was not included in the partition function of the spectroscopic predictions.}
 \tablefoottext{i}{Linewidth (\textit{FWHM}).}
 \tablefoottext{j}{Velocity offset with respect to the assumed systemic velocity given in Table~\ref{t:sources}.}
 \tablefoottext{k}{Column density ratio, with $N_{\rm ref}$ the column density of the previous reference species marked with a $\star$.}
 }

\begin{longtable}{lccrccrcccrrr}
 \caption{\label{t:coldens_an02}
 Parameters of our best-fit LTE model of AN02.
} \\ 
 \hline\hline
 \multicolumn{1}{c}{Molecule} & \multicolumn{1}{c}{\hspace*{-1ex}Tag\tablefootmark{a}} & \multicolumn{1}{c}{St.\tablefootmark{b}} & \multicolumn{1}{c}{\hspace*{-2ex}$N_{\rm det}$\tablefootmark{c}} & \multicolumn{1}{c}{Size\tablefootmark{d}} & \multicolumn{1}{c}{$T_{\mathrm{rot}}$\tablefootmark{e}} & \multicolumn{1}{c}{$N$\tablefootmark{f}} & \multicolumn{1}{c}{$C_{\rm{vib}}$\tablefootmark{g}} & \multicolumn{1}{c}{\hspace*{-1ex}$C_{\rm{conf}}$\tablefootmark{h}} & \multicolumn{1}{c}{\hspace*{-1ex}$\Delta V$\tablefootmark{i}} & \multicolumn{1}{c}{\hspace*{-1ex}$V_{\mathrm{off}}$\tablefootmark{j}} & \multicolumn{1}{c}{$\frac{N_{\rm ref}}{N}$\tablefootmark{k}} & \multicolumn{1}{c}{$\frac{N}{N_{\rm ref}}$\tablefootmark{k}} \\ 
  & & & & \multicolumn{1}{c}{$''$} & \multicolumn{1}{c}{K} & \multicolumn{1}{c}{cm$^{-2}$} & & & \multicolumn{2}{c}{\hspace*{-1.5ex} km~s$^{-1}$} & & \\ 
 \hline
 \endfirsthead
 \caption{continued.} \\ 
 \hline\hline
 \multicolumn{1}{c}{Molecule} & \multicolumn{1}{c}{\hspace*{-1ex}Tag\tablefootmark{a}} & \multicolumn{1}{c}{St.\tablefootmark{b}} & \multicolumn{1}{c}{\hspace*{-2ex}$N_{\rm det}$\tablefootmark{c}} & \multicolumn{1}{c}{Size\tablefootmark{d}} & \multicolumn{1}{c}{$T_{\mathrm{rot}}$\tablefootmark{e}} & \multicolumn{1}{c}{$N$\tablefootmark{f}} & \multicolumn{1}{c}{$C_{\rm{vib}}$\tablefootmark{g}} & \multicolumn{1}{c}{\hspace*{-1ex}$C_{\rm{conf}}$\tablefootmark{h}} & \multicolumn{1}{c}{\hspace*{-1ex}$\Delta V$\tablefootmark{i}} & \multicolumn{1}{c}{\hspace*{-1ex}$V_{\mathrm{off}}$\tablefootmark{j}} & \multicolumn{1}{c}{$\frac{N_{\rm ref}}{N}$\tablefootmark{k}} & \multicolumn{1}{c}{$\frac{N}{N_{\rm ref}}$\tablefootmark{k}} \\ 
 \hline
 \endhead
 \hline
 \endfoot
\hline 
 C$_2$H$_3$CN, $\varv=0$$^\star$ & \hspace*{-1ex} 53902 & d & \hspace*{-2ex} 49 & 0.70 &  220 & 3.9 (17) &  1.02 & \hspace*{-1ex}  1.00 & \hspace*{-1ex} 4.5 & \hspace*{-1ex} $-0.1$ & \ldots & 1 \\ 
 & & & & 0.70 &  220 & 2.6 (17) &  1.02 & \hspace*{-1ex}  1.00 & \hspace*{-1ex} 3.3 & \hspace*{-1ex} $3.8$ & 1.5 & \ldots \\ 
 \hspace*{4ex}$\varv_{11}=1$ & \hspace*{-1ex} 53903 & d & \hspace*{-2ex} 31 & 0.70 &  220 & 3.9 (17) &  1.02 & \hspace*{-1ex}  1.00 & \hspace*{-1ex} 4.5 & \hspace*{-1ex} $-0.1$ & \ldots & 1 \\ 
 & & & & 0.70 &  220 & 2.6 (17) &  1.02 & \hspace*{-1ex}  1.00 & \hspace*{-1ex} 3.3 & \hspace*{-1ex} $3.8$ & 1.5 & \ldots \\ 
 \hspace*{4ex}$\varv_{15}=1$ & \hspace*{-1ex} 53904 & d & \hspace*{-2ex} 22 & 0.70 &  220 & 3.9 (17) &  1.02 & \hspace*{-1ex}  1.00 & \hspace*{-1ex} 4.5 & \hspace*{-1ex} $-0.1$ & \ldots & 1 \\ 
 & & & & 0.70 &  220 & 2.6 (17) &  1.02 & \hspace*{-1ex}  1.00 & \hspace*{-1ex} 3.3 & \hspace*{-1ex} $3.8$ & 1.5 & \ldots \\ 
 \hspace*{4ex}$\varv_{11}=2$ & \hspace*{-1ex} 53905 & d & \hspace*{-2ex} 13 & 0.70 &  220 & 3.9 (17) &  1.02 & \hspace*{-1ex}  1.00 & \hspace*{-1ex} 4.5 & \hspace*{-1ex} $-0.1$ & \ldots & 1 \\ 
 & & & & 0.70 &  220 & 2.6 (17) &  1.02 & \hspace*{-1ex}  1.00 & \hspace*{-1ex} 3.3 & \hspace*{-1ex} $3.8$ & 1.5 & \ldots \\
\hline 
 \end{longtable}
 \tablefoot{
 \tablefoottext{a}{Entry number in our Weeds local database. See note b of Table~\ref{t:spectrobib}.}
 \tablefoottext{b}{Status: d: detection, t: tentative detection, c: contributes to detected signal close to or above 3$\sigma$ level and included in LTE model, n: nondetection.}
 \tablefoottext{c}{Number of detected lines. One line of a given species may mean a group of transitions of this species that are blended together. Transitions detected in absorption are not counted.}
 \tablefoottext{d}{Source diameter (\textit{FWHM}).}
 \tablefoottext{e}{Rotational temperature.}
 \tablefoottext{f}{Total column density of the molecule. $X$ ($Y$) means $X \times 10^Y$.}
 \tablefoottext{g}{Correction factor that was applied to the column density to account for the contribution of vibrationally or torsionally excited states, in the cases where this contribution was not included in the partition function of the spectroscopic predictions. In most cases, this factor was estimated in the harmonic approximation.}
 \tablefoottext{h}{Correction factor that was applied to the column density to account for the contribution of conformers, in the cases where this contribution was not included in the partition function of the spectroscopic predictions.}
 \tablefoottext{i}{Linewidth (\textit{FWHM}).}
 \tablefoottext{j}{Velocity offset with respect to the assumed systemic velocity given in Table~\ref{t:sources}.}
 \tablefoottext{k}{Column density ratio, with $N_{\rm ref}$ the column density of the previous reference species marked with a $\star$.}
 }

\begin{longtable}{lccrccrcccrrr}
 \caption{\label{t:coldens_an03}
 Parameters of our best-fit LTE model of AN03.
} \\ 
 \hline\hline
 \multicolumn{1}{c}{Molecule} & \multicolumn{1}{c}{\hspace*{-1ex}Tag\tablefootmark{a}} & \multicolumn{1}{c}{St.\tablefootmark{b}} & \multicolumn{1}{c}{\hspace*{-2ex}$N_{\rm det}$\tablefootmark{c}} & \multicolumn{1}{c}{Size\tablefootmark{d}} & \multicolumn{1}{c}{$T_{\mathrm{rot}}$\tablefootmark{e}} & \multicolumn{1}{c}{$N$\tablefootmark{f}} & \multicolumn{1}{c}{$C_{\rm{vib}}$\tablefootmark{g}} & \multicolumn{1}{c}{\hspace*{-1ex}$C_{\rm{conf}}$\tablefootmark{h}} & \multicolumn{1}{c}{\hspace*{-1ex}$\Delta V$\tablefootmark{i}} & \multicolumn{1}{c}{\hspace*{-1ex}$V_{\mathrm{off}}$\tablefootmark{j}} & \multicolumn{1}{c}{$\frac{N_{\rm ref}}{N}$\tablefootmark{k}} & \multicolumn{1}{c}{$\frac{N}{N_{\rm ref}}$\tablefootmark{k}} \\ 
  & & & & \multicolumn{1}{c}{$''$} & \multicolumn{1}{c}{K} & \multicolumn{1}{c}{cm$^{-2}$} & & & \multicolumn{2}{c}{\hspace*{-1.5ex} km~s$^{-1}$} & & \\ 
 \hline
 \endfirsthead
 \caption{continued.} \\ 
 \hline\hline
 \multicolumn{1}{c}{Molecule} & \multicolumn{1}{c}{\hspace*{-1ex}Tag\tablefootmark{a}} & \multicolumn{1}{c}{St.\tablefootmark{b}} & \multicolumn{1}{c}{\hspace*{-2ex}$N_{\rm det}$\tablefootmark{c}} & \multicolumn{1}{c}{Size\tablefootmark{d}} & \multicolumn{1}{c}{$T_{\mathrm{rot}}$\tablefootmark{e}} & \multicolumn{1}{c}{$N$\tablefootmark{f}} & \multicolumn{1}{c}{$C_{\rm{vib}}$\tablefootmark{g}} & \multicolumn{1}{c}{\hspace*{-1ex}$C_{\rm{conf}}$\tablefootmark{h}} & \multicolumn{1}{c}{\hspace*{-1ex}$\Delta V$\tablefootmark{i}} & \multicolumn{1}{c}{\hspace*{-1ex}$V_{\mathrm{off}}$\tablefootmark{j}} & \multicolumn{1}{c}{$\frac{N_{\rm ref}}{N}$\tablefootmark{k}} & \multicolumn{1}{c}{$\frac{N}{N_{\rm ref}}$\tablefootmark{k}} \\ 
 \hline
 \endhead
 \hline
 \endfoot
\hline 
 C$_2$H$_3$CN, $\varv=0$$^\star$ & \hspace*{-1ex} 53902 & d & \hspace*{-2ex} 63 & 0.70 &  190 & 8.0 (17) &  1.00 & \hspace*{-1ex}  1.00 & \hspace*{-1ex} 4.5 & \hspace*{-1ex} $-0.6$ & \ldots & 1 \\ 
 \hspace*{4ex}$\varv_{11}=1$ & \hspace*{-1ex} 53903 & d & \hspace*{-2ex} 37 & 0.70 &  190 & 8.0 (17) &  1.00 & \hspace*{-1ex}  1.00 & \hspace*{-1ex} 4.5 & \hspace*{-1ex} $-0.6$ & \ldots & 1 \\ 
 \hspace*{4ex}$\varv_{15}=1$ & \hspace*{-1ex} 53904 & d & \hspace*{-2ex} 36 & 0.70 &  190 & 8.0 (17) &  1.00 & \hspace*{-1ex}  1.00 & \hspace*{-1ex} 4.5 & \hspace*{-1ex} $-0.6$ & \ldots & 1 \\ 
 \hspace*{4ex}$\varv_{11}=2$ & \hspace*{-1ex} 53905 & d & \hspace*{-2ex} 20 & 0.70 &  190 & 8.0 (17) &  1.00 & \hspace*{-1ex}  1.00 & \hspace*{-1ex} 4.5 & \hspace*{-1ex} $-0.6$ & \ldots & 1 \\ 
 \hspace*{4ex}$\varv_{10}=1$ & \hspace*{-1ex} 53906 & d & \hspace*{-2ex} 5 & 0.70 &  190 & 8.0 (17) &  1.00 & \hspace*{-1ex}  1.00 & \hspace*{-1ex} 4.5 & \hspace*{-1ex} $-0.6$ & \ldots & 1 \\ 
 \hspace*{4ex}$\varv_{11}=1/\varv_{15}=1$ & \hspace*{-1ex} 53907 & d & \hspace*{-2ex} 11 & 0.70 &  190 & 8.0 (17) &  1.00 & \hspace*{-1ex}  1.00 & \hspace*{-1ex} 4.5 & \hspace*{-1ex} $-0.6$ & \ldots & 1 \\ 
 \hspace*{4ex}$\varv_{15}=2$ & \hspace*{-1ex} 53908 & d & \hspace*{-2ex} 3 & 0.70 &  190 & 8.0 (17) &  1.00 & \hspace*{-1ex}  1.00 & \hspace*{-1ex} 4.5 & \hspace*{-1ex} $-0.6$ & \ldots & 1 \\ 
 \hspace*{4ex}$\varv_{14}=1$ & \hspace*{-1ex} 53909 & t & \hspace*{-2ex} 2 & 0.70 &  190 & 8.0 (17) &  1.00 & \hspace*{-1ex}  1.00 & \hspace*{-1ex} 4.5 & \hspace*{-1ex} $-0.6$ & \ldots & 1 \\
\hline 
 \end{longtable}
 \tablefoot{
 \tablefoottext{a}{Entry number in our Weeds local database. See note b of Table~\ref{t:spectrobib}.}
 \tablefoottext{b}{Status: d: detection, t: tentative detection, c: contributes to detected signal close to or above 3$\sigma$ level and included in LTE model, n: nondetection.}
 \tablefoottext{c}{Number of detected lines. One line of a given species may mean a group of transitions of this species that are blended together. Transitions detected in absorption are not counted.}
 \tablefoottext{d}{Source diameter (\textit{FWHM}).}
 \tablefoottext{e}{Rotational temperature.}
 \tablefoottext{f}{Total column density of the molecule. $X$ ($Y$) means $X \times 10^Y$.}
 \tablefoottext{g}{Correction factor that was applied to the column density to account for the contribution of vibrationally or torsionally excited states, in the cases where this contribution was not included in the partition function of the spectroscopic predictions. In most cases, this factor was estimated in the harmonic approximation.}
 \tablefoottext{h}{Correction factor that was applied to the column density to account for the contribution of conformers, in the cases where this contribution was not included in the partition function of the spectroscopic predictions.}
 \tablefoottext{i}{Linewidth (\textit{FWHM}).}
 \tablefoottext{j}{Velocity offset with respect to the assumed systemic velocity given in Table~\ref{t:sources}.}
 \tablefoottext{k}{Column density ratio, with $N_{\rm ref}$ the column density of the previous reference species marked with a $\star$.}
 }

\begin{longtable}{lccrccrcccrrr}
 \caption{\label{t:coldens_edge}
 Parameters of our best-fit LTE model of the second velocity component toward AN06 (AN06c2).
} \\ 
 \hline\hline
 \multicolumn{1}{c}{Molecule} & \multicolumn{1}{c}{\hspace*{-1ex}Tag\tablefootmark{a}} & \multicolumn{1}{c}{St.\tablefootmark{b}} & \multicolumn{1}{c}{\hspace*{-2ex}$N_{\rm det}$\tablefootmark{c}} & \multicolumn{1}{c}{Size\tablefootmark{d}} & \multicolumn{1}{c}{$T_{\mathrm{rot}}$\tablefootmark{e}} & \multicolumn{1}{c}{$N$\tablefootmark{f}} & \multicolumn{1}{c}{$C_{\rm{vib}}$\tablefootmark{g}} & \multicolumn{1}{c}{\hspace*{-1ex}$C_{\rm{conf}}$\tablefootmark{h}} & \multicolumn{1}{c}{\hspace*{-1ex}$\Delta V$\tablefootmark{i}} & \multicolumn{1}{c}{\hspace*{-1ex}$V_{\mathrm{off}}$\tablefootmark{j}} & \multicolumn{1}{c}{$\frac{N_{\rm ref}}{N}$\tablefootmark{k}} & \multicolumn{1}{c}{$\frac{N}{N_{\rm ref}}$\tablefootmark{k}} \\ 
  & & & & \multicolumn{1}{c}{$''$} & \multicolumn{1}{c}{K} & \multicolumn{1}{c}{cm$^{-2}$} & & & \multicolumn{2}{c}{\hspace*{-1.5ex} km~s$^{-1}$} & & \\ 
 \hline
 \endfirsthead
 \caption{continued.} \\ 
 \hline\hline
 \multicolumn{1}{c}{Molecule} & \multicolumn{1}{c}{\hspace*{-1ex}Tag\tablefootmark{a}} & \multicolumn{1}{c}{St.\tablefootmark{b}} & \multicolumn{1}{c}{\hspace*{-2ex}$N_{\rm det}$\tablefootmark{c}} & \multicolumn{1}{c}{Size\tablefootmark{d}} & \multicolumn{1}{c}{$T_{\mathrm{rot}}$\tablefootmark{e}} & \multicolumn{1}{c}{$N$\tablefootmark{f}} & \multicolumn{1}{c}{$C_{\rm{vib}}$\tablefootmark{g}} & \multicolumn{1}{c}{\hspace*{-1ex}$C_{\rm{conf}}$\tablefootmark{h}} & \multicolumn{1}{c}{\hspace*{-1ex}$\Delta V$\tablefootmark{i}} & \multicolumn{1}{c}{\hspace*{-1ex}$V_{\mathrm{off}}$\tablefootmark{j}} & \multicolumn{1}{c}{$\frac{N_{\rm ref}}{N}$\tablefootmark{k}} & \multicolumn{1}{c}{$\frac{N}{N_{\rm ref}}$\tablefootmark{k}} \\ 
 \hline
 \endhead
 \hline
 \endfoot
\hline 
 C$_2$H$_3$CN, $\varv=0$$^\star$ & \hspace*{-1ex} 53902 & d & \hspace*{-2ex} 33 & 0.70 &  150 & 4.5 (15) &  1.00 & \hspace*{-1ex}  1.00 & \hspace*{-1ex} 3.5 & \hspace*{-1ex} $5.2$ & \ldots & 1 \\ 
 \hspace*{4ex}$\varv_{11}=1$ & \hspace*{-1ex} 53903 & c & \hspace*{-2ex} 0 & 0.70 &  150 & 4.5 (15) &  1.00 & \hspace*{-1ex}  1.00 & \hspace*{-1ex} 3.5 & \hspace*{-1ex} $5.2$ & \ldots & 1 \\ 
 \hspace*{4ex}$\varv_{15}=1$ & \hspace*{-1ex} 53904 & n & \hspace*{-2ex} 0 & 0.70 &  150 & $<$2.0 (16) &  1.00 & \hspace*{-1ex}  1.00 & \hspace*{-1ex} 3.5 & \hspace*{-1ex} $5.2$ & \ldots & $<$4.4 \\ 
 \hspace*{4ex}$\varv_{11}=2$ & \hspace*{-1ex} 53905 & n & \hspace*{-2ex} 0 & 0.70 &  150 & $<$7.0 (16) &  1.00 & \hspace*{-1ex}  1.00 & \hspace*{-1ex} 3.5 & \hspace*{-1ex} $5.2$ & \ldots & $<$16 \\ 
 \hspace*{4ex}$\varv_{10}=1$ & \hspace*{-1ex} 53906 & n & \hspace*{-2ex} 0 & 0.70 &  150 & $<$3.0 (17) &  1.00 & \hspace*{-1ex}  1.00 & \hspace*{-1ex} 3.5 & \hspace*{-1ex} $5.2$ & \ldots & $<$67 \\ 
 \hspace*{4ex}$\varv_{11}=1/\varv_{15}=1$ & \hspace*{-1ex} 53907 & n & \hspace*{-2ex} 0 & 0.70 &  150 & $<$2.5 (17) &  1.00 & \hspace*{-1ex}  1.00 & \hspace*{-1ex} 3.5 & \hspace*{-1ex} $5.2$ & \ldots & $<$56 \\ 
 \hspace*{4ex}$\varv_{15}=2$ & \hspace*{-1ex} 53908 & n & \hspace*{-2ex} 0 & 0.70 &  150 & $<$3.0 (17) &  1.00 & \hspace*{-1ex}  1.00 & \hspace*{-1ex} 3.5 & \hspace*{-1ex} $5.2$ & \ldots & $<$67 \\ 
 \hspace*{4ex}$\varv_{14}=1$ & \hspace*{-1ex} 53909 & n & \hspace*{-2ex} 0 & 0.70 &  150 & $<$1.0 (18) &  1.00 & \hspace*{-1ex}  1.00 & \hspace*{-1ex} 3.5 & \hspace*{-1ex} $5.2$ & \ldots & $<$220 \\
\hline 
 \end{longtable}
 \tablefoot{
 \tablefoottext{a}{Entry number in our Weeds local database. See note b of Table~\ref{t:spectrobib}.}
 \tablefoottext{b}{Status: d: detection, t: tentative detection, c: contributes to detected signal close to or above 3$\sigma$ level and included in LTE model, n: nondetection.}
 \tablefoottext{c}{Number of detected lines. One line of a given species may mean a group of transitions of this species that are blended together. Transitions detected in absorption are not counted.}
 \tablefoottext{d}{Source diameter (\textit{FWHM}).}
 \tablefoottext{e}{Rotational temperature.}
 \tablefoottext{f}{Total column density of the molecule. $X$ ($Y$) means $X \times 10^Y$.}
 \tablefoottext{g}{Correction factor that was applied to the column density to account for the contribution of vibrationally or torsionally excited states, in the cases where this contribution was not included in the partition function of the spectroscopic predictions. In most cases, this factor was estimated in the harmonic approximation.}
 \tablefoottext{h}{Correction factor that was applied to the column density to account for the contribution of conformers, in the cases where this contribution was not included in the partition function of the spectroscopic predictions.}
 \tablefoottext{i}{Linewidth (\textit{FWHM}).}
 \tablefoottext{j}{Velocity offset with respect to the assumed systemic velocity given in Table~\ref{t:sources}.}
 \tablefoottext{k}{Column density ratio, with $N_{\rm ref}$ the column density of the previous reference species marked with a $\star$.}
 }

\clearpage
\begin{longtable}{lccrccrcccrrr}
 \caption{\label{t:coldens_an06}
 Parameters of our best-fit LTE model of AN06.
} \\ 
 \hline\hline
 \multicolumn{1}{c}{Molecule} & \multicolumn{1}{c}{\hspace*{-1ex}Tag\tablefootmark{a}} & \multicolumn{1}{c}{St.\tablefootmark{b}} & \multicolumn{1}{c}{\hspace*{-2ex}$N_{\rm det}$\tablefootmark{c}} & \multicolumn{1}{c}{Size\tablefootmark{d}} & \multicolumn{1}{c}{$T_{\mathrm{rot}}$\tablefootmark{e}} & \multicolumn{1}{c}{$N$\tablefootmark{f}} & \multicolumn{1}{c}{$C_{\rm{vib}}$\tablefootmark{g}} & \multicolumn{1}{c}{\hspace*{-1ex}$C_{\rm{conf}}$\tablefootmark{h}} & \multicolumn{1}{c}{\hspace*{-1ex}$\Delta V$\tablefootmark{i}} & \multicolumn{1}{c}{\hspace*{-1ex}$V_{\mathrm{off}}$\tablefootmark{j}} & \multicolumn{1}{c}{$\frac{N_{\rm ref}}{N}$\tablefootmark{k}} & \multicolumn{1}{c}{$\frac{N}{N_{\rm ref}}$\tablefootmark{k}} \\ 
  & & & & \multicolumn{1}{c}{$''$} & \multicolumn{1}{c}{K} & \multicolumn{1}{c}{cm$^{-2}$} & & & \multicolumn{2}{c}{\hspace*{-1.5ex} km~s$^{-1}$} & & \\ 
 \hline
 \endfirsthead
 \caption{continued.} \\ 
 \hline\hline
 \multicolumn{1}{c}{Molecule} & \multicolumn{1}{c}{\hspace*{-1ex}Tag\tablefootmark{a}} & \multicolumn{1}{c}{St.\tablefootmark{b}} & \multicolumn{1}{c}{\hspace*{-2ex}$N_{\rm det}$\tablefootmark{c}} & \multicolumn{1}{c}{Size\tablefootmark{d}} & \multicolumn{1}{c}{$T_{\mathrm{rot}}$\tablefootmark{e}} & \multicolumn{1}{c}{$N$\tablefootmark{f}} & \multicolumn{1}{c}{$C_{\rm{vib}}$\tablefootmark{g}} & \multicolumn{1}{c}{\hspace*{-1ex}$C_{\rm{conf}}$\tablefootmark{h}} & \multicolumn{1}{c}{\hspace*{-1ex}$\Delta V$\tablefootmark{i}} & \multicolumn{1}{c}{\hspace*{-1ex}$V_{\mathrm{off}}$\tablefootmark{j}} & \multicolumn{1}{c}{$\frac{N_{\rm ref}}{N}$\tablefootmark{k}} & \multicolumn{1}{c}{$\frac{N}{N_{\rm ref}}$\tablefootmark{k}} \\ 
 \hline
 \endhead
 \hline
 \endfoot
\hline 
 C$_2$H$_3$CN, $\varv=0$$^\star$ & \hspace*{-1ex} 53902 & d & \hspace*{-2ex} 31 & 0.70 &  130 & 5.5 (15) &  1.00 & \hspace*{-1ex}  1.00 & \hspace*{-1ex} 3.5 & \hspace*{-1ex} $0.3$ & \ldots & 1 \\ 
 \hspace*{4ex}$\varv_{11}=1$ & \hspace*{-1ex} 53903 & c & \hspace*{-2ex} 0 & 0.70 &  130 & 5.5 (15) &  1.00 & \hspace*{-1ex}  1.00 & \hspace*{-1ex} 3.5 & \hspace*{-1ex} $0.3$ & \ldots & 1 \\ 
 \hspace*{4ex}$\varv_{15}=1$ & \hspace*{-1ex} 53904 & n & \hspace*{-2ex} 0 & 0.70 &  130 & $<$2.0 (16) &  1.00 & \hspace*{-1ex}  1.00 & \hspace*{-1ex} 3.5 & \hspace*{-1ex} $0.3$ & \ldots & $<$3.6 \\ 
 \hspace*{4ex}$\varv_{11}=2$ & \hspace*{-1ex} 53905 & n & \hspace*{-2ex} 0 & 0.70 &  130 & $<$9.0 (16) &  1.00 & \hspace*{-1ex}  1.00 & \hspace*{-1ex} 3.5 & \hspace*{-1ex} $0.3$ & \ldots & $<$16 \\ 
 \hspace*{4ex}$\varv_{10}=1$ & \hspace*{-1ex} 53906 & n & \hspace*{-2ex} 0 & 0.70 &  130 & $<$6.0 (17) &  1.00 & \hspace*{-1ex}  1.00 & \hspace*{-1ex} 3.5 & \hspace*{-1ex} $0.3$ & \ldots & $<$110 \\ 
 \hspace*{4ex}$\varv_{11}=1/\varv_{15}=1$ & \hspace*{-1ex} 53907 & n & \hspace*{-2ex} 0 & 0.70 &  130 & $<$2.5 (17) &  1.00 & \hspace*{-1ex}  1.00 & \hspace*{-1ex} 3.5 & \hspace*{-1ex} $0.3$ & \ldots & $<$45 \\ 
 \hspace*{4ex}$\varv_{15}=2$ & \hspace*{-1ex} 53908 & n & \hspace*{-2ex} 0 & 0.70 &  130 & $<$8.0 (17) &  1.00 & \hspace*{-1ex}  1.00 & \hspace*{-1ex} 3.5 & \hspace*{-1ex} $0.3$ & \ldots & $<$150 \\ 
 \hspace*{4ex}$\varv_{14}=1$ & \hspace*{-1ex} 53909 & n & \hspace*{-2ex} 0 & 0.70 &  130 & $<$2.0 (18) &  1.00 & \hspace*{-1ex}  1.00 & \hspace*{-1ex} 3.5 & \hspace*{-1ex} $0.3$ & \ldots & $<$360 \\
\hline 
 \end{longtable}
 \tablefoot{
 \tablefoottext{a}{Entry number in our Weeds local database. See note b of Table~\ref{t:spectrobib}.}
 \tablefoottext{b}{Status: d: detection, t: tentative detection, c: contributes to detected signal close to or above 3$\sigma$ level and included in LTE model, n: nondetection.}
 \tablefoottext{c}{Number of detected lines. One line of a given species may mean a group of transitions of this species that are blended together. Transitions detected in absorption are not counted.}
 \tablefoottext{d}{Source diameter (\textit{FWHM}).}
 \tablefoottext{e}{Rotational temperature.}
 \tablefoottext{f}{Total column density of the molecule. $X$ ($Y$) means $X \times 10^Y$.}
 \tablefoottext{g}{Correction factor that was applied to the column density to account for the contribution of vibrationally or torsionally excited states, in the cases where this contribution was not included in the partition function of the spectroscopic predictions. In most cases, this factor was estimated in the harmonic approximation.}
 \tablefoottext{h}{Correction factor that was applied to the column density to account for the contribution of conformers, in the cases where this contribution was not included in the partition function of the spectroscopic predictions.}
 \tablefoottext{i}{Linewidth (\textit{FWHM}).}
 \tablefoottext{j}{Velocity offset with respect to the assumed systemic velocity given in Table~\ref{t:sources}.}
 \tablefoottext{k}{Column density ratio, with $N_{\rm ref}$ the column density of the previous reference species marked with a $\star$.}
 }

\clearpage
\begin{longtable}{lrccccccrr}
 \caption{\label{t:uplim_n2b}
 Column density upper limits toward  N2b.
} \\ 
 \hline\hline
 \multicolumn{1}{c}{Molecule\tablefootmark{a}} & \multicolumn{1}{c}{Tag\tablefootmark{b}} & \multicolumn{1}{c}{Size\tablefootmark{c}} & \multicolumn{1}{c}{$T_{\mathrm{rot}}$\tablefootmark{d}} & \multicolumn{1}{c}{$N$\tablefootmark{e}} & \multicolumn{1}{c}{$C_{\rm{vib}}$\tablefootmark{f}} & \multicolumn{1}{c}{$C_{\rm{conf}}$\tablefootmark{g}} & \multicolumn{1}{c}{$\Delta V$\tablefootmark{h}} & \multicolumn{1}{c}{$V_{\mathrm{off}}$\tablefootmark{i}} & \multicolumn{1}{c}{Ref.\tablefootmark{j}} \\ 
  & & \multicolumn{1}{c}{$''$} & \multicolumn{1}{c}{K} & \multicolumn{1}{c}{cm$^{-2}$} & & & \multicolumn{2}{c}{km~s$^{-1}$} & \\ 
 \hline
 \endfirsthead
 \caption{continued.} \\ 
 \hline\hline
 \multicolumn{1}{c}{Molecule\tablefootmark{a}} & \multicolumn{1}{c}{Tag\tablefootmark{b}} & \multicolumn{1}{c}{Size\tablefootmark{c}} & \multicolumn{1}{c}{$T_{\mathrm{rot}}$\tablefootmark{d}} & \multicolumn{1}{c}{$N$\tablefootmark{e}} & \multicolumn{1}{c}{$C_{\rm{vib}}$\tablefootmark{f}} & \multicolumn{1}{c}{$C_{\rm{conf}}$\tablefootmark{g}} & \multicolumn{1}{c}{$\Delta V$\tablefootmark{h}} & \multicolumn{1}{c}{$V_{\mathrm{off}}$\tablefootmark{i}} & \multicolumn{1}{c}{Ref.\tablefootmark{j}} \\ 
 \hline
 \endhead
 \hline
 \endfoot
\hline 
\multicolumn{9}{c}{\textit{O-bearing molecules}} \\ 
 HCO $^\star$ & 29004 & 0.70 &  130 & $<$1.0 (16) &  1.00 &  1.00 & 4.0 & $0.0$ & H$_2$CO \\ 
 H$_2$COH$^+$ $^\star$ & 31504 & 0.70 &  130 & $<$6.0 (15) &  1.00 &  1.00 & 4.0 & $0.0$ & H$_2$CO \\ 
 CH$_2$OH & 31515 & 0.50 &  140 & $<$1.8 (17) &  1.00 &  1.00 & 3.5 & $0.0$ & CH$_3$OH \\ 
 CH$_3$O $^\star$ & 31518 & 0.50 &  140 & $<$1.0 (17) &  1.00 &  1.00 & 3.5 & $0.0$ & CH$_3$OH \\ 
 CH$_2$CHO & 43523 & 0.80 &  160 & $<$8.4 (15) &  1.05 &  1.00 & 4.0 & $0.3$ & CH$_3$CHO \\ 
 syn-C$_2$H$_3$OH $^\star$ & 44506 & 0.50 &  135 & $<$4.2 (16) &  1.02 &  1.02 & 3.5 & $0.0$ & C$_2$H$_5$OH \\ 
 anti-C$_2$H$_3$OH $^\star$ & 44507 & 0.50 &  135 & $<$8.4 (17) &  1.02 &  58.4 & 3.5 & $0.0$ & C$_2$H$_5$OH \\ 
 c-HCOOH $^\star$ & 46507 & 0.50 &  150 & $<$2.0 (21) &  1.00 & 485469 & 3.5 & $0.0$ & \ldots \\ 
 trans-C$_2$H$_3$CHO $^\star$ & 56519 & 0.80 &  160 & $<$2.1 (15) &  1.41 &  1.00 & 4.0 & $0.3$ & CH$_3$CHO \\ 
 CH$_3$CHCO $^\star$ & 56701 & 0.70 &  130 & $<$7.0 (15) &  1.00 &  1.00 & 3.8 & $0.0$ & CH$_2$CO \\ 
 c-CH(CH$_3$)CH$_2$O $^\star$ & 58514 & 0.50 &  150 & $<$3.7 (16) &  1.23 &  1.00 & 3.5 & $0.0$ & \ldots \\ 
 gauche-C$_2$H$_5$CHO & 58519 & 0.80 &  160 & $<$4.0 (17) &  1.91 &  1.05 & 4.0 & $0.3$ & CH$_3$CHO \\ 
 C$_2$H$_5$OCH$_3$ $^\star$ & 60901 & 0.80 &  130 & $<$9.9 (16) &  1.63 &  1.01 & 3.5 & $-0.5$ & CH$_3$OCH$_3$ \\ 
 HOCHCHOH $^\star$ & 60920 & 0.50 &  150 & $<$1.3 (16) &  1.26 &  1.00 & 3.5 & $1.0$ & a-(CH2OH)$_2$ \\ 
 CH$_3$OCH$_2$OH $^\star$ & 62527 & 0.50 &  140 & $<$1.0 (18) &  1.68 &  1.00 & 3.5 & $0.0$ & CH$_3$OH \\ 
 s-cis-C$_2$H$_3$COOH & 72502 & 0.50 &  150 & $<$7.0 (16) &  1.75 &  1.00 & 3.5 & $0.0$ & \ldots \\ 
 s-trans-C$_2$H$_3$COOH & 72503 & 0.50 &  150 & $<$3.2 (16) &  1.75 &  1.00 & 3.5 & $0.0$ & \ldots \\ 
 HOCHCHCHO $^\star$ & 72504 & 0.80 &  160 & $<$1.0 (16) &  1.27 &  1.00 & 4.0 & $0.3$ & CH$_3$CHO \\ 
 cis-trans-C$_2$H$_3$OCHO & 72505 & 0.50 &  150 & $<$4.8 (16) &  2.41 &  1.00 & 3.5 & $0.0$ & C$_2$H$_5$OCHO \\ 
 CH$_3$C(O)CHO & 72901 & 0.50 &  140 & $<$2.4 (17) &  2.36 &  1.00 & 3.5 & $-0.2$ & CH$_3$COCH$_3$ \\ 
 gauche-(CH$_3$)$_2$CHCHO & 72902 & 0.80 &  160 & $<$3.8 (16) &  2.76 &  1.25 & 4.0 & $0.3$ & CH$_3$CHO \\ 
 cis-gauche-C$_3$H$_7$CHO & 72903 & 0.80 &  160 & $<$9.2 (16) &  2.82 &  3.28 & 4.0 & $0.3$ & CH$_3$CHO \\ 
 cis-trans-C$_3$H$_7$CHO & 72904 & 0.80 &  160 & $<$4.7 (16) &  3.44 &  1.71 & 4.0 & $0.3$ & CH$_3$CHO \\ 
 CH$_2$C(OH)CHO & 72905 & 0.50 &  170 & $<$2.2 (16) &  1.46 &  1.00 & 3.5 & $1.0$ & CH$_2$(OH)CHO \\ 
 CH$_3$C(O)CH$_2$OH $^\star$ & 74003 & 0.50 &  140 & $<$4.4 (16) &  4.36 &  1.00 & 3.5 & $-0.2$ & CH$_3$OCH$_3$ \\ 
 OCHCOOH & 74516 & 0.50 &  150 & $<$2.1 (16) &  1.38 &  1.02 & 3.5 & $0.0$ & \ldots \\ 
 CH$_3$OCH$_2$CHO & 74517 & 0.50 &  150 & $<$1.3 (16) &  6.03 &  1.04 & 3.5 & $0.0$ & \ldots \\ 
 CH$_3$CH(OH)CHO & 74519 & 0.50 &  170 & $<$1.9 (16) &  2.49 &  1.00 & 3.5 & $1.0$ & CH$_2$(OH)CHO \\ 
 C$_2$H$_5$COOH & 74905 & 0.50 &  150 & $<$9.6 (16) &  2.84 &  1.13 & 3.5 & $0.0$ & \ldots \\ 
 tert-C$_4$H$_9$OH & 74908 & 0.50 &  135 & $<$6.5 (16) &  1.30 &  1.00 & 3.5 & $0.0$ & C$_2$H$_5$OH \\ 
 aG'g-CH$_3$CHOHCH$_2$OH & 76513 & 0.50 &  135 & $<$5.0 (16) &  1.79 &  1.99 & 3.5 & $0.0$ & C$_2$H$_5$OH \\ 
 CH$_2$(OH)COOH & 76514 & 0.50 &  170 & $<$2.7 (16) &  2.28 &  1.00 & 3.5 & $1.0$ & CH$_2$(OH)CHO \\ 
 gG'a-CH$_3$CHOHCH$_2$OH & 76515 & 0.50 &  135 & $<$7.8 (16) &  1.78 &  4.37 & 3.5 & $0.0$ & C$_2$H$_5$OH \\ 
 g'G'g-CH$_3$CHOHCH$_2$OH & 76516 & 0.50 &  135 & $<$1.0 (17) &  1.72 &  6.77 & 3.5 & $0.0$ & C$_2$H$_5$OH \\ 
 a'GG'g-HOCH$_2$CH$_2$CH$_2$OH & 76518 & 0.50 &  135 & $<$2.9 (16) &  2.09 &  1.39 & 3.5 & $0.0$ & C$_2$H$_5$OH \\ 
 gGG'g-HOCH$_2$CH$_2$CH$_2$OH & 76519 & 0.50 &  135 & $<$7.7 (16) &  1.98 &  4.35 & 3.5 & $0.0$ & C$_2$H$_5$OH \\ 
 g'Ga-CH$_3$CHOHCH$_2$OH & 76520 & 0.50 &  135 & $<$2.5 (17) &  1.84 &  19.0 & 3.5 & $0.0$ & C$_2$H$_5$OH \\ 
 gG'g'-CH$_3$CHOHCH$_2$OH & 76521 & 0.50 &  135 & $<$3.6 (17) &  1.72 &  23.1 & 3.5 & $0.0$ & C$_2$H$_5$OH \\ 
 aGg'-CH$_3$CHOHCH$_2$OH & 76522 & 0.50 &  135 & $<$1.5 (18) &  1.79 &  82.0 & 3.5 & $0.0$ & C$_2$H$_5$OH \\ 
 g'Gg-CH$_3$CHOHCH$_2$OH & 76523 & 0.50 &  135 & $<$3.9 (18) &  1.79 &   108 & 3.5 & $0.0$ & C$_2$H$_5$OH \\ 
 CH$_3$OCH$_2$CH$_2$OH $^\star$ & 76801 & 0.50 &  140 & $<$3.0 (16) &  1.00 &  1.01 & 3.5 & $0.0$ & CH$_3$OH \\ 
 c-C$_6$H$_5$OH & 94501 & 0.50 &  140 & $<$3.6 (16) &  1.21 &  1.00 & 3.5 & $0.0$ & CH$_3$OH \\ 
\hline 
\multicolumn{9}{c}{\textit{O- and N-bearing molecules}} \\ 
 HCNO $^\star$ & 43509 & 0.60 &  180 & $<$1.0 (15) &  2.07 &  1.00 & 3.5 & $0.0$ & HNCO \\ 
 HOCN $^\star$ & 43510 & 0.60 &  180 & $<$1.1 (15) &  1.06 &  1.00 & 3.5 & $0.0$ & HNCO \\ 
 NH$_2$CO & 44904 & 0.60 &  170 & $<$1.0 (16) &  1.00 &  1.00 & 3.5 & $0.0$ & NH$_2$CHO \\ 
 HC(O)CN $^\star$ & 55501 & 0.50 &  150 & $<$8.5 (15) &  1.21 &  1.00 & 3.5 & $0.0$ & \ldots \\ 
 CH$_3$CNO & 57510 & 0.70 &  140 & $<$6.4 (14) &  1.28 &  1.00 & 3.5 & $-0.5$ & CH$_3$NCO \\ 
 CH$_3$OCN & 57511 & 0.70 &  140 & $<$2.5 (15) &  1.65 &  1.00 & 3.5 & $-0.5$ & CH$_3$NCO \\ 
 HOCH$_2$CN $^\star$ & 57701 & 0.50 &  150 & $<$9.1 (15) &  1.00 &  1.00 & 3.5 & $0.0$ & \ldots \\ 
 NH$_2$C(O)NH$_2$ $^\star$ & 60517 & 0.50 &  150 & $<$3.2 (15) &  1.79 &  1.04 & 3.5 & $0.0$ & \ldots \\ 
 NH$_2$CH$_2$CH$_2$OH $^\star$ & 61004 & 0.50 &  150 & $<$1.3 (16) &  1.00 &  1.05 & 3.5 & $0.0$ & \ldots \\ 
 NCCHCO & 67502 & 0.50 &  150 & $<$4.2 (15) &  1.39 &  1.00 & 3.5 & $0.0$ & \ldots \\ 
 c-(C$_2$H$_3$O)-CN & 69518 & 0.70 &  130 & $<$2.4 (15) &  1.20 &  1.00 & 3.5 & $0.0$ & c-C$_2$H$_4$O \\ 
 trans-C$_2$H$_3$NCO & 69901 & 0.70 &  140 & $<$9.7 (15) &  2.43 &  1.00 & 3.5 & $-0.5$ & CH$_3$NCO \\ 
 cis-C$_2$H$_3$NCO & 69902 & 0.70 &  140 & $<$2.0 (17) &  2.47 &  1.00 & 3.5 & $-0.5$ & CH$_3$NCO \\ 
 CHCC(O)NH$_2$ & 69916 & 0.50 &  160 & $<$1.0 (16) &  1.16 &  1.00 & 3.5 & $1.0$ & CH$_3$C(O)NH$_2$ \\ 
 NCC(O)NH$_2$ & 70504 & 0.50 &  150 & $<$9.2 (15) &  1.54 &  1.00 & 3.5 & $0.0$ & \ldots \\ 
 gauche-HOCH$_2$CH$_2$CN & 71401 & 0.50 &  150 & $<$2.2 (16) &  1.00 &  1.12 & 3.5 & $0.0$ & \ldots \\ 
 gauche-CH$_3$OCH$_2$CN & 71402 & 0.50 &  150 & $<$1.3 (16) &  1.00 &  1.01 & 3.5 & $0.0$ & \ldots \\ 
 cis-C$_2$H$_5$NCO $^\star$ & 71508 & 0.70 &  140 & $<$1.8 (16) &  5.17 &  1.00 & 3.5 & $-0.5$ & CH$_3$NCO \\ 
 syn-C$_2$H$_3$C(O)NH$_2$ & 71903 & 0.50 &  160 & $<$1.8 (16) &  2.26 &  1.01 & 3.5 & $1.0$ & CH$_3$C(O)NH$_2$ \\ 
 C$_2$H$_5$C(O)NH$_2$ & 73801 & 0.50 &  150 & $<$3.0 (16) &  1.00 &  1.00 & 3.5 & $0.0$ & \ldots \\ 
 trans-sc-C$_2$H$_5$NHCHO & 73902 & 0.50 &  160 & $<$2.6 (16) &  3.69 &  1.01 & 3.5 & $0.0$ & CH$_3$NHCHO \\ 
 NH$_2$CH$_2$C(O)NH$_2$ & 74904 & 0.50 &  150 & $<$1.4 (16) &  1.23 &  1.00 & 3.5 & $0.0$ & \ldots \\ 
 NH$_2$CH$_2$COOH I & 75511 & 0.60 &  140 & $<$1.1 (17) &  2.34 &  1.00 & 3.5 & $0.6$ & NH$_2$CH$_2$CN \\ 
 NH$_2$CH$_2$COOH II & 75512 & 0.60 &  140 & $<$2.5 (18) &  2.34 &   807 & 3.5 & $0.6$ & NH$_2$CH$_2$CN \\ 
 g'Gg'-CH$_3$CH(NH$_2$)CH$_2$OH & 75518 & 0.50 &  150 & $<$1.1 (16) &  1.85 &  1.14 & 3.5 & $0.0$ & \ldots \\ 
 gG'g-CH$_3$CH(NH$_2$)CH$_2$OH & 75519 & 0.50 &  150 & $<$1.5 (18) &  1.85 &  7.94 & 3.5 & $0.0$ & \ldots \\ 
 syn-HOCH$_2$C(O)NH$_2$ $^\star$ & 75901 & 0.50 &  150 & $<$6.6 (15) &  2.53 &  1.04 & 3.5 & $0.0$ & \ldots \\ 
 OC(CN)$_2$ & 80601 & 0.50 &  150 & $<$5.9 (17) &  1.98 &  1.00 & 3.5 & $0.0$ & \ldots \\ 
 NCCH$_2$C(O)NH$_2$ & 84902 & 0.50 &  150 & $<$4.1 (17) &  3.40 &  1.00 & 3.5 & $0.0$ & \ldots \\ 
 CH$_3$CH(NH$_2$)COOH I & 89502 & 0.50 &  150 & $<$1.3 (17) &  4.14 &  1.29 & 3.5 & $0.0$ & \ldots \\ 
 CH$_3$CH(NH$_2$)COOH II & 89503 & 0.50 &  150 & $<$6.8 (16) &  4.14 &  11.0 & 3.5 & $0.0$ & \ldots \\ 
\hline 
\multicolumn{9}{c}{\textit{N-bearing molecules}} \\ 
 CH$_2$NH$_2$$^+$ & 30519 & 0.70 &  140 & $<$7.0 (16) &  1.00 &  1.00 & 3.5 & $0.0$ & CH$_2$NH \\ 
 CH$_2$CN $^\star$ & 40601 & 0.70 &  150 & $<$3.2 (15) &  1.05 &  1.00 & 3.5 & $0.0$ & CH$_3$CN \\ 
 CH$_2$CNH $^\star$ & 41503 & 0.50 &  150 & $<$2.1 (16) &  1.03 &  1.00 & 3.5 & $0.0$ & \ldots \\ 
 CH$_3$CNH$^+$ & 42504 & 0.70 &  150 & $<$7.2 (15) &  1.03 &  1.00 & 3.5 & $0.0$ & CH$_3$CN \\ 
 C$_2$H$_3$NH$_2$ $^\star$ & 43504 & 0.50 &  140 & $<$3.1 (16) &  1.04 &  1.00 & 3.5 & $0.0$ & CH$_3$NH$_2$ \\ 
 E-CH$_3$CHNH $^\star$ & 43907 & 0.50 &  150 & $<$2.1 (16) &  1.21 &  1.00 & 3.5 & $0.0$ & \ldots \\ 
 Z-CH$_3$CHNH $^\star$ & 43908 & 0.50 &  150 & $<$1.1 (16) &  1.23 &  1.00 & 3.5 & $0.0$ & \ldots \\ 
 anti-C$_2$H$_5$NH$_2$ $^\star$ & 45402 & 0.50 &  140 & $<$1.0 (17) &  1.19 &  2.15 & 3.5 & $0.0$ & CH$_3$NH$_2$ \\ 
 CH$_3$NHCH$_3$ & 45601 & 0.50 &  140 & $<$1.2 (17) &  1.23 &  1.00 & 3.5 & $0.0$ & CH$_3$NH$_2$ \\ 
 C$_3$N $^\star$ & 50511 & 0.70 &  160 & $<$1.0 (15) &  1.26 &  1.00 & 4.0 & $-0.4$ & HC$_3$N \\ 
 HNC$_3$ $^\star$ & 51528 & 0.70 &  160 & $<$7.8 (13) &  1.56 &  1.00 & 4.0 & $-0.4$ & HC$_3$N \\ 
 HC$_3$NH$^+$ $^\star$ & 52503 & 0.70 &  160 & $<$1.1 (15) &  1.55 &  1.00 & 4.0 & $-0.4$ & HC$_3$N \\ 
 C$_2$H$_3$NC & 53007 & 1.00 &  200 & $<$2.5 (15) &  1.49 &  1.00 & 4.5 & $0.0$ & C$_2$H$_3$CN \\ 
 Z-CHCCHNH $^\star$ & 53523 & 0.50 &  150 & $<$6.5 (15) &  1.24 &  1.06 & 3.5 & $0.0$ & \ldots \\ 
 E-CHCCHNH & 53524 & 0.50 &  150 & $<$2.3 (17) &  1.23 &  18.4 & 3.5 & $0.0$ & \ldots \\ 
 C$_2$H$_3$CNH$^+$ & 54518 & 1.00 &  200 & $<$5.9 (15) &  1.47 &  1.00 & 4.5 & $0.0$ & C$_2$H$_3$CN \\ 
 Z-NHCHCN & 54805 & 0.50 &  150 & $<$1.8 (16) &  1.18 &  1.00 & 3.5 & $0.0$ & \ldots \\ 
 CH$_2$NCN & 54806 & 0.50 &  150 & $<$9.1 (14) &  1.14 &  1.00 & 3.5 & $0.0$ & \ldots \\ 
 C$_2$H$_5$NC & 55507 & 0.50 &  150 & $<$3.0 (15) &  1.48 &  1.00 & 3.5 & $0.0$ & \ldots \\ 
 E-C$_2$H$_5$CHNH & 57801 & 0.50 &  150 & $<$6.5 (16) &  1.00 &  1.08 & 3.5 & $0.0$ & \ldots \\ 
 Z-C$_2$H$_5$CHNH & 57802 & 0.50 &  150 & $<$9.5 (16) &  1.00 &  13.6 & 3.5 & $0.0$ & \ldots \\ 
 trans-iso-C$_3$H$_7$NH$_2$ & 59925 & 0.50 &  140 & $<$1.3 (17) &  1.31 &  1.43 & 3.5 & $0.0$ & CH$_3$NH$_2$ \\ 
 Trans-trans-normal-C$_3$H$_7$NH$_2$ & 59928 & 0.50 &  140 & $<$1.8 (17) &  1.66 &  4.44 & 3.5 & $0.0$ & CH$_3$NH$_2$ \\ 
 CH$_3$CCNC & 65505 & 0.50 &  120 & $<$7.7 (14) &  1.54 &  1.00 & 3.5 & $0.5$ & CH$_3$C$_3$N \\ 
 CH$_2$CCHCN $^\star$ & 65506 & 0.50 &  120 & $<$1.9 (15) &  1.27 &  1.00 & 3.5 & $0.5$ & CH$_3$C$_3$N \\ 
 CHCCH$_2$CN $^\star$ & 65514 & 0.50 &  120 & $<$2.6 (15) &  1.30 &  1.00 & 3.5 & $0.5$ & CH$_3$C$_3$N \\ 
 CH$_2$CCHNC & 65519 & 0.50 &  120 & $<$2.3 (15) &  1.36 &  1.00 & 3.5 & $0.5$ & CH$_3$C$_3$N \\ 
 NCCH$_2$CN $^\star$ & 66801 & 0.50 &  150 & $<$5.6 (15) &  1.39 &  1.00 & 3.5 & $0.0$ & \ldots \\ 
 CNCH$_2$NC & 66902 & 0.50 &  150 & $<$5.9 (15) &  1.97 &  1.00 & 3.5 & $0.0$ & \ldots \\ 
 E-CH$_3$CHCHCN $^\star$ & 67513 & 0.50 &  140 & $<$2.4 (15) &  1.60 &  1.00 & 3.5 & $-0.5$ & normal-C$_3$H$_7$CN \\ 
 syn-CH$_2$CHCH$_2$CN $^\star$ & 67602 & 0.50 &  140 & $<$7.0 (15) &  1.69 &  1.03 & 3.5 & $-0.5$ & normal-C$_3$H$_7$CN \\ 
 gauche-CH$_2$CHCH$_2$CN $^\star$ & 67603 & 0.50 &  140 & $<$1.1 (17) &  1.69 &  32.6 & 3.5 & $-0.5$ & normal-C$_3$H$_7$CN \\ 
 CH$_2$NCH$_2$CN & 68506 & 0.60 &  140 & $<$8.0 (15) & \ldots &  1.00 & 3.5 & $0.6$ & NH$_2$CH$_2$CN \\ 
 Z-CH$_3$C(NH)CN & 68512 & 0.60 &  140 & $<$2.1 (16) &  1.59 &  1.00 & 3.5 & $0.6$ & NH$_2$CH$_2$CN \\ 
 CH$_3$CH(NH$_2$)CN & 70505 & 0.60 &  140 & $<$1.1 (16) &  1.55 &  1.00 & 3.5 & $0.6$ & NH$_2$CH$_2$CN \\ 
 NH$_2$CH$_2$CH$_2$CN I & 70513 & 0.60 &  140 & $<$1.4 (16) &  1.89 &  1.06 & 3.5 & $0.6$ & NH$_2$CH$_2$CN \\ 
 NH$_2$CH$_2$CH$_2$CN II & 70514 & 0.60 &  140 & $<$6.0 (16) &  1.89 &  1.06 & 3.5 & $0.6$ & NH$_2$CH$_2$CN \\ 
 HC$_5$NH$^+$ $^\star$ & 76524 & 0.50 &  160 & $<$1.6 (15) &  3.92 &  1.00 & 3.5 & $0.0$ & HC$_5$N \\ 
 C$_2$H$_3$C$_3$N $^\star$ & 77506 & 0.50 &  120 & $<$1.7 (15) &  1.84 &  1.00 & 3.5 & $0.5$ & CH$_3$C$_3$N \\ 
 CH$_3$CHCCHCN & 79506 & 0.50 &  120 & $<$3.5 (15) &  2.07 &  1.00 & 3.5 & $0.5$ & CH$_3$C$_3$N \\ 
 C$_2$H$_5$C$_3$N & 79507 & 0.50 &  120 & $<$1.3 (15) &  2.18 &  1.00 & 3.5 & $0.5$ & CH$_3$C$_3$N \\ 
 NH$_2$CH(CN)$_2$ & 81801 & 0.60 &  140 & $<$7.7 (15) &  1.93 &  1.00 & 3.5 & $0.6$ & NH$_2$CH$_2$CN \\ 
 anti-anti-normal-C$_4$H$_9$CN & 83501 & 0.50 &  140 & $<$2.9 (16) &  3.96 &  2.43 & 3.5 & $-0.5$ & normal-C$_3$H$_7$CN \\ 
 gauche-anti-normal-C$_4$H$_9$CN & 83502 & 0.50 &  140 & $<$2.6 (16) &  3.96 &  1.08 & 3.5 & $-0.5$ & normal-C$_3$H$_7$CN \\ 
 anti-gauche-normal-C$_4$H$_9$CN & 83503 & 0.50 &  140 & $<$1.8 (16) &  3.96 &  2.24 & 3.5 & $-0.5$ & normal-C$_3$H$_7$CN \\ 
 tert-C$_4$H$_9$CN & 83504 & 0.50 &  140 & $<$2.3 (15) &  1.91 &  1.00 & 3.5 & $-0.5$ & normal-C$_3$H$_7$CN \\ 
 anti-C$_2$H$_5$CH(CN)CH$_3$ & 83505 & 0.50 &  140 & $<$8.5 (15) &  2.77 &  1.13 & 3.5 & $-0.5$ & normal-C$_3$H$_7$CN \\ 
 anti-(CH$_3$)$_2$CHCH$_2$CN & 83506 & 0.50 &  140 & $<$1.4 (16) &  2.82 &  1.00 & 3.5 & $-0.5$ & normal-C$_3$H$_7$CN \\ 
 gauche-(CH$_3$)$_2$CHCH$_2$CN & 83507 & 0.50 &  140 & $<$4.9 (16) &  2.74 &  1.00 & 3.5 & $-0.5$ & normal-C$_3$H$_7$CN \\ 
 HC$_7$N $^\star$ & 99501 & 0.50 &  160 & $<$2.5 (15) & 12.26 &  1.00 & 3.5 & $0.0$ & HC$_5$N \\ 
 c-C$_6$H$_5$CN $^\star$ & 103501 & 0.50 &  150 & $<$4.2 (15) &  1.69 &  1.00 & 3.5 & $0.0$ & \ldots \\ 
\hline 
\multicolumn{9}{c}{\textit{S-bearing molecules}} \\ 
 c-C$_2$H$_4$S & 60509 & 0.50 &  150 & $<$1.2 (16) &  1.00 &  1.00 & 3.5 & $-0.3$ & CH$_3$SH \\ 
 syn-C$_2$H$_3$SH & 60521 & 0.50 &  150 & $<$1.0 (17) &  1.13 &  1.00 & 3.5 & $-0.3$ & CH$_3$SH \\ 
 anti-C$_2$H$_3$SH & 60522 & 0.50 &  150 & $<$1.0 (17) &  1.13 &  1.00 & 3.5 & $-0.3$ & CH$_3$SH \\ 
 CH$_3$CHS $^\star$ & 60701 & 0.50 &  150 & $<$8.2 (15) &  1.02 &  1.00 & 3.5 & $-0.3$ & CH$_3$SH \\ 
 gauche-C$_2$H$_5$SH $^\star$ & 62523 & 0.50 &  150 & $<$3.1 (16) &  1.31 &  1.17 & 3.5 & $-0.3$ & CH$_3$SH \\ 
 anti-C$_2$H$_5$SH & 62524 & 0.50 &  150 & $<$2.5 (17) &  1.45 &  6.81 & 3.5 & $-0.3$ & CH$_3$SH \\ 
 AA-CH$_3$SCH$_3$ $^\star$ & 62902 & 0.50 &  150 & $<$4.8 (16) &  1.61 &  1.00 & 3.5 & $-0.3$ & CH$_3$SH \\ 
 trans-HCSSH & 78506 & 0.50 &  150 & $<$8.6 (15) &  1.07 &  1.00 & 3.5 & $-0.3$ & CH$_3$SH \\ 
 cis-HCSSH & 78507 & 0.50 &  150 & $<$8.6 (16) &  1.07 &  1.00 & 3.5 & $-0.3$ & CH$_3$SH \\ 
\hline 
\multicolumn{9}{c}{\textit{S- and O-bearing molecules}} \\ 
 trans-HC(O)SH $^\star$ & 62515 & 0.50 &  150 & $<$1.5 (16) &  1.04 &  1.11 & 3.5 & $0.0$ & \ldots \\ 
 cis-HC(O)SH & 62516 & 0.50 &  150 & $<$4.2 (16) &  1.04 &  10.2 & 3.5 & $0.0$ & \ldots \\ 
\hline 
\multicolumn{9}{c}{\textit{S- and N-bearing molecules}} \\ 
 HCNS $^\star$ & 59510 & 0.50 &   40 & $<$6.0 (13) &  1.00 &  1.00 & 3.5 & $0.0$ & HNCS \\ 
 NH$_2$CHS & 61523 & 0.60 &  170 & $<$2.1 (15) &  1.06 &  1.00 & 3.5 & $0.0$ & NH$_2$CHO \\ 
\hline 
\multicolumn{9}{c}{\textit{Cl-bearing molecules}} \\ 
 CH$_3$Cl $^\star$ & 50007 & 0.50 &  150 & $<$1.5 (16) &  1.00 &  1.00 & 3.5 & $0.0$ & \ldots \\ 
\hline 
\multicolumn{9}{c}{\textit{P-bearing molecules}} \\ 
 CH$_3$CP & 58507 & 0.50 &  150 & $<$5.3 (15) &  1.06 &  1.00 & 3.5 & $0.0$ & \ldots \\ 
 C$_2$H$_5$CP & 72509 & 0.50 &  150 & $<$1.9 (16) &  1.55 &  1.00 & 3.5 & $0.0$ & \ldots \\ 
\hline 
\multicolumn{9}{c}{\textit{Hydrocarbons}} \\ 
 C$_2$H$_3$$^+$ $^\star$ & 27514 & 0.50 &  150 & $<$5.0 (17) &  1.01 &  1.00 & 3.5 & $0.0$ & \ldots \\ 
 CH$_2$CHCH$_3$ $^\star$ & 42516 & 0.50 &  150 & $<$6.1 (17) &  1.22 &  1.00 & 3.5 & $0.0$ & \ldots \\ 
 C$_3$H$_8$ & 44013 & 0.50 &  150 & $<$8.0 (18) &  1.00 &  1.00 & 3.5 & $0.0$ & \ldots \\ 
 C$_2$H$_5$CCH $^\star$ & 54519 & 0.80 &  150 & $<$5.7 (16) &  1.43 &  1.00 & 4.5 & $-0.8$ & CH$_3$CCH \\ 
 (CH$_3$)$_2$CCH$_2$ $^\star$ & 56526 & 0.50 &  150 & $<$2.3 (17) &  1.50 &  1.00 & 3.5 & $0.0$ & \ldots \\ 
 H$_2$C(CCH)$_2$ $^\star$ & 64520 & 0.80 &  150 & $<$3.1 (17) &  1.57 &  1.00 & 4.5 & $-0.8$ & CH$_3$CCH \\ 
\hline 
 \end{longtable}
 \tablefoot{
 \tablefoottext{a}{Molecules that have already been detected in the interstellar medium are marked with a star.}
 \tablefoottext{b}{Entry number in our Weeds local database. See note b of Table~\ref{t:spectrobib}.}
 \tablefoottext{c}{Source diameter (\textit{FWHM}).}
 \tablefoottext{d}{Rotational temperature.}
 \tablefoottext{e}{Upper limit to the total column density of the molecule. $X$ ($Y$) means $X \times 10^Y$.}
 \tablefoottext{f}{Correction factor that was applied to the column density to account for the contribution of vibrationally or torsionally excited states, in the cases where this contribution was not included in the partition function of the spectroscopic predictions. In most cases, this factor was estimated in the harmonic approximation.}
 \tablefoottext{g}{Correction factor that was applied to the column density to account for the contribution of conformers, in the cases where this contribution was not included in the partition function of the spectroscopic predictions.}
 \tablefoottext{h}{Linewidth (\textit{FWHM}).}
 \tablefoottext{i}{Velocity offset with respect to the assumed systemic velocity given in Table~\ref{t:sources}.}
 \tablefoottext{j}{Molecule that was used to fix the LTE parameters except for the column density. Ellipsis dots indicate that we used typical parameters of the specific source.}
 }

\begin{longtable}{lrccccccrr}
 \caption{\label{t:uplim_an02}
 Column density upper limits toward  AN02.
} \\ 
 \hline\hline
 \multicolumn{1}{c}{Molecule\tablefootmark{a}} & \multicolumn{1}{c}{Tag\tablefootmark{b}} & \multicolumn{1}{c}{Size\tablefootmark{c}} & \multicolumn{1}{c}{$T_{\mathrm{rot}}$\tablefootmark{d}} & \multicolumn{1}{c}{$N$\tablefootmark{e}} & \multicolumn{1}{c}{$C_{\rm{vib}}$\tablefootmark{f}} & \multicolumn{1}{c}{$C_{\rm{conf}}$\tablefootmark{g}} & \multicolumn{1}{c}{$\Delta V$\tablefootmark{h}} & \multicolumn{1}{c}{$V_{\mathrm{off}}$\tablefootmark{i}} & \multicolumn{1}{c}{Ref.\tablefootmark{j}} \\ 
  & & \multicolumn{1}{c}{$''$} & \multicolumn{1}{c}{K} & \multicolumn{1}{c}{cm$^{-2}$} & & & \multicolumn{2}{c}{km~s$^{-1}$} & \\ 
 \hline
 \endfirsthead
 \caption{continued.} \\ 
 \hline\hline
 \multicolumn{1}{c}{Molecule\tablefootmark{a}} & \multicolumn{1}{c}{Tag\tablefootmark{b}} & \multicolumn{1}{c}{Size\tablefootmark{c}} & \multicolumn{1}{c}{$T_{\mathrm{rot}}$\tablefootmark{d}} & \multicolumn{1}{c}{$N$\tablefootmark{e}} & \multicolumn{1}{c}{$C_{\rm{vib}}$\tablefootmark{f}} & \multicolumn{1}{c}{$C_{\rm{conf}}$\tablefootmark{g}} & \multicolumn{1}{c}{$\Delta V$\tablefootmark{h}} & \multicolumn{1}{c}{$V_{\mathrm{off}}$\tablefootmark{i}} & \multicolumn{1}{c}{Ref.\tablefootmark{j}} \\ 
 \hline
 \endhead
 \hline
 \endfoot
\hline 
\multicolumn{9}{c}{\textit{O-bearing molecules}} \\ 
 HCO $^\star$ & 29004 & 0.70 &  170 & $<$1.8 (16) &  1.00 &  1.00 & 4.0 & $0.0$ & H$_2$CO \\ 
 & & 0.70 &  170 & $<$9.5 (15) &  1.00 &  1.00 & 3.0 & $4.0$ &  \\ 
 H$_2$COH$^+$ $^\star$ & 31504 & 0.70 &  170 & $<$1.2 (16) &  1.00 &  1.00 & 4.0 & $0.0$ & H$_2$CO \\ 
 & & 0.70 &  170 & $<$6.0 (15) &  1.00 &  1.00 & 3.0 & $4.0$ &  \\ 
 CH$_2$OH & 31515 & 0.70 &  170 & $<$1.5 (17) &  1.00 &  1.00 & 4.0 & $0.0$ & CH$_3$OH \\ 
 & & 0.70 &  170 & $<$7.5 (16) &  1.00 &  1.00 & 3.0 & $4.0$ &  \\ 
 CH$_3$O $^\star$ & 31518 & 0.70 &  170 & $<$5.0 (16) &  1.00 &  1.00 & 4.0 & $0.0$ & CH$_3$OH \\
\hline 
 \end{longtable}
 \tablefoot{
 \tablefoottext{a}{Molecules that have already been detected in the interstellar medium are marked with a star.}
 \tablefoottext{b}{Entry number in our Weeds local database. See note b of Table~\ref{t:spectrobib}.}
 \tablefoottext{c}{Source diameter (\textit{FWHM}).}
 \tablefoottext{d}{Rotational temperature.}
 \tablefoottext{e}{Upper limit to the total column density of the molecule. $X$ ($Y$) means $X \times 10^Y$.}
 \tablefoottext{f}{Correction factor that was applied to the column density to account for the contribution of vibrationally or torsionally excited states, in the cases where this contribution was not included in the partition function of the spectroscopic predictions. In most cases, this factor was estimated in the harmonic approximation.}
 \tablefoottext{g}{Correction factor that was applied to the column density to account for the contribution of conformers, in the cases where this contribution was not included in the partition function of the spectroscopic predictions.}
 \tablefoottext{h}{Linewidth (\textit{FWHM}).}
 \tablefoottext{i}{Velocity offset with respect to the assumed systemic velocity given in Table~\ref{t:sources}.}
 \tablefoottext{j}{Molecule that was used to fix the LTE parameters except for the column density. Ellipsis dots indicate that we used typical parameters of the specific source.}
 }

\clearpage
\begin{longtable}{lrccccccrr}
 \caption{\label{t:uplim_an03}
 Column density upper limits toward  AN03.
} \\ 
 \hline\hline
 \multicolumn{1}{c}{Molecule\tablefootmark{a}} & \multicolumn{1}{c}{Tag\tablefootmark{b}} & \multicolumn{1}{c}{Size\tablefootmark{c}} & \multicolumn{1}{c}{$T_{\mathrm{rot}}$\tablefootmark{d}} & \multicolumn{1}{c}{$N$\tablefootmark{e}} & \multicolumn{1}{c}{$C_{\rm{vib}}$\tablefootmark{f}} & \multicolumn{1}{c}{$C_{\rm{conf}}$\tablefootmark{g}} & \multicolumn{1}{c}{$\Delta V$\tablefootmark{h}} & \multicolumn{1}{c}{$V_{\mathrm{off}}$\tablefootmark{i}} & \multicolumn{1}{c}{Ref.\tablefootmark{j}} \\ 
  & & \multicolumn{1}{c}{$''$} & \multicolumn{1}{c}{K} & \multicolumn{1}{c}{cm$^{-2}$} & & & \multicolumn{2}{c}{km~s$^{-1}$} & \\ 
 \hline
 \endfirsthead
 \caption{continued.} \\ 
 \hline\hline
 \multicolumn{1}{c}{Molecule\tablefootmark{a}} & \multicolumn{1}{c}{Tag\tablefootmark{b}} & \multicolumn{1}{c}{Size\tablefootmark{c}} & \multicolumn{1}{c}{$T_{\mathrm{rot}}$\tablefootmark{d}} & \multicolumn{1}{c}{$N$\tablefootmark{e}} & \multicolumn{1}{c}{$C_{\rm{vib}}$\tablefootmark{f}} & \multicolumn{1}{c}{$C_{\rm{conf}}$\tablefootmark{g}} & \multicolumn{1}{c}{$\Delta V$\tablefootmark{h}} & \multicolumn{1}{c}{$V_{\mathrm{off}}$\tablefootmark{i}} & \multicolumn{1}{c}{Ref.\tablefootmark{j}} \\ 
 \hline
 \endhead
 \hline
 \endfoot
\hline 
\multicolumn{9}{c}{\textit{O-bearing molecules}} \\ 
 HCO $^\star$ & 29004 & 0.70 &  200 & $<$2.5 (16) &  1.00 &  1.00 & 4.7 & $0.1$ & H$_2$CO \\ 
 H$_2$COH$^+$ $^\star$ & 31504 & 0.70 &  200 & $<$1.8 (16) &  1.00 &  1.00 & 4.7 & $0.1$ & H$_2$CO \\ 
 CH$_2$OH & 31515 & 0.70 &  190 & $<$2.0 (17) &  1.00 &  1.00 & 4.7 & $0.0$ & CH$_3$OH \\ 
 CH$_3$O $^\star$ & 31518 & 0.70 &  190 & $<$1.0 (17) &  1.01 &  1.00 & 4.7 & $0.0$ & CH$_3$OH \\ 
 CH$_2$CHO & 43523 & 1.00 &  170 & $<$8.5 (15) &  1.06 &  1.00 & 4.7 & $0.3$ & CH$_3$CHO \\ 
 syn-C$_2$H$_3$OH $^\star$ & 44506 & 0.70 &  160 & $<$6.5 (16) &  1.05 &  1.03 & 4.7 & $0.3$ & C$_2$H$_5$OH \\ 
 anti-C$_2$H$_3$OH $^\star$ & 44507 & 0.70 &  160 & $<$4.9 (17) &  1.05 &  31.5 & 4.7 & $0.3$ & C$_2$H$_5$OH \\ 
 c-HCOOH $^\star$ & 46507 & 0.70 &  150 & $<$2.0 (21) &  1.00 & 485469 & 4.5 & $0.4$ & t-HCOOH \\ 
 trans-C$_2$H$_3$CHO $^\star$ & 56519 & 1.00 &  170 & $<$4.4 (15) &  1.47 &  1.01 & 4.7 & $0.3$ & CH$_3$CHO \\ 
 CH$_3$CHCO $^\star$ & 56701 & 0.70 &  200 & $<$1.2 (16) &  1.00 &  1.00 & 4.7 & $-0.2$ & CH$_2$CO \\ 
 c-CH(CH$_3$)CH$_2$O $^\star$ & 58514 & 0.70 &  170 & $<$4.9 (16) &  1.32 &  1.00 & 5.0 & $0.0$ & \ldots \\ 
 gauche-C$_2$H$_5$CHO & 58519 & 1.00 &  140 & $<$3.9 (17) &  1.63 &  1.03 & 4.7 & $0.3$ & s-C$_2$H$_5$CHO \\ 
 C$_2$H$_5$OCH$_3$ $^\star$ & 60901 & 1.00 &  140 & $<$1.8 (17) &  1.76 &  1.02 & 4.7 & $-0.2$ & CH$_3$OCH$_3$ \\ 
 HOCHCHOH $^\star$ & 60920 & 0.50 &  150 & $<$3.8 (16) &  1.26 &  1.00 & 4.5 & $0.4$ & a-(CH2OH)$_2$ \\ 
 CH$_3$OCH$_2$OH $^\star$ & 62527 & 0.70 &  190 & $<$5.0 (18) &  2.47 &  1.01 & 4.7 & $0.0$ & CH$_3$OH \\ 
 s-cis-C$_2$H$_3$COOH & 72502 & 0.70 &  170 & $<$1.2 (17) &  1.99 &  1.00 & 5.0 & $0.0$ & \ldots \\ 
 s-trans-C$_2$H$_3$COOH & 72503 & 0.70 &  170 & $<$6.0 (16) &  1.99 &  1.00 & 5.0 & $0.0$ & \ldots \\ 
 HOCHCHCHO $^\star$ & 72504 & 1.00 &  170 & $<$1.7 (16) &  1.32 &  1.00 & 4.7 & $0.3$ & CH$_3$CHO \\ 
 cis-trans-C$_2$H$_3$OCHO & 72505 & 0.70 &  200 & $<$1.8 (17) &  3.70 &  1.00 & 4.7 & $0.0$ & C$_2$H$_5$OCHO \\ 
 CH$_3$C(O)CHO & 72901 & 1.00 &  140 & $<$2.4 (17) &  2.36 &  1.00 & 4.7 & $-0.2$ & CH$_3$COCH$_3$ \\ 
 gauche-(CH$_3$)$_2$CHCHO & 72902 & 1.00 &  170 & $<$4.7 (16) &  3.09 &  1.27 & 4.7 & $0.3$ & CH$_3$CHO \\ 
 cis-gauche-C$_3$H$_7$CHO & 72903 & 1.00 &  170 & $<$1.0 (17) &  3.15 &  3.25 & 4.7 & $0.3$ & CH$_3$CHO \\ 
 cis-trans-C$_3$H$_7$CHO & 72904 & 1.00 &  170 & $<$6.8 (16) &  3.88 &  1.75 & 4.7 & $0.3$ & CH$_3$CHO \\ 
 CH$_2$C(OH)CHO & 72905 & 0.70 &  150 & $<$1.3 (16) &  1.32 &  1.00 & 5.0 & $0.0$ & CH$_2$(OH)CHO \\ 
 CH$_3$C(O)CH$_2$OH $^\star$ & 74003 & 0.70 &  150 & $<$1.5 (17) &  4.92 &  1.00 & 5.0 & $0.0$ & CH$_3$OCH$_3$ \\ 
 OCHCOOH & 74516 & 0.70 &  170 & $<$4.7 (16) &  1.52 &  1.03 & 5.0 & $0.0$ & \ldots \\ 
 CH$_3$OCH$_2$CHO & 74517 & 0.70 &  170 & $<$2.1 (16) &  7.95 &  1.07 & 5.0 & $0.0$ & \ldots \\ 
 CH$_3$CH(OH)CHO & 74519 & 0.70 &  150 & $<$3.1 (16) &  2.06 &  1.00 & 5.0 & $0.0$ & CH$_2$(OH)CHO \\ 
 C$_2$H$_5$COOH & 74905 & 0.70 &  170 & $<$1.4 (17) &  3.40 &  1.18 & 5.0 & $0.0$ & \ldots \\ 
 tert-C$_4$H$_9$OH & 74908 & 0.70 &  160 & $<$1.1 (17) &  1.52 &  1.00 & 4.7 & $0.3$ & C$_2$H$_5$OH \\ 
 aG'g-CH$_3$CHOHCH$_2$OH & 76513 & 0.70 &  160 & $<$1.0 (17) &  2.32 &  2.24 & 4.7 & $0.3$ & C$_2$H$_5$OH \\ 
 CH$_2$(OH)COOH & 76514 & 0.70 &  150 & $<$3.5 (16) &  1.95 &  1.00 & 5.0 & $0.0$ & CH$_2$(OH)CHO \\ 
 gG'a-CH$_3$CHOHCH$_2$OH & 76515 & 0.70 &  160 & $<$2.0 (17) &  2.31 &  4.35 & 4.7 & $0.3$ & C$_2$H$_5$OH \\ 
 g'G'g-CH$_3$CHOHCH$_2$OH & 76516 & 0.70 &  160 & $<$2.8 (17) &  2.19 &  6.29 & 4.7 & $0.3$ & C$_2$H$_5$OH \\ 
 a'GG'g-HOCH$_2$CH$_2$CH$_2$OH & 76518 & 0.70 &  160 & $<$6.2 (16) &  2.73 &  1.51 & 4.7 & $0.3$ & C$_2$H$_5$OH \\ 
 gGG'g-HOCH$_2$CH$_2$CH$_2$OH & 76519 & 0.70 &  160 & $<$1.2 (17) &  2.55 &  3.95 & 4.7 & $0.3$ & C$_2$H$_5$OH \\ 
 g'Ga-CH$_3$CHOHCH$_2$OH & 76520 & 0.70 &  160 & $<$7.2 (17) &  2.40 &  15.1 & 4.7 & $0.3$ & C$_2$H$_5$OH \\ 
 gG'g'-CH$_3$CHOHCH$_2$OH & 76521 & 0.70 &  160 & $<$4.7 (17) &  2.19 &  17.7 & 4.7 & $0.3$ & C$_2$H$_5$OH \\ 
 aGg'-CH$_3$CHOHCH$_2$OH & 76522 & 0.70 &  160 & $<$1.8 (18) &  2.32 &  51.6 & 4.7 & $0.3$ & C$_2$H$_5$OH \\ 
 g'Gg-CH$_3$CHOHCH$_2$OH & 76523 & 0.70 &  160 & $<$5.3 (18) &  2.32 &  65.2 & 4.7 & $0.3$ & C$_2$H$_5$OH \\ 
 CH$_3$OCH$_2$CH$_2$OH $^\star$ & 76801 & 0.70 &  190 & $<$1.0 (17) &  1.00 &  1.03 & 4.7 & $0.0$ & CH$_3$OH \\ 
 c-C$_6$H$_5$OH & 94501 & 0.70 &  190 & $<$1.6 (17) &  1.59 &  1.00 & 4.7 & $0.0$ & CH$_3$OH \\ 
\hline 
\multicolumn{9}{c}{\textit{O- and N-bearing molecules}} \\ 
 HCNO $^\star$ & 43509 & 0.90 &  220 & $<$1.4 (15) &  2.47 &  1.00 & 5.0 & $-0.2$ & HNCO \\ 
 HOCN $^\star$ & 43510 & 0.90 &  220 & $<$2.2 (15) &  1.12 &  1.00 & 5.0 & $-0.2$ & HNCO \\ 
 NH$_2$CO & 44904 & 0.90 &  170 & $<$1.0 (16) &  1.00 &  1.00 & 5.0 & $0.5$ & NH$_2$CHO \\ 
 HC(O)CN $^\star$ & 55501 & 0.70 &  170 & $<$9.1 (15) &  1.30 &  1.00 & 5.0 & $0.0$ & \ldots \\ 
 CH$_3$CNO & 57510 & 0.70 &  150 & $<$1.3 (15) &  1.33 &  1.00 & 4.5 & $-0.3$ & CH$_3$NCO \\ 
 CH$_3$OCN & 57511 & 0.70 &  150 & $<$5.3 (15) &  1.77 &  1.00 & 4.5 & $-0.3$ & CH$_3$NCO \\ 
 HOCH$_2$CN $^\star$ & 57701 & 0.70 &  170 & $<$2.7 (16) &  1.00 &  1.01 & 5.0 & $0.0$ & \ldots \\ 
 NH$_2$C(O)NH$_2$ $^\star$ & 60517 & 0.70 &  170 & $<$2.5 (15) &  1.94 &  1.07 & 5.0 & $0.0$ & \ldots \\ 
 NH$_2$CH$_2$CH$_2$OH $^\star$ & 61004 & 0.70 &  170 & $<$2.2 (16) &  1.00 &  1.08 & 5.0 & $0.0$ & \ldots \\ 
 NCCHCO & 67502 & 0.70 &  170 & $<$7.6 (15) &  1.53 &  1.00 & 5.0 & $0.0$ & \ldots \\ 
 c-(C$_2$H$_3$O)-CN & 69518 & 0.70 &  150 & $<$5.2 (15) &  1.30 &  1.00 & 5.0 & $0.4$ & c-C$_2$H$_4$O \\ 
 trans-C$_2$H$_3$NCO & 69901 & 0.70 &  150 & $<$1.9 (16) &  2.65 &  1.00 & 4.5 & $-0.3$ & CH$_3$NCO \\ 
 cis-C$_2$H$_3$NCO & 69902 & 0.70 &  150 & $<$3.2 (17) &  2.69 &  1.00 & 4.5 & $-0.3$ & CH$_3$NCO \\ 
 CHCC(O)NH$_2$ & 69916 & 0.70 &  180 & $<$1.5 (16) &  1.25 &  1.00 & 4.7 & $1.0$ & CH$_3$C(O)NH$_2$ \\ 
 NCC(O)NH$_2$ & 70504 & 0.70 &  170 & $<$1.4 (16) &  1.77 &  1.00 & 5.0 & $0.0$ & \ldots \\ 
 gauche-HOCH$_2$CH$_2$CN & 71401 & 0.70 &  170 & $<$4.6 (16) &  1.00 &  1.15 & 5.0 & $0.0$ & \ldots \\ 
 gauche-CH$_3$OCH$_2$CN & 71402 & 0.70 &  170 & $<$3.1 (16) &  1.00 &  1.02 & 5.0 & $0.0$ & \ldots \\ 
 cis-C$_2$H$_5$NCO $^\star$ & 71508 & 0.70 &  150 & $<$4.6 (16) &  5.80 &  1.00 & 4.5 & $-0.3$ & CH$_3$NCO \\ 
 syn-C$_2$H$_3$C(O)NH$_2$ & 71903 & 0.70 &  180 & $<$2.7 (16) &  2.67 &  1.01 & 4.7 & $1.0$ & CH$_3$C(O)NH$_2$ \\ 
 C$_2$H$_5$C(O)NH$_2$ & 73801 & 0.70 &  170 & $<$5.0 (16) &  1.00 &  1.00 & 5.0 & $0.0$ & \ldots \\ 
 trans-sc-C$_2$H$_5$NHCHO & 73902 & 0.70 &  180 & $<$4.7 (16) &  4.62 &  1.01 & 4.5 & $1.0$ & CH$_3$NHCHO \\ 
 NH$_2$CH$_2$C(O)NH$_2$ & 74904 & 0.70 &  170 & $<$1.8 (16) &  1.31 &  1.00 & 5.0 & $0.0$ & \ldots \\ 
 NH$_2$CH$_2$COOH I & 75511 & 0.70 &  150 & $<$2.1 (17) &  2.56 &  1.00 & 5.0 & $0.0$ & NH$_2$CH$_2$CN \\ 
 NH$_2$CH$_2$COOH II & 75512 & 0.70 &  150 & $<$4.0 (18) &  2.56 &   517 & 5.0 & $0.0$ & NH$_2$CH$_2$CN \\ 
 g'Gg'-CH$_3$CH(NH$_2$)CH$_2$OH & 75518 & 0.70 &  150 & $<$1.1 (16) &  1.85 &  1.14 & 5.0 & $0.0$ & \ldots \\ 
 gG'g-CH$_3$CH(NH$_2$)CH$_2$OH & 75519 & 0.70 &  150 & $<$2.2 (18) &  1.85 &  7.94 & 5.0 & $0.0$ & \ldots \\ 
 syn-HOCH$_2$C(O)NH$_2$ $^\star$ & 75901 & 0.70 &  170 & $<$2.3 (16) &  3.06 &  1.06 & 5.0 & $0.0$ & \ldots \\ 
 OC(CN)$_2$ & 80601 & 0.70 &  170 & $<$1.2 (18) &  2.40 &  1.00 & 5.0 & $0.0$ & \ldots \\ 
 NCCH$_2$C(O)NH$_2$ & 84902 & 0.70 &  170 & $<$1.0 (18) &  4.18 &  1.00 & 5.0 & $0.0$ & \ldots \\ 
 CH$_3$CH(NH$_2$)COOH I & 89502 & 0.70 &  170 & $<$1.9 (17) &  5.42 &  1.40 & 5.0 & $0.0$ & \ldots \\ 
 CH$_3$CH(NH$_2$)COOH II & 89503 & 0.70 &  170 & $<$1.2 (17) &  5.42 &  9.22 & 5.0 & $0.0$ & \ldots \\ 
\hline 
\multicolumn{9}{c}{\textit{N-bearing molecules}} \\ 
 CH$_2$NH$_2$$^+$ & 30519 & 0.70 &  170 & $<$3.0 (17) &  1.00 &  1.00 & 5.0 & $0.4$ & CH$_2$NH \\ 
 CH$_2$CN $^\star$ & 40601 & 0.70 &  180 & $<$8.8 (15) &  1.10 &  1.00 & 4.5 & $-0.4$ & CH$_3$CN \\ 
 CH$_2$CNH $^\star$ & 41503 & 0.70 &  170 & $<$5.3 (16) &  1.06 &  1.00 & 5.0 & $0.0$ & \ldots \\ 
 CH$_3$CNH$^+$ & 42504 & 0.70 &  180 & $<$1.9 (16) &  1.06 &  1.00 & 4.5 & $-0.4$ & CH$_3$CN \\ 
 C$_2$H$_3$NH$_2$ $^\star$ & 43504 & 0.70 &  150 & $<$4.2 (16) &  1.06 &  1.00 & 5.0 & $0.0$ & CH$_3$NH$_2$ \\ 
 E-CH$_3$CHNH $^\star$ & 43907 & 0.70 &  170 & $<$7.7 (16) &  1.28 &  1.00 & 5.0 & $0.0$ & \ldots \\ 
 Z-CH$_3$CHNH $^\star$ & 43908 & 0.70 &  170 & $<$2.6 (16) &  1.30 &  1.00 & 5.0 & $0.0$ & \ldots \\ 
 anti-C$_2$H$_5$NH$_2$ $^\star$ & 45402 & 0.70 &  150 & $<$1.3 (17) &  1.23 &  2.19 & 5.0 & $0.0$ & CH$_3$NH$_2$ \\ 
 CH$_3$NHCH$_3$ & 45601 & 0.70 &  150 & $<$2.3 (17) &  1.28 &  1.00 & 5.0 & $0.0$ & CH$_3$NH$_2$ \\ 
 C$_3$N $^\star$ & 50511 & 0.70 &  200 & $<$2.4 (15) &  1.48 &  1.00 & 5.0 & $-0.6$ & HC$_3$N \\ 
 HNC$_3$ $^\star$ & 51528 & 0.70 &  200 & $<$2.0 (14) &  1.97 &  1.00 & 5.0 & $-0.6$ & HC$_3$N \\ 
 HC$_3$NH$^+$ $^\star$ & 52503 & 0.70 &  200 & $<$3.0 (15) &  2.01 &  1.00 & 5.0 & $-0.6$ & HC$_3$N \\ 
 C$_2$H$_3$NC & 53007 & 0.70 &  190 & $<$1.7 (15) &  1.43 &  1.00 & 4.5 & $-0.6$ & C$_2$H$_3$CN \\ 
 Z-CHCCHNH $^\star$ & 53523 & 0.70 &  170 & $<$1.0 (16) &  1.37 &  1.08 & 5.0 & $0.0$ & \ldots \\ 
 E-CHCCHNH & 53524 & 0.70 &  170 & $<$2.2 (17) &  1.36 &  13.4 & 5.0 & $0.0$ & \ldots \\ 
 C$_2$H$_3$CNH$^+$ & 54518 & 0.70 &  190 & $<$1.1 (16) &  1.41 &  1.00 & 4.5 & $-0.6$ & C$_2$H$_3$CN \\ 
 Z-NHCHCN & 54805 & 0.70 &  170 & $<$2.3 (16) &  1.25 &  1.00 & 5.0 & $0.0$ & \ldots \\ 
 CH$_2$NCN & 54806 & 0.70 &  170 & $<$1.8 (15) &  1.20 &  1.00 & 5.0 & $0.0$ & \ldots \\ 
 C$_2$H$_5$NC & 55507 & 0.70 &  170 & $<$5.0 (15) &  1.67 &  1.00 & 5.0 & $0.0$ & \ldots \\ 
 E-C$_2$H$_5$CHNH & 57801 & 0.70 &  170 & $<$8.9 (16) &  1.00 &  1.11 & 5.0 & $0.0$ & \ldots \\ 
 Z-C$_2$H$_5$CHNH & 57802 & 0.70 &  170 & $<$1.6 (17) &  1.00 &  10.4 & 5.0 & $0.0$ & \ldots \\ 
 trans-iso-C$_3$H$_7$NH$_2$ & 59925 & 0.70 &  150 & $<$2.5 (17) &  1.39 &  1.47 & 5.0 & $0.0$ & CH$_3$NH$_2$ \\ 
 Trans-trans-normal-C$_3$H$_7$NH$_2$ & 59928 & 0.70 &  150 & $<$4.1 (17) &  1.80 &  4.61 & 5.0 & $0.0$ & CH$_3$NH$_2$ \\ 
 CH$_3$CCNC & 65505 & 0.70 &  150 & $<$1.2 (15) &  2.00 &  1.00 & 5.0 & $0.0$ & CH$_3$C$_3$N \\ 
 CH$_2$CCHCN $^\star$ & 65506 & 0.70 &  150 & $<$4.5 (15) &  1.49 &  1.00 & 5.0 & $0.0$ & CH$_3$C$_3$N \\ 
 CHCCH$_2$CN $^\star$ & 65514 & 0.70 &  150 & $<$4.6 (15) &  1.54 &  1.00 & 5.0 & $0.0$ & CH$_3$C$_3$N \\ 
 CH$_2$CCHNC & 65519 & 0.70 &  150 & $<$3.8 (15) &  1.66 &  1.00 & 5.0 & $0.0$ & CH$_3$C$_3$N \\ 
 NCCH$_2$CN $^\star$ & 66801 & 0.70 &  170 & $<$1.2 (16) &  1.55 &  1.00 & 5.0 & $0.0$ & \ldots \\ 
 CNCH$_2$NC & 66902 & 0.70 &  150 & $<$9.8 (15) &  1.97 &  1.00 & 5.0 & $0.0$ & \ldots \\ 
 E-CH$_3$CHCHCN $^\star$ & 67513 & 0.70 &  160 & $<$3.7 (15) &  1.87 &  1.00 & 4.2 & $-0.6$ & normal-C$_3$H$_7$CN \\ 
 syn-CH$_2$CHCH$_2$CN $^\star$ & 67602 & 0.70 &  160 & $<$1.2 (16) &  1.97 &  1.05 & 4.2 & $-0.6$ & normal-C$_3$H$_7$CN \\ 
 gauche-CH$_2$CHCH$_2$CN $^\star$ & 67603 & 0.70 &  160 & $<$1.3 (17) &  1.97 &  21.5 & 4.2 & $-0.6$ & normal-C$_3$H$_7$CN \\ 
 CH$_2$NCH$_2$CN & 68506 & 0.70 &  150 & $<$2.0 (16) & \ldots &  1.00 & 5.0 & $-0.3$ & NH$_2$CH$_2$CN \\ 
 Z-CH$_3$C(NH)CN & 68512 & 0.70 &  150 & $<$3.4 (16) &  1.71 &  1.00 & 5.0 & $-0.3$ & NH$_2$CH$_2$CN \\ 
 CH$_3$CH(NH$_2$)CN & 70505 & 0.70 &  150 & $<$2.0 (16) &  1.68 &  1.01 & 5.0 & $-0.3$ & NH$_2$CH$_2$CN \\ 
 NH$_2$CH$_2$CH$_2$CN I & 70513 & 0.70 &  150 & $<$2.9 (16) &  2.07 &  1.07 & 5.0 & $-0.3$ & NH$_2$CH$_2$CN \\ 
 NH$_2$CH$_2$CH$_2$CN II & 70514 & 0.70 &  150 & $<$1.1 (17) &  2.07 &  1.07 & 5.0 & $-0.3$ & NH$_2$CH$_2$CN \\ 
 HC$_5$NH$^+$ $^\star$ & 76524 & 0.70 &  200 & $<$3.4 (15) &  6.84 &  1.00 & 5.0 & $-0.6$ & HC$_5$N \\ 
 C$_2$H$_3$C$_3$N $^\star$ & 77506 & 0.70 &  150 & $<$3.2 (15) &  2.48 &  1.00 & 5.0 & $0.0$ & CH$_3$C$_3$N \\ 
 CH$_3$CHCCHCN & 79506 & 0.70 &  150 & $<$4.5 (15) &  2.98 &  1.00 & 5.0 & $0.0$ & CH$_3$C$_3$N \\ 
 C$_2$H$_5$C$_3$N & 79507 & 0.70 &  150 & $<$2.8 (15) &  3.15 &  1.00 & 5.0 & $0.0$ & CH$_3$C$_3$N \\ 
 NH$_2$CH(CN)$_2$ & 81801 & 0.70 &  150 & $<$8.6 (15) &  2.14 &  1.00 & 5.0 & $-0.3$ & NH$_2$CH$_2$CN \\ 
 anti-anti-normal-C$_4$H$_9$CN & 83501 & 0.70 &  160 & $<$5.6 (16) &  5.43 &  2.59 & 4.2 & $-0.6$ & normal-C$_3$H$_7$CN \\ 
 gauche-anti-normal-C$_4$H$_9$CN & 83502 & 0.70 &  160 & $<$6.0 (16) &  5.43 &  1.11 & 4.2 & $-0.6$ & normal-C$_3$H$_7$CN \\ 
 anti-gauche-normal-C$_4$H$_9$CN & 83503 & 0.70 &  160 & $<$3.7 (16) &  5.43 &  2.30 & 4.2 & $-0.6$ & normal-C$_3$H$_7$CN \\ 
 tert-C$_4$H$_9$CN & 83504 & 0.70 &  160 & $<$3.7 (15) &  2.44 &  1.00 & 4.2 & $-0.6$ & normal-C$_3$H$_7$CN \\ 
 anti-C$_2$H$_5$CH(CN)CH$_3$ & 83505 & 0.70 &  160 & $<$1.8 (16) &  3.69 &  1.19 & 4.2 & $-0.6$ & normal-C$_3$H$_7$CN \\ 
 anti-(CH$_3$)$_2$CHCH$_2$CN & 83506 & 0.70 &  160 & $<$3.7 (16) &  3.69 &  1.00 & 4.2 & $-0.6$ & normal-C$_3$H$_7$CN \\ 
 gauche-(CH$_3$)$_2$CHCH$_2$CN & 83507 & 0.70 &  160 & $<$5.4 (16) &  3.59 &  1.00 & 4.2 & $-0.6$ & normal-C$_3$H$_7$CN \\ 
 HC$_7$N $^\star$ & 99501 & 0.70 &  200 & $<$1.1 (16) & 28.40 &  1.00 & 5.0 & $-0.6$ & HC$_5$N \\ 
 c-C$_6$H$_5$CN $^\star$ & 103501 & 0.70 &  170 & $<$7.0 (15) &  1.99 &  1.00 & 5.0 & $0.0$ & \ldots \\ 
\hline 
\multicolumn{9}{c}{\textit{S-bearing molecules}} \\ 
 c-C$_2$H$_4$S & 60509 & 0.70 &  180 & $<$2.0 (16) &  1.01 &  1.00 & 5.0 & $0.0$ & CH$_3$SH \\ 
 syn-C$_2$H$_3$SH & 60521 & 0.70 &  180 & $<$1.8 (17) &  1.22 &  1.00 & 5.0 & $0.0$ & CH$_3$SH \\ 
 anti-C$_2$H$_3$SH & 60522 & 0.70 &  180 & $<$1.1 (17) &  1.22 &  1.00 & 5.0 & $0.0$ & CH$_3$SH \\ 
 CH$_3$CHS $^\star$ & 60701 & 0.70 &  180 & $<$1.6 (16) &  1.05 &  1.00 & 5.0 & $0.0$ & CH$_3$SH \\ 
 gauche-C$_2$H$_5$SH $^\star$ & 62523 & 0.70 &  180 & $<$6.4 (16) &  1.52 &  1.21 & 5.0 & $0.0$ & CH$_3$SH \\ 
 anti-C$_2$H$_5$SH & 62524 & 0.70 &  180 & $<$3.5 (17) &  1.72 &  5.86 & 5.0 & $0.0$ & CH$_3$SH \\ 
 AA-CH$_3$SCH$_3$ $^\star$ & 62902 & 0.70 &  180 & $<$7.9 (16) &  1.96 &  1.00 & 5.0 & $0.0$ & CH$_3$SH \\ 
 trans-HCSSH & 78506 & 0.70 &  180 & $<$9.0 (15) &  1.13 &  1.00 & 5.0 & $0.0$ & CH$_3$SH \\ 
 cis-HCSSH & 78507 & 0.70 &  180 & $<$1.5 (17) &  1.13 &  1.00 & 5.0 & $0.0$ & CH$_3$SH \\ 
\hline 
\multicolumn{9}{c}{\textit{S- and O-bearing molecules}} \\ 
 trans-HC(O)SH $^\star$ & 62515 & 0.70 &  170 & $<$2.1 (16) &  1.07 &  1.14 & 5.0 & $0.0$ & \ldots \\ 
 cis-HC(O)SH & 62516 & 0.70 &  170 & $<$6.9 (16) &  1.07 &  8.08 & 5.0 & $0.0$ & \ldots \\ 
\hline 
\multicolumn{9}{c}{\textit{S- and N-bearing molecules}} \\ 
 HCNS $^\star$ & 59510 & 1.00 &  120 & $<$2.2 (14) &  1.09 &  1.00 & 5.0 & $0.0$ & HNCS \\ 
 NH$_2$CHS & 61523 & 0.90 &  170 & $<$2.7 (15) &  1.06 &  1.00 & 5.0 & $0.5$ & NH$_2$CHO \\ 
\hline 
\multicolumn{9}{c}{\textit{Cl-bearing molecules}} \\ 
 CH$_3$Cl $^\star$ & 50007 & 0.70 &  180 & $<$1.1 (16) &  1.00 &  1.00 & 5.0 & $0.0$ & \ldots \\ 
\hline 
\multicolumn{9}{c}{\textit{P-bearing molecules}} \\ 
 CH$_3$CP & 58507 & 0.70 &  170 & $<$6.5 (15) &  1.08 &  1.00 & 5.0 & $0.0$ & \ldots \\ 
 C$_2$H$_5$CP & 72509 & 0.70 &  170 & $<$3.5 (16) &  1.76 &  1.00 & 5.0 & $0.0$ & \ldots \\ 
\hline 
\multicolumn{9}{c}{\textit{Hydrocarbons}} \\ 
 C$_2$H$_3$$^+$ $^\star$ & 27514 & 0.70 &  170 & $<$2.5 (17) &  1.02 &  1.00 & 5.0 & $0.0$ & \ldots \\ 
 CH$_2$CHCH$_3$ $^\star$ & 42516 & 0.70 &  170 & $<$7.8 (17) &  1.30 &  1.00 & 5.0 & $0.0$ & \ldots \\ 
 C$_3$H$_8$ & 44013 & 0.70 &  170 & $<$1.2 (19) &  1.00 &  1.00 & 5.0 & $0.0$ & \ldots \\ 
 C$_2$H$_5$CCH $^\star$ & 54519 & 1.00 &  150 & $<$7.1 (16) &  1.43 &  1.00 & 5.0 & $-0.1$ & CH$_3$CCH \\ 
 (CH$_3$)$_2$CCH$_2$ $^\star$ & 56526 & 0.70 &  170 & $<$4.3 (17) &  1.71 &  1.00 & 5.0 & $0.0$ & \ldots \\ 
 H$_2$C(CCH)$_2$ $^\star$ & 64520 & 1.00 &  150 & $<$4.7 (17) &  1.57 &  1.00 & 5.0 & $-0.1$ & CH$_3$CCH \\ 
\hline 
 \end{longtable}
 \tablefoot{
 \tablefoottext{a}{Molecules that have already been detected in the interstellar medium are marked with a star.}
 \tablefoottext{b}{Entry number in our Weeds local database. See note b of Table~\ref{t:spectrobib}.}
 \tablefoottext{c}{Source diameter (\textit{FWHM}).}
 \tablefoottext{d}{Rotational temperature.}
 \tablefoottext{e}{Upper limit to the total column density of the molecule. $X$ ($Y$) means $X \times 10^Y$.}
 \tablefoottext{f}{Correction factor that was applied to the column density to account for the contribution of vibrationally or torsionally excited states, in the cases where this contribution was not included in the partition function of the spectroscopic predictions. In most cases, this factor was estimated in the harmonic approximation.}
 \tablefoottext{g}{Correction factor that was applied to the column density to account for the contribution of conformers, in the cases where this contribution was not included in the partition function of the spectroscopic predictions.}
 \tablefoottext{h}{Linewidth (\textit{FWHM}).}
 \tablefoottext{i}{Velocity offset with respect to the assumed systemic velocity given in Table~\ref{t:sources}.}
 \tablefoottext{j}{Molecule that was used to fix the LTE parameters except for the column density. Ellipsis dots indicate that we used typical parameters of the specific source.}
 }

\begin{longtable}{lrccccccrr}
 \caption{\label{t:uplim_edge}
 Column density upper limits toward the second velocity component toward AN06 (AN06c2).
} \\ 
 \hline\hline
 \multicolumn{1}{c}{Molecule\tablefootmark{a}} & \multicolumn{1}{c}{Tag\tablefootmark{b}} & \multicolumn{1}{c}{Size\tablefootmark{c}} & \multicolumn{1}{c}{$T_{\mathrm{rot}}$\tablefootmark{d}} & \multicolumn{1}{c}{$N$\tablefootmark{e}} & \multicolumn{1}{c}{$C_{\rm{vib}}$\tablefootmark{f}} & \multicolumn{1}{c}{$C_{\rm{conf}}$\tablefootmark{g}} & \multicolumn{1}{c}{$\Delta V$\tablefootmark{h}} & \multicolumn{1}{c}{$V_{\mathrm{off}}$\tablefootmark{i}} & \multicolumn{1}{c}{Ref.\tablefootmark{j}} \\ 
  & & \multicolumn{1}{c}{$''$} & \multicolumn{1}{c}{K} & \multicolumn{1}{c}{cm$^{-2}$} & & & \multicolumn{2}{c}{km~s$^{-1}$} & \\ 
 \hline
 \endfirsthead
 \caption{continued.} \\ 
 \hline\hline
 \multicolumn{1}{c}{Molecule\tablefootmark{a}} & \multicolumn{1}{c}{Tag\tablefootmark{b}} & \multicolumn{1}{c}{Size\tablefootmark{c}} & \multicolumn{1}{c}{$T_{\mathrm{rot}}$\tablefootmark{d}} & \multicolumn{1}{c}{$N$\tablefootmark{e}} & \multicolumn{1}{c}{$C_{\rm{vib}}$\tablefootmark{f}} & \multicolumn{1}{c}{$C_{\rm{conf}}$\tablefootmark{g}} & \multicolumn{1}{c}{$\Delta V$\tablefootmark{h}} & \multicolumn{1}{c}{$V_{\mathrm{off}}$\tablefootmark{i}} & \multicolumn{1}{c}{Ref.\tablefootmark{j}} \\ 
 \hline
 \endhead
 \hline
 \endfoot
\hline 
\multicolumn{9}{c}{\textit{O-bearing molecules}} \\ 
 HCO $^\star$ & 29004 & 0.70 &  100 & $<$8.0 (15) &  1.00 &  1.00 & 3.3 & $4.9$ & H$_2$CO \\ 
 H$_2$COH$^+$ $^\star$ & 31504 & 0.70 &  100 & $<$6.0 (15) &  1.00 &  1.00 & 3.3 & $4.9$ & H$_2$CO \\ 
 CH$_2$OH & 31515 & 0.50 &  150 & $<$2.5 (17) &  1.00 &  1.00 & 3.5 & $5.2$ & CH$_3$OH \\ 
 CH$_3$O $^\star$ & 31518 & 0.50 &  150 & $<$1.0 (17) &  1.00 &  1.00 & 3.5 & $5.2$ & CH$_3$OH \\ 
 CH$_2$CHO & 43523 & 0.50 &  135 & $<$5.1 (15) &  1.02 &  1.00 & 3.3 & $5.2$ & CH$_3$CHO \\ 
 syn-C$_2$H$_3$OH $^\star$ & 44506 & 0.50 &  135 & $<$4.2 (16) &  1.02 &  1.02 & 3.3 & $5.2$ & C$_2$H$_5$OH \\ 
 anti-C$_2$H$_3$OH $^\star$ & 44507 & 0.50 &  135 & $<$5.4 (17) &  1.02 &  58.4 & 3.3 & $5.2$ & C$_2$H$_5$OH \\ 
 c-HCOOH $^\star$ & 46507 & 0.50 &  150 & $<$9.8 (20) &  1.00 & 485469 & 3.5 & $5.2$ & \ldots \\ 
 trans-C$_2$H$_3$CHO $^\star$ & 56519 & 0.50 &  135 & $<$2.6 (15) &  1.27 &  1.00 & 3.3 & $5.2$ & CH$_3$CHO \\ 
 CH$_3$CHCO $^\star$ & 56701 & 0.50 &  150 & $<$9.0 (15) &  1.00 &  1.00 & 3.3 & $5.0$ & CH$_2$CO \\ 
 c-CH(CH$_3$)CH$_2$O $^\star$ & 58514 & 0.50 &  150 & $<$3.1 (16) &  1.23 &  1.00 & 3.5 & $5.2$ & \ldots \\ 
 gauche-C$_2$H$_5$CHO & 58519 & 0.50 &  135 & $<$8.1 (17) &  1.57 &  1.02 & 3.3 & $5.2$ & CH$_3$CHO \\ 
 C$_2$H$_5$OCH$_3$ $^\star$ & 60901 & 0.80 &  135 & $<$7.8 (16) &  1.70 &  1.02 & 3.0 & $5.2$ & CH$_3$OCH$_3$ \\ 
 HOCHCHOH $^\star$ & 60920 & 0.50 &  150 & $<$1.9 (16) &  1.26 &  1.00 & 3.5 & $5.2$ & \ldots \\ 
 CH$_3$OCH$_2$OH $^\star$ & 62527 & 0.50 &  150 & $<$2.4 (18) &  1.81 &  1.00 & 3.5 & $5.2$ & CH$_3$OH \\ 
 s-cis-C$_2$H$_3$COOH & 72502 & 0.50 &  150 & $<$5.3 (16) &  1.75 &  1.00 & 3.5 & $5.2$ & \ldots \\ 
 s-trans-C$_2$H$_3$COOH & 72503 & 0.50 &  150 & $<$3.2 (16) &  1.75 &  1.00 & 3.5 & $5.2$ & \ldots \\ 
 HOCHCHCHO $^\star$ & 72504 & 0.50 &  135 & $<$9.2 (15) &  1.15 &  1.00 & 3.3 & $5.2$ & CH$_3$CHO \\ 
 cis-trans-C$_2$H$_3$OCHO & 72505 & 0.50 &  150 & $<$7.2 (16) &  2.41 &  1.00 & 3.5 & $5.2$ & \ldots \\ 
 CH$_3$C(O)CHO & 72901 & 0.50 &  140 & $<$3.1 (17) &  2.36 &  1.00 & 3.3 & $5.2$ & CH$_3$COCH$_3$ \\ 
 gauche-(CH$_3$)$_2$CHCHO & 72902 & 0.50 &  135 & $<$2.0 (16) &  2.11 &  1.19 & 3.3 & $5.2$ & CH$_3$CHO \\ 
 cis-gauche-C$_3$H$_7$CHO & 72903 & 0.50 &  135 & $<$7.4 (16) &  2.15 &  3.43 & 3.3 & $5.2$ & CH$_3$CHO \\ 
 cis-trans-C$_3$H$_7$CHO & 72904 & 0.50 &  135 & $<$3.2 (16) &  2.56 &  1.58 & 3.3 & $5.2$ & CH$_3$CHO \\ 
 CH$_2$C(OH)CHO & 72905 & 0.50 &  150 & $<$1.6 (16) &  1.32 &  1.00 & 3.5 & $5.2$ & \ldots \\ 
 CH$_3$C(O)CH$_2$OH $^\star$ & 74003 & 0.50 &  140 & $<$4.4 (16) &  4.36 &  1.00 & 3.3 & $5.2$ & CH$_3$OCH$_3$ \\ 
 OCHCOOH & 74516 & 0.50 &  150 & $<$2.1 (16) &  1.38 &  1.02 & 3.5 & $5.2$ & \ldots \\ 
 CH$_3$OCH$_2$CHO & 74517 & 0.50 &  150 & $<$1.3 (16) &  6.03 &  1.04 & 3.5 & $5.2$ & \ldots \\ 
 CH$_3$CH(OH)CHO & 74519 & 0.50 &  150 & $<$1.5 (16) &  2.06 &  1.00 & 3.5 & $5.2$ & \ldots \\ 
 C$_2$H$_5$COOH & 74905 & 0.50 &  150 & $<$9.6 (16) &  2.84 &  1.13 & 3.5 & $5.2$ & \ldots \\ 
 tert-C$_4$H$_9$OH & 74908 & 0.50 &  135 & $<$1.0 (17) &  1.30 &  1.00 & 3.3 & $5.2$ & C$_2$H$_5$OH \\ 
 aG'g-CH$_3$CHOHCH$_2$OH & 76513 & 0.50 &  135 & $<$5.0 (16) &  1.79 &  1.99 & 3.3 & $5.2$ & C$_2$H$_5$OH \\ 
 CH$_2$(OH)COOH & 76514 & 0.50 &  150 & $<$2.3 (16) &  1.95 &  1.00 & 3.5 & $5.2$ & \ldots \\ 
 gG'a-CH$_3$CHOHCH$_2$OH & 76515 & 0.50 &  135 & $<$4.7 (16) &  1.78 &  4.37 & 3.3 & $5.2$ & C$_2$H$_5$OH \\ 
 g'G'g-CH$_3$CHOHCH$_2$OH & 76516 & 0.50 &  135 & $<$1.4 (17) &  1.72 &  6.77 & 3.3 & $5.2$ & C$_2$H$_5$OH \\ 
 a'GG'g-HOCH$_2$CH$_2$CH$_2$OH & 76518 & 0.50 &  135 & $<$2.9 (16) &  2.09 &  1.39 & 3.3 & $5.2$ & C$_2$H$_5$OH \\ 
 gGG'g-HOCH$_2$CH$_2$CH$_2$OH & 76519 & 0.50 &  135 & $<$7.7 (16) &  1.98 &  4.35 & 3.3 & $5.2$ & C$_2$H$_5$OH \\ 
 g'Ga-CH$_3$CHOHCH$_2$OH & 76520 & 0.50 &  135 & $<$2.8 (17) &  1.84 &  19.0 & 3.3 & $5.2$ & C$_2$H$_5$OH \\ 
 gG'g'-CH$_3$CHOHCH$_2$OH & 76521 & 0.50 &  135 & $<$3.6 (17) &  1.72 &  23.1 & 3.3 & $5.2$ & C$_2$H$_5$OH \\ 
 aGg'-CH$_3$CHOHCH$_2$OH & 76522 & 0.50 &  135 & $<$1.3 (18) &  1.79 &  82.0 & 3.3 & $5.2$ & C$_2$H$_5$OH \\ 
 g'Gg-CH$_3$CHOHCH$_2$OH & 76523 & 0.50 &  135 & $<$3.9 (18) &  1.79 &   108 & 3.3 & $5.2$ & C$_2$H$_5$OH \\ 
 CH$_3$OCH$_2$CH$_2$OH $^\star$ & 76801 & 0.50 &  150 & $<$3.0 (16) &  1.00 &  1.01 & 3.5 & $5.2$ & CH$_3$OH \\ 
 c-C$_6$H$_5$OH & 94501 & 0.50 &  150 & $<$5.7 (16) &  1.27 &  1.00 & 3.5 & $5.2$ & CH$_3$OH \\ 
\hline 
\multicolumn{9}{c}{\textit{O- and N-bearing molecules}} \\ 
 HCNO $^\star$ & 43509 & 1.50 &  130 & $<$3.0 (14) &  1.65 &  1.00 & 5.0 & $6.0$ & HNCO \\ 
 HOCN $^\star$ & 43510 & 1.50 &  130 & $<$4.1 (14) &  1.01 &  1.00 & 5.0 & $6.0$ & HNCO \\ 
 NH$_2$CO & 44904 & 0.80 &  200 & $<$7.0 (15) &  1.00 &  1.00 & 5.0 & $5.2$ & NH$_2$CHO \\ 
 HC(O)CN $^\star$ & 55501 & 0.50 &  150 & $<$7.3 (15) &  1.21 &  1.00 & 3.5 & $5.2$ & \ldots \\ 
 CH$_3$CNO & 57510 & 0.70 &  140 & $<$5.1 (14) &  1.28 &  1.00 & 3.0 & $5.2$ & CH$_3$NCO \\ 
 CH$_3$OCN & 57511 & 0.70 &  140 & $<$2.5 (15) &  1.65 &  1.00 & 3.0 & $5.2$ & CH$_3$NCO \\ 
 HOCH$_2$CN $^\star$ & 57701 & 0.50 &  150 & $<$1.8 (16) &  1.00 &  1.00 & 3.5 & $5.2$ & \ldots \\ 
 NH$_2$C(O)NH$_2$ $^\star$ & 60517 & 0.50 &  150 & $<$3.7 (15) &  1.79 &  1.04 & 3.5 & $5.2$ & \ldots \\ 
 NH$_2$CH$_2$CH$_2$OH $^\star$ & 61004 & 0.50 &  150 & $<$1.3 (16) &  1.00 &  1.05 & 3.5 & $5.2$ & \ldots \\ 
 NCCHCO & 67502 & 0.50 &  150 & $<$4.9 (15) &  1.39 &  1.00 & 3.5 & $5.2$ & \ldots \\ 
 c-(C$_2$H$_3$O)-CN & 69518 & 0.50 &  130 & $<$3.6 (15) &  1.20 &  1.00 & 3.0 & $5.2$ & c-C$_2$H$_4$O \\ 
 trans-C$_2$H$_3$NCO & 69901 & 0.70 &  140 & $<$8.5 (15) &  2.43 &  1.00 & 3.0 & $5.2$ & CH$_3$NCO \\ 
 cis-C$_2$H$_3$NCO & 69902 & 0.70 &  140 & $<$2.0 (17) &  2.47 &  1.00 & 3.0 & $5.2$ & CH$_3$NCO \\ 
 CHCC(O)NH$_2$ & 69916 & 0.50 &  150 & $<$1.0 (16) &  1.12 &  1.00 & 3.5 & $5.2$ & \ldots \\ 
 NCC(O)NH$_2$ & 70504 & 0.50 &  150 & $<$9.2 (15) &  1.54 &  1.00 & 3.5 & $5.2$ & \ldots \\ 
 gauche-HOCH$_2$CH$_2$CN & 71401 & 0.50 &  150 & $<$2.2 (16) &  1.00 &  1.12 & 3.5 & $5.2$ & \ldots \\ 
 gauche-CH$_3$OCH$_2$CN & 71402 & 0.50 &  150 & $<$1.3 (16) &  1.00 &  1.01 & 3.5 & $5.2$ & \ldots \\ 
 cis-C$_2$H$_5$NCO $^\star$ & 71508 & 0.70 &  140 & $<$2.1 (16) &  5.17 &  1.00 & 3.0 & $5.2$ & CH$_3$NCO \\ 
 syn-C$_2$H$_3$C(O)NH$_2$ & 71903 & 0.50 &  150 & $<$1.7 (16) &  2.08 &  1.01 & 3.5 & $5.2$ & \ldots \\ 
 C$_2$H$_5$C(O)NH$_2$ & 73801 & 0.50 &  150 & $<$1.8 (16) &  1.00 &  1.00 & 3.5 & $5.2$ & \ldots \\ 
 trans-sc-C$_2$H$_5$NHCHO & 73902 & 0.50 &  150 & $<$3.3 (16) &  3.31 &  1.00 & 3.5 & $5.2$ & \ldots \\ 
 NH$_2$CH$_2$C(O)NH$_2$ & 74904 & 0.50 &  150 & $<$9.9 (15) &  1.23 &  1.00 & 3.5 & $5.2$ & \ldots \\ 
 NH$_2$CH$_2$COOH I & 75511 & 0.50 &  150 & $<$1.2 (17) &  2.56 &  1.00 & 3.5 & $5.2$ & \ldots \\ 
 NH$_2$CH$_2$COOH II & 75512 & 0.50 &  150 & $<$1.7 (18) &  2.56 &   517 & 3.5 & $5.2$ & \ldots \\ 
 g'Gg'-CH$_3$CH(NH$_2$)CH$_2$OH & 75518 & 0.50 &  150 & $<$1.1 (16) &  1.85 &  1.14 & 3.5 & $5.2$ & \ldots \\ 
 gG'g-CH$_3$CH(NH$_2$)CH$_2$OH & 75519 & 0.50 &  150 & $<$1.0 (18) &  1.85 &  7.94 & 3.5 & $5.2$ & \ldots \\ 
 syn-HOCH$_2$C(O)NH$_2$ $^\star$ & 75901 & 0.50 &  150 & $<$9.2 (15) &  2.53 &  1.04 & 3.5 & $5.2$ & \ldots \\ 
 OC(CN)$_2$ & 80601 & 0.50 &  150 & $<$5.9 (17) &  1.98 &  1.00 & 3.5 & $5.2$ & \ldots \\ 
 NCCH$_2$C(O)NH$_2$ & 84902 & 0.50 &  150 & $<$5.8 (17) &  3.40 &  1.00 & 3.5 & $5.2$ & \ldots \\ 
 CH$_3$CH(NH$_2$)COOH I & 89502 & 0.50 &  150 & $<$1.6 (17) &  4.14 &  1.29 & 3.5 & $5.2$ & \ldots \\ 
 CH$_3$CH(NH$_2$)COOH II & 89503 & 0.50 &  150 & $<$1.0 (17) &  4.14 &  11.0 & 3.5 & $5.2$ & \ldots \\ 
\hline 
\multicolumn{9}{c}{\textit{N-bearing molecules}} \\ 
 CH$_2$NH$_2$$^+$ & 30519 & 0.50 &  150 & $<$9.0 (16) &  1.00 &  1.00 & 3.5 & $5.2$ & CH$_2$NH \\ 
 CH$_2$CN $^\star$ & 40601 & 0.50 &  140 & $<$3.1 (15) &  1.04 &  1.00 & 3.2 & $5.2$ & CH$_3$CN \\ 
 CH$_2$CNH $^\star$ & 41503 & 0.50 &  150 & $<$2.1 (16) &  1.03 &  1.00 & 3.5 & $5.2$ & \ldots \\ 
 CH$_3$CNH$^+$ & 42504 & 0.50 &  140 & $<$1.4 (16) &  1.02 &  1.00 & 3.2 & $5.2$ & CH$_3$CN \\ 
 C$_2$H$_3$NH$_2$ $^\star$ & 43504 & 0.50 &  150 & $<$4.2 (16) &  1.06 &  1.00 & 3.5 & $5.2$ & CH$_3$NH$_2$ \\ 
 E-CH$_3$CHNH $^\star$ & 43907 & 0.50 &  150 & $<$2.1 (16) &  1.21 &  1.00 & 3.5 & $5.2$ & \ldots \\ 
 Z-CH$_3$CHNH $^\star$ & 43908 & 0.50 &  150 & $<$8.6 (15) &  1.23 &  1.00 & 3.5 & $5.2$ & \ldots \\ 
 anti-C$_2$H$_5$NH$_2$ $^\star$ & 45402 & 0.50 &  150 & $<$9.4 (16) &  1.23 &  2.19 & 3.5 & $5.2$ & CH$_3$NH$_2$ \\ 
 CH$_3$NHCH$_3$ & 45601 & 0.50 &  150 & $<$1.3 (17) &  1.28 &  1.00 & 3.5 & $5.2$ & CH$_3$NH$_2$ \\ 
 C$_3$N $^\star$ & 50511 & 1.00 &  130 & $<$4.5 (14) &  1.13 &  1.00 & 4.5 & $5.2$ & HC$_3$N \\ 
 HNC$_3$ $^\star$ & 51528 & 1.00 &  130 & $<$6.6 (13) &  1.33 &  1.00 & 4.5 & $5.2$ & HC$_3$N \\ 
 HC$_3$NH$^+$ $^\star$ & 52503 & 1.00 &  130 & $<$9.2 (14) &  1.31 &  1.00 & 4.5 & $5.2$ & HC$_3$N \\ 
 C$_2$H$_3$NC & 53007 & 0.70 &  150 & $<$1.3 (15) &  1.23 &  1.00 & 3.5 & $5.2$ & C$_2$H$_3$CN \\ 
 Z-CHCCHNH $^\star$ & 53523 & 0.50 &  150 & $<$7.9 (15) &  1.24 &  1.06 & 3.5 & $5.2$ & \ldots \\ 
 E-CHCCHNH & 53524 & 0.50 &  150 & $<$2.3 (17) &  1.23 &  18.4 & 3.5 & $5.2$ & \ldots \\ 
 C$_2$H$_3$CNH$^+$ & 54518 & 0.70 &  150 & $<$4.8 (15) &  1.20 &  1.00 & 3.5 & $5.2$ & C$_2$H$_3$CN \\ 
 Z-NHCHCN & 54805 & 0.50 &  150 & $<$1.5 (16) &  1.18 &  1.00 & 3.5 & $5.2$ & \ldots \\ 
 CH$_2$NCN & 54806 & 0.50 &  150 & $<$1.1 (15) &  1.14 &  1.00 & 3.5 & $5.2$ & \ldots \\ 
 C$_2$H$_5$NC & 55507 & 0.50 &  150 & $<$3.0 (15) &  1.48 &  1.00 & 3.5 & $5.2$ & \ldots \\ 
 E-C$_2$H$_5$CHNH & 57801 & 0.50 &  150 & $<$5.4 (16) &  1.00 &  1.08 & 3.5 & $5.2$ & \ldots \\ 
 Z-C$_2$H$_5$CHNH & 57802 & 0.50 &  150 & $<$9.5 (16) &  1.00 &  13.6 & 3.5 & $5.2$ & \ldots \\ 
 trans-iso-C$_3$H$_7$NH$_2$ & 59925 & 0.50 &  150 & $<$1.4 (17) &  1.39 &  1.47 & 3.5 & $5.2$ & \ldots \\ 
 Trans-trans-normal-C$_3$H$_7$NH$_2$ & 59928 & 0.50 &  150 & $<$2.1 (17) &  1.80 &  4.61 & 3.5 & $5.2$ & \ldots \\ 
 CH$_3$CCNC & 65505 & 0.50 &  150 & $<$8.0 (14) &  2.00 &  1.00 & 3.5 & $5.2$ & \ldots \\ 
 CH$_2$CCHCN $^\star$ & 65506 & 0.50 &  150 & $<$2.2 (15) &  1.49 &  1.00 & 3.5 & $5.2$ & \ldots \\ 
 CHCCH$_2$CN $^\star$ & 65514 & 0.50 &  150 & $<$2.8 (15) &  1.54 &  1.00 & 3.5 & $5.2$ & \ldots \\ 
 CH$_2$CCHNC & 65519 & 0.50 &  150 & $<$2.5 (15) &  1.66 &  1.00 & 3.5 & $5.2$ & \ldots \\ 
 NCCH$_2$CN $^\star$ & 66801 & 0.50 &  150 & $<$5.6 (15) &  1.39 &  1.00 & 3.5 & $5.2$ & \ldots \\ 
 CNCH$_2$NC & 66902 & 0.50 &  150 & $<$7.9 (15) &  1.97 &  1.00 & 3.5 & $5.2$ & \ldots \\ 
 E-CH$_3$CHCHCN $^\star$ & 67513 & 0.50 &  130 & $<$2.2 (15) &  1.48 &  1.00 & 3.5 & $5.2$ & normal-C$_3$H$_7$CN \\ 
 syn-CH$_2$CHCH$_2$CN $^\star$ & 67602 & 0.50 &  130 & $<$6.4 (15) &  1.57 &  1.02 & 3.5 & $5.2$ & normal-C$_3$H$_7$CN \\ 
 gauche-CH$_2$CHCH$_2$CN $^\star$ & 67603 & 0.50 &  130 & $<$1.3 (17) &  1.57 &  42.2 & 3.5 & $5.2$ & normal-C$_3$H$_7$CN \\ 
 CH$_2$NCH$_2$CN & 68506 & 0.50 &  150 & $<$1.5 (16) & \ldots &  1.00 & 3.5 & $5.2$ & \ldots \\ 
 Z-CH$_3$C(NH)CN & 68512 & 0.50 &  150 & $<$2.7 (16) &  1.71 &  1.00 & 3.5 & $5.2$ & \ldots \\ 
 CH$_3$CH(NH$_2$)CN & 70505 & 0.50 &  150 & $<$1.2 (16) &  1.68 &  1.01 & 3.5 & $5.2$ & \ldots \\ 
 NH$_2$CH$_2$CH$_2$CN I & 70513 & 0.50 &  150 & $<$1.5 (16) &  2.07 &  1.07 & 3.5 & $5.2$ & \ldots \\ 
 NH$_2$CH$_2$CH$_2$CN II & 70514 & 0.50 &  150 & $<$6.6 (16) &  2.07 &  1.07 & 3.5 & $5.2$ & \ldots \\ 
 HC$_5$NH$^+$ $^\star$ & 76524 & 1.00 &  130 & $<$6.6 (14) &  2.65 &  1.00 & 4.5 & $5.2$ & HC$_3$N \\ 
 C$_2$H$_3$C$_3$N $^\star$ & 77506 & 0.50 &  150 & $<$2.5 (15) &  2.48 &  1.00 & 3.5 & $5.2$ & \ldots \\ 
 CH$_3$CHCCHCN & 79506 & 0.50 &  150 & $<$4.5 (15) &  2.98 &  1.00 & 3.5 & $5.2$ & \ldots \\ 
 C$_2$H$_5$C$_3$N & 79507 & 0.50 &  150 & $<$2.5 (15) &  3.15 &  1.00 & 3.5 & $5.2$ & \ldots \\ 
 NH$_2$CH(CN)$_2$ & 81801 & 0.50 &  150 & $<$8.6 (15) &  2.14 &  1.00 & 3.5 & $5.2$ & \ldots \\ 
 anti-anti-normal-C$_4$H$_9$CN & 83501 & 0.50 &  130 & $<$2.8 (16) &  3.39 &  2.35 & 3.5 & $5.2$ & normal-C$_3$H$_7$CN \\ 
 gauche-anti-normal-C$_4$H$_9$CN & 83502 & 0.50 &  130 & $<$2.2 (16) &  3.39 &  1.06 & 3.5 & $5.2$ & normal-C$_3$H$_7$CN \\ 
 anti-gauche-normal-C$_4$H$_9$CN & 83503 & 0.50 &  130 & $<$1.5 (16) &  3.39 &  2.20 & 3.5 & $5.2$ & normal-C$_3$H$_7$CN \\ 
 tert-C$_4$H$_9$CN & 83504 & 0.50 &  130 & $<$2.1 (15) &  1.71 &  1.00 & 3.5 & $5.2$ & normal-C$_3$H$_7$CN \\ 
 anti-C$_2$H$_5$CH(CN)CH$_3$ & 83505 & 0.50 &  130 & $<$7.2 (15) &  2.41 &  1.11 & 3.5 & $5.2$ & normal-C$_3$H$_7$CN \\ 
 anti-(CH$_3$)$_2$CHCH$_2$CN & 83506 & 0.50 &  130 & $<$1.2 (16) &  2.47 &  1.00 & 3.5 & $5.2$ & normal-C$_3$H$_7$CN \\ 
 gauche-(CH$_3$)$_2$CHCH$_2$CN & 83507 & 0.50 &  130 & $<$4.3 (16) &  2.41 &  1.00 & 3.5 & $5.2$ & normal-C$_3$H$_7$CN \\ 
 HC$_7$N $^\star$ & 99501 & 1.00 &  130 & $<$1.3 (15) &  6.64 &  1.00 & 4.5 & $5.2$ & HC$_3$N \\ 
 c-C$_6$H$_5$CN $^\star$ & 103501 & 0.50 &  150 & $<$4.2 (15) &  1.69 &  1.00 & 3.5 & $5.2$ & \ldots \\ 
\hline 
\multicolumn{9}{c}{\textit{S-bearing molecules}} \\ 
 c-C$_2$H$_4$S & 60509 & 0.50 &  150 & $<$1.8 (16) &  1.00 &  1.00 & 3.5 & $5.2$ & CH$_3$SH \\ 
 syn-C$_2$H$_3$SH & 60521 & 0.50 &  150 & $<$7.9 (16) &  1.13 &  1.00 & 3.5 & $5.2$ & CH$_3$SH \\ 
 anti-C$_2$H$_3$SH & 60522 & 0.50 &  150 & $<$1.0 (17) &  1.13 &  1.00 & 3.5 & $5.2$ & CH$_3$SH \\ 
 CH$_3$CHS $^\star$ & 60701 & 0.50 &  150 & $<$7.2 (15) &  1.02 &  1.00 & 3.5 & $5.2$ & CH$_3$SH \\ 
 gauche-C$_2$H$_5$SH $^\star$ & 62523 & 0.50 &  150 & $<$3.1 (16) &  1.31 &  1.17 & 3.5 & $5.2$ & CH$_3$SH \\ 
 anti-C$_2$H$_5$SH & 62524 & 0.50 &  150 & $<$3.0 (17) &  1.45 &  6.81 & 3.5 & $5.2$ & CH$_3$SH \\ 
 AA-CH$_3$SCH$_3$ $^\star$ & 62902 & 0.50 &  150 & $<$4.8 (16) &  1.61 &  1.00 & 3.5 & $5.2$ & CH$_3$SH \\ 
 trans-HCSSH & 78506 & 0.50 &  150 & $<$1.1 (16) &  1.07 &  1.00 & 3.5 & $5.2$ & CH$_3$SH \\ 
 cis-HCSSH & 78507 & 0.50 &  150 & $<$8.6 (16) &  1.07 &  1.00 & 3.5 & $5.2$ & CH$_3$SH \\ 
\hline 
\multicolumn{9}{c}{\textit{S- and O-bearing molecules}} \\ 
 trans-HC(O)SH $^\star$ & 62515 & 0.50 &  150 & $<$1.5 (16) &  1.04 &  1.11 & 3.5 & $5.2$ & \ldots \\ 
 cis-HC(O)SH & 62516 & 0.50 &  150 & $<$6.4 (16) &  1.04 &  10.2 & 3.5 & $5.2$ & \ldots \\ 
\hline 
\multicolumn{9}{c}{\textit{S- and N-bearing molecules}} \\ 
 HCNS $^\star$ & 59510 & 0.50 &  120 & $<$1.6 (14) &  1.09 &  1.00 & 3.5 & $3.5$ & HNCS \\ 
 NH$_2$CHS & 61523 & 0.80 &  200 & $<$1.8 (15) &  1.12 &  1.00 & 5.0 & $5.2$ & NH$_2$CHO \\ 
\hline 
\multicolumn{9}{c}{\textit{Cl-bearing molecules}} \\ 
 CH$_3$Cl $^\star$ & 50007 & 0.50 &  150 & $<$6.0 (15) &  1.00 &  1.00 & 3.5 & $5.2$ & \ldots \\ 
\hline 
\multicolumn{9}{c}{\textit{P-bearing molecules}} \\ 
 CH$_3$CP & 58507 & 0.50 &  150 & $<$4.2 (15) &  1.06 &  1.00 & 3.5 & $5.2$ & \ldots \\ 
 C$_2$H$_5$CP & 72509 & 0.50 &  150 & $<$1.5 (16) &  1.55 &  1.00 & 3.5 & $5.2$ & \ldots \\ 
\hline 
\multicolumn{9}{c}{\textit{Hydrocarbons}} \\ 
 C$_2$H$_3$$^+$ $^\star$ & 27514 & 0.50 &  150 & $<$6.1 (16) &  1.01 &  1.00 & 3.5 & $5.2$ & \ldots \\ 
 CH$_2$CHCH$_3$ $^\star$ & 42516 & 0.50 &  150 & $<$3.7 (17) &  1.22 &  1.00 & 3.5 & $5.2$ & \ldots \\ 
 C$_3$H$_8$ & 44013 & 0.50 &  150 & $<$8.0 (18) &  1.00 &  1.00 & 3.5 & $5.2$ & \ldots \\ 
 C$_2$H$_5$CCH $^\star$ & 54519 & 1.00 &  130 & $<$5.1 (16) &  1.28 &  1.00 & 4.5 & $5.2$ & CH$_3$CCH \\ 
 (CH$_3$)$_2$CCH$_2$ $^\star$ & 56526 & 0.50 &  150 & $<$2.3 (17) &  1.50 &  1.00 & 3.5 & $5.2$ & \ldots \\ 
 H$_2$C(CCH)$_2$ $^\star$ & 64520 & 1.00 &  130 & $<$2.1 (17) &  1.39 &  1.00 & 4.5 & $5.2$ & CH$_3$CCH \\ 
\hline 
 \end{longtable}
 \tablefoot{
 \tablefoottext{a}{Molecules that have already been detected in the interstellar medium are marked with a star.}
 \tablefoottext{b}{Entry number in our Weeds local database. See note b of Table~\ref{t:spectrobib}.}
 \tablefoottext{c}{Source diameter (\textit{FWHM}).}
 \tablefoottext{d}{Rotational temperature.}
 \tablefoottext{e}{Upper limit to the total column density of the molecule. $X$ ($Y$) means $X \times 10^Y$.}
 \tablefoottext{f}{Correction factor that was applied to the column density to account for the contribution of vibrationally or torsionally excited states, in the cases where this contribution was not included in the partition function of the spectroscopic predictions. In most cases, this factor was estimated in the harmonic approximation.}
 \tablefoottext{g}{Correction factor that was applied to the column density to account for the contribution of conformers, in the cases where this contribution was not included in the partition function of the spectroscopic predictions.}
 \tablefoottext{h}{Linewidth (\textit{FWHM}).}
 \tablefoottext{i}{Velocity offset with respect to the assumed systemic velocity given in Table~\ref{t:sources}.}
 \tablefoottext{j}{Molecule that was used to fix the LTE parameters except for the column density. Ellipsis dots indicate that we used typical parameters of the specific source.}
 }

\begin{longtable}{lrccccccrr}
 \caption{\label{t:uplim_an06}
 Column density upper limits toward  AN06.
} \\ 
 \hline\hline
 \multicolumn{1}{c}{Molecule\tablefootmark{a}} & \multicolumn{1}{c}{Tag\tablefootmark{b}} & \multicolumn{1}{c}{Size\tablefootmark{c}} & \multicolumn{1}{c}{$T_{\mathrm{rot}}$\tablefootmark{d}} & \multicolumn{1}{c}{$N$\tablefootmark{e}} & \multicolumn{1}{c}{$C_{\rm{vib}}$\tablefootmark{f}} & \multicolumn{1}{c}{$C_{\rm{conf}}$\tablefootmark{g}} & \multicolumn{1}{c}{$\Delta V$\tablefootmark{h}} & \multicolumn{1}{c}{$V_{\mathrm{off}}$\tablefootmark{i}} & \multicolumn{1}{c}{Ref.\tablefootmark{j}} \\ 
  & & \multicolumn{1}{c}{$''$} & \multicolumn{1}{c}{K} & \multicolumn{1}{c}{cm$^{-2}$} & & & \multicolumn{2}{c}{km~s$^{-1}$} & \\ 
 \hline
 \endfirsthead
 \caption{continued.} \\ 
 \hline\hline
 \multicolumn{1}{c}{Molecule\tablefootmark{a}} & \multicolumn{1}{c}{Tag\tablefootmark{b}} & \multicolumn{1}{c}{Size\tablefootmark{c}} & \multicolumn{1}{c}{$T_{\mathrm{rot}}$\tablefootmark{d}} & \multicolumn{1}{c}{$N$\tablefootmark{e}} & \multicolumn{1}{c}{$C_{\rm{vib}}$\tablefootmark{f}} & \multicolumn{1}{c}{$C_{\rm{conf}}$\tablefootmark{g}} & \multicolumn{1}{c}{$\Delta V$\tablefootmark{h}} & \multicolumn{1}{c}{$V_{\mathrm{off}}$\tablefootmark{i}} & \multicolumn{1}{c}{Ref.\tablefootmark{j}} \\ 
 \hline
 \endhead
 \hline
 \endfoot
\hline 
\multicolumn{9}{c}{\textit{O-bearing molecules}} \\ 
 HCO $^\star$ & 29004 & 0.70 &  125 & $<$9.0 (15) &  1.00 &  1.00 & 3.0 & $0.0$ & H$_2$CO \\ 
 H$_2$COH$^+$ $^\star$ & 31504 & 0.50 &  130 & $<$9.0 (15) &  1.00 &  1.00 & 3.0 & $0.0$ & H$_2$CO \\ 
 CH$_2$OH & 31515 & 0.50 &  130 & $<$1.5 (17) &  1.00 &  1.00 & 3.0 & $0.0$ & CH$_3$OH \\ 
 CH$_3$O $^\star$ & 31518 & 0.50 &  130 & $<$1.0 (17) &  1.00 &  1.00 & 3.0 & $0.0$ & CH$_3$OH \\ 
 CH$_2$CHO & 43523 & 0.50 &  130 & $<$4.1 (15) &  1.02 &  1.00 & 3.3 & $0.0$ & CH$_3$CHO \\ 
 syn-C$_2$H$_3$OH $^\star$ & 44506 & 0.50 &  130 & $<$4.1 (16) &  1.02 &  1.01 & 3.3 & $0.0$ & C$_2$H$_5$OH \\ 
 anti-C$_2$H$_3$OH $^\star$ & 44507 & 0.50 &  130 & $<$4.9 (17) &  1.02 &  68.1 & 3.3 & $0.0$ & C$_2$H$_5$OH \\ 
 c-HCOOH $^\star$ & 46507 & 0.50 &  130 & $<$7.3 (21) &  1.00 & 3638774 & 3.3 & $0.0$ & \ldots \\ 
 trans-C$_2$H$_3$CHO $^\star$ & 56519 & 0.50 &  130 & $<$1.9 (15) &  1.25 &  1.00 & 3.3 & $0.0$ & CH$_3$CHO \\ 
 CH$_3$CHCO $^\star$ & 56701 & 0.50 &  150 & $<$7.0 (15) &  1.00 &  1.00 & 3.0 & $0.0$ & CH$_2$CO \\ 
 c-CH(CH$_3$)CH$_2$O $^\star$ & 58514 & 0.50 &  130 & $<$2.3 (16) &  1.15 &  1.00 & 3.3 & $0.0$ & \ldots \\ 
 gauche-C$_2$H$_5$CHO & 58519 & 0.50 &  130 & $<$6.2 (17) &  1.52 &  1.02 & 3.3 & $0.0$ & CH$_3$CHO \\ 
 C$_2$H$_5$OCH$_3$ $^\star$ & 60901 & 0.50 &  120 & $<$9.2 (16) &  1.51 &  1.01 & 3.0 & $0.0$ & CH$_3$OCH$_3$ \\ 
 HOCHCHOH $^\star$ & 60920 & 0.50 &  130 & $<$1.1 (16) &  1.17 &  1.00 & 3.3 & $0.0$ & \ldots \\ 
 CH$_3$OCH$_2$OH $^\star$ & 62527 & 0.50 &  130 & $<$1.2 (18) &  1.56 &  1.00 & 3.0 & $0.0$ & CH$_3$OH \\ 
 s-cis-C$_2$H$_3$COOH & 72502 & 0.50 &  130 & $<$4.7 (16) &  1.55 &  1.00 & 3.3 & $0.0$ & \ldots \\ 
 s-trans-C$_2$H$_3$COOH & 72503 & 0.50 &  130 & $<$3.1 (16) &  1.55 &  1.00 & 3.3 & $0.0$ & \ldots \\ 
 HOCHCHCHO $^\star$ & 72504 & 0.50 &  130 & $<$7.9 (15) &  1.14 &  1.00 & 3.3 & $0.0$ & CH$_3$CHO \\ 
 cis-trans-C$_2$H$_3$OCHO & 72505 & 0.50 &  150 & $<$4.8 (16) &  2.41 &  1.00 & 2.7 & $0.0$ & C$_2$H$_5$OCHO \\ 
 CH$_3$C(O)CHO & 72901 & 0.50 &  130 & $<$2.1 (17) &  2.14 &  1.00 & 3.3 & $0.0$ & CH$_3$COCH$_3$ \\ 
 gauche-(CH$_3$)$_2$CHCHO & 72902 & 0.50 &  130 & $<$2.1 (16) &  2.01 &  1.18 & 3.3 & $0.0$ & CH$_3$CHO \\ 
 cis-gauche-C$_3$H$_7$CHO & 72903 & 0.50 &  130 & $<$5.0 (16) &  2.04 &  3.48 & 3.3 & $0.0$ & CH$_3$CHO \\ 
 cis-trans-C$_3$H$_7$CHO & 72904 & 0.50 &  130 & $<$2.6 (16) &  2.42 &  1.55 & 3.3 & $0.0$ & CH$_3$CHO \\ 
 CH$_2$C(OH)CHO & 72905 & 0.50 &  130 & $<$6.0 (15) &  1.21 &  1.00 & 3.3 & $0.0$ & \ldots \\ 
 CH$_3$C(O)CH$_2$OH $^\star$ & 74003 & 0.50 &  130 & $<$3.9 (16) &  3.87 &  1.00 & 3.3 & $0.0$ & CH$_3$OCH$_3$ \\ 
 OCHCOOH & 74516 & 0.50 &  130 & $<$1.4 (16) &  1.26 &  1.01 & 3.3 & $0.0$ & \ldots \\ 
 CH$_3$OCH$_2$CHO & 74517 & 0.50 &  130 & $<$9.3 (15) &  4.55 &  1.02 & 3.3 & $0.0$ & \ldots \\ 
 CH$_3$CH(OH)CHO & 74519 & 0.50 &  130 & $<$1.1 (16) &  1.72 &  1.00 & 3.3 & $0.0$ & \ldots \\ 
 C$_2$H$_5$COOH & 74905 & 0.50 &  130 & $<$5.2 (16) &  2.37 &  1.09 & 3.3 & $0.0$ & \ldots \\ 
 tert-C$_4$H$_9$OH & 74908 & 0.50 &  130 & $<$6.3 (16) &  1.26 &  1.00 & 3.3 & $0.0$ & C$_2$H$_5$OH \\ 
 aG'g-CH$_3$CHOHCH$_2$OH & 76513 & 0.50 &  130 & $<$4.3 (16) &  1.70 &  1.94 & 3.3 & $0.0$ & C$_2$H$_5$OH \\ 
 CH$_2$(OH)COOH & 76514 & 0.50 &  130 & $<$1.9 (16) &  1.69 &  1.00 & 3.3 & $0.0$ & \ldots \\ 
 gG'a-CH$_3$CHOHCH$_2$OH & 76515 & 0.50 &  130 & $<$4.5 (16) &  1.70 &  4.39 & 3.3 & $0.0$ & C$_2$H$_5$OH \\ 
 g'G'g-CH$_3$CHOHCH$_2$OH & 76516 & 0.50 &  130 & $<$1.0 (17) &  1.64 &  6.92 & 3.3 & $0.0$ & C$_2$H$_5$OH \\ 
 a'GG'g-HOCH$_2$CH$_2$CH$_2$OH & 76518 & 0.50 &  130 & $<$1.9 (16) &  1.99 &  1.37 & 3.3 & $0.0$ & C$_2$H$_5$OH \\ 
 gGG'g-HOCH$_2$CH$_2$CH$_2$OH & 76519 & 0.50 &  130 & $<$7.6 (16) &  1.88 &  4.47 & 3.3 & $0.0$ & C$_2$H$_5$OH \\ 
 g'Ga-CH$_3$CHOHCH$_2$OH & 76520 & 0.50 &  130 & $<$2.5 (17) &  1.75 &  20.2 & 3.3 & $0.0$ & C$_2$H$_5$OH \\ 
 gG'g'-CH$_3$CHOHCH$_2$OH & 76521 & 0.50 &  130 & $<$3.6 (17) &  1.64 &  24.7 & 3.3 & $0.0$ & C$_2$H$_5$OH \\ 
 aGg'-CH$_3$CHOHCH$_2$OH & 76522 & 0.50 &  130 & $<$1.3 (18) &  1.70 &  92.2 & 3.3 & $0.0$ & C$_2$H$_5$OH \\ 
 g'Gg-CH$_3$CHOHCH$_2$OH & 76523 & 0.50 &  130 & $<$3.6 (18) &  1.70 &   123 & 3.3 & $0.0$ & C$_2$H$_5$OH \\ 
 CH$_3$OCH$_2$CH$_2$OH $^\star$ & 76801 & 0.50 &  130 & $<$1.5 (16) &  1.00 &  1.01 & 3.0 & $0.0$ & CH$_3$OH \\ 
 c-C$_6$H$_5$OH & 94501 & 0.50 &  130 & $<$4.7 (16) &  1.16 &  1.00 & 3.0 & $0.0$ & CH$_3$OH \\ 
\hline 
\multicolumn{9}{c}{\textit{O- and N-bearing molecules}} \\ 
 HCNO $^\star$ & 43509 & 0.50 &  130 & $<$4.9 (14) &  1.65 &  1.00 & 3.5 & $0.5$ & HNCO \\ 
 HOCN $^\star$ & 43510 & 0.50 &  130 & $<$6.1 (14) &  1.01 &  1.00 & 3.5 & $0.5$ & HNCO \\ 
 NH$_2$CO & 44904 & 0.50 &  130 & $<$4.0 (15) &  1.00 &  1.00 & 3.3 & $0.0$ & \ldots \\ 
 HC(O)CN $^\star$ & 55501 & 0.50 &  130 & $<$5.7 (15) &  1.14 &  1.00 & 3.3 & $0.0$ & \ldots \\ 
 CH$_3$CNO & 57510 & 0.50 &  130 & $<$5.0 (14) &  1.24 &  1.00 & 3.0 & $0.0$ & CH$_3$NCO \\ 
 CH$_3$OCN & 57511 & 0.50 &  130 & $<$2.8 (15) &  1.54 &  1.00 & 3.0 & $0.0$ & CH$_3$NCO \\ 
 HOCH$_2$CN $^\star$ & 57701 & 0.50 &  130 & $<$8.0 (15) &  1.00 &  1.00 & 3.3 & $0.0$ & \ldots \\ 
 NH$_2$C(O)NH$_2$ $^\star$ & 60517 & 0.50 &  130 & $<$3.4 (15) &  1.66 &  1.03 & 3.3 & $0.0$ & \ldots \\ 
 NH$_2$CH$_2$CH$_2$OH $^\star$ & 61004 & 0.50 &  130 & $<$9.2 (15) &  1.00 &  1.02 & 3.3 & $0.0$ & \ldots \\ 
 NCCHCO & 67502 & 0.50 &  130 & $<$3.8 (15) &  1.28 &  1.00 & 3.3 & $0.0$ & \ldots \\ 
 c-(C$_2$H$_3$O)-CN & 69518 & 0.50 &  130 & $<$2.4 (15) &  1.20 &  1.00 & 3.0 & $0.0$ & c-C$_2$H$_4$O \\ 
 trans-C$_2$H$_3$NCO & 69901 & 0.50 &  130 & $<$6.7 (15) &  2.23 &  1.00 & 3.0 & $0.0$ & CH$_3$NCO \\ 
 cis-C$_2$H$_3$NCO & 69902 & 0.50 &  130 & $<$2.7 (17) &  2.27 &  1.00 & 3.0 & $0.0$ & CH$_3$NCO \\ 
 CHCC(O)NH$_2$ & 69916 & 0.50 &  130 & $<$7.3 (15) &  1.04 &  1.00 & 3.3 & $0.0$ & \ldots \\ 
 NCC(O)NH$_2$ & 70504 & 0.50 &  130 & $<$5.4 (15) &  1.36 &  1.00 & 3.3 & $0.0$ & \ldots \\ 
 gauche-HOCH$_2$CH$_2$CN & 71401 & 0.50 &  130 & $<$1.6 (16) &  1.00 &  1.08 & 3.3 & $0.0$ & \ldots \\ 
 gauche-CH$_3$OCH$_2$CN & 71402 & 0.50 &  130 & $<$1.3 (16) &  1.00 &  1.01 & 3.3 & $0.0$ & \ldots \\ 
 cis-C$_2$H$_5$NCO $^\star$ & 71508 & 0.50 &  130 & $<$1.7 (16) &  4.60 &  1.00 & 3.0 & $0.0$ & CH$_3$NCO \\ 
 syn-C$_2$H$_3$C(O)NH$_2$ & 71903 & 0.50 &  130 & $<$1.3 (16) &  1.78 &  1.00 & 3.3 & $0.0$ & \ldots \\ 
 C$_2$H$_5$C(O)NH$_2$ & 73801 & 0.50 &  130 & $<$2.0 (16) &  1.00 &  1.00 & 3.3 & $0.0$ & \ldots \\ 
 trans-sc-C$_2$H$_5$NHCHO & 73902 & 0.50 &  130 & $<$1.9 (16) &  2.65 &  1.00 & 3.3 & $0.0$ & \ldots \\ 
 NH$_2$CH$_2$C(O)NH$_2$ & 74904 & 0.50 &  130 & $<$8.2 (15) &  1.18 &  1.00 & 3.3 & $0.0$ & \ldots \\ 
 NH$_2$CH$_2$COOH I & 75511 & 0.50 &  130 & $<$8.6 (16) &  2.14 &  1.00 & 3.3 & $0.0$ & \ldots \\ 
 NH$_2$CH$_2$COOH II & 75512 & 0.50 &  130 & $<$3.8 (18) &  2.14 &  1348 & 3.3 & $0.0$ & \ldots \\ 
 g'Gg'-CH$_3$CH(NH$_2$)CH$_2$OH & 75518 & 0.50 &  130 & $<$6.9 (15) &  1.56 &  1.11 & 3.3 & $0.0$ & \ldots \\ 
 gG'g-CH$_3$CH(NH$_2$)CH$_2$OH & 75519 & 0.50 &  130 & $<$1.3 (18) &  1.56 &  10.4 & 3.3 & $0.0$ & \ldots \\ 
 syn-HOCH$_2$C(O)NH$_2$ $^\star$ & 75901 & 0.50 &  130 & $<$6.5 (15) &  2.11 &  1.03 & 3.3 & $0.0$ & \ldots \\ 
 OC(CN)$_2$ & 80601 & 0.50 &  130 & $<$4.0 (17) &  1.66 &  1.00 & 3.3 & $0.0$ & \ldots \\ 
 NCCH$_2$C(O)NH$_2$ & 84902 & 0.50 &  130 & $<$2.8 (17) &  2.77 &  1.00 & 3.3 & $0.0$ & \ldots \\ 
 CH$_3$CH(NH$_2$)COOH I & 89502 & 0.50 &  130 & $<$9.6 (16) &  3.19 &  1.20 & 3.3 & $0.0$ & \ldots \\ 
 CH$_3$CH(NH$_2$)COOH II & 89503 & 0.50 &  130 & $<$6.8 (16) &  3.19 &  14.2 & 3.3 & $0.0$ & \ldots \\ 
\hline 
\multicolumn{9}{c}{\textit{N-bearing molecules}} \\ 
 CH$_2$NH$_2$$^+$ & 30519 & 0.50 &  130 & $<$8.0 (16) &  1.00 &  1.00 & 3.5 & $0.0$ & CH$_2$NH \\ 
 CH$_2$CN $^\star$ & 40601 & 0.60 &  140 & $<$3.1 (15) &  1.04 &  1.00 & 3.5 & $0.2$ & CH$_3$CN \\ 
 CH$_2$CNH $^\star$ & 41503 & 0.50 &  130 & $<$1.5 (16) &  1.02 &  1.00 & 3.3 & $0.0$ & \ldots \\ 
 CH$_3$CNH$^+$ & 42504 & 0.60 &  140 & $<$1.0 (16) &  1.02 &  1.00 & 3.5 & $0.2$ & CH$_3$CN \\ 
 C$_2$H$_3$NH$_2$ $^\star$ & 43504 & 0.50 &  130 & $<$3.1 (16) &  1.03 &  1.00 & 3.5 & $0.0$ & CH$_3$NH$_2$ \\ 
 E-CH$_3$CHNH $^\star$ & 43907 & 0.50 &  130 & $<$1.6 (16) &  1.15 &  1.00 & 3.3 & $0.0$ & \ldots \\ 
 Z-CH$_3$CHNH $^\star$ & 43908 & 0.50 &  130 & $<$7.0 (15) &  1.16 &  1.00 & 3.3 & $0.0$ & \ldots \\ 
 anti-C$_2$H$_5$NH$_2$ $^\star$ & 45402 & 0.50 &  130 & $<$6.0 (16) &  1.15 &  2.10 & 3.5 & $0.0$ & CH$_3$NH$_2$ \\ 
 CH$_3$NHCH$_3$ & 45601 & 0.50 &  130 & $<$8.3 (16) &  1.18 &  1.00 & 3.5 & $0.0$ & CH$_3$NH$_2$ \\ 
 C$_3$N $^\star$ & 50511 & 1.00 &  130 & $<$3.4 (14) &  1.13 &  1.00 & 3.5 & $0.4$ & HC$_3$N \\ 
 HNC$_3$ $^\star$ & 51528 & 1.00 &  130 & $<$4.7 (13) &  1.33 &  1.00 & 3.5 & $0.4$ & HC$_3$N \\ 
 HC$_3$NH$^+$ $^\star$ & 52503 & 1.00 &  130 & $<$6.6 (14) &  1.31 &  1.00 & 3.5 & $0.4$ & HC$_3$N \\ 
 C$_2$H$_3$NC & 53007 & 0.70 &  130 & $<$9.2 (14) &  1.15 &  1.00 & 3.5 & $0.3$ & C$_2$H$_3$CN \\ 
 Z-CHCCHNH $^\star$ & 53523 & 0.50 &  130 & $<$6.1 (15) &  1.17 &  1.04 & 3.3 & $0.5$ & \ldots \\ 
 E-CHCCHNH & 53524 & 0.50 &  130 & $<$2.9 (17) &  1.17 &  28.0 & 3.3 & $0.5$ & \ldots \\ 
 C$_2$H$_3$CNH$^+$ & 54518 & 0.70 &  130 & $<$4.5 (15) &  1.13 &  1.00 & 3.5 & $0.3$ & C$_2$H$_3$CN \\ 
 Z-NHCHCN & 54805 & 0.50 &  130 & $<$1.1 (16) &  1.12 &  1.00 & 3.3 & $0.0$ & \ldots \\ 
 CH$_2$NCN & 54806 & 0.50 &  130 & $<$8.7 (14) &  1.09 &  1.00 & 3.3 & $0.0$ & \ldots \\ 
 C$_2$H$_5$NC & 55507 & 0.50 &  130 & $<$2.2 (15) &  1.32 &  1.00 & 3.3 & $0.0$ & \ldots \\ 
 E-C$_2$H$_5$CHNH & 57801 & 0.50 &  130 & $<$3.7 (16) &  1.00 &  1.05 & 3.3 & $0.0$ & \ldots \\ 
 Z-C$_2$H$_5$CHNH & 57802 & 0.50 &  130 & $<$1.4 (17) &  1.00 &  19.6 & 3.3 & $0.0$ & \ldots \\ 
 trans-iso-C$_3$H$_7$NH$_2$ & 59925 & 0.50 &  130 & $<$1.0 (17) &  1.24 &  1.38 & 3.5 & $0.0$ & CH$_3$NH$_2$ \\ 
 Trans-trans-normal-C$_3$H$_7$NH$_2$ & 59928 & 0.50 &  130 & $<$1.4 (17) &  1.53 &  4.26 & 3.5 & $0.0$ & CH$_3$NH$_2$ \\ 
 CH$_3$CCNC & 65505 & 0.50 &  130 & $<$5.9 (14) &  1.68 &  1.00 & 3.3 & $0.0$ & \ldots \\ 
 CH$_2$CCHCN $^\star$ & 65506 & 0.50 &  130 & $<$2.0 (15) &  1.34 &  1.00 & 3.3 & $0.0$ & \ldots \\ 
 CHCCH$_2$CN $^\star$ & 65514 & 0.50 &  130 & $<$2.1 (15) &  1.37 &  1.00 & 3.3 & $0.0$ & \ldots \\ 
 CH$_2$CCHNC & 65519 & 0.50 &  130 & $<$2.2 (15) &  1.45 &  1.00 & 3.3 & $0.0$ & \ldots \\ 
 NCCH$_2$CN $^\star$ & 66801 & 0.50 &  130 & $<$5.0 (15) &  1.26 &  1.00 & 3.3 & $0.0$ & \ldots \\ 
 CNCH$_2$NC & 66902 & 0.50 &  130 & $<$5.0 (15) &  1.66 &  1.00 & 3.3 & $0.0$ & \ldots \\ 
 E-CH$_3$CHCHCN $^\star$ & 67513 & 0.50 &  130 & $<$1.5 (15) &  1.48 &  1.00 & 3.5 & $0.0$ & normal-C$_3$H$_7$CN \\ 
 syn-CH$_2$CHCH$_2$CN $^\star$ & 67602 & 0.50 &  130 & $<$6.4 (15) &  1.57 &  1.02 & 3.5 & $0.0$ & normal-C$_3$H$_7$CN \\ 
 gauche-CH$_2$CHCH$_2$CN $^\star$ & 67603 & 0.50 &  130 & $<$1.3 (17) &  1.57 &  42.2 & 3.5 & $0.0$ & normal-C$_3$H$_7$CN \\ 
 CH$_2$NCH$_2$CN & 68506 & 0.50 &  130 & $<$1.0 (16) & \ldots &  1.00 & 3.3 & $0.0$ & \ldots \\ 
 Z-CH$_3$C(NH)CN & 68512 & 0.50 &  130 & $<$1.5 (16) &  1.48 &  1.00 & 3.3 & $0.0$ & \ldots \\ 
 CH$_3$CH(NH$_2$)CN & 70505 & 0.50 &  130 & $<$7.2 (15) &  1.43 &  1.00 & 3.3 & $0.0$ & \ldots \\ 
 NH$_2$CH$_2$CH$_2$CN I & 70513 & 0.50 &  130 & $<$1.3 (16) &  1.74 &  1.05 & 3.3 & $0.0$ & \ldots \\ 
 NH$_2$CH$_2$CH$_2$CN II & 70514 & 0.50 &  130 & $<$5.5 (16) &  1.74 &  1.05 & 3.3 & $0.0$ & \ldots \\ 
 HC$_5$NH$^+$ $^\star$ & 76524 & 0.50 &  130 & $<$7.9 (14) &  2.65 &  1.00 & 3.3 & $0.0$ & \ldots \\ 
 C$_2$H$_3$C$_3$N $^\star$ & 77506 & 0.50 &  130 & $<$2.0 (15) &  2.03 &  1.00 & 3.3 & $0.0$ & \ldots \\ 
 CH$_3$CHCCHCN & 79506 & 0.50 &  130 & $<$2.8 (15) &  2.33 &  1.00 & 3.3 & $0.0$ & \ldots \\ 
 C$_2$H$_5$C$_3$N & 79507 & 0.50 &  130 & $<$2.0 (15) &  2.46 &  1.00 & 3.3 & $0.0$ & \ldots \\ 
 NH$_2$CH(CN)$_2$ & 81801 & 0.50 &  130 & $<$5.3 (15) &  1.75 &  1.00 & 3.3 & $0.0$ & \ldots \\ 
 anti-anti-normal-C$_4$H$_9$CN & 83501 & 0.50 &  130 & $<$2.0 (16) &  3.39 &  2.35 & 3.5 & $0.0$ & normal-C$_3$H$_7$CN \\ 
 gauche-anti-normal-C$_4$H$_9$CN & 83502 & 0.50 &  130 & $<$2.2 (16) &  3.39 &  1.06 & 3.5 & $0.0$ & normal-C$_3$H$_7$CN \\ 
 anti-gauche-normal-C$_4$H$_9$CN & 83503 & 0.50 &  130 & $<$1.5 (16) &  3.39 &  2.20 & 3.5 & $0.0$ & normal-C$_3$H$_7$CN \\ 
 tert-C$_4$H$_9$CN & 83504 & 0.50 &  130 & $<$2.1 (15) &  1.71 &  1.00 & 3.5 & $0.0$ & normal-C$_3$H$_7$CN \\ 
 anti-C$_2$H$_5$CH(CN)CH$_3$ & 83505 & 0.50 &  130 & $<$7.2 (15) &  2.41 &  1.11 & 3.5 & $0.0$ & normal-C$_3$H$_7$CN \\ 
 anti-(CH$_3$)$_2$CHCH$_2$CN & 83506 & 0.50 &  130 & $<$1.2 (16) &  2.47 &  1.00 & 3.5 & $0.0$ & normal-C$_3$H$_7$CN \\ 
 gauche-(CH$_3$)$_2$CHCH$_2$CN & 83507 & 0.50 &  130 & $<$4.3 (16) &  2.41 &  1.00 & 3.5 & $0.0$ & normal-C$_3$H$_7$CN \\ 
 HC$_7$N $^\star$ & 99501 & 1.00 &  130 & $<$1.3 (15) &  6.64 &  1.00 & 3.5 & $0.4$ & HC$_3$N \\ 
 c-C$_6$H$_5$CN $^\star$ & 103501 & 0.50 &  130 & $<$3.7 (15) &  1.47 &  1.00 & 3.3 & $0.0$ & \ldots \\ 
\hline 
\multicolumn{9}{c}{\textit{S-bearing molecules}} \\ 
 c-C$_2$H$_4$S & 60509 & 0.50 &  130 & $<$1.2 (16) &  1.00 &  1.00 & 3.5 & $0.0$ & CH$_3$SH \\ 
 syn-C$_2$H$_3$SH & 60521 & 0.50 &  130 & $<$6.5 (16) &  1.08 &  1.00 & 3.5 & $0.0$ & CH$_3$SH \\ 
 anti-C$_2$H$_3$SH & 60522 & 0.50 &  130 & $<$9.7 (16) &  1.08 &  1.00 & 3.5 & $0.0$ & CH$_3$SH \\ 
 CH$_3$CHS $^\star$ & 60701 & 0.50 &  130 & $<$7.1 (15) &  1.01 &  1.00 & 3.5 & $0.0$ & CH$_3$SH \\ 
 gauche-C$_2$H$_5$SH $^\star$ & 62523 & 0.50 &  130 & $<$2.8 (16) &  1.20 &  1.15 & 3.5 & $0.0$ & CH$_3$SH \\ 
 anti-C$_2$H$_5$SH & 62524 & 0.50 &  130 & $<$2.0 (17) &  1.30 &  7.85 & 3.5 & $0.0$ & CH$_3$SH \\ 
 AA-CH$_3$SCH$_3$ $^\star$ & 62902 & 0.50 &  130 & $<$2.8 (16) &  1.42 &  1.00 & 3.5 & $0.0$ & CH$_3$SH \\ 
 trans-HCSSH & 78506 & 0.50 &  130 & $<$7.3 (15) &  1.04 &  1.00 & 3.5 & $0.0$ & CH$_3$SH \\ 
 cis-HCSSH & 78507 & 0.50 &  130 & $<$1.0 (17) &  1.04 &  1.00 & 3.5 & $0.0$ & CH$_3$SH \\ 
\hline 
\multicolumn{9}{c}{\textit{S- and O-bearing molecules}} \\ 
 trans-HC(O)SH $^\star$ & 62515 & 0.50 &  130 & $<$1.4 (16) &  1.02 &  1.08 & 3.3 & $0.0$ & \ldots \\ 
 cis-HC(O)SH & 62516 & 0.50 &  130 & $<$10.0 (16) &  1.02 &  13.9 & 3.3 & $0.0$ & \ldots \\ 
\hline 
\multicolumn{9}{c}{\textit{S- and N-bearing molecules}} \\ 
 HCNS $^\star$ & 59510 & 0.50 &  120 & $<$9.8 (13) &  1.09 &  1.00 & 3.5 & $-1.0$ & HNCS \\ 
 NH$_2$CHS & 61523 & 0.50 &  130 & $<$1.5 (15) &  1.02 &  1.00 & 3.3 & $0.0$ & \ldots \\ 
\hline 
\multicolumn{9}{c}{\textit{Cl-bearing molecules}} \\ 
 CH$_3$Cl $^\star$ & 50007 & 0.50 &  130 & $<$4.0 (15) &  1.00 &  1.00 & 3.3 & $0.0$ & \ldots \\ 
\hline 
\multicolumn{9}{c}{\textit{P-bearing molecules}} \\ 
 CH$_3$CP & 58507 & 0.50 &  130 & $<$3.6 (15) &  1.03 &  1.00 & 3.3 & $0.0$ & \ldots \\ 
 C$_2$H$_5$CP & 72509 & 0.50 &  130 & $<$1.1 (16) &  1.37 &  1.00 & 3.3 & $0.0$ & \ldots \\ 
\hline 
\multicolumn{9}{c}{\textit{Hydrocarbons}} \\ 
 C$_2$H$_3$$^+$ $^\star$ & 27514 & 0.50 &  130 & $<$2.0 (16) &  1.00 &  1.00 & 3.3 & $0.0$ & \ldots \\ 
 CH$_2$CHCH$_3$ $^\star$ & 42516 & 0.50 &  130 & $<$2.3 (17) &  1.15 &  1.00 & 3.3 & $0.0$ & \ldots \\ 
 C$_3$H$_8$ & 44013 & 0.50 &  130 & $<$8.0 (18) &  1.00 &  1.00 & 3.3 & $0.0$ & \ldots \\ 
 C$_2$H$_5$CCH $^\star$ & 54519 & 1.00 &  130 & $<$3.8 (16) &  1.28 &  1.00 & 4.0 & $0.0$ & CH$_3$CCH \\ 
 (CH$_3$)$_2$CCH$_2$ $^\star$ & 56526 & 0.50 &  130 & $<$1.7 (17) &  1.33 &  1.00 & 3.3 & $0.0$ & \ldots \\ 
 H$_2$C(CCH)$_2$ $^\star$ & 64520 & 1.00 &  130 & $<$1.7 (17) &  1.39 &  1.00 & 4.0 & $0.0$ & CH$_3$CCH \\ 
\hline 
 \end{longtable}
 \tablefoot{
 \tablefoottext{a}{Molecules that have already been detected in the interstellar medium are marked with a star.}
 \tablefoottext{b}{Entry number in our Weeds local database. See note b of Table~\ref{t:spectrobib}.}
 \tablefoottext{c}{Source diameter (\textit{FWHM}).}
 \tablefoottext{d}{Rotational temperature.}
 \tablefoottext{e}{Upper limit to the total column density of the molecule. $X$ ($Y$) means $X \times 10^Y$.}
 \tablefoottext{f}{Correction factor that was applied to the column density to account for the contribution of vibrationally or torsionally excited states, in the cases where this contribution was not included in the partition function of the spectroscopic predictions. In most cases, this factor was estimated in the harmonic approximation.}
 \tablefoottext{g}{Correction factor that was applied to the column density to account for the contribution of conformers, in the cases where this contribution was not included in the partition function of the spectroscopic predictions.}
 \tablefoottext{h}{Linewidth (\textit{FWHM}).}
 \tablefoottext{i}{Velocity offset with respect to the assumed systemic velocity given in Table~\ref{t:sources}.}
 \tablefoottext{j}{Molecule that was used to fix the LTE parameters except for the column density. Ellipsis dots indicate that we used typical parameters of the specific source.}
 }

%\clearpage
%\twocolumn
\section{Additional figures}
\label{a:morefigures}

\begin{figure}[h!]
\centerline{\resizebox{0.5\hsize}{!}{\includegraphics[angle=0]{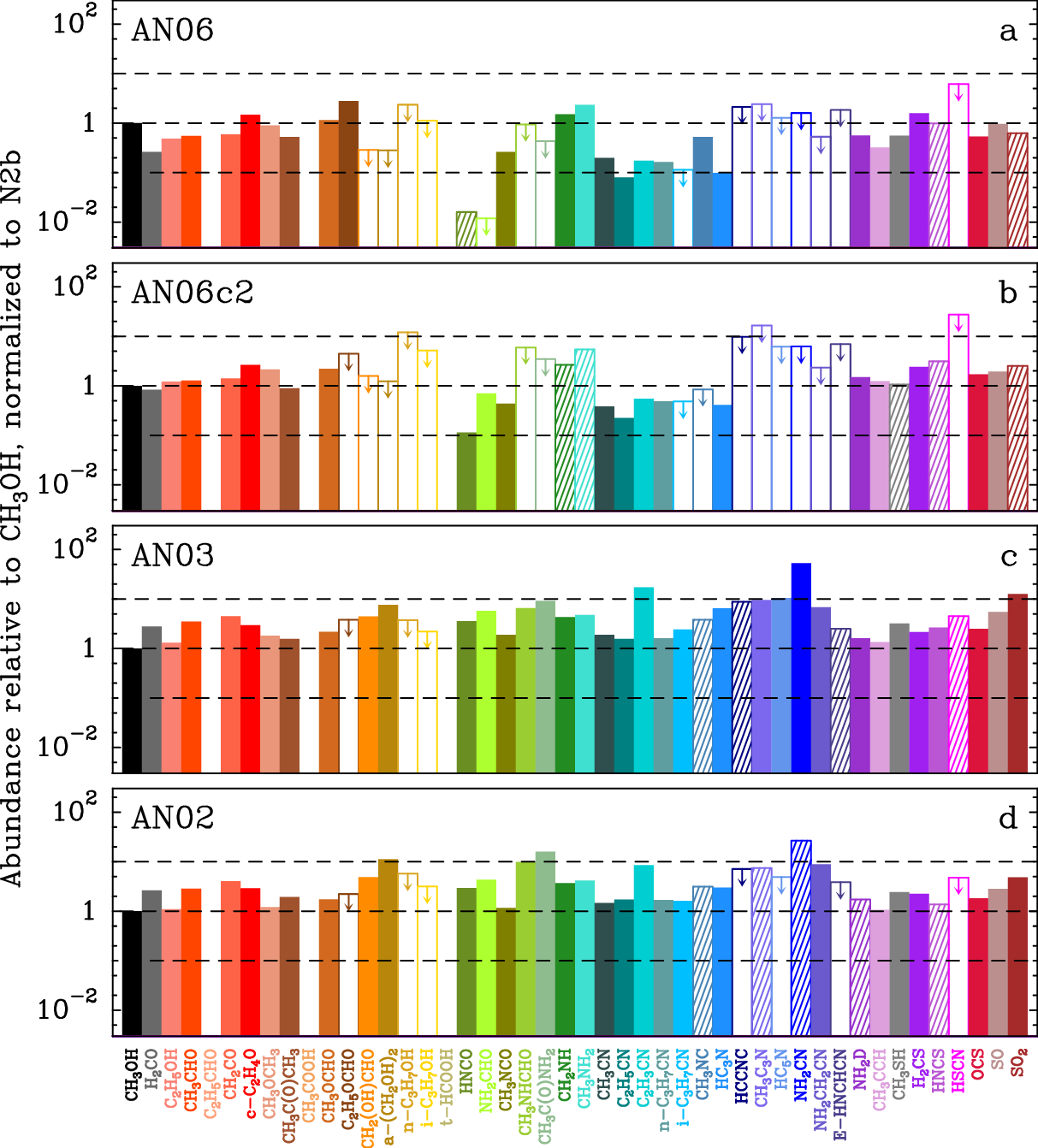}}}
\caption{Same as for Fig.~\ref{f:chemcomp_ch3oh} but 
normalized to N2b. The dashed lines indicate values of 0.1, 1, and 10.}
\label{f:chemcomp_ch3oh_normn2b}
\end{figure}

\begin{figure*}[h!]
\centerline{\resizebox{0.77\hsize}{!}{\includegraphics[angle=0]{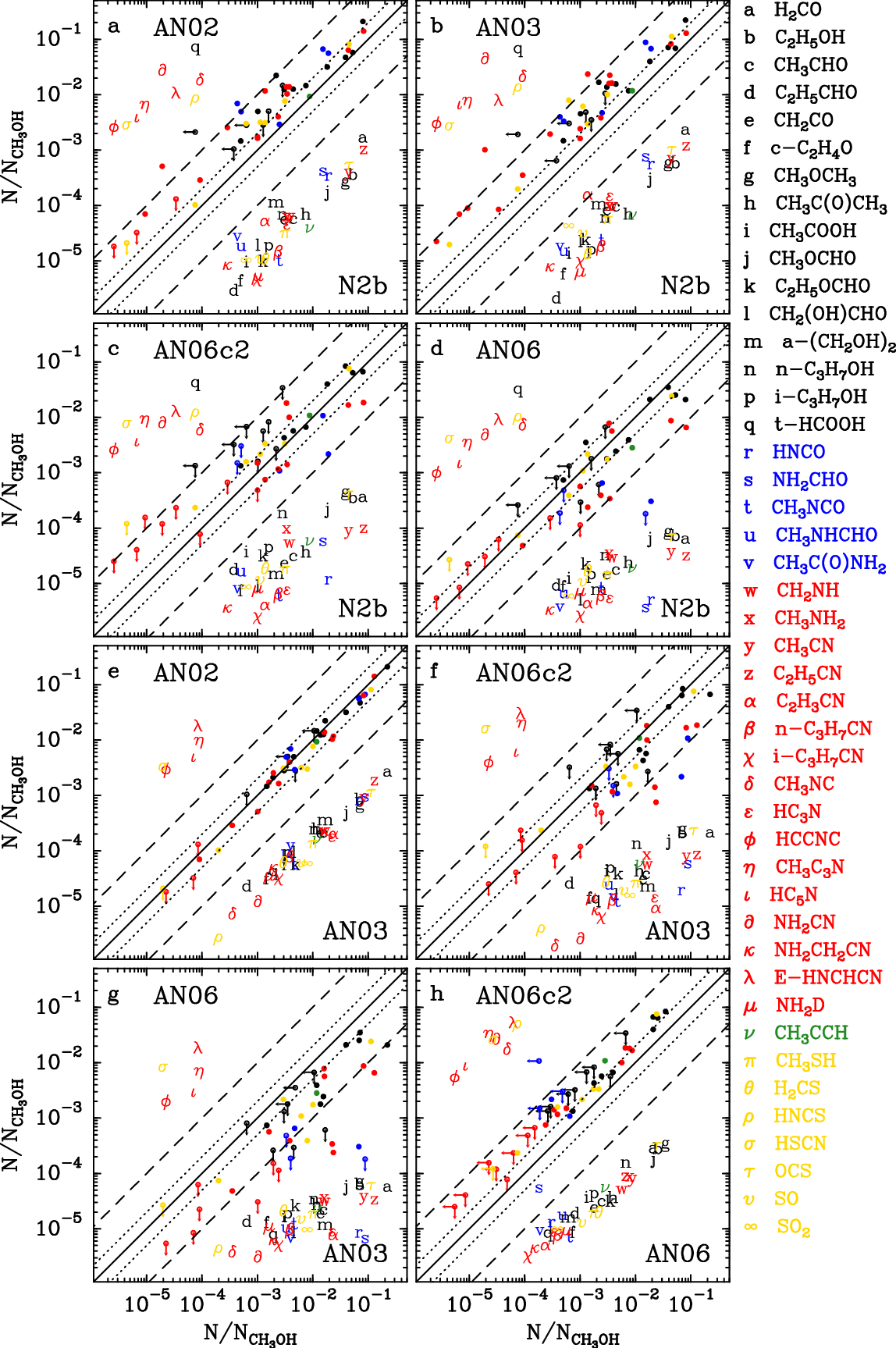}}}
\caption{Same as Fig.~\ref{f:correl_normch3oh} but with colors coding for the
class of molecules (black for O-bearing, blue for O+N-bearing, red for 
N-bearing, green for pure hydrocarbon, and yellow for S-bearing).}
\label{f:correl_normch3oh_type}
\end{figure*}

\begin{figure*}[h!]
\centerline{\resizebox{0.8\hsize}{!}{\includegraphics[angle=0]{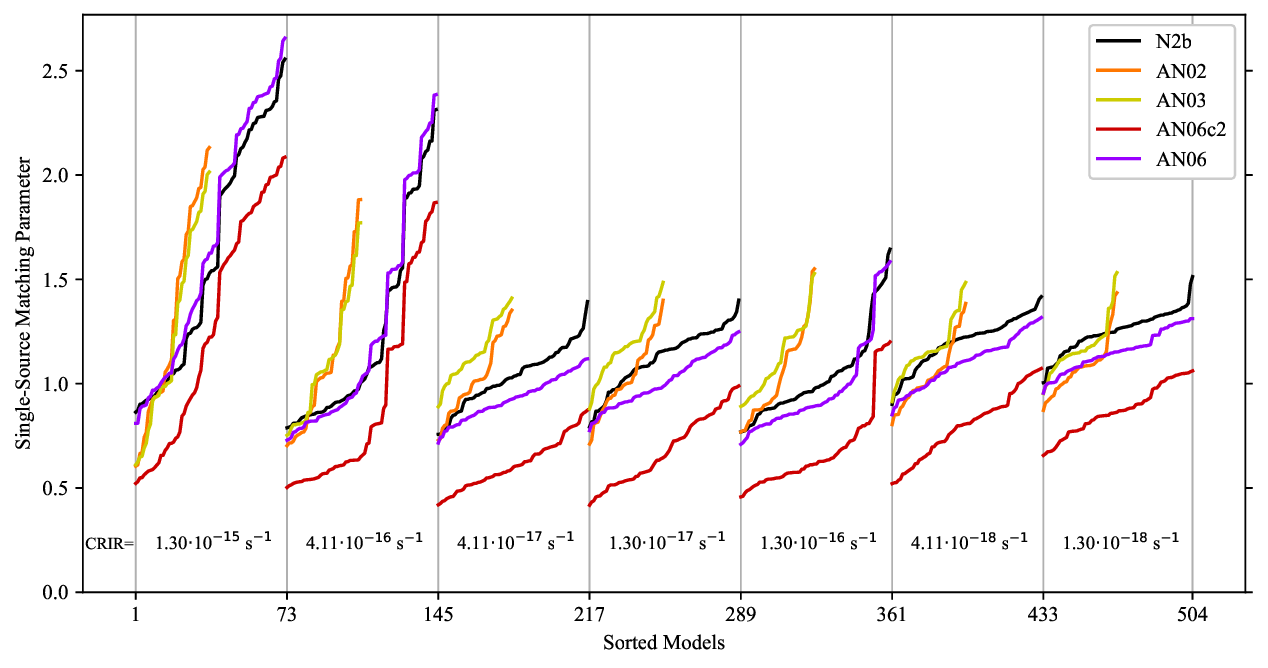}}}
\caption{Matching parameters for all tested models, using the 
density-restricted setup. Each curve shows the matching parameter for a single 
source as a function of all the models in the grid with which it is compared, 
using the same fixed CRIR. The results are arranged in bands, one for each 
CRIR. The bands are arranged left to right in order of best to worst overall 
matching parameter. Within each band, the models contributing to each curve 
are arranged in order of increasing (i.e.~worse) matching parameter, as 
determined independently for each source comparison.}
\label{f:model_match}
\end{figure*}

\begin{figure*}[h!]
\centerline{\resizebox{0.82\hsize}{!}{\includegraphics[angle=0]{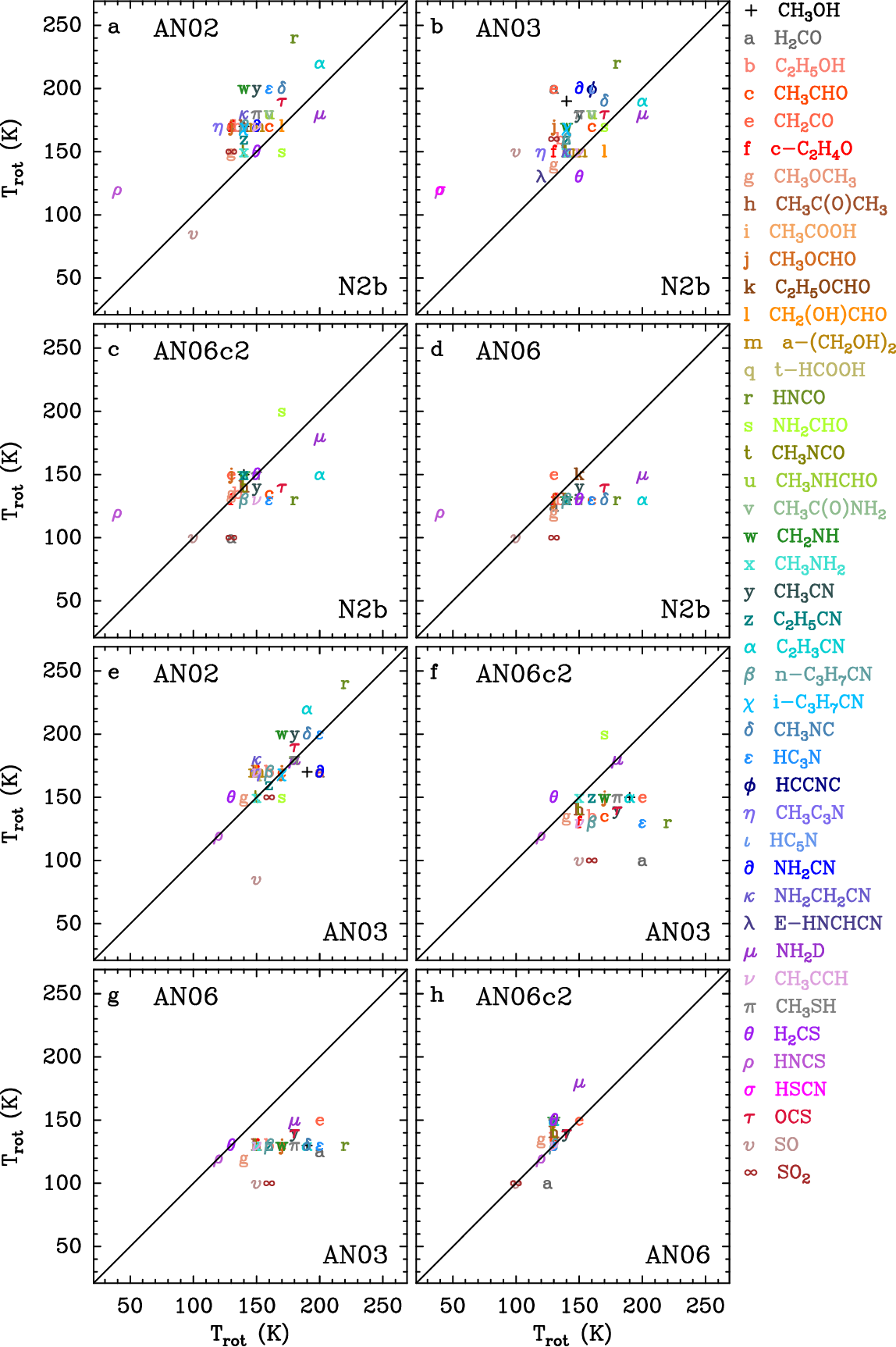}}}
\caption{Correlation plots of rotational temperatures for 
various pairs of positions. The x- and y-axes of each panel correspond to the 
positions written in the bottom right and top left corners, respectively. The 
color coding of the molecules is the same as in Fig.~\ref{f:correl_normch3oh}, 
plus black for methanol, and is indicated on 
the right. The plain line indicates the one-to-one relation.}
\label{f:correl_temp}
\end{figure*}

\begin{figure}[h!]
\centerline{\resizebox{0.45\hsize}{!}{\includegraphics[angle=0]{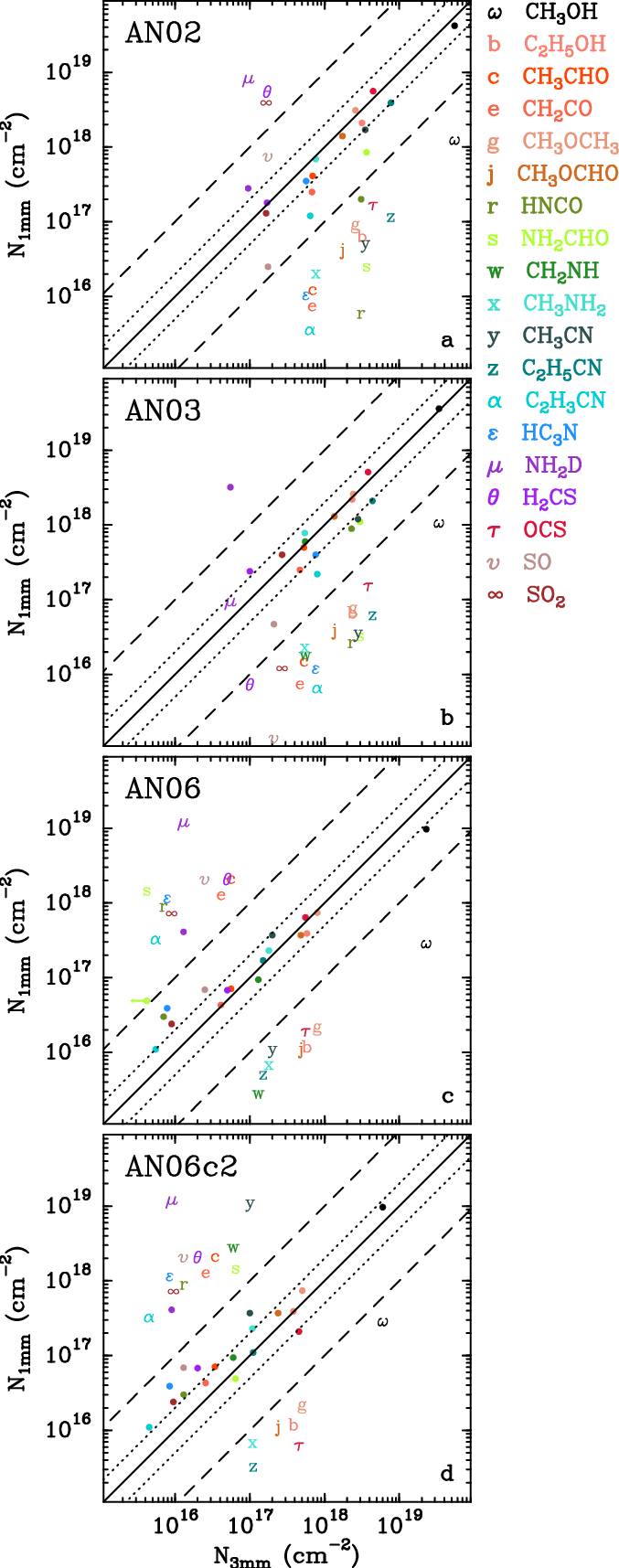}}}
\caption{Comparison of the column densities obtained from the ReMoCA survey 
toward AN02, AN03, AN06, and AN06c2 to those derived from an ALMA survey of 
Sgr~B2(N) at 1.2~mm by \citet{Moeller25}. The ReMoCA column densities 
correspond to peak column densities while the 1.2~mm column densities were 
derived from spectra averaged over the shaded blue polygons shown in Fig.~2 of 
\citet{Moeller25}. Panels c and d show the same 1.2~mm column 
densities except for C$_2$H$_5$CN and OCS which were modeled with two 
velocity components similar to the components that we extracted from the 
ReMoCA survey. The dashed, dotted, and plain lines and the color coding and 
labels of the molecules are the same as in Fig.~\ref{f:correl_normch3oh}.} 
\label{f:correl_chemcomp_1mm}
\end{figure}

Figure~\ref{f:chemcomp_ch3oh_normn2b} shows the chemical composition (relative 
to methanol) of AN06, AN06c2, AN03, and AN02, normalized to the composition
of N2b. Figure~\ref{f:correl_normch3oh_type} shows the same as 
Fig.~\ref{f:correl_normch3oh} but with a different color scheme.
Figure~\ref{f:model_match} shows the matching parameters for all tested 
chemical models. Figure~\ref{f:correl_temp} displays correlation plots of 
rotational temperatures for various pairs of positions in Sgr~B2(N2). 
Figure~\ref{f:correl_chemcomp_1mm} compares the ReMoCA column densities to the
column densities derived by \citet{Moeller25}.

\twocolumn
\end{appendix}

\end{document}